\RequirePackage{etoolbox}
\patchcmd{\bibliographystyle}{#1}{sn-basic-unsort}{}{}

\documentclass[sn-basic, Numbered]{sn-jnl}

\usepackage{graphicx}%
\usepackage{multirow}%
\usepackage{amsmath,amssymb,amsfonts}%
\usepackage{amsthm}%
\usepackage{mathrsfs}%
\usepackage[title]{appendix}%
\usepackage{xcolor}%
\usepackage{textcomp}%
\usepackage{manyfoot}%
\usepackage{booktabs}%
\usepackage{algorithm}%
\usepackage{algorithmicx}%
\usepackage{algpseudocode}%
\usepackage{listings}%
\usepackage {longtable}

\usepackage{ltablex}
\usepackage{lscape}
\usepackage{setspace}
\usepackage{array}
\usepackage{bibentry}
\usepackage{pgfplots}
\pgfplotsset{compat=1.18}
\usepackage{enumitem}
\setlist[enumerate]{left=0pt, label={[\arabic*]}, labelsep=0.5em}

\newcolumntype{P}[1]{>{\centering\arraybackslash}p{#1}}
\newcolumntype{L}[1]{>{\raggedright\arraybackslash}p{#1}}

\newcommand{\tabularxcaption}[2]{%
    \begingroup
    \setbox\tabcapbox\vbox{\tablecaptionfont\raggedright%
    {\bfseries 
    \mbox{#1}\nobreak}{\hskip2mm}\mbox{#2}\vphantom{y}\par\vskip\belowcaptionskip}%
    \box\tabcapbox%
    \endgroup
}
\usepackage{url}

\usepackage{geometry}
\usepackage{tikz}
\usetikzlibrary{positioning, arrows.meta, calc,fit,shapes}

\theoremstyle{thmstyleone}%
\theoremstyle{thmstyletwo}%

\theoremstyle{thmstylethree}%

\renewcommand{\orcidlogo}{%
  \includegraphics[width=10pt]{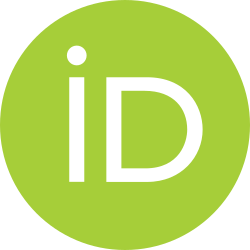}%
}

\begin{document}

\title[Ethical Aspects of the Use of Social Robots]{Ethical Aspects of the Use of Social Robots in Elderly Care }
\subtitle{A Systematic Qualitative Review}

\author[1]{\fnm{Marianne} \sur{Leineweber}}
\author[1]{\fnm{Clara Victoria} \sur{Keusgen}\orcid{https://orcid.org/0009-0002-5052-2280}}
\author[1]{\fnm{Marc} \sur{Bubeck}\orcid{https://orcid.org/0009-0003-0667-6191}}
\author[1]{\fnm{Joschka} \sur{Haltaufderheide}\orcid{https://orcid.org/0000-0002-5014-4593}}
\author*[1]{\fnm{Robert} \sur{Ranisch}\orcid{https://orcid.org/0000-0002-1676-1694}}\email{ranisch@uni-potsdam.de}\equalcont{Robert Ranisch and Corinna Klingler contributed equally as last authors.}
\author[1]{\fnm{Corinna} \sur{Klingler}\orcid{https://orcid.org/0000-0002-8148-5999}}\equalcont{Robert Ranisch and Corinna Klingler contributed equally as last authors.}

\affil[1]{\orgdiv{Juniorprofessorship for Medical Ethics with a focus on Digitization}, \orgname{Faculty for Health Sciences Brandenburg, University of Potsdam}, \orgaddress{\street{Am M\"uhlenberg 9}, \city{Potsdam}, \postcode{14476}, \state{Brandenburg}, \country{Germany}}}


 \abstract{\textbf{Background:} The use of social robotics in elderly care is increasingly discussed as one way of meeting emerging care needs due to scarce resources. While many potential benefits are associated with robotic care technologies, there is a variety of ethical challenges. To support steps towards a responsible implementation and use, this review develops an overview on ethical aspects of the use of social robots in elderly care from a decision-makers’ perspective. \\
\textbf{Methods:} Electronic databases were queried using a comprehensive search strategy based on the key concepts of “ethical aspects”, “social robotics” and “elderly care”. Abstract and title screening was conducted by two authors independently. Full-text screening was conducted by one author following a joint consolidation phase. Data was extracted using MAXQDA24 by one author, based on a consolidated coding framework. Analysis was performed through modified qualitative content analysis.\\
\textbf{Results:} A total of 1,518 publications were screened, and 248 publications were included. We have organized our analysis in a scheme of ethical hazards, ethical opportunities and unsettled questions, identifying at least 60 broad ethical aspects affecting three different stakeholder groups. While some ethical issues are well-known and broadly discussed our analysis shows a plethora of potentially relevant aspects, often only marginally recognized, that are worthy of consideration from a practical perspective.\\
\textbf{Discussion:} The findings highlight the need for a contextual and detailed evaluation of implementation scenarios. To make use of the vast knowledge of the ethical discourse, we hypothesize that decision-makers need to understand the specific nature of this discourse to be able to engage in careful ethical deliberation.}




\maketitle
\section*{Background}

Demographic change and the persistent care worker shortage have prompted discussion on the use of assistive robotic technologies as potential solutions for expected gaps in elderly care \cite{Haltaufderheide_etal__2023,Berner_etal__2020}. As a result, extensive efforts are being made in the area of research, research funding and development of robotic technologies \cite{Wright__2024}. The industry is more and more investing in the development of care robots \cite{Buxbaum_Sen__2018}. In some countries like Japan, where a high percentage of the population comprises older citizens, the government already provides subsidies to care facilities for the purchase of robots \cite{Savage__2022}, shifting from a developmental to an implementation and use focus.

Robotic systems designed to support care work can be grouped roughly into two overlapping categories of (physically) assistive and socially assistive robots \cite{Maalouf_etal__2018,Mataric_Scassellati__2016}. While the former support tasks such as bringing, carrying, or lifting activities of the caregivers or physical activities of the care recipients, the latter facilitate social interaction. Systems of that kind are known as social robots, socially assistive robots or companion robots. They can support social interaction\footnote{E.g., Giraff, a telepresence robot able to connect users with their loved ones or care providers virtually \cite{Ienca_etal__2016}.}  or even interact with the elderly themselves \cite{Kehl__2018b}, thereby simulating emotions or use natural language.\footnote{E.g., Pepper, a humanoid robot able to talk and used, for example, as a fitness instructor or storyteller in care facilities \cite{Ghafurian_etal__2022,Pandey_Gelin__2018}}
. As such they gain access to particularly sensitive areas of human life. For that reason, the use of social robotics in elderly care is considered particularly ethically sensitive, while also providing novel opportunities to connect older people and improve their well-being \cite{Kachouie_etal__2014}.

While there are many potential benefits associated with the use of social robotics, there are also a variety of ethical challenges discussed in the literature: Advocates often emphasize that the use of robotics may mitigate upcoming resource shortages by relieving caregivers, especially in particularly time-intensive or physically demanding tasks \cite{Haltaufderheide_etal__2023,Sharkey_Sharkey__2012}. More cautious perspectives on the other hand, argue that the “human touch of care” may be lost and care relationships thereby impoverished \cite{Kehl__2018c} or that robots cannot consider the individually varying needs of older people \cite{Remmers__2020}. In addition, questions of autonomy and dependency \cite{Boada_etal__2021}, adequate data protection \cite{Felzmann_etal__2016}, as well as issues related to deceptions \cite{Sharkey_Sharkey__2021} are frequently discussed. 

Consequently, the research discourse notes a variety of ethical challenges and opportunities that might vary depending on the type of robot, implementation scenario and broader sociotechnical arrangement and the ethical perspective \cite{Vandemeulebroucke_etal__2018}. Given that it is to be expected that social robots will become a part of nursing care practices for the elderly, decision-makers on the micro-, meso- and macro-level will have to consider and navigate these ethical issues arising. While this topic is critically discussed in the literature \cite{Haltaufderheide_etal__2020,Hung_etal__2022,Wachsmuth__2018} and increasingly addressed by national ethics bodies \cite{BioethikkommissionbeimBundeskanzleramtOsterreich__2019,DeutscherEthikrat__2020}, no pertinent standards for the use of social robots in care settings have been developed. For both the development of guidelines and individual decision-making, knowing about relevant ethical aspects arising in this context and being able to engage with the existing ethical knowledge will be paramount. Against this background, this systematic review aims to provide a broad overview of relevant ethical aspects concerning the use of social robotics in elderly care as discussed in the literature following the question: What are the ethical aspects of using social robots in elderly care? In the following, we will outline the methods for realizing the review, including the search, the screening process and the analysis of the data. Then, we will present our results regarding our research question and discuss their implications.

\section*{Methods}

The review is reported following the RESERVE guideline for systematic reviews in ethics \cite{Kahrass_etal__2023}. A protocol was agreed upon by the authors and has been registered with PROSPERO \cite{Klingler_etal__2023}. Unlike other reviews on this topic that primarily focus on ethical challenges \cite{Boada_etal__2021} or specific aspects and concepts \cite{Vandemeulebroucke_etal__2018}, our aim was to include all relevant aspects – including the opportunities or benefits of technologies – deemed significant for ethical decision-making. The goal of this work is thereby purely descriptive. It does not evaluate or weigh up the aspects discussed in the literature as no accepted criteria or procedures for synthesizing normative data have been developed \cite{Klingler_Mertz__2021}. We assume that decision-makers will need to take further evaluative steps and apply arguments raised to their context as discussed below. Nevertheless, this overview serves as a valuable resource for decision-makers at both local and policy levels, providing a clear synthesis of the ethical aspects that should be considered. 

\subsection*{Search Strategy}

We assumed that relevant discussions would take place within the ethical, nursing, medical, and technical literature. Accordingly, we included the following databases covering the relevant thematic areas: PubMed/Medline, CINAHL, the Technical Information Library (TIB-Portal) and BELIT. To ensure comprehensive coverage, we included databases containing grey literature, such as policy briefs and statements from ethics councils, which are available in the TIB-Portal and BELIT.

Search strings were constructed around the three core concepts: a) ethical aspects, b) social robotics, and c) elderly care. In doing so, we followed Droste et al. \cite{Droste_etal__2010}, who argue that the PICO framework is unsuitable for ethics-related topics. Instead, we narrowed the scope by specifying the ethical dimension, the intervention, and the user group or care setting. The search string development was informed by a preliminary search and literature screening to identify relevant terminology. The search string for PubMed is presented in Table 1.

\begin{table}[ht]
\centering
\caption{Search string components for the systematic review}
\label{tab:search-strings}
\begin{tabular}{@{}lp{9cm}@{}}
\toprule
Concept & Corresponding part of search string \\ \midrule
a) Ethical aspects & (ethics[MeSH Terms] OR human rights[MeSH Terms] OR ethic*[Text Word] OR moral*[Text Word]) AND \\[0.5em]
b) Social robotics & (robotics[MeSH Terms] OR robot*[Text Word] OR (social*[Text Word] AND assistiv*[Text Word]) OR (social*[Text Word] AND interactiv*[Text Word])) AND \\[0.5em]
c) Elderly care & (aged[MeSH Terms] OR aged[Text Word] OR geriatr*[Text Word] OR elder*[Text Word] OR senior*[Text Word] OR nursing home*[Text Word] OR dement*[Text Word]) \\ 
\bottomrule
\end{tabular}
\end{table}

Search strings for the other databases were modeled after the PubMED-string (see Appendix D) and adjusted for database-specific features such as language or literature type. The selection of databases and the refinement of search strings were carried out in consultation with a library scientist from the University Library Potsdam. The search was conducted in February and March of 2023.

\subsection*{Inclusion/exclusion criteria }

As at least two terms used in the research question are contested and only vaguely specified in the literature, we developed operationalizable criteria for the purpose of this review. We understand “social robots” as robots whose primary purpose is to either support social interaction as a mediator (socially assistive robots) or serve as an actual interaction partner (socially interactive robots) – a definition employed by Kehl \cite{Kehl__2018b}, building on the work of Feil-Seifer and Matarić \cite{FeilSeifer_Mataric__2005}. 

In specifying what constitutes a robot, we are guided by Fosch-Villaronga and Drukarch, who define a “robot” as a “movable machine that performs tasks either automatically or with degree of autonomy” \cite{FoschVillaronga_Drukarch__2022}. A “machine” is operationalized as having a physical body capable of interacting with its environment. This definition is deemed appropriate as it allows us to exclude neighboring technologies that fall outside the scope of this review, such as exoskeletons, cleaning robots, or virtual avatars. To determine whether this criterion was met, respective device descriptions were checked using included literature and – where necessary – further sources describing them in more detail. 

Terms like “ethical issues” or “ethical aspects” are also inconsistently used in the literature \cite{Schofield_etal__2021}. We used the latter to refer to hazards, opportunities or unsettled questions that need to be considered (or clarified) when determining how to responsibly handle a given phenomenon. To operationalize the term “ethical aspects”, we employed a principle-based approach that presumes certain ethical principles as relevant orientation points in the field of robotics/AI: autonomy, beneficence, nonmaleficence, justice and explicability \cite{Beauchamp_Childress__2019,Floridi_etal__2018}. 

We defined ethical aspects as instances where at least one of the five principles is promoted (ethical opportunities), violated (ethical hazards), or where there is ambiguity or a conflict between different ethical claims or principles (unsettled questions). By choosing such a substantive approach, we reduced the risk of overlooking relevant literature that does not explicitly use normative terminology to describe ethical aspects. Further context-related criteria (e.g., publications had to self-identify as discussing elderly care, as we did not want to introduce an age cutoff) and formal inclusion criteria (to guarantee a certain quality of publications and due to language limitations) were defined upfront.

\subsection*{Screening procedures }

Two researchers independently screened title and abstract of identified literature using Colandr \cite{Cheng_etal__2018}. During the title/abstract screening, we adopted an inclusive approach, including all papers that addressed ethical aspects of using robots in elderly care, even if we were uncertain whether social robotics were specifically discussed. To manage the significant quantity of findings, collected volumes were not included in their entirety in the full-text screening; instead, only potentially relevant book chapters were considered. One researcher (ML) screened the title, table of contents, and summary descriptions of identified books to determine relevant chapters.

Access to full texts was sought via various libraries, and authors of relevant publications were contacted directly when necessary. We were able to retrieve 295 publications for full-text screening – a number significantly higher than expected. Accordingly, after a consolidation phase, it was decided that the full-text screening would be conducted by only one reviewer (either ML or CVK), in parallel with analysis. The first 25 publications were screened jointly by at least two reviewers, one of whom (CK) has extensive experience in conducting systematic reviews in ethics. All decisions were discussed extensively, and any uncertainties arising during the full-text screening were resolved through further discussion with CK.

\subsection*{Quality Appraisal}

As outlined by Mertz \cite{Mertz__2019}, there are many open questions regarding how to conduct quality appraisal in systematic reviews that address normative questions. Due to the lack of an appropriate methodology, we have decided to not conduct a quality appraisal as part of our review. However, we attempted to ensure a baseline level of quality for the included literature by defining formal inclusion criteria (e.g., focus on academic publications and policy documents). 

\subsection*{Data extraction}

Based on our research question and context of interest we did not conduct data extraction in a standardized, quantitative format. Instead, a qualitative document analysis was conducted using MAXQDA24, as described below. However, we did extract certain information from the papers to be able to describe our sample, such as bibliographic information (see Appendix B), the disciplinary affiliation of the first author, and the type of articles. Visualizations of the sample based on those dimensions have been made available as supplemental material (see Appendix C). 

\subsection*{Data analysis and synthesis}
The data was analyzed using qualitative content analysis adapted from Kuckartz \cite{Kuckartz__2018} and Schreier \cite{Schreier__2012}. First, a purposively selected sample – maximum variation sampling \cite{Patton__2001b} – of 10 publications was analyzed independently by two authors (CK and ML). Categories describing the material were developed inductively using the strategies of summarizing and subsumption, as outlined by Schreier \cite{Schreier__2012}. The resulting coding frames were discussed between the two authors, and a consolidated framework was developed. This framework was used to analyze a second purposively selected sample of 10 papers. 

This step was conducted by three reviewers independently (CK, CVK, and ML). The resulting preliminary coding frame was discussed with the rest of the team to ensure its validity, consistency, and clarity. Once robustness of the coding frame and a shared understanding among the authors of this paper were established (by an additional joint analysis of five papers by CK and CVK), the remaining literature was analyzed by only one reviewer (ML or CVK). The two reviewers met regularly with CK to discuss new codes or uncertainties that arose during analysis. 

After the first 100 publications, further works were only used to validate the framework. Relevant text passages were checked for representation in the coding frame but were only explicitly coded if they discussed an ethical aspect not already included in the framework or where they constituted a particularly illustrative example quote.

\section*{Results}

\begin{figure}[b!]  
\centering
\caption{\label{fig1:prisma}Flow of studies through the screening process}
\includegraphics[width=\textwidth]{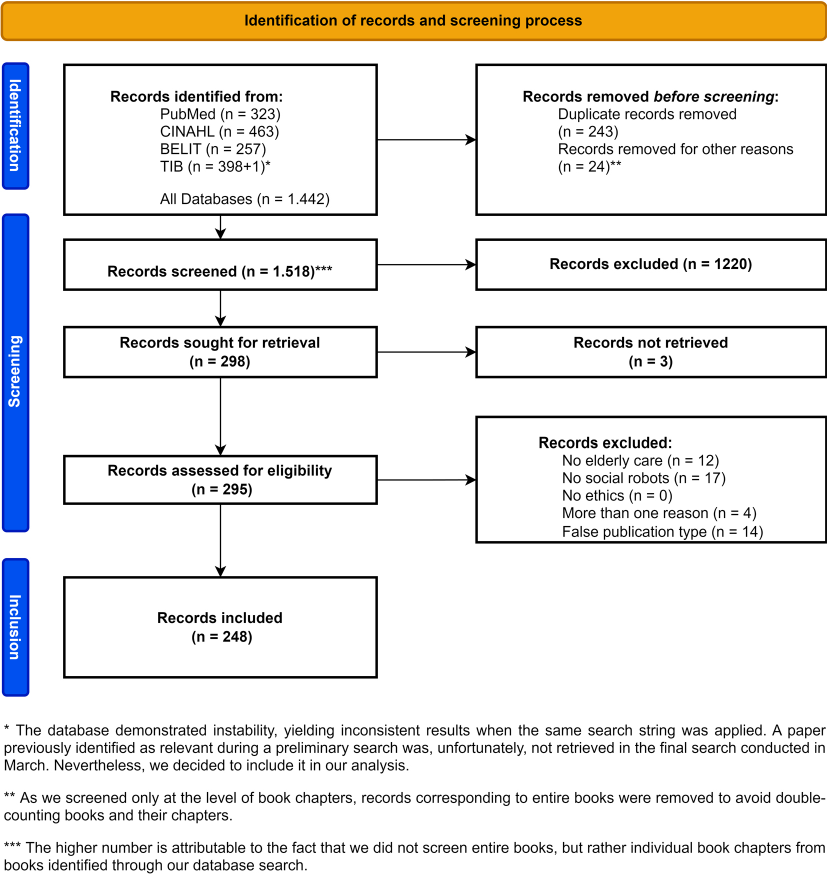}
\end{figure}

Our literature search identified 1.442 publications across the mentioned four databases. After the removal of duplicates and identification of potentially relevant book chapters, 1.518 publications remained for title-abstract screening. During this process, 1.220 publications were excluded. The remaining 298 publications were sought for retrieval, and all but three publications could be obtained. Therefore 295 publications were included in full-text screening. In this phase, 47 publications were excluded due to different reasons (see Figure 1 for full information on screening process), leaving 248 publications for analysis (see bibliographic information of included publications in Appendix B).

In the following, we use a framework of ethical opportunities and hazards to structure our reporting. This framework acknowledges that the integration of technologies often brings various benefits but also involves trade-offs, sometimes resembling a “Faustian Bargain” as Neil Postman called it, where advancements provide new possibilities while simultaneously taking away or compromising certain values or goods. Ethical opportunities and hazards are categorized according to their impact on three primary stakeholder groups: care recipients (Table 2), care providers/facilities (including informal carers) (Table 3), and society and the healthcare system (Table 4). This approach allows for a balanced examination of how these technologies contribute to the realization of goods while also highlighting their potential risks or costs.

For each main category (e.g., ethical hazards/opportunities) we identified subcategories, which we differentiated up to the third level (decreasing in abstraction) to make visible the wide variety of perspectives and potential usages of social robotics. Only the first-level subcategories are presented in the paper, but an overview of all subcategories, including example quotes for transparency, is available as supplemental material (see Appendix A). At the first level, we identified 60 subcategories, with 141 subcategories on the second level and 123 on the third. Most categories on either level pertain to the care recipient group. 

Besides these categories, our analysis revealed a number of unsettled questions which were highlighted as ethically salient but could not be clearly categorized within the framework of ethical opportunities and hazards. These include, for example, ethical arguments pointing to conflicts between different stakeholder groups, potential issues affecting all stakeholders in divergent ways, or aspects which were conceptually, normatively or empirically too ambiguous to be classified as a hazard or an opportunity. These aspects are, therefore, presented separately (Table 5). 

In the following, we will provide a short narrative overview of our findings. As the ethical aspects encountered in the literature are extensive, a more in-depth introduction of the diverse aspects will only be exemplary. For a detailed overview, consult Appendix A.

The ethical opportunities and hazards for care recipients primarily relate to the ethical dimensions of well-being and health, privacy, discrimination, independency, informed decision-making, social connection, adaptation of care to individual needs, efficiency, deception, trust, and the impact of lacking regulation (see Table 2 for a full overview of first-level categories). It is argued that engagement in relations with social robotics may positively impact perceptions of quality of life, various health parameters, and overall well-being \cite{Jecker__2021}. However, caution is warranted with regard to a satisfaction of emotional and, in some cases, physical needs, which some authors suggest may only be superficially “satisficed” rather than fully satisfied in a broader and ethically significant sense \cite{Frennert_Ostlund__2014,Misselhorn__2018}. 

Aspects of discrimination and biases are frequently discussed. Regarding the design and appearance of devices, interacting with robots that exhibit certain features (e.g., toy-like or child-like designs) may risk infantilizing or stigmatizing users \cite{Frennert_Ostlund__2014}. With respect to the interaction of social robots with care recipients, it has been positively highlighted that the consistent behavior of these devices – regardless of the age, race, or gender of care recipients – may help circumvent implicit or explicit biases in caring practices \cite{Weel_etal__2022}. 

Quality of care is another dimension subject to both opportunities and hazards. One the one hand, quality of care could improve as robots reduce caregiver workload, thereby mitigating risks of neglect or even abuse stemming from caregiver fatigue \cite{Noori_etal__2019}. One the other hand, quality may decline if robots are assigned tasks they are not (yet) capable of performing effectively or were never designed to handle \cite{Zollick_etal__2020}. Aspects of independency reflect a similar ambiguity. Social robots may enable care recipients to become less dependent on other people \cite{Haltaufderheide_etal__2020}, but they may simultaneously increase technological dependency \cite{Li_etal__2020}, leading to significant challenges when robots malfunction. 

Privacy concerns are another major theme. Health data collection and the (digital) transfer of such data by social robots frequently raise privacy concerns \cite{Bendel__2020}. However, robots lacking personal involvement or interests of their own may, in some cases, offer an advantage over human caregivers, who, by virtue of their roles, naturally intrude into private contexts, potentially infringing on privacy \cite{Sharkey_Sharkey__2014}. Similarly, while robots substituting human caregivers are often criticized for reducing social interaction and exacerbating loneliness, the opposite may hold true in some settings: 
\begin{quote}
“During a pandemic emergency in particular, the alternative to robot companionship for many older people is social isolation and loneliness. Without support, older adults are left to languish. Under these conditions, sociable robots do not rob older adults of human companionship but afford companionship where it is lacking.” \cite{Jecker__2021}
\end{quote}
On most of these dimensions, the usage of social robotics is seen to potentially produce both hazards and opportunities, where their realization often depends on specific contextual factors. However, some dimensions lean more heavily toward potential hazards, such as debates around deception regarding the robot’s nature or the relationship with it \cite{Mataric_Scassellati__2016}. While the term deception is sometimes used as a prescriptive term \cite{Danaher__2020}, implying that deception inherently constitutes a moral wrong, other sources emphasize that the risk of deception might be particularly pronounced in robots with anthropomorphic or zoomorphic design, especially when interacting with patients suffering from neurocognitive impairments such as dementia. 

\begin{table}[ht]
\centering
\caption{Ethical opportunities and hazards for care recipients in the use of social robots in elderly care }
\label{tab:ethical-categories}
\begin{tabular}{@{}lp{9.5cm}@{}}
\toprule
Main category & Subcategory (first-level) \\ \midrule

\multirow{8}{*}{Ethical opportunities for care recipients} 
& Improving/ensuring well-being/health \\ 
& Restoring privacy \\ 
& Promoting positive perception/preventing stigmatization/discrimination of care recipient \\ 
& Providing support in daily living (especially where capacities reduced) \\ 
& Sustaining independent and established lifestyle \\ 
& Furthering social connection \\ 
& Supports tailoring care approaches to needs/taste \\ 
& Support/care provided (more) efficiently \\ \midrule

\multirow{11}{*}{Ethical hazards for care recipients} 
& Reducing/not adequately securing well-being/health \\ 
& Infringing the right to privacy \\ 
& Promoting discrimination/negative images of care recipient \\ 
& Increasing dependency in everyday life \\ 
& Creating barriers for autonomous and informed decision-making \\ 
& Increasing social isolation \\ 
& Care not tailored to varying needs \\ 
& Deception about relationships/personal bonds that do not exist \\ 
& Loss of trust in carers (including social robot carer) \\ 
& Beneficial social robots not/no longer put to use \\ 
& Lack of protection due to lack of guidelines/legal regulations \\ 

\bottomrule
\end{tabular}
\end{table}

At the level of care providers and care facilities the ethical opportunities and hazards can be assigned to the dimensions of well-being/health, privacy, working conditions, relationships, nursing as profession, technological literacy, care facility management, and the impact of lacking regulations (see Table 3 for a more detailed overview). Similar to the ethical aspects relevant to care recipients, most dimensions present both hazards and opportunities. As the dimensions well-being/health, privacy and impact of lacking regulations overlap with those of the care recipients, the following examples focus on aspects specific to care providers and care facilities. 

One of the main aspects regarding care providers concerns the conditions under which care is delivered. Social robots may mitigate difficult care conditions or alleviate the burden of a high workload in professional care \cite{Wirth_etal__2020}. This could positively affect the health and well-being of care providers. However, some authors caution that a shift in tasks and roles may lead caregivers to become supervisors of their technical aids \cite{Frebel__2015} or face additional tasks (e.g., cleaning the device) \cite{Hung_etal__2019}. A more severe risk may lie in the disturbance of human care processes, as most devices lack awareness of when to interrupt or not to interrupt care interactions \cite{BioethikkommissionbeimBundeskanzleramtOsterreich__2019}. 

Another dimension that can be positively and negatively impacted by robotic technology is the relationship of care recipients and providers. Relationships can be strengthened if robots take on straining tasks, such as assisting with bathroom visits \cite{Borenstein_Pearson__2014}. In contrast, the use of social robots could also negatively impact relationships, for example, if the care recipient loses confidence in the caregiver because the social robot is used to monitor them \cite{Jenkins_Draper__2015}, or if conflicts arise between those involved in care processes over which tasks robots should be allowed to perform \cite{Tan_etal__2021}. 

A further hazard mentioned is that care providers might feel insecure or stressed when dealing with social robots, particularly if they lack the necessary technological literacy and cannot acquire relevant competencies during work hours \cite{Niemela_etal__2021}. 

Interestingly, ethical aspects relating to the care facility and its management are also highlighted, as social robots might ease or complicate their managerial tasks. For the care facility management, robots might turn out as a market advantage by supporting the acquisition of personnel and nursing home residents \cite{Bleuler_Caroni__2021}, or reducing cost of care \cite{Fruh_Gasser__2018}. Additionally, they could be used to monitor the performance of care personnel \cite{Draper_Sorell__2017} and can be used flexibly since, unlike human caregivers, robots do not tire or act out \cite{Pirhonen_etal__2020}. However, robots might also pose a market disadvantage if they are perceived as an indicator for low quality, potentially deterring prospective residents \cite{Meyer__2011}. Other challenges include organizing maintenance and support, especially if the technology is no longer supported, and the high costs associated with training staff to use these devices\cite{Hung_etal__2019}. 

\begin{table}[ht]
\centering
\caption{Ethical opportunities and hazards for care providers and facilities}
\label{tab:ethical-careproviders}
\begin{tabular}{@{}p{5cm}p{9cm}@{}}
\toprule
Main category & Subcategory (first-level) \\ \midrule

\multirow{5}{=}{Ethical opportunities for care providers and facilities \\ (including caring relatives)} 
& Positively impacting care provider’s health and well-being \\ 
& Improving working conditions for care providers \\ 
& Positively implicates relationships \\ 
& Upgrading nursing profession \\ 
& Easing management for care facilities \\ \midrule

\multirow{7}{=}{Ethical hazards for care providers and facilities (including caring relatives)} 
& Adverse effect on health/well-being of care providers \\ 
& Aggravate work situation for care providers \\ 
& Infringing on privacy of care providers \\ 
& Negatively implicates relationships \\ 
& Feeling insecure in handling social robots \\ 
& Complicating management for care facilities \\ 
& Lack of protection due to lack of guidelines/legal regulations \\ 

\bottomrule
\end{tabular}
\end{table}

The ethical opportunities and hazards for society and the (healthcare) system can be categorized into the following dimensions: ensuring provision of care, discrimination, social and moral development, economy and ecology, regulation and design of social robots, (under)standing of care/being human, justice, responsibility, efficiency, and the “rise of the robots”. 

One frequently discussed opportunity is that social robots could help ensure the provision of care in the context of demographic change and a nursing shortage by taking on tasks that would otherwise require significant human resource investment \cite{Kehl__2018}. Robots could also contribute to a more just society by being deployed to underserved populations and context, such as rural areas \cite{Bleses_Dammert__2020}. Conversely, there is the concern that the (under)standing of care in society could change in problematic ways through the use of social robots \cite{Manzeschke__2019}. It is argued that there is more to care than an input-output-process and that an important component of caregiving is lost when tasks are performed by robots, which cannot care in the way humans can (they take care, but do not care) \cite{Maur__2023}. 

Another prominently discussed hazard is that the design of social robots may be driven more by technical feasibility or engineering preferences than by the specific needs of real-world care situation, which often differ significantly from laboratory settings \cite{Paletta_etal__2019}. This misalignment could result in robots being underutilized, thereby wasting resources \cite{Boada_etal__2021}. 

A less frequently raised but interesting argument is that relying on social robots in care could curtail the need for human interaction with vulnerable individuals, reducing opportunities to develop compassion and leading to moral de-skilling. As Boada et al. state: 
\begin{quote}
“… the adoption of SARs in care, by outsourcing practices central to human existence to non-human actors, could blind us from the awareness of the constitutive vulnerability and (inter)dependence of human life, thus threatening the cultivation of virtues essential to a flourishing society. More specifically, the new technological practices could reduce the opportunities for cultivating moral skills regarding human caregiving.”\cite{Boada_etal__2021} 
\end{quote}
While the introduction of social robots has been viewed as a potential hazard to social development, in some areas, their integration may also signal moral progress. For example, the use of sex robots – a type of social robot – has been argued to help break taboos surrounding sexuality in old age \cite{Koumpis_Gees__2020}. Similarly, robots designed with gender-critical perspectives could play a role in dismantling traditional gender stereotypes, thereby fostering inclusivity and equality \cite{Weel_etal__2021}.

\begin{table}[ht]
\centering
\caption{Ethical opportunities and hazards for society and the (healthcare) system}
\label{tab:ethical-society}
\begin{tabular}{@{}p{5cm}p{9cm}@{}}
\toprule
Main category & Subcategory (first-level) \\ \midrule

\multirow{4}{=}{Ethical opportunities for society \\ and the (healthcare) system} 
& Supporting state in ensuring comprehensive nursing care for all \\ 
& Reducing discrimination/stigmatization of certain groups \\ 
& Promoting desired behavior of citizens \\ 
& Important field for economic development (location advantage) \\ \midrule

\multirow{13}{=}{Ethical hazards for society \\ and the (healthcare) system} 
& Moving forward without an adequate knowledge/discussion base \\ 
& Using a misguided approach to regulation/design of social robots \\ 
& Beneficial social robots not used due to diverse barriers \\ 
& Problematic change in (under)standing of care \\ 
& Problematic change in understanding of being human \\ 
& Damaging interpersonal and societal connectedness \\ 
& Moral de-skilling \\ 
& Fueling discrimination/stigmatization of certain groups \\ 
& Unfair distribution of benefits and burdens of robot introduction \\ 
& Erosion of responsibility/gaps in the responsibility structure \\ 
& Inefficient use of social robots \\ 
& Ecologically and socially not sustainable \\ 
& Social robots get too much power and endanger humans/\newline human society \\

\bottomrule
\end{tabular}
\end{table}

Unsettled questions in the scientific discourse relate to the cost-benefit ratio of social robots, just allocation/prioritization, discrimination, privacy, deception, autonomy and informed decision-making, responsibility, understanding of care, and social robots as autonomous agents. These issues are highlighted as being worthy of consideration from an ethical perspective but do not clearly classify as either hazard or opportunity or are conceptually, normatively or empirically too ambiguous to be classified as such. Some of the hazards and opportunities mentioned above are revisited here but framed as ethical conflicts, illustrating what is gained and lost through the use of social robots and how certain hazards and opportunities are interconnected. A prominent example is the question of how to ethically appraise the possible deception of care recipients given the potential for positive effects on well-being. As mentioned above, particularly cognitively impaired patients can mistake social robots for real humans or animals. At the same time, there are indications in the literature that certain opportunities, like reduced stress \cite{Wada_Shibata__2007} or increased communication between care recipient and relatives \cite{Sharkey_Sharkey__2011} may better be realized where at least some level of deception is accepted. In the literature it remains normatively unresolved whether the deceit is ethically acceptable given the resulting health gains \cite{Kreis__2018}. 

Further aspects mentioned in this context concern, for example, the question whether privacy risks associated with monitoring and transmitting health data – for instance to ensure well-being through immediate emergency care if needed – are justifiable \cite{Kehl__2018c}. Again, this largely depends on the normative stance one adopts. Conflicts arising from the design of robots are less frequently discussed. Weßel et al. \cite{Weel_etal__2021} argue that designing robots based on problematic gender stereotypes (e.g., portraying women as subservient and gentle) can enhance the well-being and robot acceptance of care recipients who grew up with these stereotypes, but potentially at the expense of societal progress toward gender diversity. 

Further unsettled aspects noted in the literature do not concern conflicts but rather conceptual gaps that complicate decision-making. One such gap is the question of how to operationalize the concept of responsibility in the context of social robots: Who bears responsibility for the actions of social robots, particularly when the robot causes harm or negative consequences for users \cite{FeilSeifer_Mataric__2011}? This is not a trivial question, as certain hazards only arise due to a lack of clear ascriptions of responsibilities and liabilities – for example, the risk that care recipients face prohibitive costs if a care robot damages property because designers, vendors, or care facilities are not legally assigned responsibility \cite{Schmidhuber_Stoger__2021}. 

There is also broader uncertainty regarding the concept of “good care”. How we understand (good) care will heavily influence how we evaluate the employment of social robots. Further unsettled questions are summarized in table 5.

\begin{table}[ht]
\centering
\caption{Unsettled questions regarding social robots}
\label{tab:unsettled-questions}
\begin{tabular}{@{}p{5cm}p{9cm}@{}}
\toprule
Main category & Subcategory (first-level) \\ \midrule

\multirow{12}{=}{Unsettled questions} 
& Do the benefits of social robots outweigh the costs? \\ 
& Who should receive/how to allocate scarce (robotic) resources? \\ 
& Whose interests regarding social robot usage should be prioritized? \\ 
& Design by stereotype to foster wellbeing/compliance while \newline reproducing harmful effects/reducing acceptance of diverse ways of life? \\ 
& What privacy risks are acceptable for security gains? \\ 
& Is deception about relationships acceptable to realize benefits? \\ 
& Should control/access of social robots be restricted to reduce risk of harm? \\ 
& Who should/how to decide when people are cognitively impaired? \\ 
& Who bears responsibility for the social robot and its actions? \\ 
& (When) should social robots be allowed to act without consent to protect? \\ 
& Should autonomy be bestowed onto social robots (at all/to what extent)? \\ 
& What normative status should we assign social robots? \\ 
& How do we understand the concept of care? \\ 

\bottomrule
\end{tabular}
\end{table}

\section*{Discussion}

Our results testify to a vast array of ethical aspects to consider when using social robots in care practices. This includes addressing the interests from different stakeholder groups as well as various value dimensions. As our findings indicate, most of these dimensions entail both potentially negative as well as positive ethical aspects.

From a perspective of decision-making – broadly defined as seeking normative action guidance and practical orientation – the discourse captured within this review may initially appear to lead to an impasse, offering little actionable guidance for concrete scenarios. However, it is important to understand that the actual realization of most of the identified ethical aspects is largely contingent on the specific circumstances of each situation \cite{Lee_etal__2017}. This implies that sweeping judgements about the use of social robots in elderly care are often misplaced \cite{Wahl_etal__2021} and fail to reflect the depth and complexity of ethical considerations outlined in this discourse. Secondly, it suggests that any practical application of this knowledge requires careful analysis of the specifics of the situation. This includes essential aspects such as the care setting, the type of the device used, its intended role, and the tasks assigned to it. Perhaps most importantly, it requires consideration of the characteristics, interests, and preferences of care receivers and care givers \cite{Geier_etal__2020}. 

With this, our results also testify to some key challenges arising when making ethical decisions about the use of social robots from a practical perspective. The first key challenge is about being aware of and acknowledging the different needs and interests connected to the various perspectives involved. It is well-known, for example, that social robots in elderly care are often ill-prepared to fulfill tasks care providers/recipients consider relevant or helpful \cite{Frennert_Ostlund__2014,Kehl__2018c,Kehl__2018} and decision makers do not know how and whose interests are to be taken into account. One reason is that development is often driven more by technically motivations and possibilities than actual care needs. It is, therefore, crucial that decision makers are aware of the plurality of perspectives and interests and enable greater engagement with affected user groups. This applies especially during early decisions such as in the design and testing process but is not limited to that. As a result of growing experience and becoming more acquainted with a technology, interests may evolve during later stages, necessitating periodic reassessments \cite{Flandorfer_Huang__2012}. In addition, any assessment of social robots used in care setting is highly dependent on the specifics of the situation. A monitoring function of a robot can, for example, in some contexts be perceived as enhancing independence by allowing individuals to remain at home for longer, while in other contexts, it may be viewed as a critical infringement of privacy. Such functions may accordingly be welcomed or rejected, depending on the particular circumstances \cite{Frennert_Ostlund__2014} but also depending on user preferences \cite{Buhr_etal__2024}. Participatory approaches to the design and implementation of social robots have been highlighted as a useful strategy to simultaneously ensure that the interests of the affected user groups are respected and that robots are practical, for example in terms of meeting the structural requirements \cite{Lee_etal__2017}. 

Secondly, we note the importance for decision-makers to recognize that the discourse on social robots, as captured with this review, is not always substantiated by empirical evidence. As social robotics is still in an early stage, much of the discourse exists in the realm of speculative hope and fear. Whether all identified hazards and opportunities are genuinely relevant will need to be evaluated against actual and future developments. For instance, the perpetuations of problematic (gender-) stereotypes are more likely when robots take anthropomorphic forms, but it remains uncertain whether technological development will continue in this direction. In addition, many claims about ethically relevant outcomes such as impact on quality of life, are not only highly intervention-, device-, and setting-specific but currently lack sufficient empirical support \cite{Haltaufderheide_etal__2023}. 

A third aspect that decision-makers, or others working with our results, should consider is that while the realization of many hazards and opportunities depends on how robotically assisted care practices are implemented, some opportunities will only be achieved by accepting the materialization of certain hazards. These aspects are categorized as unsettled questions, as they require to take a normative position on the hierarchies of values and the permissibility of value trade-offs. One example is the use of Paro and other similar zoomorphic robots to improve the well-being of dementia patients, which may only be achievable by accepting a degree of deception – an issue, often seen as morally problematic \cite{Sparrow__2002,Coeckelbergh__2015,Remmers__2018}. Another example concerns possible privacy breaches caused by constant monitoring through social robots, which may or may not be deemed acceptable in exchange for enhanced security, particularly of dementia patients. We do not claim to have identified all unsettled questions that require such value-balancing but consider it highly likely additional trade-offs can be reconstructed from the hazards and opportunities highlighted in specific situations. 

Finally, we note that terminology and concepts in the research discourse are often used imprecisely, complicating debates about robots, care, and old age. For instance, it is often argued that social robots can relieve caregivers, allowing them more time for social interactions with care recipients. However, this seems contradictory, as social robots are explicitly designed to take on or support social interactions themselves. Interestingly, many examples given in the texts included in the analysis focus on tasks unrelated to social interaction, such as disinfecting door handles, transporting objects \cite{Bleuler_Caroni__2021}, or lifting \cite{Parviainen_etal__2019}, which could just as effectively be performed by simpler assistance robots or non-robotic systems \cite{Zollick_etal__2020}. This raises the question of whether the development of social robots should be prioritized over non-social robotic systems that may achieve similar benefits without the associated ethical hazards. At the same time, it is important to consider that routine tasks, such as carrying objects or dressing \cite{Parviainen_Pirhonen__2017}, often provide moments of interpersonal contact, and it is worth reflecting on how care recipients might be affected if these interactions are carried out by robots. Furthermore, terms like “social robots” are not morally neutral, as they evoke specific images that may not accurately reflect the robots’ actual functions. For example, the robot Pepper might be better described as an entertainment robot, as it is primarily used for games (e.g., Bingo or Memory) or as a fitness instructor.

Another important concept shaping the ethical assessment of social robotics in elderly care – often implicitly – is the understanding of “care” or “good care” \cite{Coeckelbergh__2015}. If “good care” is equated with human care \cite{Maur__2023} or necessarily involves interpersonal relationships \cite{DeutscherEthikrat__2020}, this results in different evaluation standards for the use of social robotics than if “good care” is equated with the fulfillment of specific tasks \cite{Sparrow__2016,Suwa_etal__2020}. Misselhorn \cite{Misselhorn__2018} provides a helpful distinction in this context, between goal-oriented or practice-oriented care. She describes goal-oriented care as focusing on outcomes (e.g., bringing a needed object to the care recipient), while practice-oriented care emphasizes the social and empathic interactions inherent in caregiving activities. Robots are currently incapable of practice-oriented care, at least according to the majority opinion \cite{Bertolini_Arian__2020,Noori_etal__2019,Steinrotter__2020,Maur__2023}. Consequently, those who adopt a practice-oriented view of care are likely to reject social robots as adequate tools in elderly care. In the literature, the concept of care is repeatedly used without much reflection. For robust evaluation and meaningful discussion, it would be important to make the underlying understanding of care in this context explicit \cite{Kehl__2018b}.

The same applies with regard to underlying perceptions of images of old age which warrant caution \cite{Neven__2010}. How old age is perceived not only influences how the prospect of care being provided by robots is evaluated but also determines the type of robots that are designed and implemented. A recurring issue is that the technical perspective on social robotics in elderly care is often characterized by deficit-oriented images of old age, reducing older people in need of care to technical problems requiring solutions \cite{Frennert_Ostlund__2014,Kamphof__2015}. A further problem is that older people are frequently viewed as a uniform group, neglecting the fundamentally different preferences, values, and priorities that exist – just as in any other human group. Such perspectives not only objectify older adults but also neglect the creativity and potential for self-determined agency in old age \cite{Remmers__2019}. Moreover, they fail to account for individual needs, which can vary widely depending on personal experiences, cultural backgrounds, and specific circumstances. Acknowledging and addressing these diverse needs is essential to developing social robots that genuinely enhance quality of life of older adults, while respecting their own values. In our opinion, it is desirable for the development of social robots in elderly care to focus not only on compensating for lost abilities (“problem-solving”) but also on enabling older people to participate productively in society \cite{Remmers__2019}.

\section*{Limitations}

Our findings come with certain limitations we want to highlight. First, the operationalization of key terms, particularly the ethical aspects that largely determined the scope of our review, warrants caution. We based our ethical framework on principlism, adapted from Beauchamp and Childress \cite{Beauchamp_Childress__2019} for the context of robots \cite{Floridi_etal__2018}. Although this approach is debated, it is widely applied in normative systematic reviews \cite{Mertz_etal__2017} and proved effective for our purposes. We deliberately avoided relying solely on authors’ self-identification of ethical aspects, as non-ethicists often do not use normative terminology, potentially leading to relevant points being overlooked. However, our approach may have had exclusionary effects as well – such as missing discussions clarifying normative concepts like responsibility or good care. 

Second, our findings might be less reliable due to the interpretative nature of qualitative systematic reviews and the fact that most included literature was analyzed by only one reviewer. This decision was made to manage the unexpectedly high number of publications. To mitigate the risks of solo interpretation, the first 25 publications, which represented a variety of disciplines and formats, were analyzed by two to three reviewers. This subset helped establish a robust coding frame, which was then applied to the remaining publications. Additionally, regular meetings were held between the two reviewers and the project lead to address any challenges, uncertainties, or emerging codes.

\section*{Conclusion}
With this work, we provide a comprehensive overview of the ethical hazards and opportunities that should be considered in practical evaluations of the use of social robots. Our findings demonstrate that the research discourse provides a vast array of potential ethical aspects. Its structure implies, however, that decision makers need to be aware of its complexity and depth, as well as its empirical and conceptual weaknesses and the various affected interests and needs involved to be able make ethically informed decisions.

\section*{Statements and declarations}
\subsection*{Acknowledgements}
We gratefully acknowledge support by Marie Pelzer and Dr Laura Gawinski. We would also like to thank the Library of the University of Potsdam for their support in the development of the search strings, which was fundamental to the thoroughness and accuracy of our literature review.

\subsection*{Funding}

This publication is part of the project Ethics Guidelines for Socially Assistive Robots in Elderly Care: An Empirical-Participatory Approach (E-cARE) funded by the German Ministry of Health under grant no. 2521FSB008

\subsection*{Competing interests}

The authors declare to have no competing interests.

\subsection*{Ethics approval}
The study did not involve human paticipants, animals, or their data or biological material. Ethics approval was not deemed necessary.


\clearpage
\begin{appendices}

\clearpage
\section*{Appendix A: Categories with example quotes}\label{secA1}

\begin{tikzpicture}[
  every node/.style={align=left,},
  category/.style={text width=6cm, align=left, inner sep=4pt, font=\small},
  root/.style={text width=10cm, align=left, inner sep=4pt},
  quote/.style={rounded corners, thick, text width=6cm, align=justify, inner sep=4pt, font=\footnotesize},
  arrow/.style={-{Stealth[length=3mm]}, thick},
  connector/.style={thick},
  spacing/.store in=\myspacing, spacing=1cm 
]

\node[root] (cat1) at (0, 0) {Ethical opportunities for care recipients};

\node[category, anchor=west] (sub1) at ($(cat1.south west)+(0.5,-\myspacing+0.5cm)$) {Improving/ensuring well-being/health};
\node[quote, anchor=west] (q1) at ($(sub1.east)+(1.5cm,0)$) {“Research shows that when older adults form relationships with robots, they generally report better health and well-being (Broekens et al. 2009).”(Jecker 2021, p. S36)};
\draw[arrow] (sub1.east) -- (q1.west);
\node[draw=none, fit=(sub1)(q1), inner sep=0pt] (pair1) {};

\node[category, anchor=north west] (sub2) at ($(pair1.south west)+(0.5,-\myspacing)$) {Supporting therapeutic/rehabilitative procedures};
\node[quote, anchor=west] (q2) at ($(sub2.east)+(1cm,0)$) {“Yet, social robots cannot only entertain and engender social interaction, they may also be used to initiate therapeutic processes that enhance self-understanding.” (Misselhorn et al. 2013, p. 128)};
\draw[arrow] (sub2.east) -- (q2.west);
\node[draw=none, fit=(sub2)(q2), inner sep=0pt] (pair2) {};

\node[category, anchor=north west] (sub3) at ($(pair2.south west)+(0.5,-\myspacing)$) {(Re-)activation of (lost) capacities};
\node[quote, anchor=west] (q3) at ($(sub3.east)+(0.5cm,0)$) {“Social assistive-robot-based systems with the abilities to perform activities, play cognitive games, and socialize, can be used to stimulate the physical, cognitive, and social conditions of older adults and restrain deterioration of their cognitive state [20].”(Mishra et al. 2021, p. 492)};
\draw[arrow] (sub3.east) -- (q3.west);
\node[draw=none, fit=(sub3)(q3), inner sep=0pt] (pair3) {};

\node[category, anchor=north west] (sub4) at ($(pair3.south west)+(0,-\myspacing)$) {Remote doctor’s appointment as facilitation};
\node[quote, anchor=west] (q4) at ($(sub4.east)+(0.5cm,0)$) {“Nowadays a large number of companies are developing robots to monitor the health and safety of old citizens. Some robots like RP-7 are teleoperated so the elderly can interact with their family doctors instantly.” (Noori et al. 2019 Robot-Care for the Older People: Ethically Justified or Not?, S. 44)};
\draw[arrow] (sub4.east) -- (q4.west);
\node[draw=none, fit=(sub4)(q4), inner sep=0pt] (pair4) {};

\node[category, anchor=north west] (sub5) at ($(pair4.south west)+(-0.5,-\myspacing)$) {Reducing risk of mistreatment by care providers};
\node[quote, anchor=west] (q5) at ($(sub5.east)+(1cm,0)$) {“The primary aim of robots is to make senior citizens independent, which sometimes they dont [sic] feel comfortable with caregivers or some caregivers dont [sic] treat/care them according to the human rights act.”(Noori et al. 2019 Robot-Care for the Older People: Ethically Justified or Not?, p. 44)};
\draw[arrow] (sub5.east) -- (q5.west);
\node[draw=none, fit=(sub5)(q5), inner sep=0pt] (pair5) {};

\node[category, anchor=north west] (sub6) at ($(pair5.south west)+(0.5,-\myspacing-0.5cm)$) {Reducing risk of abuse/neglect (e.g. stemming from caregiver fatigue)};
\node[quote, anchor=west] (q6) at ($(sub6.east)+(0.5cm,0)$) {“Psychologically, assuming that robotics technology has been programmed to act ethically, employing robotics and autonomous systems to deal with some of the caregiving tasks for the older people could potentially reduce the likelihood of psychological and emotional abuse and neglect which often stem from caregiver fatigue (O’Brolchain, 2017).” (Tan et al. 2021 Tensions and antagonistic interactions of risks and ethics of using robotics and autonomous systems in long-term care, p. 2)};
\draw[arrow] (sub6.east) -- (q6.west);
\node[draw=none, fit=(sub6)(q6), inner sep=0pt] (pair6) {};

\node[category, anchor=north west] (sub7) at ($(pair6.south west)+(0,-\myspacing)$) {Controlling care personnel and reducing problematic practices};
\node[quote, anchor=west] (q7) at ($(sub7.east)+(0.5cm,0)$) {“Employing the robot to ‘‘police’’ care may discourage poor care practices, with benefits to older people.” (Draper and Sorell 2017 Ethical values and social care robots for older people: an international qualitative study, p. 63)};
\draw[arrow] (sub7.east) -- (q7.west);
\node[draw=none, fit=(sub7)(q7), inner sep=0pt] (pair7) {};

\node[category, anchor=north west] (sub8) at ($(pair7.south west)+(-0.5,-\myspacing)$) {Addressing physical well-being and safety of care recipients};
\node[quote, anchor=west] (q8) at ($(sub8.east)+(1cm,0)$) {“Robots are an opportunity for the health and social care sector. They increase individual autonomy, relieve staff of repetitive tasks and contribute to the safety of residents, for example, by detecting falls.” (Translation by the authors) (Früh and Gasser 2018 Erfahrungen aus dem Einsatz von Pflegerobotern für Menschen im Alter, p. 45; in: Bendel 2018 Pflegeroboter)};
\draw[arrow] (sub8.east) -- (q8.west);
\node[draw=none, fit=(sub8)(q8), inner sep=0pt] (pair8) {};

\node[category, anchor=north west] (sub9) at ($(pair8.south west)+(+0.5,0)$) {\textit{Cont. on next page}};
\node[quote, anchor=west] (q9) at ($(sub9.east)+(1.5cm,0)$) {};
\node[draw=none, fit=(sub9)(q9), inner sep=0pt] (pair9) {};

\node[category, anchor=north west] (sub10) at ($(pair9.south west)+(-0.5,0)$) {\textit{Cont. on next page}};
\node[quote, anchor=west] (q10) at ($(sub10.east)+(2cm,0)$) {};
\node[draw=none, fit=(sub10)(q10), inner sep=0pt] (pair10) {};

\node[category, anchor=north west] (sub11) at ($(pair10.south west)+(-0.5,0)$) {\textit{Cont. on next page}};
\node[quote, anchor=west] (q11) at ($(sub11.east)+(2.5cm,0)$) {};
\node[draw=none, fit=(sub11)(q11), inner sep=0pt] (pair11) {};

\draw[connector] (cat1.west) |- ([xshift=-2mm]sub1.west);
\draw[connector] (sub1.west) |- ([xshift=-2mm]sub2.west);
\draw[connector] (sub2.west) |- ([xshift=-2mm]sub3.west);
\draw[connector] (sub2.west) |- ([xshift=-2mm]sub4.west);
\draw[connector] (sub1.west) |- ([xshift=-2mm]sub5.west);
\draw[connector] (sub5.west) |- ([xshift=-2mm]sub6.west);
\draw[connector] (sub5.west) |- ([xshift=-2mm]sub7.west);
\draw[connector] (sub1.west) |- ([xshift=-2mm]sub8.west);
\draw[connector] (sub8.west) |- ([xshift=-2mm]sub9.west);
\draw[connector] (sub1.west) |- ([xshift=-2mm]sub10.west);
\draw[connector] (cat1.west) |- ([xshift=-2mm]sub11.west);

\end{tikzpicture}

\clearpage

\newdimen\spacing
\spacing=1cm 

\begin{tikzpicture}[
  every node/.style={align=left,},
  category/.style={text width=6cm, align=left, inner sep=4pt, font=\small},
  root/.style={text width=10cm, align=left, inner sep=4pt},
  quote/.style={rounded corners, thick, text width=6cm, align=justify, inner sep=4pt, font=\footnotesize},
  arrow/.style={-{Stealth[length=3mm]}, thick},
  connector/.style={thick},
  spacing/.store in=\myspacing, spacing=1.0cm 
]

\node[root] (cat1) at (0, 0) {Ethical opportunities for care recipients (continued)};

\node[category, anchor=west] (sub1) at ($(cat1.south west)+(0.5,-\myspacing+0.5cm)$) {Improving/ensuring well-being/health (continued)};
\node[quote, anchor=west] (q1) at ($(sub1.east)+(1.5cm,0)$) {};
\node[draw=none, fit=(sub1)(q1), inner sep=0pt] (pair1) {};

\node[category, anchor=north west] (sub2) at ($(pair1.south west)+(0.5,-\myspacing+0.5cm)$){Addressing physical well-being and safety of care recipients (continued)};
\node[quote, anchor=west] (q2) at ($(sub2.east)+(1.5cm,0)$) {};
\node[draw=none, fit=(sub2)(q2), inner sep=0pt] (pair2) {};

\node[category, anchor=north west] (sub3) at ($(pair2.south west)+(0.5,-\myspacing)$) {(Reducing risk to catch infectious diseases (e. g. from care providers)};
\node[quote, anchor=west] (q3) at ($(sub3.east)+(0.5cm,0)$) {“First, sociable robots can be sanitized and offer a safe means of interacting with older people during a pandemic (Armitage and Nellums 2020).” (Jecker 2021 You’ve got a friend in me: sociable robots for older adults in an age of global pandemics, p. S36)};
\draw[arrow] (sub3.east) -- (q3.west);
\node[draw=none, fit=(sub3)(q3), inner sep=0pt] (pair3) {};

\node[category, anchor=north west] (sub4) at ($(pair3.south west)+(0.,-\myspacing-0.5cm)$) {Motivate or execute behavior to reduce (health) risks};
\node[quote, anchor=west] (q4) at ($(sub4.east)+(0.5cm,0)$) {“A robot could register potential dangers; for example, turn off a forgotten hotplate; prevent an elderly person from climbing onto a wobbly chair to reach an object; but also simply take high-calorie food away from a person who should be watching their weight for health reasons or prevent them from pouring themselves a third glass of wine.” (Translation by the authors) (Bioethikkommission beim Bundeskanzleramt Österreich 2018 Roboter in der Betreuung alter Menschen (Stellungnahme), p. 368)};
\draw[arrow] (sub4.east) -- (q4.west);
\node[draw=none, fit=(sub4)(q4), inner sep=0pt] (pair4) {};

\node[category, anchor=north west] (sub5) at ($(pair4.south west)+(-0.5,-\myspacing-0.5cm)$) {Monitoring health and alarming others in crisis situations};
\node[quote, anchor=west] (q5) at ($(sub5.east)+(1cm,0)$) {“In both cases, a robot is used in the household - in the case of SRS the Care-O-Bot, in the case of Robo M.D. the Nao (Aldebaran), which detects the senior’s state of health and passes it on to a medical center or reports critical situations to relatives, neighbors, family doctors, emergency doctors and/or relatives.” (Translation by the authors) (Kapitel Meyer 2011 Akzeptanz ausgewählter Anwendungsszenarien, p. 89; in: Meyer 2011 Mein Freund der Roboter - Servicerobotik für ältere Menschen - eine Antwort auf den demographischen Wandel?)};
\draw[arrow] (sub5.east) -- (q5.west);
\node[draw=none, fit=(sub5)(q5), inner sep=0pt] (pair5) {};

\node[category, anchor=north west] (sub6) at ($(pair5.south west)+(0.5,-\myspacing)$) {As instructor in fitness sessions};
\node[quote, anchor=west] (q6) at ($(sub6.east)+(0.5cm,0)$) {"Based on the recommendation of her physiotherapist, the robot demonstrates some soft and slow stretching motions that Mrs. Smith can replicate." (Gelin 2017 The Domestic Robot: Ethical and Technical Concerns, p. 208; in: Ferreira et al. 2017 A World with Robots - Intelligent Systems, Control and Automation: Science and Engineering)};
\draw[arrow] (sub6.east) -- (q6.west);
\node[draw=none, fit=(sub6)(q6), inner sep=0pt] (pair6) {};

\node[category, anchor=north west] (sub7) at ($(pair6.south west)+(0.0,-\myspacing)$) {Motivating movement/exercise/free-time activities};
\node[quote, anchor=west] (q7) at ($(sub7.east)+(0.5cm,0)$) {"Her Care-O-bot® knows that she should be encouraged to move about, and suggests several times a day that she walks with it to look out of the window at either the garden or the street below."  (Draper and Sorell 2017 Ethical values and social care robots for older people: an international qualitative study, p. 53)};
\draw[arrow] (sub7.east) -- (q7.west);
\node[draw=none, fit=(sub7)(q7), inner sep=0pt] (pair7) {};

\node[category, anchor=north west] (sub8) at ($(pair7.south west)+(-0.0,-\myspacing)$) {Sex robots allow staying sexually active};
\node[quote, anchor=west] (q8) at ($(sub8.east)+(0.5cm,0)$) {"[…] finally, after responding to objections, I conclude that sex robots are a reasonable way to support later-life sexuality for persons with disabilities."  (Jecker 2021 Nothing to be ashamed of - sex robots for older adults with disabilities, p. 26)};
\draw[arrow] (sub8.east) -- (q8.west);
\node[draw=none, fit=(sub8)(q8), inner sep=0pt] (pair8) {};

\node[category, anchor=north west] (sub9) at ($(pair8.south west)+(+0.0,0)$) {\textit{Cont. on next page}};
\node[quote, anchor=west] (q9) at ($(sub9.east)+(0.5cm,0)$) {};
\node[draw=none, fit=(sub9)(q9), inner sep=0pt] (pair9) {};

\node[category, anchor=north west] (sub10) at ($(pair9.south west)+(-0.5,0)$) {\textit{Cont. on next page}};
\node[quote, anchor=west] (q10) at ($(sub10.east)+(1.5cm,0)$) {};
\node[draw=none, fit=(sub10)(q10), inner sep=0pt] (pair10) {};

\node[category, anchor=north west] (sub11) at ($(pair10.south west)+(-0.5,0)$) {\textit{Cont. on next page}};
\node[quote, anchor=west] (q11) at ($(sub11.east)+(1.5cm,0)$) {};
\node[draw=none, fit=(sub11)(q11), inner sep=0pt] (pair11) {};

\draw[connector] (cat1.west) |- ([xshift=-2mm]sub1.west);
\draw[connector] (sub1.west) |- ([xshift=-2mm]sub2.west);
\draw[connector] (sub2.west) |- ([xshift=-2mm]sub3.west);
\draw[connector] (sub2.west) |- ([xshift=-2mm]sub4.west);
\draw[connector] (sub2.west) |- ([xshift=-2mm]sub5.west);
\draw[connector] (sub2.west) |- ([xshift=-2mm]sub6.west);
\draw[connector] (sub2.west) |- ([xshift=-2mm]sub7.west);
\draw[connector] (sub2.west) |- ([xshift=-2mm]sub8.west);
\draw[connector] (sub2.west) |- ([xshift=-2mm]sub9.west);
\draw[connector] (sub1.west) |- ([xshift=-2mm]sub10.west);
\draw[connector] (cat1.west) |- ([xshift=-2mm]sub11.west);

\end{tikzpicture}
\clearpage

\newdimen\spacing
\spacing=1cm 

\begin{tikzpicture}[
  every node/.style={align=left,},
  category/.style={text width=6cm, align=left, inner sep=4pt, font=\small},
  root/.style={text width=10cm, align=left, inner sep=4pt},
  quote/.style={rounded corners, thick, text width=6cm, align=justify, inner sep=4pt, font=\footnotesize},
  arrow/.style={-{Stealth[length=3mm]}, thick},
  connector/.style={thick},
  spacing/.store in=\myspacing, spacing=1.0cm 
]

\node[root] (cat1) at (0, 0) {Ethical opportunities for care recipients (continued)};

\node[category, anchor=west] (sub1) at ($(cat1.south west)+(0.5,-\myspacing+0.5cm)$) {Improving/ensuring well-being/health (continued)};
\node[quote, anchor=west] (q1) at ($(sub1.east)+(1.5cm,0)$) {};
\node[draw=none, fit=(sub1)(q1), inner sep=0pt] (pair1) {};

\node[category, anchor=north west] (sub2) at ($(pair1.south west)+(0.5,-\myspacing+0.5cm)$){Addressing physical well-being and safety of care recipients (continued)};
\node[quote, anchor=west] (q2) at ($(sub2.east)+(1.5cm,0)$) {};
\node[draw=none, fit=(sub2)(q2), inner sep=0pt] (pair2) {};

\node[category, anchor=north west] (sub3) at ($(pair2.south west)+(0.5,-\myspacing)$) {Positive physiological reactions (e. g. of organs)};
\node[quote, anchor=west] (q3) at ($(sub3.east)+(0.5cm,0)$) {“We consider that the interaction with Paro and with other people (through Paro) physiologically improved the reactions of the residents’ vital organs.” (Wada und Shibata 2007 Living With Seal Robots – Its Sociopsychological and Physiological Influences on the Elderly at a Care House, p. 978)};
\draw[arrow] (sub3.east) -- (q3.west);
\node[draw=none, fit=(sub3)(q3), inner sep=0pt] (pair3) {};

\node[category, anchor=north west] (sub4) at ($(pair3.south west)+(-0.5,-\myspacing)$) {Addressing emotional well-being of care recipients};
\node[quote, anchor=west] (q4) at ($(sub4.east)+(1.0cm,0)$) {“Khosla and Chu proposed the design and implementation of a human-like assistive robot Matilda to improve the emotional wellbeing of the elderly in residential care facilities.” (Wang et al. 2019 Affective Interaction Technology of Companion Robots for the Elderly: A Review, p. 79)};
\draw[arrow] (sub4.east) -- (q4.west);
\node[draw=none, fit=(sub4)(q4), inner sep=0pt] (pair4) {};

\node[category, anchor=north west] (sub5) at ($(pair4.south west)+(0.5,-\myspacing)$) {Reducing boredom};
\node[quote, anchor=west] (q5) at ($(sub5.east)+(0.5cm,0)$) {“Positive expectations of the caregivers were that the robotic cushion [“Qoobo”] would be more practical than the already used SAR, that it could improve mood, would be a pastime and boredom would be reduced.” (Lehmann et al. 2021 Using a Socially Assistive Robot in a Nursing Home: Caregivers’ Expectations and Concerns, p. 148)
};
\draw[arrow] (sub5.east) -- (q5.west);
\node[draw=none, fit=(sub5)(q5), inner sep=0pt] (pair5) {};

\node[category, anchor=north west] (sub6) at ($(pair5.south west)+(0.0,-\myspacing)$) {Reducing worry/fear };
\node[quote, anchor=west] (q6) at ($(sub6.east)+(0.5cm,0)$) {“The main reported benefits of Paro at home were in reducing anxiety and enhancing mood. Caregivers reported that touching Paro’s soft texture was soothing for some care recipients and that it could serve as a distraction from distressing events.” (Liang et al. 2017 A Pilot Randomized Trial of a Companion Robot for People With Dementia Living in the Community, p. 875)
};
\draw[arrow] (sub6.east) -- (q6.west);
\node[draw=none, fit=(sub6)(q6), inner sep=0pt] (pair6) {};

\node[category, anchor=north west] (sub7) at ($(pair6.south west)+(0.0,-\myspacing-1cm)$) {Reducing feelings of loneliness/enhancing feelings of attachment};
\node[quote, anchor=west] (q7) at ($(sub7.east)+(0.5cm,0)$) {“Of particular interest here is the case in which a 76-year-old male resident in a nursing home who is recovering from depression is given a My Real Baby to interact with for 4 months. As a natural animal lover who considers all the neighborhood animals his friends (and believes that he communicates with them), he responds warmly to the introduction of My Real Baby. He explains to the researchers that he not only enjoys the robots presence as it alleviates feelings of loneliness, but that he also talks to the robot as if it were his former wife; its presence and responsiveness gives him some sort of permission to vent his private feelings of attachment, regret, and confusion.” (Misselhorn et al. 2013 Ethical Considerations Regarding the Use of Social Robots in the Fourth Age, p. 128)
};
\draw[arrow] (sub7.east) -- (q7.west);
\node[draw=none, fit=(sub7)(q7), inner sep=0pt] (pair7) {};

\node[category, anchor=north west] (sub8) at ($(pair7.south west)+(-0.0,-\myspacing)$) {Reducing stress};
\node[quote, anchor=west] (q8) at ($(sub8.east)+(0.5cm,0)$) {“An interesting possibility that could be explored in future research is that some of the benefits that result from playing with, or ‘‘nurturing’’ robot pets are a consequence of increased levels of oxytocin, which as discussed earlier, appears to reduce stress.” (Sharkey and Sharkey 2012 Granny and the robots: ethical issues in robot care for the elderly, p. 34) 
};
\draw[arrow] (sub8.east) -- (q8.west);
\node[draw=none, fit=(sub8)(q8), inner sep=0pt] (pair8) {};

\node[category, anchor=north west] (sub9) at ($(pair8.south west)+(+0.0,0)$) {\textit{Cont. on next page}};
\node[quote, anchor=west] (q9) at ($(sub9.east)+(0.5cm,0)$) {};
\node[draw=none, fit=(sub9)(q9), inner sep=0pt] (pair9) {};

\node[category, anchor=north west] (sub10) at ($(pair9.south west)+(-0.5,0)$) {\textit{Cont. on next page}};
\node[quote, anchor=west] (q10) at ($(sub10.east)+(1cm,0)$) {};
\node[draw=none, fit=(sub10)(q10), inner sep=0pt] (pair10) {};

\node[category, anchor=north west] (sub11) at ($(pair10.south west)+(-0.5,0)$) {\textit{Cont. on next page}};
\node[quote, anchor=west] (q11) at ($(sub11.east)+(1.5cm,0)$) {};
\node[draw=none, fit=(sub11)(q11), inner sep=0pt] (pair11) {};

\draw[connector] (cat1.west) |- ([xshift=-2mm]sub1.west);
\draw[connector] (sub1.west) |- ([xshift=-2mm]sub2.west);
\draw[connector] (sub2.west) |- ([xshift=-2mm]sub3.west);
\draw[connector] (sub1.west) |- ([xshift=-2mm]sub4.west);
\draw[connector] (sub4.west) |- ([xshift=-2mm]sub5.west);
\draw[connector] (sub4.west) |- ([xshift=-2mm]sub6.west);
\draw[connector] (sub4.west) |- ([xshift=-2mm]sub7.west);
\draw[connector] (sub4.west) |- ([xshift=-2mm]sub8.west);
\draw[connector] (sub4.west) |- ([xshift=-2mm]sub9.west);
\draw[connector] (sub1.west) |- ([xshift=-2mm]sub10.west);
\draw[connector] (cat1.west) |- ([xshift=-2mm]sub11.west);

\end{tikzpicture}

\clearpage

\newdimen\spacing
\spacing=1cm 

\begin{tikzpicture}[
  every node/.style={align=left,},
  category/.style={text width=6cm, align=left, inner sep=4pt, font=\small},
  root/.style={text width=10cm, align=left, inner sep=4pt},
  quote/.style={rounded corners, thick, text width=6cm, align=justify, inner sep=4pt, font=\footnotesize},
  arrow/.style={-{Stealth[length=3mm]}, thick},
  connector/.style={thick},
  spacing/.store in=\myspacing, spacing=1.0cm 
]

\node[root] (cat1) at (0, 0) {Ethical opportunities for care recipients (continued)};

\node[category, anchor=west] (sub1) at ($(cat1.south west)+(0.5,-\myspacing+0.5cm)$) {Improving/ensuring well-being/health (continued)};
\node[quote, anchor=west] (q1) at ($(sub1.east)+(1.5cm,0)$) {};
\node[draw=none, fit=(sub1)(q1), inner sep=0pt] (pair1) {};

\node[category, anchor=north west] (sub2) at ($(pair1.south west)+(0.5,-\myspacing+0.5cm)$){Addressing emotional well-being of care recipients (continued)};
\node[quote, anchor=west] (q2) at ($(sub2.east)+(1.5cm,0)$) {};
\node[draw=none, fit=(sub2)(q2), inner sep=0pt] (pair2) {};

\node[category, anchor=north west] (sub3) at ($(pair2.south west)+(0.5,-\myspacing)$) {Enhancing the mood and other positive neuropsychological effects};
\node[quote, anchor=west] (q3) at ($(sub3.east)+(0.5cm,0)$) {“In nursing homes in Japan, "Qoobo" was found to have positive effects on psychological well-being, such as patient satisfaction, happiness, and mood changes [34].” (Lehmann et al. 2021 Using a Socially Assistive Robot in a Nursing Home: Caregivers’ Expectations and Concerns, p. 150)
};
\draw[arrow] (sub3.east) -- (q3.west);
\node[draw=none, fit=(sub3)(q3), inner sep=0pt] (pair3) {};

\node[category, anchor=north west] (sub4) at ($(pair3.south west)+(-0.0,-\myspacing-0.5cm)$) {Reassurance/confirmation of worth by social robot};
\node[quote, anchor=west] (q4) at ($(sub4.east)+(0.5cm,0)$) {“The residents’ reassurance of worth could be strengthened by both direct and indirect social robots. Direct social robots could address people using their favorite ‘titles’ (whether from their previous profession or private life) and by their name (reassurance of worth) and ask questions about daily life in the old times or recent history, thus eliciting positive reminiscence.” (Pirhonen et al. 2020 Can robots tackle late-life loneliness? Scanning of future opportunities and challenges in assisted living facilities, p. 8)
};
\draw[arrow] (sub4.east) -- (q4.west);
\node[draw=none, fit=(sub4)(q4), inner sep=0pt] (pair4) {};

\node[category, anchor=north west] (sub5) at ($(pair4.south west)+(0.0,-\myspacing)$) {Sex robots can reduce shame/negative feelings (e. g. around sexuality)};
\node[quote, anchor=west] (q5) at ($(sub5.east)+(0.5cm,0)$) {“Sex robots could be used to practice sexual dysfunction without fear or shame […].” (Translation by the authors)
(Döring 2018 Sollten Pflegeroboter auch sexuelle Assistenzfunktionen bieten?, p. 260; in: Bendel 2018 Pflegeroboter)
};
\draw[arrow] (sub5.east) -- (q5.west);
\node[draw=none, fit=(sub5)(q5), inner sep=0pt] (pair5) {};

\node[category, anchor=north west] (sub6) at ($(pair5.south west)+(0.0,-\myspacing-0.5cm)$) {Social robots experienced as less pressurizing than care providers };
\node[quote, anchor=west] (q6) at ($(sub6.east)+(0.5cm,0)$) {“The situation was also unpleasant for the person in need of care, as they felt forced to eat. The calm and consistently monotonous nature of Care-O-bot's communication not only relieved the care staff of the repetitive and mentally stressful request to eat, but also made the patient feel less coerced.” (Translation by the authors) (Scorna 2015 Servicerobotik in der Altenpflege – Eine empirische Untersuchung des Einsatzes der Serviceroboter in der stationären Altenpflege am Beispiel von PARO und Care-O-bot, p. 91; in: Weber et al. 2015 Technisierung des Alltags – Beitrag für ein gutes Leben?)
};
\draw[arrow] (sub6.east) -- (q6.west);
\node[draw=none, fit=(sub6)(q6), inner sep=0pt] (pair6) {};

\node[category, anchor=north west] (sub7) at ($(pair6.south west)+(-1.0,-\myspacing-0.5cm)$) {Restoring privacy};
\node[quote, anchor=west] (q7) at ($(sub7.east)+(1.5cm,0)$) {“While they repeatedly acknowledge the importance of respecting older persons’ privacy, the AEG [Alzheimer Europe's Guidelines and Position on the Ethical Use of Assistive Technologies for/by People with Dementia] also mention a different view of the relationship between SARs and privacy, pointing out how relying on robot assistance may actually enhance privacy and not interfere with it.” (Battistuzzi et al. 2018 Embedding Ethics in the Design of Culturally Competent Socially Assistive Robots, p. 1999)
};
\draw[arrow] (sub7.east) -- (q7.west);
\node[draw=none, fit=(sub7)(q7), inner sep=0pt] (pair7) {};

\node[category, anchor=north west] (sub8) at ($(pair7.south west)+(0.5,-\myspacing)$) {Intimate situations have no longer to be shared with a person};
\node[quote, anchor=west] (q8) at ($(sub8.east)+(1cm,0)$) {“In the area of intimate care or in other very personal areas, support from machines is less invasive of privacy for some people than support from humans.” (Translation by the authors) (Bauberger 2020 Medizin und Pflege, p. 71; in: Bauberger 2020 Welche KI? Künstliche Intelligenz demokratisch gestalten) 
};
\draw[arrow] (sub8.east) -- (q8.west);
\node[draw=none, fit=(sub8)(q8), inner sep=0pt] (pair8) {};

\node[category, anchor=north west] (sub9) at ($(pair8.south west)+(+0.0,0)$) {\textit{Cont. on next page}};
\node[quote, anchor=west] (q9) at ($(sub9.east)+(0.5cm,0)$) {};
\node[draw=none, fit=(sub9)(q9), inner sep=0pt] (pair9) {};


\node[category, anchor=north west] (sub11) at ($(pair9.south west)+(-0.5,0)$) {\textit{Cont. on next page}};
\node[quote, anchor=west] (q11) at ($(sub11.east)+(1.5cm,0)$) {};
\node[draw=none, fit=(sub11)(q11), inner sep=0pt] (pair11) {};

\draw[connector] (cat1.west) |- ([xshift=-2mm]sub1.west);
\draw[connector] (sub1.west) |- ([xshift=-2mm]sub2.west);
\draw[connector] (sub2.west) |- ([xshift=-2mm]sub3.west);
\draw[connector] (sub2.west) |- ([xshift=-2mm]sub4.west);
\draw[connector] (sub2.west) |- ([xshift=-2mm]sub5.west);
\draw[connector] (sub2.west) |- ([xshift=-2mm]sub6.west);
\draw[connector] (cat1.west) |- ([xshift=-2mm]sub7.west);
\draw[connector] (sub7.west) |- ([xshift=-2mm]sub8.west);
\draw[connector] (sub7.west) |- ([xshift=-2mm]sub9.west);
\draw[connector] (cat1.west) |- ([xshift=-2mm]sub11.west);

\end{tikzpicture}

\clearpage

\newdimen\spacing
\spacing=1cm 

\begin{tikzpicture}[
  every node/.style={align=left,},
  category/.style={text width=6cm, align=left, inner sep=4pt, font=\small},
  root/.style={text width=10cm, align=left, inner sep=4pt},
  quote/.style={rounded corners, thick, text width=6cm, align=justify, inner sep=4pt, font=\footnotesize},
  arrow/.style={-{Stealth[length=3mm]}, thick},
  connector/.style={thick},
  spacing/.store in=\myspacing, spacing=1.0cm 
]

\node[root] (cat1) at (0, 0) {Ethical opportunities for care recipients (continued)};

\node[category, anchor=west] (sub1) at ($(cat1.south west)+(0.5,-\myspacing+0.5cm)$) {Restoring privacy (continued)};
\node[quote, anchor=west] (q1) at ($(sub1.east)+(1.5cm,0)$) {};
\node[draw=none, fit=(sub1)(q1), inner sep=0pt] (pair1) {};

\node[category, anchor=north west] (sub2) at ($(pair1.south west)+(0.5,-\myspacing)$){Care providers might be nosey/judgy, social robots are not};
\node[quote, anchor=west] (q2) at ($(sub2.east)+(1.0cm,0)$) {“There is a shared understanding of the potential normative implications of a carer’s seeing an unexpected person sharing the householder’s bed, overhearing a phone call to the betting office or alcohol retailer, or reading aloud a letter from a solicitor about changes to a will. If confronted by a wife or child asking questions about these events, a human servant/carer/companion has to make a normative judgement about the relative importance of infidelity, gambling, alcohol use and disinheriting a family member compared with some prior agreement to maintain confidentiality. For the robot there is no such tension. This could be regarded by some older people as a potential advantage of having robot as a carer. The robot is not nosey—it has no personal interest in finding certain things out, items of information are merely data.” (Draper and Sorell 2017 Ethical values and social care robots for older people: an international qualitative study, p. 64)
};
\draw[arrow] (sub2.east) -- (q2.west);
\node[draw=none, fit=(sub2)(q2), inner sep=0pt] (pair2) {};

\node[category, anchor=north west] (sub3) at ($(pair2.south west)+(-0.5,-\myspacing)$) {Promoting positive perception/preventing stigmatization/discrimination of care recipient};
\node[quote, anchor=west] (q3) at ($(sub3.east)+(1.5cm,0)$) {“Studies indicate that in some situations it may be less stigmatising for the older user to receive robotic care or help instead of human [34].” (Frennert and Östlund 2014 Review: Seven Matters of Concern of Social Robots and Older People, pp. 301, 302)
};
\draw[arrow] (sub3.east) -- (q3.west);
\node[draw=none, fit=(sub3)(q3), inner sep=0pt] (pair3) {};

\node[category, anchor=north west] (sub4) at ($(pair3.south west)+(0.5,-\myspacing-1cm)$) {Sex robots produce counter (non-ageist) narratives about sexuality};
\node[quote, anchor=west] (q4) at ($(sub4.east)+(1cm,0)$) {“One way this can occur is by generating counternarratives that include sex robots. Robots designed to integrate with the life narratives of older users may bear little resemblance to contemporary sex robots that dominate film and television shows, such as the young gynoids, Ave and Kyoko, of the 2014 film Ex Machina or those featured in the 1976 television series The Bionic Women and the Austin Powers film series. These images tilt towards users who are young, able-bodied, heterosexual and male. By contrast, sex robots designed for older people might resemble a deceased spouse in appearance or manner. For an older user, a desirable sex robot might evoke film stars of a bygone era, like Sidney Portier or Bette Davis.” (Jecker 2020 Nothing to be ashamed of: sex robots for older adults  with disabilities, p. 29)
};
\draw[arrow] (sub4.east) -- (q4.west);
\node[draw=none, fit=(sub4)(q4), inner sep=0pt] (pair4) {};

\node[category, anchor=north west] (sub5) at ($(pair4.south west)+(-0.5,-\myspacing)$) {Providing support in daily living (esp. where capacities reduced)};
\node[quote, anchor=west] (q5) at ($(sub5.east)+(1.5cm,0)$) {“The older person becomes something that needs to be governed and the robot assists the person in her everyday life [31]. Examples of robots that fit into this paradigm are Fraunhofer’s Care-O-Bot […]” (Frennert and Östlund 2014 Review: Seven Matters of Concern of Social Robots and Older People, p. 302)
};
\draw[arrow] (sub5.east) -- (q5.west);
\node[draw=none, fit=(sub5)(q5), inner sep=0pt] (pair5) {};

\node[category, anchor=north west] (sub6) at ($(pair5.south west)+(0.5,-\myspacing)$) {By acting as memory aid};
\node[quote, anchor=west] (q6) at ($(sub6.east)+(1cm,0)$) {“For example, Fraunhofer IPA’s Care-o-bot (now at its 4th generation, Care-O-bot 4), has been successfully tested to assist the specific memory deficits of older adults with dementia […].” (Ienca et al. 2016 Social and Assistive Robotics in Dementia Care: Ethical Recommendations for Research and Practice, p. 566)
};
\draw[arrow] (sub6.east) -- (q6.west);
\node[draw=none, fit=(sub6)(q6), inner sep=0pt] (pair6) {};

\node[category, anchor=north west] (sub7) at ($(pair6.south west)+(0.5,-\myspacing)$) {Reminding of important appointments/activities (e. g. taking medication)};
\node[quote, anchor=west] (q7) at ($(sub7.east)+(0.5cm,0)$) {“Cognitive assistance care robots may meet the need for senior citizens to live independently at home for a longer time. A robot can assist someone to remember appointments, to take medication or to eat on time […].” (Van Est et al. 2016 Robotisation as Rationalisation - In Search for a Human Robot Future, p. 45)
};
\draw[arrow] (sub7.east) -- (q7.west);
\node[draw=none, fit=(sub7)(q7), inner sep=0pt] (pair7) {};

\node[category, anchor=north west] (sub8) at ($(pair7.south west)+(+0.0,0)$) {\textit{Cont. on next page}};
\node[quote, anchor=west] (q8) at ($(sub8.east)+(0.5cm,0)$) {};
\node[draw=none, fit=(sub8)(q8), inner sep=0pt] (pair8) {};

\node[category, anchor=north west] (sub9) at ($(pair8.south west)+(-0.5,0)$) {\textit{Cont. on next page}};
\node[quote, anchor=west] (q9) at ($(sub9.east)+(1cm,0)$) {};
\node[draw=none, fit=(sub9)(q9), inner sep=0pt] (pair9) {};

\node[category, anchor=north west] (sub10) at ($(pair9.south west)+(-0.5,0)$) {\textit{Cont. on next page}};
\node[quote, anchor=west] (q10) at ($(sub10.east)+(1.5cm,0)$) {};
\node[draw=none, fit=(sub10)(q10), inner sep=0pt] (pair10) {};

\draw[connector] (cat1.west) |- ([xshift=-2mm]sub1.west);
\draw[connector] (sub1.west) |- ([xshift=-2mm]sub2.west);
\draw[connector] (cat1.west) |- ([xshift=-2mm]sub3.west);
\draw[connector] (sub3.west) |- ([xshift=-2mm]sub4.west);
\draw[connector] (cat1.west) |- ([xshift=-2mm]sub5.west);
\draw[connector] (sub5.west) |- ([xshift=-2mm]sub6.west);
\draw[connector] (sub6.west) |- ([xshift=-2mm]sub7.west);
\draw[connector] (sub6.west) |- ([xshift=-2mm]sub8.west);
\draw[connector] (sub5.west) |- ([xshift=-2mm]sub9.west);
\draw[connector] (cat1.west) |- ([xshift=-2mm]sub10.west);

\end{tikzpicture}

\clearpage

\newdimen\spacing
\spacing=1cm 

\begin{tikzpicture}[
  every node/.style={align=left,},
  category/.style={text width=6cm, align=left, inner sep=4pt, font=\small},
  root/.style={text width=10cm, align=left, inner sep=4pt},
  quote/.style={rounded corners, thick, text width=6cm, align=justify, inner sep=4pt, font=\footnotesize},
  arrow/.style={-{Stealth[length=3mm]}, thick},
  connector/.style={thick},
  spacing/.store in=\myspacing, spacing=1.0cm 
]

\node[root] (cat1) at (0, 0) {Ethical opportunities for care recipients (continued)};

\node[category, anchor=west] (sub1) at ($(cat1.south west)+(0.5,-\myspacing+0.5cm)$) {Providing support in daily living (esp. where capacities reduced) (continued)};
\node[quote, anchor=west] (q1) at ($(sub1.east)+(1.5cm,0)$) {};
\node[draw=none, fit=(sub1)(q1), inner sep=0pt] (pair1) {};

\node[category, anchor=north west] (sub2) at ($(pair1.south west)+(0.5,-\myspacing+0.5cm)$){By acting as memory aid (continued)};
\node[quote, anchor=west] (q2) at ($(sub2.east)+(1.0cm,0)$) {
};
\node[draw=none, fit=(sub2)(q2), inner sep=0pt] (pair2) {};

\node[category, anchor=north west] (sub3) at ($(pair2.south west)+(0.5,-\myspacing)$) {Reminding how things get done};
\node[quote, anchor=west] (q3) at ($(sub3.east)+(0.5cm,0)$) {“Cognitive assistance care robots may meet the need for senior citizens to live independently at home for a longer time. A robot can assist someone to remember appointments, to take medication or to eat on time […].” (Van Est et al. 2016 Robotisation as Rationalisation - In Search for a Human Robot Future, p. 45)
};
\draw[arrow] (sub3.east) -- (q3.west);
\node[draw=none, fit=(sub3)(q3), inner sep=0pt] (pair3) {};

\node[category, anchor=north west] (sub4) at ($(pair3.south west)+(0.0,-\myspacing)$) {Reminding of further relevant aspects of daily living};
\node[quote, anchor=west] (q4) at ($(sub4.east)+(0.5cm,0)$) {“Small social robots that help you learn a foreign language, remind you of medication, save birthdays or shopping lists are unusual but relatively harmless in the eyes of the test persons.” (Translation by the authors) (Meyer 2011 Akzeptanzbedingungen für Roboter-Assistenten, p. 106; in: Meyer 2011 Mein Freund der Roboter - Servicerobotik für ältere Menschen - eine Antwort auf den demographischen Wandel?)
};
\draw[arrow] (sub4.east) -- (q4.west);
\node[draw=none, fit=(sub4)(q4), inner sep=0pt] (pair4) {};

\node[category, anchor=north west] (sub5) at ($(pair4.south west)+(-0.5,-\myspacing-0.5cm)$) {By carrying out certain tasks for/with the care recipient};
\node[quote, anchor=west] (q5) at ($(sub5.east)+(1cm,0)$) {“It would be ideal if the user and the robot could perform tasks together. When performing a task together, the robot would only need to perform that part of the task the user cannot perform. For example, when the user is still able to make a cup of coffee but unable to carry the cup of coffee to the living room, the robot should do that latter part of the task for the user.” (Bedaf et al. 2019 What are the preferred characteristics of a service robot for the elderly? A multi-country focus group study with older adults and caregivers, p. 8)
};
\draw[arrow] (sub5.east) -- (q5.west);
\node[draw=none, fit=(sub5)(q5), inner sep=0pt] (pair5) {};

\node[category, anchor=north west] (sub6) at ($(pair5.south west)+(0.5,-\myspacing-0.5cm)$) {Supporting eating/feeding};
\node[quote, anchor=west] (q6) at ($(sub6.east)+(0.5cm,0)$) {“Other ‘class 2 medical robots’ are assistive robots (typically mobile robots and mobile manipulators) that help impaired people to everyday functions, such as eating, drinking, washing, etc. or for hospital operations. Examples of such robots are: the ‘Care-0-Bot 3’ robot […].” (Tzafestas 2016 Socialised Roboethics, p. 110; in: Tzafestas 2016 Roboethics - A Navigating Overview)
};
\draw[arrow] (sub6.east) -- (q6.west);
\node[draw=none, fit=(sub6)(q6), inner sep=0pt] (pair6) {};

\node[category, anchor=north west] (sub7) at ($(pair6.south west)+(0.0,-\myspacing-0.5cm)$) {Supporting having sex};
\node[quote, anchor=west] (q7) at ($(sub7.east)+(0.5cm,0)$) {“I am simply arguing that sexbots are a plausible solution that could help residents who still have sexual drives but are currently unable to find a suitable outlet for them.” (Lancaster 2023 Granny and the Sexbots: An ethical appraisal of the use of sexbots in residential care institutions for elderly people, p. 200; in: Loh and Loh 2023 Social Robotics and the Good Life - The Normative Side of Forming Emotional Bonds With Robots)
};
\draw[arrow] (sub7.east) -- (q7.west);
\node[draw=none, fit=(sub7)(q7), inner sep=0pt] (pair7) {};

\node[category, anchor=north west] (sub8) at ($(pair7.south west)+(0.0,-\myspacing)$) {Supporting with getting up and moving around};
\node[quote, anchor=west] (q8) at ($(sub8.east)+(0.5cm,0)$) {“However, general support in moving around the residence was viewed more positively. A robot which was remotely operated and monitored by care professionals could be used in the homes of older people as an assistant for moving, walking and getting up.” (Parviainen et al. 2019 Social Robots and Human Touch in Care: The Perceived Usefulness of Robot Assistance among Healthcare Professionals, p. 10) 
};
\draw[arrow] (sub8.east) -- (q8.west);
\node[draw=none, fit=(sub8)(q8), inner sep=0pt] (pair8) {};

\node[category, anchor=north west] (sub9) at ($(pair8.south west)+(+0.0,0)$) {\textit{Cont. on next page}};
\node[quote, anchor=west] (q9) at ($(sub9.east)+(0.5cm,0)$) {};
\node[draw=none, fit=(sub9)(q9), inner sep=0pt] (pair9) {};

\node[category, anchor=north west] (sub10) at ($(pair9.south west)+(-0.5,0)$) {\textit{Cont. on next page}};
\node[quote, anchor=west] (q10) at ($(sub10.east)+(1cm,0)$) {};
\node[draw=none, fit=(sub10)(q10), inner sep=0pt] (pair10) {};

\node[category, anchor=north west] (sub11) at ($(pair10.south west)+(-0.5,0)$) {\textit{Cont. on next page}};
\node[quote, anchor=west] (q11) at ($(sub11.east)+(1.5cm,0)$) {};
\node[draw=none, fit=(sub11)(q11), inner sep=0pt] (pair11) {};

\draw[connector] (cat1.west) |- ([xshift=-2mm]sub1.west);
\draw[connector] (sub1.west) |- ([xshift=-2mm]sub2.west);
\draw[connector] (sub2.west) |- ([xshift=-2mm]sub3.west);
\draw[connector] (sub2.west) |- ([xshift=-2mm]sub4.west);
\draw[connector] (sub1.west) |- ([xshift=-2mm]sub5.west);
\draw[connector] (sub5.west) |- ([xshift=-2mm]sub6.west);
\draw[connector] (sub5.west) |- ([xshift=-2mm]sub7.west);
\draw[connector] (sub5.west) |- ([xshift=-2mm]sub8.west);
\draw[connector] (sub5.west) |- ([xshift=-2mm]sub9.west);
\draw[connector] (sub1.west) |- ([xshift=-2mm]sub10.west);
\draw[connector] (cat1.west) |- ([xshift=-2mm]sub11.west);

\end{tikzpicture}

\clearpage

\newdimen\spacing
\spacing=1cm 

\begin{tikzpicture}[
  every node/.style={align=left,},
  category/.style={text width=6cm, align=left, inner sep=4pt, font=\small},
  root/.style={text width=10cm, align=left, inner sep=4pt},
  quote/.style={rounded corners, thick, text width=6cm, align=justify, inner sep=4pt, font=\footnotesize},
  arrow/.style={-{Stealth[length=3mm]}, thick},
  connector/.style={thick},
  spacing/.store in=\myspacing, spacing=1.0cm 
]

\node[root] (cat1) at (0, 0) {Ethical opportunities for care recipients (continued)};

\node[category, anchor=west] (sub1) at ($(cat1.south west)+(0.5,-\myspacing+0.5cm)$) {Providing support in daily living (esp. where capacities reduced) (continued)};
\node[quote, anchor=west] (q1) at ($(sub1.east)+(1.5cm,0)$) {};
\node[draw=none, fit=(sub1)(q1), inner sep=0pt] (pair1) {};

\node[category, anchor=north west] (sub2) at ($(pair1.south west)+(0.5,-\myspacing+0.5cm)$){By carrying out certain tasks for/with the care recipient (continued)};
\node[quote, anchor=west] (q2) at ($(sub2.east)+(1.0cm,0)$) {
};
\node[draw=none, fit=(sub2)(q2), inner sep=0pt] (pair2) {};

\node[category, anchor=north west] (sub3) at ($(pair2.south west)+(0.5,-\myspacing)$) {Supporting with doing household chores};
\node[quote, anchor=west] (q3) at ($(sub3.east)+(0.5cm,0)$) {“It has been demonstrated that most older people could accept robots as functional helpers in the home doing household chores […].” (Frennert and Östlund 2014 Review: Seven Matters of Concern of Social Robots and Older People, p. 304)
};
\draw[arrow] (sub3.east) -- (q3.west);
\node[draw=none, fit=(sub3)(q3), inner sep=0pt] (pair3) {};

\node[category, anchor=north west] (sub4) at ($(pair3.south west)+(0.0,-\myspacing-1cm)$) {Accompanying to/support with appointments and shopping};
\node[quote, anchor=west] (q4) at ($(sub4.east)+(0.5cm,0)$) {“The idea of using service robots as shopping guides is aimed at the problems associated with shopping. The basic idea is that the service robot waits in the entrance area of a shopping center, asks the customer what they want and then guides them to the goods they are looking for. If the barcode of the product is held under the built-in scanner, it can also answer questions about the prices of the desired items or the ingredients of the food. If desired, the shopping assistant can guide the customer to the checkout.” (Translation by the authors) (Kapitel Meyer 2011 Akzeptanz ausgewählter Anwendungsszenarien, p. 72; in: Meyer 2011 Mein Freund der Roboter - Servicerobotik für ältere Menschen - eine Antwort auf den demographischen Wandel?)
};
\draw[arrow] (sub4.east) -- (q4.west);
\node[draw=none, fit=(sub4)(q4), inner sep=0pt] (pair4) {};

\node[category, anchor=north west] (sub5) at ($(pair4.south west)+(-0.0,-\myspacing)$) {Getting relevant things for care recipients};
\node[quote, anchor=west] (q5) at ($(sub5.east)+(0.5cm,0)$) {“The third generation of the robot, Care-O-bot III, was already able to move independently and collision-free between people and provide them with household items or drinks of their choice, among other things.” (Translation by the authors) (Kreis 2018 Umsorgen, überwachen, unterhalten - sind Pflegeroboter ethisch vertretbar? p. 218; in: Bendel 2018 Pflegeroboter)
};
\draw[arrow] (sub5.east) -- (q5.west);
\node[draw=none, fit=(sub5)(q5), inner sep=0pt] (pair5) {};

\node[category, anchor=north west] (sub6) at ($(pair5.south west)+(0.0,-\myspacing)$) {Providing relevant public information (e. g. weather)  };
\node[quote, anchor=west] (q6) at ($(sub6.east)+(0.5cm,0)$) {“Lio and P-Care can answer questions about the weather or the latest news and thus serve as conversation partners for the residents.” (Translation by the authors) (Früh and Gasser 2018 Erfahrungen aus dem Einsatz von Pflegerobotern für Menschen im Alter, p. 39; in: Bendel 2018 Pflegeroboter)
};
\draw[arrow] (sub6.east) -- (q6.west);
\node[draw=none, fit=(sub6)(q6), inner sep=0pt] (pair6) {};

\node[category, anchor=north west] (sub7) at ($(pair6.south west)+(0.0,-\myspacing-0.5cm)$) {Carrying out further tasks (e. g. guarding the home against intruders)};
\node[quote, anchor=west] (q7) at ($(sub7.east)+(0.5cm,0)$) {“Aido, a robot from the company InGen Dynamics (InGen Dynamics n.d.), can move around the house and is primarily intended for interaction with people (Aido User Manual n.d.). On the one hand, it has various entertainment functions, but can also be used to monitoring the house or as a personal assistant.” (Translation by the authors) (Bleuler and Caroni 2021 Roboter in der Pflege - Welche Aufgaben können Roboter heute schon übernehmen?, pp. 445, 446; in: Bendel 2021 Soziale Roboter)
};
\draw[arrow] (sub7.east) -- (q7.west);
\node[draw=none, fit=(sub7)(q7), inner sep=0pt] (pair7) {};

\node[category, anchor=north west] (sub8) at ($(pair7.south west)+(-1.0,-\myspacing)$) {Sustaining independent and established lifestyle };
\node[quote, anchor=west] (q8) at ($(sub8.east)+(1.5cm,0)$) {“On the other hand, robots that remind people to take their medication or monitor them, for example, are designed precisely for cognitively impaired people so that they can live more independently despite their limitations.” (Translation by the authors) (Bleuler and Caroni 2021 Roboter in der Pflege - Welche Aufgaben können Roboter heute schon übernehmen?, p. 444; in: Bendel 2021 Soziale Roboter) 
};
\draw[arrow] (sub8.east) -- (q8.west);
\node[draw=none, fit=(sub8)(q8), inner sep=0pt] (pair8) {};

\node[category, anchor=north west] (sub9) at ($(pair8.south west)+(0.5,0)$) {\textit{Cont. on next page}};
\node[quote, anchor=west] (q9) at ($(sub9.east)+(1.cm,0)$) {};
\node[draw=none, fit=(sub9)(q9), inner sep=0pt] (pair9) {};

\node[category, anchor=north west] (sub10) at ($(pair9.south west)+(-0.5,0)$) {\textit{Cont. on next page}};
\node[quote, anchor=west] (q10) at ($(sub10.east)+(1.5cm,0)$) {};
\node[draw=none, fit=(sub10)(q10), inner sep=0pt] (pair10) {};

\draw[connector] (cat1.west) |- ([xshift=-2mm]sub1.west);
\draw[connector] (sub1.west) |- ([xshift=-2mm]sub2.west);
\draw[connector] (sub2.west) |- ([xshift=-2mm]sub3.west);
\draw[connector] (sub2.west) |- ([xshift=-2mm]sub4.west);
\draw[connector] (sub2.west) |- ([xshift=-2mm]sub5.west);
\draw[connector] (sub2.west) |- ([xshift=-2mm]sub6.west);
\draw[connector] (sub2.west) |- ([xshift=-2mm]sub7.west);
\draw[connector] (cat1.west) |- ([xshift=-2mm]sub8.west);
\draw[connector] (sub8.west) |- ([xshift=-2mm]sub9.west);
\draw[connector] (cat1.west) |- ([xshift=-2mm]sub10.west);

\end{tikzpicture}

\clearpage

\newdimen\spacing
\spacing=1cm 

\begin{tikzpicture}[
  every node/.style={align=left,},
  category/.style={text width=6cm, align=left, inner sep=4pt, font=\small},
  root/.style={text width=10cm, align=left, inner sep=4pt},
  quote/.style={rounded corners, thick, text width=6cm, align=justify, inner sep=4pt, font=\footnotesize},
  arrow/.style={-{Stealth[length=3mm]}, thick},
  connector/.style={thick},
  spacing/.store in=\myspacing, spacing=1.0cm 
]

\node[root] (cat1) at (0, 0) {Ethical opportunities for care recipients (continued)};

\node[category, anchor=west] (sub1) at ($(cat1.south west)+(0.5,-\myspacing+0.5cm)$) {Sustaining independent and established lifestyle ) (continued)};
\node[quote, anchor=west] (q1) at ($(sub1.east)+(1.5cm,0)$) {};
\node[draw=none, fit=(sub1)(q1), inner sep=0pt] (pair1) {};

\node[category, anchor=west] (sub2) at ($(pair1.south west)+(0.5,-\myspacing)$) {Allow sustaining established routines/relations of daily living};
\node[quote, anchor=west] (q2) at ($(sub2.east)+(1.cm,0)$) {“They are potentially important not only in relieving loneliness but in helping their users maintain a normal routine in the face of frailty […].” (Draper and Sorell 2017 Ethical values and social care robots for older people: an international qualitative study, p. 50)};
\draw[arrow] (sub2.east) -- (q2.west);
\node[draw=none, fit=(sub2)(q2), inner sep=0pt] (pair2) {};

\node[category, anchor=west] (sub3) at ($(pair2.south west)+(0.,-\myspacing-1cm)$) {Remaining in own home for longer};
\node[quote, anchor=west] (q3) at ($(sub3.east)+(1.cm,0)$) {“To avoid moving to a nursing home or other care facility, Margaret has purchased a home-care robot that can: remind her to take her medications; fetch items for her if she is too tired or is already in bed; help with simple cleaning tasks; and, can facilitate her staying in contact with her family, friends and healthcare provider via video chat.”(van Wynsberghe 2015 Designing Care Robots with Care, p. 9; in: van Wynsberghe 2015 Healthcare Robots - Ethics, Design and Implementation) };
\draw[arrow] (sub3.east) -- (q3.west);
\node[draw=none, fit=(sub3)(q3), inner sep=0pt] (pair3) {};

\node[category, anchor=west] (sub4) at ($(pair3.south west)+(0.,-\myspacing-0.5cm)$) {Reducing dependencies on other people};
\node[quote, anchor=west] (q4) at ($(sub4.east)+(1.cm,0)$) {“The option to rely on assistive technology systems instead of being dependent on human caregivers can thus prima facie be understood as a valuable option of gaining a degree of independence.”(Wiertz 2020 Trusting robots? – On the concept of trust and on forms of human vulnerability, p. 53; in: Haltaufderheide et al. 2020 Aging between Participation and Simulation - Ethical Dimensions of Social Assistive Technologies)};
\draw[arrow] (sub4.east) -- (q4.west);
\node[draw=none, fit=(sub4)(q4), inner sep=0pt] (pair4) {};

\node[category, anchor=west] (sub5) at ($(pair4.south west)+(0.,-\myspacing)$) {Social robots are flexible and can help at any time (e. g. at night)};
\node[quote, anchor=west] (q5) at ($(sub5.east)+(1.cm,0)$) {“The robots can also be called around the clock to provide assistance to people in the middle of the night or early in the morning.” (Translation by the authors)(Früh and Gasser 2018 Erfahrungen aus dem Einsatz von Pflegerobotern für Menschen im Alter, p. 45; in: Bendel 2018 Pflegeroboter)};
\draw[arrow] (sub5.east) -- (q5.west);
\node[draw=none, fit=(sub5)(q5), inner sep=0pt] (pair5) {};

\node[category, anchor=west] (sub6) at ($(pair5.south west)+(-0.5,-\myspacing-1cm)$) {Furthering social connection};
\node[quote, anchor=west] (q6) at ($(sub6.east)+(1.5cm,0)$) {“Social networks change over the lifespan, with the average size of private networks already decreasing from a maximum in the fourth decade of life [23]. In order to possibly compensate for the impending loss of resources, there could be an increasing number of technical possibilities offered by social robotics.” (Translation by the authors) (Zöllick et al. 2022 Potenziale und Herausforderungen von sozialen Robotern für Beziehungen älterer Menschen: eine Bestandsaufnahme mittels 'rapid review'", p. 299) };
\draw[arrow] (sub6.east) -- (q6.west);
\node[draw=none, fit=(sub6)(q6), inner sep=0pt] (pair6) {};

\node[category, anchor=west] (sub7) at ($(pair6.south west)+(0.5,-\myspacing)$) {Social robots offer opportunities for social interaction with them};
\node[quote, anchor=west] (q7) at ($(sub7.east)+(1.cm,0)$) {“The performance of social or socio-emotional robots initially consists less in performing certain activities than in entering into social interaction with people.” (Translation by the authors) (Manzeschke 2019 Roboter in der Pflege - Von Menschen, Maschinen und anderen hilfreichen Wesen, p. 4)};
\draw[arrow] (sub7.east) -- (q7.west);
\node[draw=none, fit=(sub7)(q7), inner sep=0pt] (pair7) {};

\node[category, anchor=north west] (sub8) at ($(pair7.south west)+(0.5,0)$) {\textit{Cont. on next page}};
\node[quote, anchor=west] (q8) at ($(sub8.east)+(1.cm,0)$) {};
\node[draw=none, fit=(sub8)(q8), inner sep=0pt] (pair8) {};

\node[category, anchor=north west] (sub9) at ($(pair8.south west)+(-0.5,0)$) {\textit{Cont. on next page}};
\node[quote, anchor=west] (q9) at ($(sub9.east)+(1.cm,0)$) {};
\node[draw=none, fit=(sub9)(q9), inner sep=0pt] (pair9) {};

\node[category, anchor=north west] (sub10) at ($(pair9.south west)+(-0.5,0)$) {\textit{Cont. on next page}};
\node[quote, anchor=west] (q10) at ($(sub10.east)+(1.5cm,0)$) {};
\node[draw=none, fit=(sub10)(q10), inner sep=0pt] (pair10) {};

\draw[connector] (cat1.west) |- ([xshift=-2mm]sub1.west);
\draw[connector] (sub1.west) |- ([xshift=-2mm]sub2.west);
\draw[connector] (sub1.west) |- ([xshift=-2mm]sub3.west);
\draw[connector] (sub1.west) |- ([xshift=-2mm]sub4.west);
\draw[connector] (sub1.west) |- ([xshift=-2mm]sub5.west);
\draw[connector] (cat1.west) |- ([xshift=-2mm]sub6.west);
\draw[connector] (sub6.west) |- ([xshift=-2mm]sub7.west);
\draw[connector] (sub7.west) |- ([xshift=-2mm]sub8.west);
\draw[connector] (sub6.west) |- ([xshift=-2mm]sub9.west);
\draw[connector] (cat1.west) |- ([xshift=-2mm]sub10.west);
\end{tikzpicture}

\clearpage

\newdimen\spacing
\spacing=1cm 

\begin{tikzpicture}[
  every node/.style={align=left,},
  category/.style={text width=6cm, align=left, inner sep=4pt, font=\small},
  root/.style={text width=10cm, align=left, inner sep=4pt},
  quote/.style={rounded corners, thick, text width=6cm, align=justify, inner sep=4pt, font=\footnotesize},
  arrow/.style={-{Stealth[length=3mm]}, thick},
  connector/.style={thick},
  spacing/.store in=\myspacing, spacing=1.0cm 
]

\node[root] (cat1) at (0, 0) {Ethical opportunities for care recipients (continued)};

\node[category, anchor=west] (sub1) at ($(cat1.south west)+(0.5,-\myspacing+0.5cm)$) {Furthering social connection (continued)};
\node[quote, anchor=west] (q1) at ($(sub1.east)+(1.5cm,0)$) {};
\node[draw=none, fit=(sub1)(q1), inner sep=0pt] (pair1) {};

\node[category, anchor=west] (sub2) at ($(sub1.south west)+(0.5,-\myspacing+0.5cm)$) {Social robots offer opportunities for social interaction with them (continued)};
\node[quote, anchor=west] (q2) at ($(sub2.east)+(1cm,0)$) {};
\node[draw=none, fit=(sub2)(q2), inner sep=0pt] (pair2) {};

\node[category, anchor=west] (sub3) at ($(pair2.south west)+(0.5,-\myspacing)$) {Social robots as companions};
\node[quote, anchor=west] (q3) at ($(sub3.east)+(0.5cm,0)$) {“During a pandemic emergency in particular, the alternative to robot companionship for many older people is social isolation and loneliness. Without support, older adults are left to languish. Under these conditions, sociable robots do not rob older adults of human companionship but afford companionship where it is lacking.”(Jecker 2021 You’ve got a friend in me: sociable robots for older adults in an age of global pandemics, p. S39)};
\draw[arrow] (sub3.east) -- (q3.west);
\node[draw=none, fit=(sub3)(q3), inner sep=0pt] (pair3) {};

\node[category, anchor=west] (sub4) at ($(pair3.south west)+(0.,-\myspacing-0.5cm)$) {Social robots as entertainers};
\node[quote, anchor=west] (q4) at ($(sub4.east)+(0.5cm,0)$) {“Various functions are used to entertain and activate residents. The assistance robot cheers them up with jokes and short stories.” (Translation by the authors)(Wirth et al. 2020 Erfahrungen aus dem Einsatz von Assistenzrobotern für Menschen im Alter, p. 261; in: Buxbaum 2020 Mensch-Roboter-Kollaboration) };
\draw[arrow] (sub4.east) -- (q4.west);
\node[draw=none, fit=(sub4)(q4), inner sep=0pt] (pair4) {};

\node[category, anchor=west] (sub5) at ($(pair4.south west)+(0.,-\myspacing)$) {Sex robots as (safe) sexual contacts };
\node[quote, anchor=west] (q5) at ($(sub5.east)+(0.5cm,0)$) {“[… ] the advantage of sex robots is that they can meet sexual needs without violating the rights of any other person.”(Jecker 2021 Sex robots for older adults with disabilities: reply to critics, p. 1)};
\draw[arrow] (sub5.east) -- (q5.west);
\node[draw=none, fit=(sub5)(q5), inner sep=0pt] (pair5) {};

\node[category, anchor=west] (sub6) at ($(pair5.south west)+(-0.5,-\myspacing-1cm)$) {Social robots facilitate interaction with other human beings};
\node[quote, anchor=west] (q6) at ($(sub6.east)+(1.cm,0)$) {“Nevertheless, the developments in robotics - if viewed within these limits - give rise to justified hope for an improvement in the care situation. Such systems can help to maintain communication and the ability of people to meet in situations where other options are not available [...].” (Translation of the authors)(Gräb-Schmidt and Stritzelberger 2018 Ethische Herausforderungen durch autonome Systeme und Robotik im Bereich Pflege, p. 368)};
\draw[arrow] (sub6.east) -- (q6.west);
\node[draw=none, fit=(sub6)(q6), inner sep=0pt] (pair6) {};

\node[category, anchor=west] (sub7) at ($(pair6.south west)+(0.5,-\myspacing-2cm)$) {By translating if care recipient and other actors don’t speak same language};
\node[quote, anchor=west] (q7) at ($(sub7.east)+(0.5cm,0)$) {“Due to progressive dementia, Adam, who is in need of care, speaks almost exclusively his native language and does not even realize that he is doing so. This means that the care staff can no longer understand the person in need of care. The robot recognizes such situations through speech recognition, which is supported by artificial intelligence. It then translates into the other language in such situations. In this way, the person in need of care knows that the meal is ready and can let the carer know when he is thirsty. It also gives the senior the opportunity to continue speaking his native language.” (Translation by the authors)(Carros et al. 2022 Roboter als intelligente Assistenten in Betreuung und Pflege – Grenzen und Perspektiven im Praxiseinsatz, p. 808; in: Pfannstiel 2022 Künstliche Intelligenz im Gesundheitswesen)};
\draw[arrow] (sub7.east) -- (q7.west);
\node[draw=none, fit=(sub7)(q7), inner sep=0pt] (pair7) {};

\node[category, anchor=north west] (sub8) at ($(pair7.south west)+(0.5,0)$) {\textit{Cont. on next page}};
\node[quote, anchor=west] (q8) at ($(sub8.east)+(1.cm,0)$) {};
\node[draw=none, fit=(sub8)(q8), inner sep=0pt] (pair8) {};

\node[category, anchor=north west] (sub9) at ($(pair8.south west)+(-0.5,0)$) {\textit{Cont. on next page}};
\node[quote, anchor=west] (q9) at ($(sub9.east)+(1.cm,0)$) {};
\node[draw=none, fit=(sub9)(q9), inner sep=0pt] (pair9) {};

\node[category, anchor=north west] (sub10) at ($(pair9.south west)+(-0.5,0)$) {\textit{Cont. on next page}};
\node[quote, anchor=west] (q10) at ($(sub10.east)+(1.5cm,0)$) {};
\node[draw=none, fit=(sub10)(q10), inner sep=0pt] (pair10) {};

\draw[connector] (cat1.west) |- ([xshift=-2mm]sub1.west);
\draw[connector] (sub1.west) |- ([xshift=-2mm]sub2.west);
\draw[connector] (sub2.west) |- ([xshift=-2mm]sub3.west);
\draw[connector] (sub2.west) |- ([xshift=-2mm]sub4.west);
\draw[connector] (sub2.west) |- ([xshift=-2mm]sub5.west);
\draw[connector] (sub1.west) |- ([xshift=-2mm]sub6.west);
\draw[connector] (sub6.west) |- ([xshift=-2mm]sub7.west);
\draw[connector] (sub7.west) |- ([xshift=-2mm]sub8.west);
\draw[connector] (sub6.west) |- ([xshift=-2mm]sub9.west);
\draw[connector] (cat1.west) |- ([xshift=-2mm]sub10.west);

\end{tikzpicture}

\clearpage

\newdimen\spacing
\spacing=1cm 

\begin{tikzpicture}[
  every node/.style={align=left,},
  category/.style={text width=6cm, align=left, inner sep=4pt, font=\small},
  root/.style={text width=10cm, align=left, inner sep=4pt},
  quote/.style={rounded corners, thick, text width=6cm, align=justify, inner sep=4pt, font=\footnotesize},
  arrow/.style={-{Stealth[length=3mm]}, thick},
  connector/.style={thick},
  spacing/.store in=\myspacing, spacing=1.0cm 
]

\node[root] (cat1) at (0, 0) {Ethical opportunities for care recipients (continued)};

\node[category, anchor=west] (sub1) at ($(cat1.south west)+(0.5,-\myspacing+0.5cm)$) {Furthering social connection (continued)};
\node[quote, anchor=west] (q1) at ($(sub1.east)+(1.5cm,0)$) {};
\node[draw=none, fit=(sub1)(q1), inner sep=0pt] (pair1) {};

\node[category, anchor=west] (sub2) at ($(sub1.south west)+(0.5,-\myspacing+0.5cm)$) {Social robots facilitate interaction with other human beings (continued)};
\node[quote, anchor=west] (q2) at ($(sub2.east)+(1cm,0)$) {};
\node[draw=none, fit=(sub2)(q2), inner sep=0pt] (pair2) {};

\node[category, anchor=west] (sub3) at ($(pair2.south west)+(0.5,-\myspacing)$) {Care providers have more time for social interaction};
\node[quote, anchor=west] (q3) at ($(sub3.east)+(0.5cm,0)$) {“Due to the assumption of various smaller tasks, the interviewees assume that robots will relieve the nursing staff in their everyday work and thus give them more freedom. This will benefit nursing home residents, who will be able to have other activities and conversations with nursing and care staff or benefit from less stressed staff (Lipp \& Massen, 2019).” (Translation by the authors) (Carros et al. 2022 Roboter als intelligente Assistenten in Betreuung und Pflege – Grenzen und Perspektiven im Praxiseinsatz, pp. 811, 812; in: Pfannstiel 2022 Künstliche Intelligenz im Gesundheitswesen)};
\draw[arrow] (sub3.east) -- (q3.west);
\node[draw=none, fit=(sub3)(q3), inner sep=0pt] (pair3) {};

\node[category, anchor=west] (sub4) at ($(pair3.south west)+(0.,-\myspacing-0.5cm)$) {Providing something to talk about (communication facilitator)};
\node[quote, anchor=west] (q4) at ($(sub4.east)+(0.5cm,0)$) {“On the one hand, social robots can promote communication with them by initiating conversations. On the other hand, when used in a therapeutic group setting, they promote interpersonal communication about them, which in turn can strengthen social integration [20].”(Zöllick et al. 2022 Potenziale und Herausforderungen von sozialen Robotern für Beziehungen älterer Menschen: eine Bestandsaufnahme mittels "rapid review", pp. 300, 302) };
\draw[arrow] (sub4.east) -- (q4.west);
\node[draw=none, fit=(sub4)(q4), inner sep=0pt] (pair4) {};

\node[category, anchor=west] (sub5) at ($(pair4.south west)+(0.,-\myspacing-0.5cm)$) {Providing information on chances for interaction (e. g. events)};
\node[quote, anchor=west] (q5) at ($(sub5.east)+(0.5cm,0)$) {“Frank’s daughter has suggested that the Care-O-bot should be used to encourage Frank join a virtual fishing forum on the internet. She is worried that he only has one friend who is older and poorly and may die leaving Frank with no one else to talk to about fishing.” (Draper and Sorell 2017 Ethical values and social care robots for older people: an international qualitative study, p. 53)};
\draw[arrow] (sub5.east) -- (q5.west);
\node[draw=none, fit=(sub5)(q5), inner sep=0pt] (pair5) {};

\node[category, anchor=west] (sub6) at ($(pair5.south west)+(0.,-\myspacing)$) {Attracting visitors (particularly grandchildren)};
\node[quote, anchor=west] (q6) at ($(sub6.east)+(0.5cm,0)$) {“A robot that facilitates conversation may function as an attractor for visitors.” (Sharkey and Sharkey 2012 Granny and the robots: ethical issues in robot care for the elderly, p. 35)};
\node[draw=none, fit=(sub6)(q6), inner sep=0pt] (pair6) {};

\node[category, anchor=west] (sub7) at ($(pair6.south west)+(0.,-\myspacing-1cm)$) {Promoting positive/friendly communication};
\node[quote, anchor=west] (q7) at ($(sub7.east)+(0.5cm,0)$) {“For instance, when Wada and Shibata (2006) videoed interactions between a Paro robot seal and a group of elderly care home residents, they found evidence that the social interactions between the residents themselves increased, at the same time that physiological indicators showed reduced stress levels. It seems that Paro even encouraged positive communication, and resulted in a reduction of the ‘‘backbiting’’ that had previously characterised some of their interactions.” (Sharkey and Sharkey 2012 Granny and the robots: ethical issues in robot care for the elderly, p. 35)};
\draw[arrow] (sub7.east) -- (q7.west);
\node[draw=none, fit=(sub7)(q7), inner sep=0pt] (pair7) {};

\node[category, anchor=west] (sub8) at ($(pair7.south west)+(0.,-\myspacing-1cm)$) {Allowing easy exchange via virtual interfaces};
\node[quote, anchor=west] (q8) at ($(sub8.east)+(0.5cm,0)$) {“Telepresence is a potential technology to alleviate loneliness by allowing friends and family members to be virtually present for the elderly person (e.g. [4]). […] For elderly persons, robotic telepresence provides benefits compared to non-mobile video connection [7]: because of the remote control of the robot, the elderly user can interact with it in a natural manner, with little additional learning.”(Niemelä et al. 2021 Towards Ethical Guidelines of Using Telepresence Robots in Residential Care, p. 431)};
\draw[arrow] (sub8.east) -- (q8.west);
\node[draw=none, fit=(sub8)(q8), inner sep=0pt] (pair8) {};

\node[category, anchor=north west] (sub9) at ($(pair8.south west)+(-1.,0)$) {\textit{Cont. on next page}};
\node[quote, anchor=west] (q9) at ($(sub9.east)+(1.cm,0)$) {};
\node[draw=none, fit=(sub9)(q9), inner sep=0pt] (pair9) {};

\draw[connector] (cat1.west) |- ([xshift=-2mm]sub1.west);
\draw[connector] (sub1.west) |- ([xshift=-2mm]sub2.west);
\draw[connector] (sub2.west) |- ([xshift=-2mm]sub3.west);
\draw[connector] (sub2.west) |- ([xshift=-2mm]sub4.west);
\draw[connector] (sub2.west) |- ([xshift=-2mm]sub5.west);
\draw[connector] (sub2.west) |- ([xshift=-2mm]sub6.west);
\draw[connector] (sub2.west) |- ([xshift=-2mm]sub7.west);
\draw[connector] (sub2.west) |- ([xshift=-2mm]sub8.west);
\draw[connector] (cat1.west) |- ([xshift=-2mm]sub9.west);

\end{tikzpicture}

\clearpage

\newdimen\spacing
\spacing=1cm 

\begin{tikzpicture}[
  every node/.style={align=left,},
  category/.style={text width=6cm, align=left, inner sep=4pt, font=\small},
  root/.style={text width=10cm, align=left, inner sep=4pt},
  quote/.style={rounded corners, thick, text width=6cm, align=justify, inner sep=4pt, font=\footnotesize},
  arrow/.style={-{Stealth[length=3mm]}, thick},
  connector/.style={thick},
  spacing/.store in=\myspacing, spacing=1.0cm 
]

\node[root] (cat1) at (0, 0) {Ethical opportunities for care recipients (continued)};

\node[category, anchor=west] (sub1) at ($(cat1.south west)+(0.5,-\myspacing-1cm)$) {Supports tailoring care approaches to needs/taste};
\node[quote, anchor=west] (q1) at ($(sub1.east)+(1.5cm,0)$) {“Second, sociable robots that leverage recent advances in artificial intelligence (AI) are up to the challenge. They display increasingly sophisticated emotional intelligence; interact in ways that seem lifelike, such as recognizing voices, faces and emotions; interpret speech and gestures; respond appropriately to complex verbal and nonverbal cues; make eye contact; speak conversationally; and adapt to people’s needs by learning from feedback, rewards, and criticisms.”(Jecker 2021 You’ve got a friend in me: sociable robots for older adults in an age of global pandemics, p. S36)};
\draw[arrow] (sub1.east) -- (q1.west);
\node[draw=none, fit=(sub1)(q1), inner sep=0pt] (pair1) {};

\node[category, anchor=west] (sub2) at ($(pair1.south west)+(0.5,-\myspacing-1.5cm)$) {Social robots will learn to read care recipient and react adequately};
\node[quote, anchor=west] (q2) at ($(sub2.east)+(1.cm,0)$) {“Future robot fitness coaches will be able to correctly interpret the mood of their ‘pupil’ in order to find the right approach to motivate them. To do this, the fitness robot must be able to recognize people, recognize naturally spoken language and process gestures. This development could also be harnessed for older people and used in rehabilitation or prevention.” (Translation by the authors)(Meyer 2011 Akzeptanz ausgewählter Anwendungsszenarien, p. 87; in: Meyer 2011 Mein Freund der Roboter - Servicerobotik für ältere Menschen - eine Antwort auf den demographischen Wandel?)};
\draw[arrow] (sub2.east) -- (q2.west);
\node[draw=none, fit=(sub2)(q2), inner sep=0pt] (pair2) {};

\node[category, anchor=west] (sub3) at ($(pair2.south west)+(0.,-\myspacing-0.5cm)$) {Simplified by broadening scope of action (more approaches to choose from)};
\node[quote, anchor=west] (q3) at ($(sub3.east)+(1.cm,0)$) {“It is important however to observe the choice of the care recipient. Some people might prefer a human caregiver, while others may prefer the support of robots, depending on which one gives them a greater sense of self-worth.” (Van Est et al. 2016 Robotisation as Rationalisation - In Search for a Human Robot Future, p. 45)};
\draw[arrow] (sub3.east) -- (q3.west);
\node[draw=none, fit=(sub3)(q3), inner sep=0pt] (pair3) {};

\node[category, anchor=west] (sub4) at ($(pair3.south west)+(-0.5,-\myspacing-1cm)$) {Support/care provided (more) efficiently};
\node[quote, anchor=west] (q4) at ($(sub4.east)+(1.5cm,0)$) {“In view of the expected increase in the number of very old people living alone and those remaining alone in (rural) regions, telematics-supported systems are seen as promising technologies for improving care provision. The article shows the potential benefits that the implementation of telepresence robotics can have for both relatives and people in need of help in the case of dementia.” (Translation by the authors)(Bleses and Dammert 2020 Neue Technologien aus Sicht der Pflegewissenschaft, p. 56; in: Hanika 2020 Künstliche Intelligenz, Robotik und autonome Systeme in der Gesundheitsversorgung)};
\draw[arrow] (sub4.east) -- (q4.west);
\node[draw=none, fit=(sub4)(q4), inner sep=0pt] (pair4) {};

\node[category, anchor=west] (sub5) at ($(pair4.south west)+(0.5,-\myspacing-0.5cm)$) {Social robots can save costs (e. g. compared to smart homes)};
\node[quote, anchor=west] (q5) at ($(sub5.east)+(1.cm,0)$) {“Gecko Systems claim that the CareBot provides a more cost-effective version of a “smart home”, because the robot can follow the occupant from room to room, and reduce the need for sensors and wires throughout the house.”(Sharkey and Sharkey 2012 Granny and the robots: ethical issues in robot care for the elderly, p. 31)};
\draw[arrow] (sub5.east) -- (q5.west);
\node[draw=none, fit=(sub5)(q5), inner sep=0pt] (pair5) {};

\node[category, anchor=west] (sub6) at ($(pair5.south west)+(0.,-\myspacing-1.5cm)$) {Pet robots easier to maintain than real animals};
\node[quote, anchor=west] (q6) at ($(sub6.east)+(1.cm,0)$) {“In contrast to real animals, Paro has several advantages: for example, the seal cannot bite patients who grip it hard, does not trigger allergic reactions due to its antiseptic fur, does not cause dirt, for example through excrement, and does not need food. Paro is intended to offer patients who have problems caring for themselves an easy-care alternative to real animals (Calo et al. 2011: 21).” (Translation by the authors)(Lange and Bauer 2021 - Eine Robbe für Oma - Die zukünftige Dauerausstellung Robotik im Deutschen Museum, p. 231; in: Inthorn und Seising 2021 Digitale Patientenversorgung - Zur Computerisierung von Diagnostik, Therapie und Pflege)};
\draw[arrow] (sub6.east) -- (q6.west);
\node[draw=none, fit=(sub6)(q6), inner sep=0pt] (pair6) {};

\draw[connector] (cat1.west) |- ([xshift=-2mm]sub1.west);
\draw[connector] (sub1.west) |- ([xshift=-2mm]sub2.west);
\draw[connector] (sub1.west) |- ([xshift=-2mm]sub3.west);
\draw[connector] (cat1.west) |- ([xshift=-2mm]sub4.west);
\draw[connector] (sub4.west) |- ([xshift=-2mm]sub5.west);
\draw[connector] (sub4.west) |- ([xshift=-2mm]sub6.west);

\end{tikzpicture}

\clearpage
\begin{tikzpicture}[
  every node/.style={align=left,},
  category/.style={text width=6cm, align=left, inner sep=4pt, font=\small},
  root/.style={text width=10cm, align=left, inner sep=4pt},
  quote/.style={rounded corners, thick, text width=6cm, align=justify, inner sep=4pt, font=\footnotesize},
  arrow/.style={-{Stealth[length=3mm]}, thick},
  connector/.style={thick},
  spacing/.store in=\myspacing, spacing=1.0cm 
]

\node[root] (cat1) at (0, 0) {Ethical hazards for care recipients (Main category)};

\node[category, anchor=west] (sub1) at ($(cat1.south west)+(0.5,-\myspacing)$) {Reducing/not adequately securing well-being/health};
\node[quote, anchor=west] (q1) at ($(sub1.east)+(1.5cm,0)$) {“Sullins suggests that robotic companions can only satisfice, but not truly satisfy, our physical and emotional needs.”(Frennert and Östlund 2014 Review: Seven Matters of Concern of Social Robots and Older People, p. 304)};
\draw[arrow] (sub1.east) -- (q1.west);
\node[draw=none, fit=(sub1)(q1), inner sep=0pt] (pair1) {};"

\node[category, anchor=west] (sub2) at ($(pair1.south west)+(0.5,-\myspacing-0.5cm)$) {Facilitating harmful behavior of care recipient};
\node[quote, anchor=west] (q2) at ($(sub2.east)+(1.cm,0)$) {“Louis has discovered that he can use the interface on the Care-O-bot to visit online gambling sites and enjoys playing poker in the evening.”(Draper and Sorell 2017 Ethical values and social care robots for older people: an international qualitative study, p. 53)};
\draw[arrow] (sub2.east) -- (q2.west);
\node[draw=none, fit=(sub2)(q2), inner sep=0pt] (pair2) {};

\node[category, anchor=west] (sub3) at ($(pair2.south west)+(0.5,-\myspacing-0.5cm)$) {Social robot assists in dangerous actions/addictive behavior};
\node[quote, anchor=west] (q3) at ($(sub3.east)+(0.5cm,0)$) {“Frank, an elderly gentleman suffering from mild dementia, is given a robot by his son to help him with his everyday life. He only reluctantly accepts this housemate until he discovers that this robot can help him to resume his dormant activity as a jewel thief in an enormously effective way.” (Translation by the authors) (Manzeschke 2019 Roboter in der Pflege - Von Menschen, Maschinen und anderen hilfreichen Wesen, p. 7)};
\draw[arrow] (sub3.east) -- (q3.west);
\node[draw=none, fit=(sub3)(q3), inner sep=0pt] (pair3) {};

\node[category, anchor=west] (sub4) at ($(pair3.south west)+(0.,-\myspacing-0.5cm)$) {Becoming lazy/moving less as social robot does it all};
\node[quote, anchor=west] (q4) at ($(sub4.east)+(0.5cm,0)$) {“Marie is reluctant to get up from her chair because she is afraid of falling and walking is uncomfortable. She also uses the Care-O-bot® to get drinks for her from the kitchen, even though the nurse has suggested that she should go to the kitchen with the Care-O-bot® but let it carry the drinks back to her chair for her.”(Draper and Sorell 2017 Ethical values and social care robots for older people: an international qualitative study, p. 53)};
\draw[arrow] (sub4.east) -- (q4.west);
\node[draw=none, fit=(sub4)(q4), inner sep=0pt] (pair4) {};

\node[category, anchor=west] (sub5) at ($(pair4.south west)+(-0.5,-\myspacing)$) {Reduction in the quality of care provided};
\node[quote, anchor=west] (q5) at ($(sub5.east)+(1.cm,0)$) {“The introduction of robotic care determined by limited stakeholders leaves customised care in danger and instead could potentially turn the quality of elderly care into ‘elderly management’.” (Poulsen and Burmeister 2019 Overcoming carer shortages with care robots: Dynamic value trade-offs in run-time, p. 1)};
\draw[arrow] (sub5.east) -- (q5.west);
\node[draw=none, fit=(sub5)(q5), inner sep=0pt] (pair5) {};

\node[category, anchor=west] (sub6) at ($(pair5.south west)+(0.5,-\myspacing-1.5cm)$) {Introduction of social robots leads to disinvestment in care providers};
\node[quote, anchor=west] (q6) at ($(sub6.east)+(0.5cm,0)$) {“However, should such applications be intended as a way to radically reduce or worse eliminate professional human care, with the primary aim of abiding budgetary constraints grounded on the observed increase in age-related welfare costs, that would also contrast the anthropocentric perspective on the use of technology, and violate the dignity of the user, thereby being inadmissible in light of our extant legal system.”(Bertolini and Arian 2020 Do robots care? - Towards an anthropocentric framework in the caring of frail individuals through assistive technology, p. 49; in: Haltaufderheide et al. 2020 Aging between Participation and Simulation - Ethical Dimensions of Social Assistive Technologies)};
\draw[arrow] (sub6.east) -- (q6.west);
\node[draw=none, fit=(sub6)(q6), inner sep=0pt] (pair6) {};

\node[category, anchor=west] (sub7) at ($(pair6.south west)+(0.,-\myspacing-1cm)$) {(Informal) caregivers withdraw care/are less engaged because of social robot};
\node[quote, anchor=west] (q7) at ($(sub7.east)+(0.5cm,0)$) {“The fear that robots will cause ICs to withdraw their care is one that should be taken seriously. The withdrawal of care by ICs could have the effect of shifting the burden onto FCs, or of reducing the overall degree of care the older person receives, or both. This displacement may foster resentment amongst informal carers, which may impact negatively on the older person’s care.” (Jenkins and Draper 2015 Care, Monitoring, and Companionship: Views on Care Robots from Older People and Their Carer, p. 679)};
\draw[arrow] (sub7.east) -- (q7.west);
\node[draw=none, fit=(sub7)(q7), inner sep=0pt] (pair7) {};

\node[category, anchor=north west] (sub8) at ($(pair7.south west)+(0.0,0)$) {\textit{Cont. on next page}};
\node[quote, anchor=west] (q8) at ($(sub8.east)+(1.cm,0)$) {};
\node[draw=none, fit=(sub8)(q8), inner sep=0pt] (pair8) {};

\node[category, anchor=north west] (sub9) at ($(pair8.south west)+(-0.5,0)$) {\textit{Cont. on next page}};
\node[quote, anchor=west] (q9) at ($(sub9.east)+(1.cm,0)$) {};
\node[draw=none, fit=(sub9)(q9), inner sep=0pt] (pair9) {};

\node[category, anchor=north west] (sub10) at ($(pair9.south west)+(-0.5,0)$) {\textit{Cont. on next page}};
\node[quote, anchor=west] (q10) at ($(sub10.east)+(1.5cm,0)$) {};
\node[draw=none, fit=(sub10)(q10), inner sep=0pt] (pair10) {};

\draw[connector] (cat1.west) |- ([xshift=-2mm]sub1.west);
\draw[connector] (sub1.west) |- ([xshift=-2mm]sub2.west);
\draw[connector] (sub2.west) |- ([xshift=-2mm]sub3.west);
\draw[connector] (sub2.west) |- ([xshift=-2mm]sub4.west);
\draw[connector] (sub1.west) |- ([xshift=-2mm]sub5.west);
\draw[connector] (sub5.west) |- ([xshift=-2mm]sub6.west);
\draw[connector] (sub5.west) |- ([xshift=-2mm]sub7.west);
\draw[connector] (sub5.west) |- ([xshift=-2mm]sub8.west);
\draw[connector] (sub1.west) |- ([xshift=-2mm]sub9.west);
\draw[connector] (cat1.west) |- ([xshift=-2mm]sub10.west);

\end{tikzpicture}

\clearpage
\begin{tikzpicture}[
  every node/.style={align=left,},
  category/.style={text width=6cm, align=left, inner sep=4pt, font=\small},
  root/.style={text width=10cm, align=left, inner sep=4pt},
  quote/.style={rounded corners, thick, text width=6cm, align=justify, inner sep=4pt, font=\footnotesize},
  arrow/.style={-{Stealth[length=3mm]}, thick},
  connector/.style={thick},
  spacing/.store in=\myspacing, spacing=1.0cm 
]

\node[root] (cat1) at (0, 0) {Ethical hazards for care recipients (continued)};

\node[category, anchor=west] (sub1) at ($(cat1.south west)+(0.5,-\myspacing+0.5cm)$) {Reducing/not adequately securing well-being/health (continued)};
\node[quote, anchor=west] (q1) at ($(sub1.east)+(1.5cm,0)$) {};
\node[draw=none, fit=(sub1)(q1), inner sep=0pt] (pair1) {};

\node[category, anchor=west] (sub2) at ($(sub1.south west)+(0.5,-\myspacing+0.5cm)$) {Reduction in the quality of care provided (continued)};
\node[quote, anchor=west] (q2) at ($(sub2.east)+(1cm,0)$) {};
\node[draw=none, fit=(sub2)(q2), inner sep=0pt] (pair2) {};

\node[category, anchor=west] (sub3) at ($(pair2.south west)+(0.5,-\myspacing)$) {Improper delegation of tasks to social robot};
\node[quote, anchor=west] (q3) at ($(sub3.east)+(0.5cm,0)$) {“‘The lack of human resources must not lead to technology being seen as the solution. Technology is only as good as those who deploy and use it. Anyone who uses cuddly robots to sedate dementia patients left alone has not understood their profession’ (geriatric nurse, female, aged 53).” (Translation by the authors)(Zöllick et al. 2020 Technik und Pflege – eine ambivalente Beziehung, p. 51) };
\draw[arrow] (sub3.east) -- (q3.west);
\node[draw=none, fit=(sub3)(q3), inner sep=0pt] (pair3) {};

\node[category, anchor=west] (sub4) at ($(pair3.south west)+(0.,-\myspacing-1cm)$) {Social robots are not fit to meet care tasks they are designed for};
\node[quote, anchor=west] (q4) at ($(sub4.east)+(0.5cm,0)$) {“Others don’t see technology as being able to provide social and emotional support: ‘No robot can understand the subtle nuances of words, tone of voice and dialect, the connotations in communication. They are mixed with gestures, facial expressions, environment, situation and culture and are not always easy for humans to understand. A robot can't do that. You can’t touch with plastic’. (Geriatric nurse, female, 53 years old).” (Translation by the authors)(Zöllick et al. 2020 Technik und Pflege – eine ambivalente Beziehung, p. 51) };
\draw[arrow] (sub4.east) -- (q4.west);
\node[draw=none, fit=(sub4)(q4), inner sep=0pt] (pair4) {};

\node[category, anchor=west] (sub5) at ($(pair4.south west)+(0.,-\myspacing-0.5cm)$) {Social robots can’t replace a “trustful care relationship”};
\node[quote, anchor=west] (q5) at ($(sub5.east)+(0.5cm,0)$) {“Care robots thus cannot be considered to replace caregivers in their function as a person trusted.”(Wiertz 2020 Trusting robots? – On the concept of trust and on forms of human vulnerability, p. 65; in: Haltaufderheide et al. 2020 Aging between Participation and Simulation - Ethical Dimensions of Social Assistive Technologies)};
\draw[arrow] (sub5.east) -- (q5.west);
\node[draw=none, fit=(sub5)(q5), inner sep=0pt] (pair5) {};

\node[category, anchor=west] (sub6) at ($(pair5.south west)+(0.,-\myspacing-0.5cm)$) {Social robots don’t have the same wealth of experience as nursing staff };
\node[quote, anchor=west] (q6) at ($(sub6.east)+(0.5cm,0)$) {“All these developments are associated with benefits such as physical and cognitive relief for care staff, but also risks such as the loss of experience-saturated judgment and relationship work based on this, among other things.” (Translation by the authors)(Remmers 2019 Pflege und Technik. Stand der Diskussion und zentrale ethische Fragen, p. 413)};
\draw[arrow] (sub6.east) -- (q6.west);
\node[draw=none, fit=(sub6)(q6), inner sep=0pt] (pair6) {};

\node[category, anchor=west] (sub7) at ($(pair6.south west)+(0.,-\myspacing-0.5cm)$) {Reduced sensitivity/empathy of care providers due to social robots};
\node[quote, anchor=west] (q7) at ($(sub7.east)+(0.5cm,0)$) {“Artificial systems therefore have a particularly negative impact on the development of the virtues of care staff and their relationship with those in need of care. For example, the sensitivity, empathy, commitment and personal development of care staff suffer from the use of artificial systems.” (Translation by the authors)(Misselhorn 2018 Pflegesysteme, p. 153; in: Misselhorn 2018 Grundfragen der Maschinenethik) };
\draw[arrow] (sub7.east) -- (q7.west);
\node[draw=none, fit=(sub7)(q7), inner sep=0pt] (pair7) {};

\node[category, anchor=west] (sub8) at ($(pair7.south west)+(0.,-\myspacing-1cm)$) {Being a testbed for social robots};
\node[quote, anchor=west] (q8) at ($(sub8.east)+(0.5cm,0)$) {“A study commissioned by the European Commission (Butter et el., 2008) on the use of robotics in healthcare discusses the following ethical problems, which are also relevant to the care of the elderly: [...] the use of ill and vulnerable people for experimental purposes; [...].” (Translation by the authors)(Bioethikkommission beim Bundeskanzleramt Österreich 2018 Roboter in der Betreuung alter Menschen (Stellungnahme), pp. 363, 364)};
\draw[arrow] (sub8.east) -- (q8.west);
\node[draw=none, fit=(sub8)(q8), inner sep=0pt] (pair8) {};

\node[category, anchor=north west] (sub9) at ($(pair8.south west)+(-0.5,0)$) {\textit{Cont. on next page}};
\node[quote, anchor=west] (q9) at ($(sub9.east)+(1.cm,0)$) {};
\node[draw=none, fit=(sub9)(q9), inner sep=0pt] (pair9) {};

\node[category, anchor=north west] (sub10) at ($(pair9.south west)+(-0.5,0)$) {\textit{Cont. on next page}};
\node[quote, anchor=west] (q10) at ($(sub10.east)+(1.5cm,0)$) {};
\node[draw=none, fit=(sub10)(q10), inner sep=0pt] (pair10) {};

\draw[connector] (cat1.west) |- ([xshift=-2mm]sub1.west);
\draw[connector] (sub1.west) |- ([xshift=-2mm]sub2.west);
\draw[connector] (sub2.west) |- ([xshift=-2mm]sub3.west);
\draw[connector] (sub2.west) |- ([xshift=-2mm]sub4.west);
\draw[connector] (sub2.west) |- ([xshift=-2mm]sub5.west);
\draw[connector] (sub2.west) |- ([xshift=-2mm]sub6.west);
\draw[connector] (sub2.west) |- ([xshift=-2mm]sub7.west);
\draw[connector] (sub2.west) |- ([xshift=-2mm]sub8.west);
\draw[connector] (sub1.west) |- ([xshift=-2mm]sub9.west);
\draw[connector] (cat1.west) |- ([xshift=-2mm]sub10.west);

\end{tikzpicture}

\clearpage
\begin{tikzpicture}[
  every node/.style={align=left,},
  category/.style={text width=6cm, align=left, inner sep=4pt, font=\small},
  root/.style={text width=10cm, align=left, inner sep=4pt},
  quote/.style={rounded corners, thick, text width=6cm, align=justify, inner sep=4pt, font=\footnotesize},
  arrow/.style={-{Stealth[length=3mm]}, thick},
  connector/.style={thick},
  spacing/.store in=\myspacing, spacing=1.0cm 
]

\node[root] (cat1) at (0, 0) {Ethical hazards for care recipients (continued)};

\node[category, anchor=west] (sub1) at ($(cat1.south west)+(0.5,-\myspacing+0.5cm)$) {Reducing/not adequately securing well-being/health (continued)};
\node[quote, anchor=west] (q1) at ($(sub1.east)+(1.5cm,0)$) {};
\node[draw=none, fit=(sub1)(q1), inner sep=0pt] (pair1) {};

\node[category, anchor=west] (sub2) at ($(pair1.south west)+(0.5,-\myspacing)$) {Adverse effect on emotional/psychological well-being};
\node[quote, anchor=west] (q2) at ($(sub2.east)+(1.cm,0)$) {“In some cases, interventions with PARO caused negative emotional responses, including anger, wandering, fearfulness, and agitation [17, 20, 35].” (Hung et al. 2019 The benefits of and barriers to using a social robot PARO in care settings: a scoping review, p. 6)};
\draw[arrow] (sub2.east) -- (q2.west);
\node[draw=none, fit=(sub2)(q2), inner sep=0pt] (pair2) {};

\node[category, anchor=west] (sub3) at ($(pair2.south west)+(0.5,-\myspacing)$) {Feeling uncomfortable with social robot};
\node[quote, anchor=west] (q3) at ($(sub3.east)+(0.5cm,0)$) {“Companion robots are designed to counter social isolation and loneliness. Most people, however, feel uncomfortable with such robots […].”(Kemenade et al. 2019 Do You Care for Robots That Care? Exploring the Opinions of Vocational Care Students on the Use of Healthcare Robots, p. 2)};
\draw[arrow] (sub3.east) -- (q3.west);
\node[draw=none, fit=(sub3)(q3), inner sep=0pt] (pair3) {};

\node[category, anchor=west] (sub4) at ($(pair3.south west)+(0.,-\myspacing-1cm)$) {Enhancing feelings of isolation/loneliness};
\node[quote, anchor=west] (q4) at ($(sub4.east)+(0.5cm,0)$) {“In this regards, it is plausible to assume that humans will not only stay lonely when someone places a robot in their home, but that this allows them realizing even more that in fact no-body cares about them (which is the reason for being lonely in the first place). Thus, supposedly being cared about by a robot might serve as the ultimate proof for really being lonely.”(von Maur 2023 Alice Does not Care – Or: Wy it Matters That Robots “Don’t Give a Damn”, p. 224; in: Loh und Loh 2023 Social Robotics and the Good Life - The Normative Side of Forming Emotional Bonds With Robots)};
\draw[arrow] (sub4.east) -- (q4.west);
\node[draw=none, fit=(sub4)(q4), inner sep=0pt] (pair4) {};

\node[category, anchor=west] (sub5) at ($(pair4.south west)+(0.,-\myspacing-1.5cm)$) {User interface is experienced as overwhelming/hard to control};
\node[quote, anchor=west] (q5) at ($(sub5.east)+(0.5cm,0)$) {“In addition to their curiosity about the system, some of the people with dementia had reservations about using a telepresence robot. For them, the use of this technology - often in contrast to their relatives - did not always seem comprehensible, even when the device was demonstrated to them ‘in use’ in their own home. In some cases, people with dementia specifically stated that they were ‘overwhelmed’ or ‘unable to cope’ with operating the telepresence robot.” (Translation by the authors)(Bleses and Dammert 2020 Neue Technologien aus Sicht der Pflegewissenschaft, p. 69; in: Hanika 2020 Künstliche Intelligenz, Robotik und autonome Systeme in der Gesundheitsversorgung)};
\draw[arrow] (sub5.east) -- (q5.west);
\node[draw=none, fit=(sub5)(q5), inner sep=0pt] (pair5) {};

\node[category, anchor=west] (sub6) at ($(pair5.south west)+(0.,-\myspacing-0.5)$) {Feeling of loss/jealousy when robot is away, broken or shared};
\node[quote, anchor=west] (q6) at ($(sub6.east)+(0.5cm,0)$) {“The robot, however, may have to be shared with others – other residents in a care home, for instance – or break down, […] leading at times to feelings of jealousy, disillusionment or disappointment.”(Battistuzzi et al. 2018 Embedding Ethics in the Design of Culturally Competent Socially Assistive Robots, p. 1998)};
\draw[arrow] (sub6.east) -- (q6.west);
\node[draw=none, fit=(sub6)(q6), inner sep=0pt] (pair6) {};

\node[category, anchor=west] (sub7) at ($(pair6.south west)+(0.,-\myspacing-0.5cm)$) {Being disappointed/frustrated because expectation of robot not met};
\node[quote, anchor=west] (q7) at ($(sub7.east)+(0.5cm,0)$) {“We observed that Henry, when he was first deployed at the care site was often expected to be able to engage in natural conversation and to understand human language. As this was not the case, people were quickly frustrated by failing to communicate with the robot.”(Körtner 2016 Ethical challenges in the use of social service robots for elderly people, p. 304)};
\draw[arrow] (sub7.east) -- (q7.west);
\node[draw=none, fit=(sub7)(q7), inner sep=0pt] (pair7) {};

\node[category, anchor=west] (sub8) at ($(pair7.south west)+(0.,-\myspacing)$) {Increased stress level (e. g. due to malfunction/reduced contact)};
\node[quote, anchor=west] (q8) at ($(sub8.east)+(0.5cm,0)$) {“The robot’s absence may produce distress and/or loss of therapeutic benefits. This consequence can occur especially in users who cannot understand the reason of the robot removal.” (Tzafestas 2016 Socialised Roboethics, p. 119; in: Tzafestas 2016 Roboethics - A Navigating Overview)};
\draw[arrow] (sub8.east) -- (q8.west);
\node[draw=none, fit=(sub8)(q8), inner sep=0pt] (pair8) {};

\node[category, anchor=north west] (sub9) at ($(pair8.south west)+(-0.0,0)$) {\textit{Cont. on next page}};
\node[quote, anchor=west] (q9) at ($(sub9.east)+(1.cm,0)$) {};
\node[draw=none, fit=(sub9)(q9), inner sep=0pt] (pair9) {};

\node[category, anchor=north west] (sub10) at ($(pair9.south west)+(-0.5,0)$) {\textit{Cont. on next page}};
\node[quote, anchor=west] (q10) at ($(sub10.east)+(1.5cm,0)$) {};
\node[draw=none, fit=(sub10)(q10), inner sep=0pt] (pair10) {};

\node[category, anchor=north west] (sub11) at ($(pair10.south west)+(-0.5,0)$) {\textit{Cont. on next page}};
\node[quote, anchor=west] (q11) at ($(sub11.east)+(1.5cm,0)$) {};
\node[draw=none, fit=(sub11)(q11), inner sep=0pt] (pair11) {};

\draw[connector] (cat1.west) |- ([xshift=-2mm]sub1.west);
\draw[connector] (sub1.west) |- ([xshift=-2mm]sub2.west);
\draw[connector] (sub2.west) |- ([xshift=-2mm]sub3.west);
\draw[connector] (sub2.west) |- ([xshift=-2mm]sub4.west);
\draw[connector] (sub2.west) |- ([xshift=-2mm]sub5.west);
\draw[connector] (sub2.west) |- ([xshift=-2mm]sub6.west);
\draw[connector] (sub2.west) |- ([xshift=-2mm]sub7.west);
\draw[connector] (sub2.west) |- ([xshift=-2mm]sub8.west);
\draw[connector] (sub2.west) |- ([xshift=-2mm]sub9.west);
\draw[connector] (sub1.west) |- ([xshift=-2mm]sub10.west);
\draw[connector] (cat1.west) |- ([xshift=-2mm]sub11.west);

\end{tikzpicture}

\begin{tikzpicture}[
  every node/.style={align=left,},
  category/.style={text width=6cm, align=left, inner sep=4pt, font=\small},
  root/.style={text width=10cm, align=left, inner sep=4pt},
  quote/.style={rounded corners, thick, text width=6cm, align=justify, inner sep=4pt, font=\footnotesize},
  arrow/.style={-{Stealth[length=3mm]}, thick},
  connector/.style={thick},
  spacing/.store in=\myspacing, spacing=1.0cm 
]

\node[root] (cat1) at (0, 0) {Ethical hazards for care recipients (continued)};

\node[category, anchor=west] (sub1) at ($(cat1.south west)+(0.5,-\myspacing+0.5cm)$) {Reducing/not adequately securing well-being/health (continued)};
\node[quote, anchor=west] (q1) at ($(sub1.east)+(1.5cm,0)$) {};
\node[draw=none, fit=(sub1)(q1), inner sep=0pt] (pair1) {};

\node[category, anchor=west] (sub2) at ($(pair1.south west)+(0.5,-\myspacing+0.5cm)$) {Adverse effect on emotional/psychological well-being};
\node[quote, anchor=west] (q2) at ($(sub2.east)+(1.5cm,0)$) {};
\node[draw=none, fit=(sub2)(q2), inner sep=0pt] (pair2) {};

\node[category, anchor=west] (sub3) at ($(pair2.south west)+(0.5,-\myspacing)$) {Suffering from conflicts with care providers about usage of social robot};
\node[quote, anchor=west] (q3) at ($(sub3.east)+(0.5cm,0)$) {“We will show that assisting older people and their carers may often not amount to the same thing, because there can be serious tensions between older people and their carers that can be exacerbated by the introduction of a robot.”(Jenkins and Draper 2015 Care, Monitoring, and Companionship: Views on Care Robots from Older People and Their Carer, p. 673)};
\draw[arrow] (sub3.east) -- (q3.west);
\node[draw=none, fit=(sub3)(q3), inner sep=0pt] (pair3) {};

\node[category, anchor=west] (sub4) at ($(pair3.south west)+(0.,-\myspacing-1cm)$) {Feeling humiliated/disrespected by implied images of oneself};
\node[quote, anchor=west] (q4) at ($(sub4.east)+(0.5cm,0)$) {“In the next example from the beginning of the film, it becomes clear how much Frank initially resists the imposed robot: Hunter, the son, gets the robot out of the trunk of his car. [...] Frank: ‘Are you kidding me? I'm not that bad off! I don't need to be spoon-fed by a goddamn robot!’ Hunter: ‘Dad, Dad, no one says that either! He's new... He's more like a, eh, something like a butler!’” (Translation by the authors)(Frebel 2015 Roboter gegen das Vergessen? - Technische Assistenz bei Altersdemenz im Spielfilm aus medizinethischer Sicht, p. 112)};
\draw[arrow] (sub4.east) -- (q4.west);
\node[draw=none, fit=(sub4)(q4), inner sep=0pt] (pair4) {};

\node[category, anchor=west] (sub5) at ($(pair4.south west)+(0.,-\myspacing-1cm)$) {Being ashamed/scared of how others view them when needing social robot};
\node[quote, anchor=west] (q5) at ($(sub5.east)+(0.5cm,0)$) {“Some participants described interacting with robots as potentially shameful or embarrassing, signifying to others that they may be experiencing cognitive decline. This represented a threat to a person’s self-esteem and sense of identity and even a loss of control over one’s own life (e.g., in the case of the individual fearing she would be removed from her own home).”(Coghlan et al. 2021 Dignity, Autonomy, and Style of Company: Dimensions Older Adults Consider for Robot Companions, p. 16)};
\draw[arrow] (sub5.east) -- (q5.west);
\node[draw=none, fit=(sub5)(q5), inner sep=0pt] (pair5) {};

\node[category, anchor=west] (sub6) at ($(pair5.south west)+(0.,-\myspacing-0.4cm)$) {Further negative feelings};
\node[quote, anchor=west] (q6) at ($(sub6.east)+(0.5cm,0)$) {“Some may be confused by the SAR, or have difficulty understanding how it works, what to do to make it work or what triggers its actions, which could lead to self-blame and lowered self-efficacy.”(Battistuzzi et al. 2021 Socially Assistive Robots, Older Adults and Research Ethics: The Case for Case-Based Ethics Training, p. 650) };
\draw[arrow] (sub6.east) -- (q6.west);
\node[draw=none, fit=(sub6)(q6), inner sep=0pt] (pair6) {};

\node[category, anchor=west] (sub7) at ($(pair6.south west)+(-0.5,-\myspacing-0.5cm)$) {Adverse effects on safety/physical well-being};
\node[quote, anchor=west] (q7) at ($(sub7.east)+(1.cm,0)$) {“Safe and responsible introduction of social robots to this target group is essential, as a potential lack of knowledge of and experience with new technologies may lead to situations that potentially affect psychological and/or physical safety (Borenstein et al., 2017)”.(Van Maris et al. 2020 Designing Ethical Social Robots – A Longitudinal Field Study With Older Adults, p. 2)};
\draw[arrow] (sub7.east) -- (q7.west);
\node[draw=none, fit=(sub7)(q7), inner sep=0pt] (pair7) {};

\node[category, anchor=west] (sub8) at ($(pair7.south west)+(0.5,-\myspacing-1.1cm)$) {Harms caused by limited control/overtrust in social robot };
\node[quote, anchor=west] (q8) at ($(sub8.east)+(0.5cm,0)$) {“This then can be a problem of truthfulness, but also a source of practical misunderstandings and the dangers that arise from them: if users are not able to tell what exactly the machine is capable of, they might be led to assume that the machine has understood a command that, in reality, it hasn’t; for example the command to remind the patient to take medication. If the patient relies on the machine to perform the task, a dangerous situation can arise, in which the machine’s failure to clearly advertise its capabilities and limitations can directly lead to harm.”(Matthias 2015 Robot Lies in Health Care: When Is Deception Morally Permissible?, pp. 170, 171)};
\draw[arrow] (sub8.east) -- (q8.west);
\node[draw=none, fit=(sub8)(q8), inner sep=0pt] (pair8) {};

\node[category, anchor=north west] (sub9) at ($(pair8.south west)+(-0.0,0)$) {\textit{Cont. on next page}};
\node[quote, anchor=west] (q9) at ($(sub9.east)+(1.cm,0)$) {};
\node[draw=none, fit=(sub9)(q9), inner sep=0pt] (pair9) {};

\node[category, anchor=north west] (sub10) at ($(pair9.south west)+(-0.5,0)$) {\textit{Cont. on next page}};
\node[quote, anchor=west] (q10) at ($(sub10.east)+(1.5cm,0)$) {};
\node[draw=none, fit=(sub10)(q10), inner sep=0pt] (pair10) {};

\node[category, anchor=north west] (sub11) at ($(pair10.south west)+(-0.5,0)$) {\textit{Cont. on next page}};
\node[quote, anchor=west] (q11) at ($(sub11.east)+(1.5cm,0)$) {};
\node[draw=none, fit=(sub11)(q11), inner sep=0pt] (pair11) {};

\draw[connector] (cat1.west) |- ([xshift=-2mm]sub1.west);
\draw[connector] (sub1.west) |- ([xshift=-2mm]sub2.west);
\draw[connector] (sub2.west) |- ([xshift=-2mm]sub3.west);
\draw[connector] (sub2.west) |- ([xshift=-2mm]sub4.west);
\draw[connector] (sub2.west) |- ([xshift=-2mm]sub5.west);
\draw[connector] (sub2.west) |- ([xshift=-2mm]sub6.west);
\draw[connector] (sub1.west) |- ([xshift=-2mm]sub7.west);
\draw[connector] (sub7.west) |- ([xshift=-2mm]sub8.west);
\draw[connector] (sub7.west) |- ([xshift=-2mm]sub9.west);
\draw[connector] (sub1.west) |- ([xshift=-2mm]sub10.west);
\draw[connector] (cat1.west) |- ([xshift=-2mm]sub11.west);

\end{tikzpicture}

\clearpage
\begin{tikzpicture}[
  every node/.style={align=left,},
  category/.style={text width=6cm, align=left, inner sep=4pt, font=\small},
  root/.style={text width=10cm, align=left, inner sep=4pt},
  quote/.style={rounded corners, thick, text width=6cm, align=justify, inner sep=4pt, font=\footnotesize},
  arrow/.style={-{Stealth[length=3mm]}, thick},
  connector/.style={thick},
  spacing/.store in=\myspacing, spacing=1.0cm 
]

\node[root] (cat1) at (0, 0) {Ethical hazards for care recipients (continued)};

\node[category, anchor=west] (sub1) at ($(cat1.south west)+(0.5,-\myspacing+0.5cm)$) {Reducing/not adequately securing well-being/health (continued)};
\node[quote, anchor=west] (q1) at ($(sub1.east)+(1.5cm,0)$) {};
\node[draw=none, fit=(sub1)(q1), inner sep=0pt] (pair1) {};

\node[category, anchor=west] (sub2) at ($(pair1.south west)+(0.5,-\myspacing+0.5cm)$) {Adverse effects on safety/physical well-being};
\node[quote, anchor=west] (q2) at ($(sub2.east)+(1.5cm,0)$) {};
\node[draw=none, fit=(sub2)(q2), inner sep=0pt] (pair2) {};

\node[category, anchor=west] (sub3) at ($(pair2.south west)+(0.5,-\myspacing)$) {Harms caused by malfunction};
\node[quote, anchor=west] (q3) at ($(sub3.east)+(0.5cm,0)$) {“In an aged care setting where robots interact in close, direct contact with the elderly, any malfunctioning can result in a disastrous outcome.”(Poulsen et al. 2020 Cybersecurity, value sensing robots for LGBTIQ+ elderly, and the need for revised codes of conduct, p. 2)};
\draw[arrow] (sub3.east) -- (q3.west);
\node[draw=none, fit=(sub3)(q3), inner sep=0pt] (pair3) {};

\node[category, anchor=west] (sub4) at ($(pair3.south west)+(0.,-\myspacing-1.5cm)$) {Harms caused by social robot allowed to learn/make own decisions};
\node[quote, anchor=west] (q4) at ($(sub4.east)+(0.5cm,0)$) {“Suppose the care facility uses a brand-new robotic system that has been designed by the manufacturer to be as safe as possible, tested many times and also equipped with self-learning functions that should enable the system to adapt to the special needs and sensitivities of each individual patient as optimally as possible. [...] Hacker H manages to hack the hospital network and manipulate the system’s learning process in such a way that the machine learns to injure patients.” (Translation by the authors)(Hilgendorf 2018 Recht und Ethik in der Pflegerobotik – ein Überblick, p. 380)};
\draw[arrow] (sub4.east) -- (q4.west);
\node[draw=none, fit=(sub4)(q4), inner sep=0pt] (pair4) {};

\node[category, anchor=west] (sub5) at ($(pair4.south west)+(0.,-\myspacing-0.5cm)$) {Harms caused by accidents where social robots and humans live together};
\node[quote, anchor=west] (q5) at ($(sub5.east)+(0.5cm,0)$) {“Of course it is important to make sure the robot does not become an additional fall risk for the users.”(Wiczorek et al. 2020 Domestic Robots for Older Adults: Design Approaches and Recommendations, p. 207; in: Woodcock et al. 2020 Design of Assistive Technology for Ageing Populations)};
\draw[arrow] (sub5.east) -- (q5.west);
\node[draw=none, fit=(sub5)(q5), inner sep=0pt] (pair5) {};

\node[category, anchor=west] (sub6) at ($(pair5.south west)+(0.,-\myspacing-0.5cm)$) {Harms caused by hygiene problems with social robot};
\node[quote, anchor=west] (q6) at ($(sub6.east)+(0.5cm,0)$) {“Studies highlighted that it can be difficult to keep PARO’s fur clean [11, 38], and that the fur covering is not designed to be regularly removed or machine washed, which may post a concern especially to individuals who are immunocompromised [3, 22].” (Hung et al. 2019 The benefits of and barriers to using a social robot PARO in a care setting: a scoping review, p. 6)};
\draw[arrow] (sub6.east) -- (q6.west);
\node[draw=none, fit=(sub6)(q6), inner sep=0pt] (pair6) {};

\node[category, anchor=west] (sub7) at ($(pair6.south west)+(-1.0,-\myspacing)$) {Infringing the right to privacy};
\node[quote, anchor=west] (q7) at ($(sub7.east)+(1.5cm,0)$) {“Health robots and monitoring can entail a deterioration of an elderly person’s quality of life and a disproportionate invasion of their privacy.” (The Swedish National Council on Medical Ethics 2015 Robots and Surveillance in Health Care of the Elderly – Ethical Aspects, p. 447)};
\draw[arrow] (sub7.east) -- (q7.west);
\node[draw=none, fit=(sub7)(q7), inner sep=0pt] (pair7) {};

"
\node[category, anchor=west] (sub8) at ($(pair7.south west)+(0.5,-\myspacing-0.5cm)$) {Limited capacity of social robot to understand what should (not) be shared};
\node[quote, anchor=west] (q8) at ($(sub8.east)+(1.cm,0)$) {“When discussing authority with respect to SAR, privacy is of utmost importance. A robot might not have sufficient capabilities to distinguish between privileged information and information that can be distributed. A robot may also lack the ability to distinguish between individuals who have the authority to receive information about the user and those who do not.”(Feil-Seifer and Mataric 2011 Socially Assistive Robotics - Ethical Issues Related to Technology, p. 29)};
\draw[arrow] (sub8.east) -- (q8.west);
\node[draw=none, fit=(sub8)(q8), inner sep=0pt] (pair8) {};"

\node[category, anchor=north west] (sub9) at ($(pair8.south west)+(0.5,0)$) {\textit{Cont. on next page}};
\node[quote, anchor=west] (q9) at ($(sub9.east)+(1.cm,0)$) {};
\node[draw=none, fit=(sub9)(q9), inner sep=0pt] (pair9) {};

\node[category, anchor=north west] (sub10) at ($(pair9.south west)+(-0.5,0)$) {\textit{Cont. on next page}};
\node[quote, anchor=west] (q10) at ($(sub10.east)+(1.5cm,0)$) {};
\node[draw=none, fit=(sub10)(q10), inner sep=0pt] (pair10) {};

\node[category, anchor=north west] (sub11) at ($(pair10.south west)+(-0.5,0)$) {\textit{Cont. on next page}};
\node[quote, anchor=west] (q11) at ($(sub11.east)+(1.5cm,0)$) {};
\node[draw=none, fit=(sub11)(q11), inner sep=0pt] (pair11) {};

\draw[connector] (cat1.west) |- ([xshift=-2mm]sub1.west);
\draw[connector] (sub1.west) |- ([xshift=-2mm]sub2.west);
\draw[connector] (sub2.west) |- ([xshift=-2mm]sub3.west);
\draw[connector] (sub2.west) |- ([xshift=-2mm]sub4.west);
\draw[connector] (sub2.west) |- ([xshift=-2mm]sub5.west);
\draw[connector] (sub2.west) |- ([xshift=-2mm]sub6.west);
\draw[connector] (cat1.west) |- ([xshift=-2mm]sub7.west);
\draw[connector] (sub7.west) |- ([xshift=-2mm]sub8.west);
\draw[connector] (sub8.west) |- ([xshift=-2mm]sub9.west);
\draw[connector] (sub7.west) |- ([xshift=-2mm]sub10.west);
\draw[connector] (cat1.west) |- ([xshift=-2mm]sub11.west);

\end{tikzpicture}

\clearpage
\begin{tikzpicture}[
  every node/.style={align=left,},
  category/.style={text width=6cm, align=left, inner sep=4pt, font=\small},
  root/.style={text width=10cm, align=left, inner sep=4pt},
  quote/.style={rounded corners, thick, text width=6cm, align=justify, inner sep=4pt, font=\footnotesize},
  arrow/.style={-{Stealth[length=3mm]}, thick},
  connector/.style={thick},
  spacing/.store in=\myspacing, spacing=1.0cm 
]

\node[root] (cat1) at (0, 0) {Ethical hazards for care recipients (continued)};

\node[category, anchor=west] (sub1) at ($(cat1.south west)+(0.5,-\myspacing+0.5cm)$) {Infringing the right to privacy (continued)};
\node[quote, anchor=west] (q1) at ($(sub1.east)+(1.5cm,0)$) {};
\node[draw=none, fit=(sub1)(q1), inner sep=0pt] (pair1) {};

\node[category, anchor=west] (sub2) at ($(pair1.south west)+(0.5,-\myspacing+0.5cm)$) {Limited capacity of social robot to understand what should (not) be shared (continued)};
\node[quote, anchor=west] (q2) at ($(sub2.east)+(1.5cm,0)$) {};
\node[draw=none, fit=(sub2)(q2), inner sep=0pt] (pair2) {};

\node[category, anchor=west] (sub3) at ($(pair2.south west)+(0.5,-\myspacing)$) {Sharing personal information aloud in public places};
\node[quote, anchor=west] (q3) at ($(sub3.east)+(0.5cm,0)$) {“As the user may prefer not to share the content of these messages with other people in the room, carrying out this type of tasks [reading out messages aloud, addition by the authors] should be explicitly required or authorized by users.”(Battistuzzi et al. 2018 Embedding Ethics in the Design of Culturally Competent Socially Assistive Robots, p. 2001)};
\draw[arrow] (sub3.east) -- (q3.west);
\node[draw=none, fit=(sub3)(q3), inner sep=0pt] (pair3) {};

\node[category, anchor=west] (sub4) at ($(pair3.south west)+(-0.5,-\myspacing-0.5cm)$) {Limited possibilities to control data recording/sharing};
\node[quote, anchor=west] (q4) at ($(sub4.east)+(1.cm,0)$) {“Our second concern is that the elderly care receivers may lose control of their data monitored by the robot.” (Li et al. 2020 The Complexity of Autonomy: A Consideration of the Impacts of Care Robots on the Autonomy of Elderly Care Receivers, p. 322; in: Nørskov et al. 2020 Frontiers in Artificial Intelligence and Applications)};
\draw[arrow] (sub4.east) -- (q4.west);
\node[draw=none, fit=(sub4)(q4), inner sep=0pt] (pair4) {};

\node[category, anchor=west] (sub5) at ($(pair4.south west)+(0.5,-\myspacing-1.5cm)$) {Social robot records/operates on/shares distorted data};
\node[quote, anchor=west] (q5) at ($(sub5.east)+(0.5cm,0)$) {“Such a robotic system interacts with the user to learn the user’s preferences, with the aim of providing a better service, and the user or any authorised users should be able to interact with the robotic system locally or remotely. In the process, this poses several unanswered questions pertaining to users’ privacy and data security: What if the input data is manipulated? What if the stored data is either visible to, or corrupted or tampered with by an unauthorised user? What if the sensor data is fraudulent?” (Marchang and Di Nuovo 2022 Assistive Multimodal Robotic System (AMRSys): Security and Privacy Issues, Challenges, and Possible Solutions, p. 2)};
\draw[arrow] (sub5.east) -- (q5.west);
\node[draw=none, fit=(sub5)(q5), inner sep=0pt] (pair5) {};

\node[category, anchor=west] (sub6) at ($(pair5.south west)+(-0.5,-\myspacing-0.5cm)$) {Data collection or virtual connection in moments of intimacy};
\node[quote, anchor=west] (q6) at ($(sub6.east)+(1.cm,0)$) {“An operator could drive a robot to peer round an elder’s apartment before they were dressed or when they are taking a bath. An autonomous robot could record in the same circumstances.”(Sharkey and Sharkey 2014 The Rights and Wrongs of Robot Care, p. 271; in: Lin et al. 2012 Robot Ethics: The Ethical and Social Implications of Robotics)};
\draw[arrow] (sub6.east) -- (q6.west);
\node[draw=none, fit=(sub6)(q6), inner sep=0pt] (pair6) {};

\node[category, anchor=west] (sub7) at ($(pair6.south west)+(0.,-\myspacing-0.2cm)$) {Feeling of being constantly controlled/followed by social robot };
\node[quote, anchor=west] (q7) at ($(sub7.east)+(1.cm,0)$) {“The other concern relative to the robot’s proactivity is the discomfort caused by the feeling of being permanently observed by a machine.” (Gelin 2017 The Domestic Robot: Ethical and Technical Concerns, p. 210; in: Ferreira et al. 2017 A World with Robots - Intelligent Systems, Control and Automation: Science and Engineering) };
\draw[arrow] (sub7.east) -- (q7.west);
\node[draw=none, fit=(sub7)(q7), inner sep=0pt] (pair7) {};

\node[category, anchor=west] (sub8) at ($(pair7.south west)+(0.,-\myspacing-0.5cm)$) {Not being aware that data is collected};
\node[quote, anchor=west] (q8) at ($(sub8.east)+(1.cm,0)$) {“A person with Alzheimer’s may soon forget that a robot is present and might perform acts or say things believing he is in the privacy of his own home, or thinking that he is alone with his robot friend.”(Sharkey and Sharkey 2014 The Rights and Wrongs of Robot Care, p. 271; in: Lin et al. 2012 Robot Ethics: The Ethical and Social Implications of Robotics)};
\draw[arrow] (sub8.east) -- (q8.west);
\node[draw=none, fit=(sub8)(q8), inner sep=0pt] (pair8) {};

\node[category, anchor=north west] (sub9) at ($(pair8.south west)+(0.0,0)$) {\textit{Cont. on next page}};
\node[quote, anchor=west] (q9) at ($(sub9.east)+(1.cm,0)$) {};
\node[draw=none, fit=(sub9)(q9), inner sep=0pt] (pair9) {};

\node[category, anchor=north west] (sub10) at ($(pair9.south west)+(-0.5,0)$) {\textit{Cont. on next page}};
\node[quote, anchor=west] (q10) at ($(sub10.east)+(1.5cm,0)$) {};
\node[draw=none, fit=(sub10)(q10), inner sep=0pt] (pair10) {};

\draw[connector] (cat1.west) |- ([xshift=-2mm]sub1.west);
\draw[connector] (sub1.west) |- ([xshift=-2mm]sub2.west);
\draw[connector] (sub2.west) |- ([xshift=-2mm]sub3.west);
\draw[connector] (sub1.west) |- ([xshift=-2mm]sub4.west);
\draw[connector] (sub4.west) |- ([xshift=-2mm]sub5.west);
\draw[connector] (sub1.west) |- ([xshift=-2mm]sub6.west);
\draw[connector] (sub1.west) |- ([xshift=-2mm]sub7.west);
\draw[connector] (sub1.west) |- ([xshift=-2mm]sub8.west);
\draw[connector] (sub1.west) |- ([xshift=-2mm]sub9.west);
\draw[connector] (cat1.west) |- ([xshift=-2mm]sub10.west);

\end{tikzpicture}

\clearpage
\begin{tikzpicture}[
  every node/.style={align=left,},
  category/.style={text width=6cm, align=left, inner sep=4pt, font=\small},
  root/.style={text width=10cm, align=left, inner sep=4pt},
  quote/.style={rounded corners, thick, text width=6cm, align=justify, inner sep=4pt, font=\footnotesize},
  arrow/.style={-{Stealth[length=3mm]}, thick},
  connector/.style={thick},
  spacing/.store in=\myspacing, spacing=1.0cm 
]

\node[root] (cat1) at (0, 0) {Ethical hazards for care recipients (continued)};

\node[category, anchor=west] (sub1) at ($(cat1.south west)+(0.5,-\myspacing+0.5cm)$) {Infringing the right to privacy (continued)};
\node[quote, anchor=west] (q1) at ($(sub1.east)+(1.5cm,0)$) {};
\node[draw=none, fit=(sub1)(q1), inner sep=0pt] (pair1) {};

\node[category, anchor=west] (sub2) at ($(pair1.south west)+(0.5,-\myspacing)$) {Data security/improper access to data};
\node[quote, anchor=west] (q2) at ($(sub2.east)+(1.cm,0)$) {“Furthermore, intimacy and privacy and the related informational autonomy are important issues. Robots usually have cameras and sensors of all kinds, especially when they are mobile, and can thus, in principle, photograph and monitor patients in need of care. Personal data in this area is very sensitive and must be protected. Self-learning systems pose further problems. The AI-enhanced robot could create detailed profiles that would be interesting for nefarious groups, for example for coercion and extortion.”(Bendel 2020 Care Robots with Sexual Assistance Functions, p. 5)};
\draw[arrow] (sub2.east) -- (q2.west);
\node[draw=none, fit=(sub2)(q2), inner sep=0pt] (pair2) {};

\node[category, anchor=west] (sub3) at ($(pair2.south west)+(-0.5,-\myspacing-0.5cm)$) {Promoting discrimination/negative images of care recipient};
\node[quote, anchor=west] (q3) at ($(sub3.east)+(1.5cm,0)$) {“Past research indicates that having a robot that looks like a toy such as My Real Baby (a robotic doll that becomes happy if you bounce it but gets distraught if you bounce it too much) may be stigmatising, patronising and infantilizing.”(Frennert and Östlund 2014 Review: Seven Matters of Concern of Social Robots and Older People, p. 306)};
\draw[arrow] (sub3.east) -- (q3.west);
\node[draw=none, fit=(sub3)(q3), inner sep=0pt] (pair3) {};

\node[category, anchor=west] (sub4) at ($(pair3.south west)+(0.5,-\myspacing-0.5cm)$) {Social robot might discriminate certain groups (e. g. hearing impaired)};
\node[quote, anchor=west] (q4) at ($(sub4.east)+(1.cm,0)$) {“Another conceptual study examining various legal-ethical issues of mobile servant robots raised the possibility of robots discriminating against older people, especially those with speech and hearing impairments who could not actively summon the robots to assist them (Fosch-Villaronga and Virk, 2017).”(Tan et al. 2021 Tensions and antagonistic interactions of risks and ethics of using robotics and autonomous systems in long-term care, p. 10)};
\draw[arrow] (sub4.east) -- (q4.west);
\node[draw=none, fit=(sub4)(q4), inner sep=0pt] (pair4) {};

\node[category, anchor=west] (sub5) at ($(pair4.south west)+(0.,-\myspacing)$) {Furthering image of care recipient as dependent};
\node[quote, anchor=west] (q5) at ($(sub5.east)+(1.cm,0)$) {“Ethical issues included the […] patients’ stigmatisation as being frail and dependent when using robotic devices,25, 28, 32 […] .” (Servaty et al. 2020 Implementation of robotic devices in nursing care. Barriers and facilitators: an integrative review, p. 9) };
\draw[arrow] (sub5.east) -- (q5.west);
\node[draw=none, fit=(sub5)(q5), inner sep=0pt] (pair5) {};

\node[category, anchor=west] (sub6) at ($(pair5.south west)+(0.,-\myspacing-0.5cm)$) {Furthering objectification of care recipients};
\node[quote, anchor=west] (q6) at ($(sub6.east)+(1.cm,0)$) {“Regarding social robots in care, the objectification of older people (Decker, 2008; Sparrow \& Sparrow, 2006) might occur when residents are provided with robots due to the needs of the care organization or other people instead of the older people themselves.”(Pirhonen et al. 2020 Can robots tackle late-life loneliness? Scanning of future opportunities and challenges in assisted living facilities, p. 9)};
\draw[arrow] (sub6.east) -- (q6.west);
\node[draw=none, fit=(sub6)(q6), inner sep=0pt] (pair6) {};

\node[category, anchor=west] (sub7) at ($(pair6.south west)+(0.,-\myspacing)$) {Further infantilization and deficit-orientation};
\node[quote, anchor=west] (q7) at ($(sub7.east)+(1.cm,0)$) {“Similarly, interaction with toy robots will have the effects of infantilizing old people.”(Noori et al. 2019 Robot-Care for the Older People: Ethically Justified or Not?, p. 46)};
\draw[arrow] (sub7.east) -- (q7.west);
\node[draw=none, fit=(sub7)(q7), inner sep=0pt] (pair7) {};

\node[category, anchor=west] (sub8) at ($(pair7.south west)+(-0.5,-\myspacing-0.5cm)$) {Increasing dependency in everyday life};
\node[quote, anchor=west] (q8) at ($(sub8.east)+(1.5cm,0)$) {“Finally, scholars point out that dependence and lack of competencies can be consequences of robots that provide too much help or assistance.” (Salvini 2015 On Ethical, Legal and Social Issues of Care Robots, p. 440; in: Mohammed 2015 Intelligent Assistive Robots - Recent Advances in Assistive Robotics for Everyday Activities)};
\draw[arrow] (sub8.east) -- (q8.west);
\node[draw=none, fit=(sub8)(q8), inner sep=0pt] (pair8) {};

\node[category, anchor=north west] (sub9) at ($(pair8.south west)+(0.5,0)$) {\textit{Cont. on next page}};
\node[quote, anchor=west] (q9) at ($(sub9.east)+(1.cm,0)$) {};
\node[draw=none, fit=(sub9)(q9), inner sep=0pt] (pair9) {};

\node[category, anchor=north west] (sub10) at ($(pair9.south west)+(-0.5,0)$) {\textit{Cont. on next page}};
\node[quote, anchor=west] (q10) at ($(sub10.east)+(1.5cm,0)$) {};
\node[draw=none, fit=(sub10)(q10), inner sep=0pt] (pair10) {};

\draw[connector] (cat1.west) |- ([xshift=-2mm]sub1.west);
\draw[connector] (sub1.west) |- ([xshift=-2mm]sub2.west);
\draw[connector] (cat1.west) |- ([xshift=-2mm]sub3.west);
\draw[connector] (sub3.west) |- ([xshift=-2mm]sub4.west);
\draw[connector] (sub3.west) |- ([xshift=-2mm]sub5.west);
\draw[connector] (sub3.west) |- ([xshift=-2mm]sub6.west);
\draw[connector] (sub3.west) |- ([xshift=-2mm]sub7.west);
\draw[connector] (cat1.west) |- ([xshift=-2mm]sub8.west);
\draw[connector] (sub8.west) |- ([xshift=-2mm]sub9.west);
\draw[connector] (cat1.west) |- ([xshift=-2mm]sub10.west);

\end{tikzpicture}

\clearpage
\begin{tikzpicture}[
  every node/.style={align=left,},
  category/.style={text width=6cm, align=left, inner sep=4pt, font=\small},
  root/.style={text width=10cm, align=left, inner sep=4pt},
  quote/.style={rounded corners, thick, text width=6cm, align=justify, inner sep=4pt, font=\footnotesize},
  arrow/.style={-{Stealth[length=3mm]}, thick},
  connector/.style={thick},
  spacing/.store in=\myspacing, spacing=1.0cm 
]

\node[root] (cat1) at (0, 0) {Ethical hazards for care recipients (continued)};

\node[category, anchor=west] (sub1) at ($(cat1.south west)+(0.5,-\myspacing+0.5cm)$) {Increasing dependency in everyday life (continued)};
\node[quote, anchor=west] (q1) at ($(sub1.east)+(1.5cm,0)$) {};
\node[draw=none, fit=(sub1)(q1), inner sep=0pt] (pair1) {};

\node[category, anchor=west] (sub2) at ($(pair1.south west)+(0.5,-\myspacing)$) {Accelerate admission to care homes};
\node[quote, anchor=west] (q2) at ($(sub2.east)+(1.cm,0)$) {“A tangible fear was that allowing others, including robots, to take over certain tasks and activities could result in a deterioration of mental function and the erosion of independence, potentially representing a step on the road to residential aged care.”(Coghlan et al. 2021 Dignity, Autonomy, and Style of Company: Dimensions Older Adults Consider for Robot Companions, p. 16)};
\draw[arrow] (sub2.east) -- (q2.west);
\node[draw=none, fit=(sub2)(q2), inner sep=0pt] (pair2) {};

\node[category, anchor=west] (sub3) at ($(pair2.south west)+(0.,-\myspacing-0.5cm)$) {Becoming dependent on technology};
\node[quote, anchor=west] (q3) at ($(sub3.east)+(1.cm,0)$) {“After receiving excessive assistance by delegating tasks to robots, the care receivers may become so dependent on robots concerning daily tasks they could have done by themselves that this gradually leads to deterioration of bodily functions […].”(Li et al. 2020 The Complexity of Autonomy: A Consideration of the Impacts of Care Robots on the Autonomy of Elderly Care Receivers, p. 323; in: Nørskov et al. 2020 Frontiers in Artificial Intelligence and Applications)};
\draw[arrow] (sub3.east) -- (q3.west);
\node[draw=none, fit=(sub3)(q3), inner sep=0pt] (pair3) {};

\node[category, anchor=west] (sub4) at ($(pair3.south west)+(0.5,-\myspacing-0.5cm)$) {Emotional dependency};
\node[quote, anchor=west] (q4) at ($(sub4.east)+(0.5cm,0)$) {“Attachment to and dependency on the robot: Robot users may get emotionally attached to their robots. While such attachment is desirable for increasing robot acceptance, it also has the potential to be emotionally harmful [29, 33].”(Casey et al. 2016 What People with Dementia Want: Designing MARIO an Acceptable Robot Companion, p. 321)};
\node[draw=none, fit=(sub4)(q4), inner sep=0pt] (pair4) {};

\node[category, anchor=west] (sub5) at ($(pair4.south west)+(0.,-\myspacing-0.5cm)$) {Dependent on maintenance and support};
\node[quote, anchor=west] (q5) at ($(sub5.east)+(0.5cm,0)$) {“We are vulnerable whenever our assistive device does not fulfill its function and in regard to this vulnerability we are dependent on the people responsible for the functioning and maintenance of our assistive device.”(Wiertz 2020 Trusting robots? – On the concept of trust and on forms of human vulnerability, p. 64; in: Haltaufderheide et al. 2020 Aging between Participation and Simulation - Ethical Dimensions of Social Assistive Technologies)};
\draw[arrow] (sub5.east) -- (q5.west);
\node[draw=none, fit=(sub5)(q5), inner sep=0pt] (pair5) {};

\node[category, anchor=west] (sub6) at ($(pair5.south west)+(-0.5,-\myspacing-0.5cm)$) {Causing loss of further capacities (as not practiced anymore)};
\node[quote, anchor=west] (q6) at ($(sub6.east)+(1.cm,0)$) {“Some of the elders inquired have expressed their concern that an excessively prestative robot could lead seniors to rely on robots for everything, even for the tasks they could perform well causing them to lose more quickly their remaining physical abilities […].”(Gelin 2017 The Domestic Robot: Ethical and Technical Concerns, pp. 211, 212; aus Ferreira et al. 2017 A World with Robots - Intelligent Systems, Control and Automation: Science and Engineering)};
\draw[arrow] (sub6.east) -- (q6.west);
\node[draw=none, fit=(sub6)(q6), inner sep=0pt] (pair6) {};

\node[category, anchor=west] (sub7) at ($(pair6.south west)+(-0.5,-\myspacing-0.5cm)$) {Creating barriers for autonomous and informed decision-making};
\node[quote, anchor=west] (q7) at ($(sub7.east)+(1.5cm,0)$) {“The use of robots in care facilities must also be viewed critically, as many residents are no longer in a position to give their voluntary and informed consent to such measures, for example because they suffer from mental impairments.” (Translation by the authors)(Misselhorn 2018 Pflegesysteme, p. 153; in: Misselhorn 2018 Grundfragen der Maschinenethik)};
\draw[arrow] (sub7.east) -- (q7.west);
\node[draw=none, fit=(sub7)(q7), inner sep=0pt] (pair7) {};

\node[category, anchor=west] (sub8) at ($(pair7.south west)+(0.5,-\myspacing-0.5cm)$) {Social robot is designed in a way that limits choices/control };
\node[quote, anchor=west] (q8) at ($(sub8.east)+(1.cm,0)$) {“A focus group study [20] in Belgium shows that elderly care receivers worry if they may partly lose their autonomy when robots are used in care. They insist that they should be able to turn off the robots when robots infringe on their autonomy.”(Li et al. 2020 The Complexity of Autonomy: A Consideration of the Impacts of Care Robots on the Autonomy of Elderly Care Receivers, p. 318; in: Nørskov et al. 2020 Frontiers in Artificial Intelligence and Applications)};
\draw[arrow] (sub8.east) -- (q8.west);
\node[draw=none, fit=(sub8)(q8), inner sep=0pt] (pair8) {};"

\node[category, anchor=north west] (sub9) at ($(pair8.south west)+(0.0,0)$) {\textit{Cont. on next page}};
\node[quote, anchor=west] (q9) at ($(sub9.east)+(1.cm,0)$) {};
\node[draw=none, fit=(sub9)(q9), inner sep=0pt] (pair9) {};

\node[category, anchor=north west] (sub10) at ($(pair9.south west)+(-0.5,0)$) {\textit{Cont. on next page}};
\node[quote, anchor=west] (q10) at ($(sub10.east)+(1.5cm,0)$) {};
\node[draw=none, fit=(sub10)(q10), inner sep=0pt] (pair10) {};

\draw[connector] (cat1.west) |- ([xshift=-2mm]sub1.west);
\draw[connector] (sub1.west) |- ([xshift=-2mm]sub2.west);
\draw[connector] (sub1.west) |- ([xshift=-2mm]sub3.west);
\draw[connector] (sub3.west) |- ([xshift=-2mm]sub4.west);
\draw[connector] (sub3.west) |- ([xshift=-2mm]sub5.west);
\draw[connector] (sub1.west) |- ([xshift=-2mm]sub6.west);
\draw[connector] (cat1.west) |- ([xshift=-2mm]sub7.west);
\draw[connector] (sub7.west) |- ([xshift=-2mm]sub8.west);
\draw[connector] (sub7.west) |- ([xshift=-2mm]sub9.west);
\draw[connector] (cat1.west) |- ([xshift=-2mm]sub10.west);

\end{tikzpicture}

\clearpage
\begin{tikzpicture}[
  every node/.style={align=left,},
  category/.style={text width=6cm, align=left, inner sep=4pt, font=\small},
  root/.style={text width=10cm, align=left, inner sep=4pt},
  quote/.style={rounded corners, thick, text width=6cm, align=justify, inner sep=4pt, font=\footnotesize},
  arrow/.style={-{Stealth[length=3mm]}, thick},
  connector/.style={thick},
  spacing/.store in=\myspacing, spacing=1.0cm 
]

\node[root] (cat1) at (0, 0) {Ethical hazards for care recipients (continued)};

\node[category, anchor=west] (sub1) at ($(cat1.south west)+(0.5,-\myspacing+0.5cm)$) {Creating barriers for autonomous and informed decision-making (continued)};
\node[quote, anchor=west] (q1) at ($(sub1.east)+(1.5cm,0)$) {};
\node[draw=none, fit=(sub1)(q1), inner sep=0pt] (pair1) {};

\node[category, anchor=west] (sub2) at ($(pair1.south west)+(0.5,-\myspacing)$) {Exposure to manipulation or feeling pressured to act by social robot};
\node[quote, anchor=west] (q2) at ($(sub2.east)+(1.cm,0)$) {“That is, the social nature of the interaction depends crucially on the user being tricked or fooled into ascribing emotional states to the robot seal which do not pertain to it. There is an emotional manipulation of the user by the designer of the robot.11 This is an offense against the person’s rational nature, since the workings of this mechanism depend on the suspension of rational reflection and control.”(Misselhorn et al. 2013 Ethical Considerations Regarding the Use of Social Robots in the Fourth Age, p. 129)};
\draw[arrow] (sub2.east) -- (q2.west);
\node[draw=none, fit=(sub2)(q2), inner sep=0pt] (pair2) {};

\node[category, anchor=west] (sub3) at ($(pair2.south west)+(0.,-\myspacing-2.5cm)$) {Improper delegation of decision-making authority to social robot};
\node[quote, anchor=west] (q3) at ($(sub3.east)+(1.cm,0)$) {“Such loss of freedom is related to the issue of autonomy raised in the preceding section. If robots are to be used to help elderly people to live safely in their own homes for longer, there is likely to be a need for them to function as autonomous supervisors. A simple extension would be to allow home customisation so that the robot can recognise danger signs, such as the cooker being left on, or the bath overflowing. A robot could make a first pass at warning a senior to stop doing or engaging in a potentially dangerous activity. But there are ethical problems here. If a robot can predict a dangerous situation, it could be programmed to autonomously take steps to prevent it, e.g. turning the cooker off, or even restraining the elderly person (gently) from carrying out a potentially dangerous action such as climbing up on a chair to get something from a cupboard. However, restraining a person to avoid harm could be a slippery slope towards authoritarian robotics.”(Sharkey and Sharkey 2012 Granny and the robots: ethical issues in robot care for the elderly, p. 33)};
\draw[arrow] (sub3.east) -- (q3.west);
\node[draw=none, fit=(sub3)(q3), inner sep=0pt] (pair3) {};

\node[category, anchor=west] (sub4) at ($(pair3.south west)+(0.,-\myspacing-0.5cm)$) {Care recipients might be pressured into using social robot};
\node[quote, anchor=west] (q4) at ($(sub4.east)+(1.cm,0)$) {“Similar external pressures should also be prevented at the level of domestic or residential use of robots. In particular, scenarios where family members or other informal caregivers force a patient with capacity to consent to have a service robot in the house—e.g. because they want to reduce their time-investment and caregiving-workload should be prevented.”(Ienca et al. 2016 Social and Assistive Robotics in Dementia Care: Ethical Recommendations for Research and Practice, p. 568)};
\draw[arrow] (sub4.east) -- (q4.west);
\node[draw=none, fit=(sub4)(q4), inner sep=0pt] (pair4) {};

\node[category, anchor=west] (sub5) at ($(pair4.south west)+(0.,-\myspacing-0.5cm)$) {Care recipients might not be adequately informed about social robots};
\node[quote, anchor=west] (q5) at ($(sub5.east)+(1.cm,0)$) {“Finally, the authority that explains the robot and conveys the allegedly neutral information assumes a position of power. This creates a danger of manipulation of the users. The person who explains the robot can decide to share the information they feel is important about the robot and withhold other information.”(Weßel et al. 2021 Gender Stereotyping of Robotic Systems in Elderly Care An Exploratory Analysis of Ethical Problems and Possible Solutions, p. 7)};
\draw[arrow] (sub5.east) -- (q5.west);
\node[draw=none, fit=(sub5)(q5), inner sep=0pt] (pair5) {};

\node[category, anchor=west] (sub6) at ($(pair5.south west)+(0.5,-\myspacing)$) {Social robots add an extra layer of complexity};
\node[quote, anchor=west] (q6) at ($(sub6.east)+(0.5cm,0)$) {“PARO was found to not have the option to easily turn off because its hidden switch between the split tail fins; older people with dementia did not know how to turn the robot on or off [43].” (Hung et al. 2019 The benefits of and barriers to using a social robot PARO in a care setting: a scoping review, p. 7)};
\draw[arrow] (sub6.east) -- (q6.west);
\node[draw=none, fit=(sub6)(q6), inner sep=0pt] (pair6) {};

\node[category, anchor=west] (sub7) at ($(pair6.south west)+(0.,-\myspacing-0.5cm)$) {KI-informed care approaches stay opaque };
\node[quote, anchor=west] (q7) at ($(sub7.east)+(0.5cm,0)$) {“…the technological complexity of several IATs makes it harder for patients and/or their proxy decision makers (e.g. caregivers) to adequately understand what they are consenting to.” (Wangmo et al. 2019 Ethical concerns with the use of intelligent assistive technology: findings from a qualitative study with professional stakeholders, p. 8)};
\draw[arrow] (sub7.east) -- (q7.west);
\node[draw=none, fit=(sub7)(q7), inner sep=0pt] (pair7) {};

\node[category, anchor=north west] (sub8) at ($(pair7.south west)+(-0.5,0)$) {\textit{Cont. on next page}};
\node[quote, anchor=west] (q8) at ($(sub8.east)+(1.cm,0)$) {};
\node[draw=none, fit=(sub8)(q8), inner sep=0pt] (pair8) {};

\node[category, anchor=north west] (sub9) at ($(pair8.south west)+(-0.5,0)$) {\textit{Cont. on next page}};
\node[quote, anchor=west] (q9) at ($(sub9.east)+(1.cm,0)$) {};
\node[draw=none, fit=(sub9)(q9), inner sep=0pt] (pair9) {};

\draw[connector] (cat1.west) |- ([xshift=-2mm]sub1.west);
\draw[connector] (sub1.west) |- ([xshift=-2mm]sub2.west);
\draw[connector] (sub1.west) |- ([xshift=-2mm]sub3.west);
\draw[connector] (sub1.west) |- ([xshift=-2mm]sub4.west);
\draw[connector] (sub1.west) |- ([xshift=-2mm]sub5.west);
\draw[connector] (sub5.west) |- ([xshift=-2mm]sub6.west);
\draw[connector] (sub5.west) |- ([xshift=-2mm]sub7.west);
\draw[connector] (sub1.west) |- ([xshift=-2mm]sub8.west);
\draw[connector] (cat1.west) |- ([xshift=-2mm]sub9.west);

\end{tikzpicture}

\clearpage
\begin{tikzpicture}[
  every node/.style={align=left,},
  category/.style={text width=6cm, align=left, inner sep=4pt, font=\small},
  root/.style={text width=10cm, align=left, inner sep=4pt},
  quote/.style={rounded corners, thick, text width=6cm, align=justify, inner sep=4pt, font=\footnotesize},
  arrow/.style={-{Stealth[length=3mm]}, thick},
  connector/.style={thick},
  spacing/.store in=\myspacing, spacing=1.0cm 
]

\node[root] (cat1) at (0, 0) {Ethical hazards for care recipients (continued)};

\node[category, anchor=west] (sub1) at ($(cat1.south west)+(0.5,-\myspacing+0.5cm)$) {Creating barriers for autonomous and informed decision-making (continued)};
\node[quote, anchor=west] (q1) at ($(sub1.east)+(1.5cm,0)$) {};
\node[draw=none, fit=(sub1)(q1), inner sep=0pt] (pair1) {};

\node[category, anchor=west] (sub2) at ($(pair1.south west)+(0.5,-\myspacing)$) {Design (e. g. anthropomorphic/female) leads to over-trust};
\node[quote, anchor=west] (q2) at ($(sub2.east)+(1.cm,0)$) {“For instance, deception could lead to over-trust the robot –thus endangering the user’s safety or reasonable decision-making processes–, […].”(Boada et al. 2021 The ethical issues of social assistive robotics: A critical literature review, p. 6)};
\draw[arrow] (sub2.east) -- (q2.west);
\node[draw=none, fit=(sub2)(q2), inner sep=0pt] (pair2) {};

\node[category, anchor=west] (sub3) at ($(pair2.south west)+(-0.5,-\myspacing-1cm)$) {Increasing social isolation};
\node[quote, anchor=west] (q3) at ($(sub3.east)+(1.5cm,0)$) {“Another argument put forward against the use of Paro is that, contrary to its intended use, Paro could cause or promote isolation if it were to replace, rather than supplement, human contact.” (Translation by the authors)(Lange and Bauer 2021 - Eine Robbe für Oma - Die zukünftige Dauerausstellung Robotik im Deutschen Museum, p. 232; in: Inthorn und Seising 2021 Digitale Patientenversorgung - Zur Computerisierung von Diagnostik, Therapie und Pflege)};
\draw[arrow] (sub3.east) -- (q3.west);
\node[draw=none, fit=(sub3)(q3), inner sep=0pt] (pair3) {};

\node[category, anchor=west] (sub4) at ($(pair3.south west)+(0.5,-\myspacing-1cm)$) {Social robot fosters social seclusion of care recipient (via dependent relation)};
\node[quote, anchor=west] (q4) at ($(sub4.east)+(1.cm,0)$) {“While many studies support the idea of a preference for human-versus-robotic interaction, De Graaf57 expressed the concern that the human–robot relationship may replace their human counter-part in social relations and predicted that if people share feelings with robotic others, they may become accustomed to the reduced emotional range that these machines can offer.”(Diaz-Orueta et al. 2020 Shaping technologies for older adults with dementia: Reflection on ethics and preferences, p. 3222)};
\draw[arrow] (sub4.east) -- (q4.west);
\node[draw=none, fit=(sub4)(q4), inner sep=0pt] (pair4) {};

\node[category, anchor=west] (sub5) at ($(pair4.south west)+(0.,-\myspacing)$) {Reduction of physical visits};
\node[quote, anchor=west] (q5) at ($(sub5.east)+(1.cm,0)$) {“There is evidence that robots can reduce visits by care staff at home (Broadbent 2017).” (Translation by the authors)(Bleuler and Caroni 2021 Roboter in der Pflege - Welche Aufgaben können Roboter heute schon übernehmen?, p. 445; in: Bendel 2021 Soziale Roboter)};
\draw[arrow] (sub5.east) -- (q5.west);
\node[draw=none, fit=(sub5)(q5), inner sep=0pt] (pair5) {};

\node[category, anchor=west] (sub6) at ($(pair5.south west)+(0.5,-\myspacing-0.5cm)$) {Uncomfortable feelings around social robot discourage visits};
\node[quote, anchor=west] (q6) at ($(sub6.east)+(0.5cm,0)$) {“N=15 would not like being received by a service robot as a relative.” (Translation by the authors)(Meyer 2011 Akzeptanz ausgewählter Anwendungsszenarien, p. 96; in: Meyer 2011 Mein Freund der Roboter - Servicerobotik für ältere Menschen - eine Antwort auf den demographischen Wandel?)};
\draw[arrow] (sub6.east) -- (q6.west);
\node[draw=none, fit=(sub6)(q6), inner sep=0pt] (pair6) {};

\node[category, anchor=west] (sub7) at ($(pair6.south west)+(0.,-\myspacing)$) {Don’t worry about Granny, she’s got the social robot to talk to};
\node[quote, anchor=west] (q7) at ($(sub7.east)+(0.5cm,0)$) {“The robot might indeed be a good excuse to cancel some of Anna’s visits, as she feels less guilty because her grandmother has a ‘companionship’.” (Felber et al. 2022 The concept of social dignity as a yardstick to delimit ethical use of robotic assistance in the care of older persons, p. 105)};
\draw[arrow] (sub7.east) -- (q7.west);
\node[draw=none, fit=(sub7)(q7), inner sep=0pt] (pair7) {};

\node[category, anchor=west] (sub8) at ($(pair7.south west)+(0.,-\myspacing)$) {Less control visits necessary and therefore visitors};
\node[quote, anchor=west] (q8) at ($(sub8.east)+(0.5cm,0)$) {“If care robots can be used to monitor the activities of daily living of the older people by their dependents who live apart, less social visits could be an unintended consequence in the process of leveraging the technology.”(Tan et al. 2021 Tensions and antagonistic interactions of risks and ethics of using robotics and autonomous systems in long-term care, p. 11)};
\draw[arrow] (sub8.east) -- (q8.west);
\node[draw=none, fit=(sub8)(q8), inner sep=0pt] (pair8) {};

\node[category, anchor=north west] (sub9) at ($(pair8.south west)+(0.0,0)$) {\textit{Cont. on next page}};
\node[quote, anchor=west] (q9) at ($(sub9.east)+(1.cm,0)$) {};
\node[draw=none, fit=(sub9)(q9), inner sep=0pt] (pair9) {};

\node[category, anchor=north west] (sub10) at ($(pair9.south west)+(-0.5,0)$) {\textit{Cont. on next page}};
\node[quote, anchor=west] (q10) at ($(sub10.east)+(1.5cm,0)$) {};
\node[draw=none, fit=(sub10)(q10), inner sep=0pt] (pair10) {};

\node[category, anchor=north west] (sub11) at ($(pair10.south west)+(-0.5,0)$) {\textit{Cont. on next page}};
\node[quote, anchor=west] (q11) at ($(sub11.east)+(1.5cm,0)$) {};
\node[draw=none, fit=(sub11)(q11), inner sep=0pt] (pair11) {};

\draw[connector] (cat1.west) |- ([xshift=-2mm]sub1.west);
\draw[connector] (sub1.west) |- ([xshift=-2mm]sub2.west);
\draw[connector] (cat1.west) |- ([xshift=-2mm]sub3.west);
\draw[connector] (sub3.west) |- ([xshift=-2mm]sub4.west);
\draw[connector] (sub3.west) |- ([xshift=-2mm]sub5.west);
\draw[connector] (sub5.west) |- ([xshift=-2mm]sub6.west);
\draw[connector] (sub5.west) |- ([xshift=-2mm]sub7.west);
\draw[connector] (sub5.west) |- ([xshift=-2mm]sub8.west);
\draw[connector] (sub5.west) |- ([xshift=-2mm]sub9.west);
\draw[connector] (sub3.west) |- ([xshift=-2mm]sub10.west);
\draw[connector] (cat1.west) |- ([xshift=-2mm]sub11.west);
\end{tikzpicture}

\clearpage
\begin{tikzpicture}[
  every node/.style={align=left,},
  category/.style={text width=6cm, align=left, inner sep=4pt, font=\small},
  root/.style={text width=10cm, align=left, inner sep=4pt},
  quote/.style={rounded corners, thick, text width=6cm, align=justify, inner sep=4pt, font=\footnotesize},
  arrow/.style={-{Stealth[length=3mm]}, thick},
  connector/.style={thick},
  spacing/.store in=\myspacing, spacing=1.0cm 
]

\node[root] (cat1) at (0, 0) {Ethical hazards for care recipients (continued)};

\node[category, anchor=west] (sub1) at ($(cat1.south west)+(0.5,-\myspacing+0.5cm)$) {Increasing social isolation (continued)};
\node[quote, anchor=west] (q1) at ($(sub1.east)+(1.5cm,0)$) {};
\node[draw=none, fit=(sub1)(q1), inner sep=0pt] (pair1) {};

\node[category, anchor=west] (sub2) at ($(pair1.south west)+(0.5,-\myspacing+0.5cm)$) {Reduction of physical visits (continued)};
\node[quote, anchor=west] (q2) at ($(sub2.east)+(1.5cm,0)$) {};
\node[draw=none, fit=(sub2)(q2), inner sep=0pt] (pair2) {};

\node[category, anchor=west] (sub3) at ($(pair2.south west)+(0.5,-\myspacing-0.5cm)$) {Physical visits are substituted by virtual visits};
\node[quote, anchor=west] (q3) at ($(sub3.east)+(0.5cm,0)$) {“For example, maybe elderly care receivers will not see their family and friends as frequently when they can virtually meet via the robot.” (Li et al. 2020 The Complexity of Autonomy: A Consideration of the Impacts of Care Robots on the Autonomy of Elderly Care Receivers, p. 322; in: Nørskov et al. 2020 Frontiers in Artificial Intelligence and Applications)};
\draw[arrow] (sub3.east) -- (q3.west);
\node[draw=none, fit=(sub3)(q3), inner sep=0pt] (pair3) {};

\node[category, anchor=west] (sub4) at ($(pair3.south west)+(-0.5,-\myspacing-1.5cm)$) {Reduction of human attention/contact resp. “real”/emotional component of care in care relationships};
\node[quote, anchor=west] (q4) at ($(sub4.east)+(1.cm,0)$) {“The concern that humanity could be reduced in care and nursing is nevertheless not entirely unjustified. On the one hand, working with the robotic seal Paro actually shows that people with dementia no longer need much additional care during the time with Paro on their lap because they are busy with the seal, and on the other hand, carers - especially in hospitals - are usually tied to a tight schedule that leaves little room for activity and conversation (Newerla 2017).” (Translation by the authors)(Schmidhuber 2022 Werden Roboter Menschen in der Pflege ersetzen? Ethische Überlegungen, pp. 169, 170; in: Stronegger und Platzer 2022 Technisierung der Pflege: 4. Goldegger Dialogforum Mensch und Endlichkeit) };
\draw[arrow] (sub4.east) -- (q4.west);
\node[draw=none, fit=(sub4)(q4), inner sep=0pt] (pair4) {};

\node[category, anchor=west] (sub5) at ($(pair4.south west)+(0.5,-\myspacing)$) {(Sex)robots can’t give the same love like humans};
\node[quote, anchor=west] (q5) at ($(sub5.east)+(0.5cm,0)$) {“Jecker’s illusions about sex robots originate from a failure to recognise that what ‘sex’ robots facilitate is masturbation, not intercourse. Sex robots are sex toys, not partners.”(Sparrow 2021 Sex robot fantasies, p. 33)};
\draw[arrow] (sub5.east) -- (q5.west);
\node[draw=none, fit=(sub5)(q5), inner sep=0pt] (pair5) {};

\node[category, anchor=west] (sub6) at ($(pair5.south west)+(-0.5,-\myspacing-0.5cm)$) {Promotes distance from non-robot-using humans (and vice versa)};
\node[quote, anchor=west] (q6) at ($(sub6.east)+(1.cm,0)$) {“Of course at the same time, the robot may encourage people to disengage from the non-robotic environment and to take distance from others who are not involved in the robotic adventure. There is likely to be a distance between those who interact with the robot and talk about among themselves, and those who are excluded from the interaction and robot-mediated and robot-focused conversation.”(Coeckelbergh 2016 Care robots and the future of ICT-mediated elderly care: a response to doom scenarios, p. 459)};
\draw[arrow] (sub6.east) -- (q6.west);
\node[draw=none, fit=(sub6)(q6), inner sep=0pt] (pair6) {};

\node[category, anchor=west] (sub7) at ($(pair6.south west)+(-0.5,-\myspacing-0.5cm)$) {Care not tailored to varying needs};
\node[quote, anchor=west] (q7) at ($(sub7.east)+(1.5cm,0)$) {“One participant (male) would not want to use the robot at all. This participant stated that, although he liked interacting with the robot, he found that it was not sufficient for his needs as he preferred a robot that was capable of physical assistance.”(Van Maris et al. 2020 Designing Ethical Social Robots – A Longitudinal Field Study With Older Adults, p. 10)};
\draw[arrow] (sub7.east) -- (q7.west);
\node[draw=none, fit=(sub7)(q7), inner sep=0pt] (pair7) {};

\node[category, anchor=west] (sub8) at ($(pair7.south west)+(0.5,-\myspacing)$) {Developed/designed based on stereotypes/images that ...};
\node[quote, anchor=west] (q8) at ($(sub8.east)+(1.cm,0)$) {“Sex robots and love dolls can certainly be used in old people’s and nursing homes. However, there are several points against this: […] They partly correspond to stereotypes that are perceived differently and sometimes rejected.”(Bendel 2020 Care Robots with Sexual Assistance Functions, p. 4)};
\draw[arrow] (sub8.east) -- (q8.west);
\node[draw=none, fit=(sub8)(q8), inner sep=0pt] (pair8) {};

\node[category, anchor=north west] (sub9) at ($(pair8.south west)+(0.5,0)$) {\textit{Cont. on next page}};
\node[quote, anchor=west] (q9) at ($(sub9.east)+(1.cm,0)$) {};
\node[draw=none, fit=(sub9)(q9), inner sep=0pt] (pair9) {};

\node[category, anchor=north west] (sub10) at ($(pair9.south west)+(-0.5,0)$) {\textit{Cont. on next page}};
\node[quote, anchor=west] (q10) at ($(sub10.east)+(1.5cm,0)$) {};
\node[draw=none, fit=(sub10)(q10), inner sep=0pt] (pair10) {};

\node[category, anchor=north west] (sub11) at ($(pair10.south west)+(-0.5,0)$) {\textit{Cont. on next page}};
\node[quote, anchor=west] (q11) at ($(sub11.east)+(1.5cm,0)$) {};
\node[draw=none, fit=(sub11)(q11), inner sep=0pt] (pair11) {};

\draw[connector] (cat1.west) |- ([xshift=-2mm]sub1.west);
\draw[connector] (sub1.west) |- ([xshift=-2mm]sub2.west);
\draw[connector] (sub2.west) |- ([xshift=-2mm]sub3.west);
\draw[connector] (sub1.west) |- ([xshift=-2mm]sub4.west);
\draw[connector] (sub4.west) |- ([xshift=-2mm]sub5.west);
\draw[connector] (sub1.west) |- ([xshift=-2mm]sub6.west);
\draw[connector] (cat1.west) |- ([xshift=-2mm]sub7.west);
\draw[connector] (sub7.west) |- ([xshift=-2mm]sub8.west);
\draw[connector] (sub8.west) |- ([xshift=-2mm]sub9.west);
\draw[connector] (sub7.west) |- ([xshift=-2mm]sub10.west);
\draw[connector] (cat1.west) |- ([xshift=-2mm]sub11.west);

\end{tikzpicture}

\clearpage
\begin{tikzpicture}[
  every node/.style={align=left,},
  category/.style={text width=6cm, align=left, inner sep=4pt, font=\small},
  root/.style={text width=10cm, align=left, inner sep=4pt},
  quote/.style={rounded corners, thick, text width=6cm, align=justify, inner sep=4pt, font=\footnotesize},
  arrow/.style={-{Stealth[length=3mm]}, thick},
  connector/.style={thick},
  spacing/.store in=\myspacing, spacing=1.0cm 
]

\node[root] (cat1) at (0, 0) {Ethical hazards for care recipients (continued)};

\node[category, anchor=west] (sub1) at ($(cat1.south west)+(0.5,-\myspacing+0.5cm)$) {Care not tailored to varying needs (continued)};
\node[quote, anchor=west] (q1) at ($(sub1.east)+(1.5cm,0)$) {};
\node[draw=none, fit=(sub1)(q1), inner sep=0pt] (pair1) {};

\node[category, anchor=west] (sub2) at ($(pair1.south west)+(0.5,-\myspacing+0.5cm)$) {Developed/designed based on stereotypes/images that ... (continued)};
\node[quote, anchor=west] (q2) at ($(sub2.east)+(1.5cm,0)$) {};
\node[draw=none, fit=(sub2)(q2), inner sep=0pt] (pair2) {};

\node[category, anchor=west] (sub3) at ($(pair2.south west)+(0.5,-\myspacing-0.5cm)$) {Not adequately consider diversity of lifestyles};
\node[quote, anchor=west] (q3) at ($(sub3.east)+(0.5cm,0)$) {“Moreover, there is often an imbalance between perceptions of older peoples’ technology needs and knowledge about their actual needs. The supposed user employs the technology according to the manufacturer’s idea of how the item should be used. However, this is distinct from the real user, who is actually using the technology, and may for instance change the purpose of the technology (Dekker 2015). If diversity in users is incorporated at all, it is most often based on basic social distinctions such as age and gender differences (Flandorfer 2012).”(Hoppe et al. 2020 Assistive robots in care: Expectations and perceptions of older people, p. 144; in: Haltaufderheide et al. 2020 Aging between Participation and Simulation - Ethical Dimensions of Social Assistive Technologies)};
\draw[arrow] (sub3.east) -- (q3.west);
\node[draw=none, fit=(sub3)(q3), inner sep=0pt] (pair3) {};

\node[category, anchor=west] (sub4) at ($(pair3.south west)+(0.,-\myspacing-1cm)$) {Not adequately address needs of targeted population};
\node[quote, anchor=west] (q4) at ($(sub4.east)+(0.5cm,0)$) {“Sexbots in their current form are generally designed for able-bodied heterosexual men, and are therefore not well-suited to providing sexual pleasure for frail, elderly people – most of whom are women – with limited dexterity.” (Lancaster 2023 Granny and the Sexbots: An ethical appraisal of the use of sexbots in residential care institutions for elderly people, p. 203; in: Loh und Loh 2023 Social Robotics and the Good Life - The Normative Side of Forming Emotional Bonds With Robots) };
\draw[arrow] (sub4.east) -- (q4.west);
\node[draw=none, fit=(sub4)(q4), inner sep=0pt] (pair4) {};

\node[category, anchor=west] (sub5) at ($(pair4.south west)+(-0.5,-\myspacing-1cm)$) {Care not provided culturally-sensitive};
\node[quote, anchor=west] (q5) at ($(sub5.east)+(1.cm,0)$) {“All forms of care—including SAR use—need to be sensitive to personal and cultural differences in experiences (Coeckelbergh, 2010; Körtner, 2016; Parks, 2010; Sorell \& Draper, 2014). This sensitivity must be expanded to include not only differences in care experiences of today’s older adults, but also those of tomorrow (Misselhorn et al., 2013; Sorell \& Draper, 2014).” (Vandemeulebroucke et al. 2021 Socially Assistive Robots in Aged Care: Ethical Orientations Beyond the Care-Romantic and Technology-Deterministic Gaze, p. 17)};
\draw[arrow] (sub5.east) -- (q5.west);
\node[draw=none, fit=(sub5)(q5), inner sep=0pt] (pair5) {};

\node[category, anchor=west] (sub6) at ($(pair5.south west)+(0.,-\myspacing-0.5cm)$) {Social robot cannot tailor support to individually varying needs};
\node[quote, anchor=west] (q6) at ($(sub6.east)+(1.cm,0)$) {“The now famous robotic seal ‘Paro’ is also a joyful change for some people with dementia because it is soft, can sit on the lap and reacts to being touched with noises and a glance. However, people who don’t like animals often have little use for this form of activity.” (Translation by the authors)(Schmidhuber and Stöger 2021 Ethisches und Rechtliches zur Zukunft der Robotik in der Pflege. Grundfragen für österreichische und deutsche Debatten, p. 95)  };
\draw[arrow] (sub6.east) -- (q6.west);
\node[draw=none, fit=(sub6)(q6), inner sep=0pt] (pair6) {};

\node[category, anchor=west] (sub7) at ($(pair6.south west)+(0.5,-\myspacing-0.7cm)$) {Care approach based on learning model that restricts options};
\node[quote, anchor=west] (q7) at ($(sub7.east)+(0.5cm,0)$) {“Although the system has some learning capacity, Anderson and Anderson's approach is therefore blind to cases in which the specific needs and moral values of individual care recipients become relevant.” (Translation by the authors) (Misselhorn 2019 Moralische Maschinen in der Pflege? Grundlagen und eine Roadmap für ein moralisch lernfähiges Altenpflegesystem, p. 59; in: Woopen und Jannes 2019 Roboter in der Gesellschaft – Technische Möglichkeiten und menschliche Verantwortung)};
\draw[arrow] (sub7.east) -- (q7.west);
\node[draw=none, fit=(sub7)(q7), inner sep=0pt] (pair7) {};

\node[category, anchor=north west] (sub8) at ($(pair7.south west)+(0.0,0)$) {\textit{Cont. on next page}};
\node[quote, anchor=west] (q8) at ($(sub8.east)+(1.cm,0)$) {};
\node[draw=none, fit=(sub8)(q8), inner sep=0pt] (pair8) {};

\node[category, anchor=north west] (sub9) at ($(pair8.south west)+(-0.5,0)$) {\textit{Cont. on next page}};
\node[quote, anchor=west] (q9) at ($(sub9.east)+(1.cm,0)$) {};
\node[draw=none, fit=(sub9)(q9), inner sep=0pt] (pair9) {};

\node[category, anchor=north west] (sub10) at ($(pair9.south west)+(-0.5,0)$) {\textit{Cont. on next page}};
\node[quote, anchor=west] (q10) at ($(sub10.east)+(1.5cm,0)$) {};
\node[draw=none, fit=(sub10)(q10), inner sep=0pt] (pair10) {};

\draw[connector] (cat1.west) |- ([xshift=-2mm]sub1.west);
\draw[connector] (sub1.west) |- ([xshift=-2mm]sub2.west);
\draw[connector] (sub2.west) |- ([xshift=-2mm]sub3.west);
\draw[connector] (sub2.west) |- ([xshift=-2mm]sub4.west);
\draw[connector] (sub1.west) |- ([xshift=-2mm]sub5.west);
\draw[connector] (sub1.west) |- ([xshift=-2mm]sub6.west);
\draw[connector] (sub6.west) |- ([xshift=-2mm]sub7.west);
\draw[connector] (sub6.west) |- ([xshift=-2mm]sub8.west);
\draw[connector] (sub1.west) |- ([xshift=-2mm]sub9.west);
\draw[connector] (cat1.west) |- ([xshift=-2mm]sub10.west);
\end{tikzpicture}

\clearpage
\begin{tikzpicture}[
  every node/.style={align=left,},
  category/.style={text width=6cm, align=left, inner sep=4pt, font=\small},
  root/.style={text width=10cm, align=left, inner sep=4pt},
  quote/.style={rounded corners, thick, text width=6cm, align=justify, inner sep=4pt, font=\footnotesize},
  arrow/.style={-{Stealth[length=3mm]}, thick},
  connector/.style={thick},
  spacing/.store in=\myspacing, spacing=1.0cm 
]

\node[root] (cat1) at (0, 0) {Ethical hazards for care recipients (continued)};

\node[category, anchor=west] (sub1) at ($(cat1.south west)+(0.5,-\myspacing+0.5cm)$) {Care not tailored to varying needs (continued)};
\node[quote, anchor=west] (q1) at ($(sub1.east)+(1.5cm,0)$) {};
\node[draw=none, fit=(sub1)(q1), inner sep=0pt] (pair1) {};

\node[category, anchor=west] (sub2) at ($(pair1.south west)+(0.5,-\myspacing+0.5cm)$) {Social robot cannot tailor support to individually varying needs (continued)};
\node[quote, anchor=west] (q2) at ($(sub2.east)+(1.5cm,0)$) {};
\node[draw=none, fit=(sub2)(q2), inner sep=0pt] (pair2) {};

\node[category, anchor=west] (sub3) at ($(pair2.south west)+(0.5,-\myspacing)$) {Rule-based/reproducible action clashes with tailored approaches};
\node[quote, anchor=west] (q3) at ($(sub3.east)+(0.5cm,0)$) {“It is particularly feared that technical systems that function with situation-invariant regularity will contradict the individualization and contextualization of care activities that are essential for good care.” (Translation by the authors)(Deutscher Ethikrat 2020 Robotik für gute Pflege (Stellungnahme), p. 43)};
\draw[arrow] (sub3.east) -- (q3.west);
\node[draw=none, fit=(sub3)(q3), inner sep=0pt] (pair3) {};

\node[category, anchor=west] (sub4) at ($(pair3.south west)+(-0.5,-\myspacing)$) {Care providers are less attuned to the individual needs of care recipients because of social robot};
\node[quote, anchor=west] (q4) at ($(sub4.east)+(1.cm,0)$) {“Critics fear, perhaps justifiably, that caregivers might become less attuned to the specific needs of care recipients because a technological crutch is available.”(Borenstein and Pearson 2012 Robot Caregivers: Ethical Issues across the Human Lifespan, p. 260; in: Lin et al. 2012 Robot Ethics: The Ethical and Social Implications of Robotics) };
\draw[arrow] (sub4.east) -- (q4.west);
\node[draw=none, fit=(sub4)(q4), inner sep=0pt] (pair4) {};

\node[category, anchor=west] (sub5) at ($(pair4.south west)+(-0.5,-\myspacing-1cm)$) {Deception about relationships/personal bonds that do not exist};
\node[quote, anchor=west] (q5) at ($(sub5.east)+(1.5cm,0)$) {"However, it was also demonstrated that some users formed attachments and emotional bonds with robots they were interacting with. These attachments led to misconceptions about the robots’ emotional capabilities. For example, one user felt that the robot would miss him when he was gone, something that the robot was not capable of doing.” (Matari? and Scassellati 2016 Socially Assistive Robotics, p. 1987; in: Siciliano und Khatib 2016 Springer Handbook of Robotics)};
\draw[arrow] (sub5.east) -- (q5.west);
\node[draw=none, fit=(sub5)(q5), inner sep=0pt] (pair5) {};

\node[category, anchor=west] (sub6) at ($(pair5.south west)+(0.5,-\myspacing-1cm)$) {Impoverishment of one’s life by illusioned relations};
\node[quote, anchor=west] (q6) at ($(sub6.east)+(1.cm,0)$) {“For instance, linking dignity to the capabilities approach, one perspective in this vein argues to assess SAR depending on whether a robot’s use expands or restricts the access to the basic set of capabilities to live a worthy human life. Also, dignity is under attack when these unauthentic intersubjective interactions alienate users from real life, thereby impoverishing their world.”(Boada et al. 2021 The ethical issues of social assistive robotics: A critical literature review, p. 7)};
\draw[arrow] (sub6.east) -- (q6.west);
\node[draw=none, fit=(sub6)(q6), inner sep=0pt] (pair6) {};

\node[category, anchor=west] (sub7) at ($(pair6.south west)+(-0.5,-\myspacing-0.5cm)$) {Loss of trust in care providers (including social robot carer)};
\node[quote, anchor=west] (q7) at ($(sub7.east)+(1.5cm,0)$) {“Hijacking a companion robot could lead unintended users to vocalize offensive, anti-LGBTIQ+ language, which could seriously endanger the trust the user put into the system.”(Poulsen et al. 2020 Cybersecurity, value sensing robots for LGBTIQ+ elderly, and the need for revised codes of conduct, p. 4)};
\draw[arrow] (sub7.east) -- (q7.west);
\node[draw=none, fit=(sub7)(q7), inner sep=0pt] (pair7) {};

\node[category, anchor=west] (sub8) at ($(pair7.south west)+(0.5,-\myspacing-0.5cm)$) {Loss of trust due to inadequate design of social robot};
\node[quote, anchor=west] (q8) at ($(sub8.east)+(1.cm,0)$) {“(Some people may) ‘Lose trust in the system if the robot goes behind their back ... blackmailing the patient. Depends whether the person really had taken the medication (human intervention that the robot was not aware of) […].’ (Robotics academic)”(Poulsen and Burmeister 2019 Overcoming carer shortages with care robots: Dynamic value trade-offs in run-time, pp. 9, 10) };
\draw[arrow] (sub8.east) -- (q8.west);
\node[draw=none, fit=(sub8)(q8), inner sep=0pt] (pair8) {};

\node[category, anchor=north west] (sub9) at ($(pair8.south west)+(0.0,0)$) {\textit{Cont. on next page}};
\node[quote, anchor=west] (q9) at ($(sub9.east)+(1.cm,0)$) {};
\node[draw=none, fit=(sub9)(q9), inner sep=0pt] (pair9) {};

\node[category, anchor=north west] (sub10) at ($(pair9.south west)+(-0.5,0)$) {\textit{Cont. on next page}};
\node[quote, anchor=west] (q10) at ($(sub10.east)+(1.5cm,0)$) {};
\node[draw=none, fit=(sub10)(q10), inner sep=0pt] (pair10) {};

\draw[connector] (cat1.west) |- ([xshift=-2mm]sub1.west);
\draw[connector] (sub1.west) |- ([xshift=-2mm]sub2.west);
\draw[connector] (sub2.west) |- ([xshift=-2mm]sub3.west);
\draw[connector] (sub1.west) |- ([xshift=-2mm]sub4.west);
\draw[connector] (cat1.west) |- ([xshift=-2mm]sub5.west);
\draw[connector] (sub5.west) |- ([xshift=-2mm]sub6.west);
\draw[connector] (cat1.west) |- ([xshift=-2mm]sub7.west);
\draw[connector] (sub7.west) |- ([xshift=-2mm]sub8.west);
\draw[connector] (sub7.west) |- ([xshift=-2mm]sub9.west);
\draw[connector] (cat1.west) |- ([xshift=-2mm]sub10.west);

\end{tikzpicture}

\clearpage
\begin{tikzpicture}[
  every node/.style={align=left,},
  category/.style={text width=6cm, align=left, inner sep=4pt, font=\small},
  root/.style={text width=10cm, align=left, inner sep=4pt},
  quote/.style={rounded corners, thick, text width=6cm, align=justify, inner sep=4pt, font=\footnotesize},
  arrow/.style={-{Stealth[length=3mm]}, thick},
  connector/.style={thick},
  spacing/.store in=\myspacing, spacing=1.0cm 
]

\node[root] (cat1) at (0, 0) {Ethical hazards for care recipients (continued)};

\node[category, anchor=west] (sub1) at ($(cat1.south west)+(0.5,-\myspacing+0.5cm)$) {Loss of trust in care providers (including social robot carer) (continued)};
\node[quote, anchor=west] (q1) at ($(sub1.east)+(1.5cm,0)$) {};
\node[draw=none, fit=(sub1)(q1), inner sep=0pt] (pair1) {};

\node[category, anchor=west] (sub2) at ($(pair1.south west)+(0.5,-\myspacing)$) {Loss of trust in care provider through reduced physical interaction};
\node[quote, anchor=west] (q2) at ($(sub2.east)+(1.cm,0)$) {“Conventional physical interaction, which often conveys calm and builds trust, could be omitted when using technology.” (Translation by the authors)(Deutscher Ethikrat 2020 Robotik für gute Pflege (Stellungnahme), p. 26)};
\draw[arrow] (sub2.east) -- (q2.west);
\node[draw=none, fit=(sub2)(q2), inner sep=0pt] (pair2) {};

\node[category, anchor=west] (sub3) at ($(pair2.south west)+(-0.5,-\myspacing)$) {Beneficial social robots not/no longer put to use};
\node[quote, anchor=west] (q3) at ($(sub3.east)+(1.5cm,0)$) {“Some studies described situations where PARO appeared ineffective for some individuals or lost effect over time [34, 41].” (Hung et al. 2019 The benefits of and barriers to using a social robot PARO in care setting: a scoping review, p. 7)};
\draw[arrow] (sub3.east) -- (q3.west);
\node[draw=none, fit=(sub3)(q3), inner sep=0pt] (pair3) {};

\node[category, anchor=west] (sub4) at ($(pair3.south west)+(0.5,-\myspacing-0.5cm)$) {Reception of social robot might vary over time (reduction in interest)};
\node[quote, anchor=west] (q4) at ($(sub4.east)+(1.cm,0)$) {“It was noted that when using PARO, staff should uphold a person-centred approach, as just because the resident liked PARO 1 day does not mean that he or she will enjoy it the next [26].” (Hung et al. 2019 The benefits of and barriers to using a social robot PARO in a care setting: a scoping review, p. 7)};
\draw[arrow] (sub4.east) -- (q4.west);
\node[draw=none, fit=(sub4)(q4), inner sep=0pt] (pair4) {};

\node[category, anchor=west] (sub5) at ($(pair4.south west)+(0.,-\myspacing-0.5cm)$) {Lack of acceptance prevents usage };
\node[quote, anchor=west] (q5) at ($(sub5.east)+(1.cm,0)$) {“In order to be accepted, social robots need to be perceived by users as useful and relevant to their current unmet needs [15–19]. A mismatch between needs and the solutions offered by robots are a barrier to acceptance and robot adoption.”(Casey et al. 2016 What People with Dementia Want: Designing MARIO an Acceptable Robot Companion, p. 319)};
\draw[arrow] (sub5.east) -- (q5.west);
\node[draw=none, fit=(sub5)(q5), inner sep=0pt] (pair5) {};

\node[category, anchor=west] (sub6) at ($(pair5.south west)+(0.,-\myspacing-0.5cm)$) {Costs of social robots prevent usage };
\node[quote, anchor=west] (q6) at ($(sub6.east)+(1.cm,0)$) {“[…] each unit of the Justocat costs about US \$1350, an AIBO dog costs US \$3000, and a Paro costs approximately US \$6000. Cost and affordability can therefore influence equal access to such innovations by older adults and people with dementia [25].” (Koh et al. 2021 Impacts of Low-cost Robotic Pets for Older Adults and People with Dementia: Scoping Review, p. 2)};
\draw[arrow] (sub6.east) -- (q6.west);
\node[draw=none, fit=(sub6)(q6), inner sep=0pt] (pair6) {};

\node[category, anchor=west] (sub7) at ($(pair6.south west)+(0.5,-\myspacing-3cm)$) {Environment of care recipient needs to be adapted to social robot (cost-intensive)};
\node[quote, anchor=west] (q7) at ($(sub7.east)+(0.5cm,0)$) {“Due to its size, the robot pictured in the scenario would not be able to operate in the homes of many older adults as they tend to have lots of furniture, knickknacks and rugs in their homes. The professional caregivers indicated it would also be challenging for the robot to operate in the houses of older adults on account of doorsteps, doors and small spaces (especially the kitchen, which is often small). Therefore, it was felt, the homes of older adults should be adapted to the robot. Some professional caregivers thought this would become an issue, as they thought many older adults would be unwilling to do so (thought the elderly participants stated they would not mind adapting their homes to the robot). Professional caregivers also wanted the robot to have its own place in the home and not to follow the user around everywhere. The robot’s place, they opined, should be chosen cooperatively with the user.”(Bedaf et al. 2019 What are the preferred characteristics of a service robot for the elderly? A multi-country focus group study with older adults and caregivers, p. 9)};
\draw[arrow] (sub7.east) -- (q7.west);
\node[draw=none, fit=(sub7)(q7), inner sep=0pt] (pair7) {};

\node[category, anchor=north west] (sub8) at ($(pair7.south west)+(-0.5,0)$) {\textit{Cont. on next page}};
\node[quote, anchor=west] (q8) at ($(sub8.east)+(1.cm,0)$) {};
\node[draw=none, fit=(sub8)(q8), inner sep=0pt] (pair8) {};

\node[category, anchor=north west] (sub9) at ($(pair8.south west)+(-0.5,0)$) {\textit{Cont. on next page}};
\node[quote, anchor=west] (q9) at ($(sub9.east)+(1.cm,0)$) {};
\node[draw=none, fit=(sub9)(q9), inner sep=0pt] (pair9) {};

\draw[connector] (cat1.west) |- ([xshift=-2mm]sub1.west);
\draw[connector] (sub1.west) |- ([xshift=-2mm]sub2.west);
\draw[connector] (cat1.west) |- ([xshift=-2mm]sub3.west);
\draw[connector] (sub3.west) |- ([xshift=-2mm]sub4.west);
\draw[connector] (sub3.west) |- ([xshift=-2mm]sub5.west);
\draw[connector] (sub3.west) |- ([xshift=-2mm]sub6.west);
\draw[connector] (sub6.west) |- ([xshift=-2mm]sub7.west);
\draw[connector] (sub3.west) |- ([xshift=-2mm]sub8.west);
\draw[connector] (cat1.west) |- ([xshift=-2mm]sub9.west);

\end{tikzpicture}

\clearpage
\begin{tikzpicture}[
  every node/.style={align=left,},
  category/.style={text width=6cm, align=left, inner sep=4pt, font=\small},
  root/.style={text width=10cm, align=left, inner sep=4pt},
  quote/.style={rounded corners, thick, text width=6cm, align=justify, inner sep=4pt, font=\footnotesize},
  arrow/.style={-{Stealth[length=3mm]}, thick},
  connector/.style={thick},
  spacing/.store in=\myspacing, spacing=1.0cm 
]

\node[root] (cat1) at (0, 0) {Ethical hazards for care recipients (continued)};

\node[category, anchor=west] (sub1) at ($(cat1.south west)+(0.5,-\myspacing+0.5cm)$) {Beneficial social robots not/no longer put to use (continued)};
\node[quote, anchor=west] (q1) at ($(sub1.east)+(1.5cm,0)$) {};
\node[draw=none, fit=(sub1)(q1), inner sep=0pt] (pair1) {};

\node[category, anchor=west] (sub2) at ($(pair1.south west)+(0.5,-\myspacing)$) {Further points that prevent use};
\node[quote, anchor=west] (q2) at ($(sub2.east)+(1.cm,0)$) {“And which technical systems can be used also depends on their living conditions. Humanoid robots and other (self-)driving assistance systems need room to move.” (Translation by the authors)(Hübner and Müller 2020 Roboter in der Pflege, p. 6)};
\draw[arrow] (sub2.east) -- (q2.west);
\node[draw=none, fit=(sub2)(q2), inner sep=0pt] (pair2) {};

\node[category, anchor=west] (sub3) at ($(pair2.south west)+(-0.5,-\myspacing-0.5cm)$) {Lack of protection due to lack of guidelines/legal regulations};
\node[quote, anchor=west] (q3) at ($(sub3.east)+(1.5cm,0)$) {“The lack of agreed standardized procedures and guidelines for therapeutic robots impedes the establishment of a safeguard baseline to be respected by anyone working in this area, and this leaves users unprotected.”(Fosch-Villaronga and Albo-Canals 2019 “I’ll take care of you,” said the robot, p. 81)};
\draw[arrow] (sub3.east) -- (q3.west);
\node[draw=none, fit=(sub3)(q3), inner sep=0pt] (pair3) {};

\node[category, anchor=west] (sub4) at ($(pair3.south west)+(0.5,-\myspacing-1cm)$) {Problems following from unresolved liability issues};
\node[quote, anchor=west] (q4) at ($(sub4.east)+(1.cm,0)$) {“‘What happens if the robot makes a mistake, e.?g. delivers the wrong medicine? Who is liable? Actually, it is obviously the caregivers in the case that they provide the wrong medicine. But if the robot gives a wrong drug, who is liable then?’ (GER 2018).”(Hoppe et al. 2020 Assistive robots in care: Expectations and perceptions of older people, p. 148; in: Haltaufderheide et al. 2020 Aging between Participation and Simulation - Ethical Dimensions of Social Assistive Technologies)};
\draw[arrow] (sub4.east) -- (q4.west);
\node[draw=none, fit=(sub4)(q4), inner sep=0pt] (pair4) {};

\node[category, anchor=west] (sub5) at ($(pair4.south west)+(0.5,-\myspacing-1cm)$) {Costs of false alarms have to be shouldered by care recipients};
\node[quote, anchor=west] (q5) at ($(sub5.east)+(0.5cm,0)$) {“For instance, there are no clear rules and guidelines to date in specifying which party should be held responsible should robots inflict harm on the older people (Chou et al., 2018). For example, in case of a false alarm, the care robot could call for an ambulance, but the cost to the hospital or insurer will most likely be borne by the person being cared for.”(Tan et al. 2021 Tensions and antagonistic interactions of risks and ethics of using robotics and autonomous systems in long-term care, p. 8)};
\draw[arrow] (sub5.east) -- (q5.west);
\node[draw=none, fit=(sub5)(q5), inner sep=0pt] (pair5) {};

\node[category, anchor=west] (sub6) at ($(pair5.south west)+(0.,-\myspacing-1cm)$) {No compensation for harm realized by social robot};
\node[quote, anchor=west] (q6) at ($(sub6.east)+(0.5cm,0)$) {“Recently, ethical advisory bodies have increasingly emphasized that the use of robotics (as well as AI) must not lead to (cared for) people having to bear the damage caused to them by a (partially) autonomous system (Bioethikkommission, 2020).” (Translation by the authors)(Schmidhuber 2022 Werden Roboter Menschen in der Pflege ersetzen? Ethische Überlegungen, p. 171; in: Stronegger und Platzer 2022 Technisierung der Pflege: 4. Goldegger Dialogforum Mensch und Endlichkeit)};
\draw[arrow] (sub6.east) -- (q6.west);
\node[draw=none, fit=(sub6)(q6), inner sep=0pt] (pair6) {};

\draw[connector] (cat1.west) |- ([xshift=-2mm]sub1.west);
\draw[connector] (sub1.west) |- ([xshift=-2mm]sub2.west);
\draw[connector] (cat1.west) |- ([xshift=-2mm]sub3.west);
\draw[connector] (sub3.west) |- ([xshift=-2mm]sub4.west);
\draw[connector] (sub4.west) |- ([xshift=-2mm]sub5.west);
\draw[connector] (sub4.west) |- ([xshift=-2mm]sub6.west);

\end{tikzpicture}

\clearpage

\begin{tikzpicture}[
  every node/.style={align=left,},
  category/.style={text width=6cm, align=left, inner sep=4pt, font=\small},
  root/.style={text width=10cm, align=left, inner sep=4pt},
  quote/.style={rounded corners, thick, text width=6cm, align=justify, inner sep=4pt, font=\footnotesize},
  arrow/.style={-{Stealth[length=3mm]}, thick},
  connector/.style={thick},
  spacing/.store in=\myspacing, spacing=1.5cm 
]

\node[root] (cat1) at (0, 0) {Ethical opportunities for care providers/facilities (including caring relatives) (Main category)};

\node[category, anchor=west] (sub1) at ($(cat1.south west)+(0.5,-\myspacing)$) {Positively impacting care provider’s health and well-being};
\node[quote, anchor=west] (q1) at ($(sub1.east)+(1.5cm,0)$) {“Care robots are receiving growing attention in elder care since their use has proven to have a positive impact, not only on older adults but also for those who work with them, including professional caregivers and relatives [19].”(Johansson-Pajala et al. 2020 Care Robot Orientation: What, Who and How? Potential Users’ Perceptions, p. 1104) 
};
\draw[arrow] (sub1.east) -- (q1.west);
\node[draw=none, fit=(sub1)(q1), inner sep=0pt] (pair1) {};

\node[category, anchor=west] (sub2) at ($(pair1.south west)+(0.5,-\myspacing-1cm)$) {Reduced risk of infection in care homes};
\node[quote, anchor=west] (q2) at ($(sub2.east)+(1.cm,0)$) {“Lio differs from other service robots in that it has an arm that enables it to manipulate objects. It can open and close doors, but also pick up objects from the floor or take them out of a cupboard and hand them over. These functions are particularly useful in times of a pandemic, as they make it possible to limit contact between wards or between nursing staff and patients, thereby reducing the spread of pathogens.” (Translation by the authors) (Bleuler and Caroni 2021 Roboter in der Pflege - Welche Aufgaben können Roboter heute schon übernehmen?, p. 451; in: Bendel 2021 Soziale Roboter)
 };
\draw[arrow] (sub2.east) -- (q2.west);
\node[draw=none, fit=(sub2)(q2), inner sep=0pt] (pair2) {};

\node[category, anchor=west] (sub3) at ($(pair2.south west)+(0.,-\myspacing-0.5cm)$) {Reduced risk of workload negatively affecting health};
\node[quote, anchor=west] (q3) at ($(sub3.east)+(1.cm,0)$) {“It is well known that professional carers in geriatric care are under more psychological and physical strain than carers in other areas. High time pressure and stress, together with the physical consequences of work (especially back pain), lead to high staff turnover. Robotic systems can be used here in various ways to provide support and relieve the burden on caregivers and/or informal or formal caregivers.” (Translation by the authors)(Wahl et al. 2021 Robotik und Altenpflege: Freund oder Feind?, p. 63) 
};
\draw[arrow] (sub3.east) -- (q3.west);
\node[draw=none, fit=(sub3)(q3), inner sep=0pt] (pair3) {};

\node[category, anchor=west] (sub4) at ($(pair3.south west)+(0.,-\myspacing)$) {Feeling of “being there” for care recipient despite distance (e. g. relatives)};
\node[quote, anchor=west] (q4) at ($(sub4.east)+(1.cm,0)$) {“According to current heuristic assumptions, caregiving relatives seem to experience a sense of presence through the use of a telepresence robot.” (Translation by the authors)(Bleses and Dammert 2020 Neue Technologien aus Sicht der Pflegewissenschaft, p. 71; in: Hanika 2020 Künstliche Intelligenz, Robotik und autonome Systeme in der Gesundheitsversorgung)
};
\draw[arrow] (sub4.east) -- (q4.west);
\node[draw=none, fit=(sub4)(q4), inner sep=0pt] (pair4) {};

\node[category, anchor=west] (sub5) at ($(pair4.south west)+(0.,-\myspacing)$) {Increasing the feeling of safety };
\node[quote, anchor=west] (q5) at ($(sub5.east)+(1.cm,0)$) {“Relatives report in interviews that it reassures them when they can see [thanks to the telepresence robot] that the elderly person is eating something and something ‘right’ or is wearing the emergency call button, for example.” (Translation by the authors)(Bleses and Dammert 2020 Neue Technologien aus Sicht der Pflegewissenschaft, p. 72; in: Hanika 2020 Künstliche Intelligenz, Robotik und autonome Systeme in der Gesundheitsversorgung)
};
\draw[arrow] (sub5.east) -- (q5.west);
\node[draw=none, fit=(sub5)(q5), inner sep=0pt] (pair5) {};

\node[category, anchor=west] (sub6) at ($(pair5.south west)+(0.,-\myspacing)$) {Reducing stress};
\node[quote, anchor=west] (q6) at ($(sub6.east)+(1.cm,0)$) {"While Paro was originally developed for use by elderly residents of nursing homes, investigation has shown that it also has a positive effect on the carers who showed lower levels of stress and took fewer sick leaves when they worked in environments where Paro is used."(Lau et al. 2009 R\&D in healthcare robotics, p. 29; in: Lau et al. 2009 Beyond the Surface – An Exploration in Healthcare Robotics in Japan) 
};
\draw[arrow] (sub6.east) -- (q6.west);
\node[draw=none, fit=(sub6)(q6), inner sep=0pt] (pair6) {};

\node[category, anchor=north west] (sub7) at ($(pair6.south west)+(-0.,0)$) {\textit{Cont. on next page}};
\node[quote, anchor=west] (q7) at ($(sub7.east)+(1.cm,0)$) {};
\node[draw=none, fit=(sub7)(q7), inner sep=0pt] (pair7) {};

\node[category, anchor=north west] (sub8) at ($(pair7.south west)+(-0.5,0)$) {\textit{Cont. on next page}};
\node[quote, anchor=west] (q8) at ($(sub8.east)+(1.cm,0)$) {};
\node[draw=none, fit=(sub8)(q8), inner sep=0pt] (pair8) {};

\draw[connector] (cat1.west) |- ([xshift=-2mm]sub1.west);
\draw[connector] (sub1.west) |- ([xshift=-2mm]sub2.west);
\draw[connector] (sub1.west) |- ([xshift=-2mm]sub3.west);
\draw[connector] (sub1.west) |- ([xshift=-2mm]sub4.west);
\draw[connector] (sub1.west) |- ([xshift=-2mm]sub5.west);
\draw[connector] (sub1.west) |- ([xshift=-2mm]sub6.west);
\draw[connector] (sub1.west) |- ([xshift=-2mm]sub7.west);
\draw[connector] (cat1.west) |- ([xshift=-2mm]sub8.west);

\end{tikzpicture}

\clearpage

\begin{tikzpicture}[
  every node/.style={align=left,},
  category/.style={text width=6cm, align=left, inner sep=4pt, font=\small},
  root/.style={text width=10cm, align=left, inner sep=4pt},
  quote/.style={rounded corners, thick, text width=6cm, align=justify, inner sep=4pt, font=\footnotesize},
  arrow/.style={-{Stealth[length=3mm]}, thick},
  connector/.style={thick},
  spacing/.store in=\myspacing, spacing=1.5cm 
]

\node[root] (cat1) at (0, 0) {Ethical opportunities for care providers/facilities (including caring relatives)  (continued)};

\node[category, anchor=west] (sub1) at ($(cat1.south west)+(0.5,-\myspacing+0.5cm)$) {Positively impacting care provider’s health and well-being (continued)};
\node[quote, anchor=west] (q1) at ($(sub1.east)+(1.5cm,0)$) {};
\node[draw=none, fit=(sub1)(q1), inner sep=0pt] (pair1) {};

\node[category, anchor=west] (sub2) at ($(pair1.south west)+(0.5,-\myspacing)$) {Less inhibitions in dealing with cognitive impaired people};
\node[quote, anchor=west] (q2) at ($(sub2.east)+(1.cm,0)$) {“On the one hand, the robotic seal increased the willingness of the person in need of care to cooperate with the geriatric nurse, as communication between them was improved, and on the other hand, career changers in particular were able to overcome their own inhibitions in dealing with dementia patients more quickly by working with PARO.” (Translation by the authors)(Scorna 2015 Servicerobotik in der Altenpflege – Eine empirische Untersuchung des Einsatzes der Serviceroboter in der stationären Altenpflege am Beispiel von PARO und Care-O-bot, p. 92; in: Weber et al. 2015 Technisierung des Alltags – Beitrag für ein gutes Leben?)
};
\draw[arrow] (sub2.east) -- (q2.west);
\node[draw=none, fit=(sub2)(q2), inner sep=0pt] (pair2) {};

\node[category, anchor=west] (sub3) at ($(pair2.south west)+(-0.5,-\myspacing)$) {Improving working conditions for care providers};
\node[quote, anchor=west] (q3) at ($(sub3.east)+(1.5cm,0)$) {“Another of the main arguments used to support the development of care robots is the need to improve the quality of care. […] Improving the quality of care it means providing better working conditions for care-givers [sic] […].”(Salvini 2015 On Ethical, Legal and Social Issues of Care Robots, p. 432; in: Mohammed 2015 Intelligent Assistive Robots - Recent Advances in Assistive Robotics for Everyday Activities)
};
\draw[arrow] (sub3.east) -- (q3.west);
\node[draw=none, fit=(sub3)(q3), inner sep=0pt] (pair3) {};

\node[category, anchor=west] (sub4) at ($(pair3.south west)+(0.5,-\myspacing-0.5cm)$) {Reducing workload};
\node[quote, anchor=west] (q4) at ($(sub4.east)+(1.cm,0)$) {“For example, a robot that does something that a human caregiver would otherwise do (e.g., providing encouragement for performing exercises) might have as much impact on the human caregiver as on the patient, through the reduction of tasks related to a patient or through the reduction of workplace monotony. Specifically, many SAR systems are being designed to reduce the burden and burnout of family members and other caregivers.”(Feil-Seifer and Matarić 2011 Socially Assistive Robotics - Ethical Issues Related to Technology, p. 27) 
};
\draw[arrow] (sub4.east) -- (q4.west);
\node[draw=none, fit=(sub4)(q4), inner sep=0pt] (pair4) {};

\node[category, anchor=west] (sub5) at ($(pair4.south west)+(0.5,-\myspacing)$) {Freeing up time for social interaction with care recipients};
\node[quote, anchor=west] (q5) at ($(sub5.east)+(0.5cm,0)$) {“Lio relieves the nursing staff and saves time by taking over repetitive tasks. These are errands or long walks in the corridor to remind residents of activities and appointments. This frees up more time for the specialized staff to invest in interpersonal relationships.” (Translation by the authors) (Wirth et al. 2020 Erfahrungen aus dem Einsatz von Assistenzrobotern für Menschen im Alter, p. 261; in: Buxbaum 2020 Mensch-Roboter-Kollaboration)
};
\draw[arrow] (sub5.east) -- (q5.west);
\node[draw=none, fit=(sub5)(q5), inner sep=0pt] (pair5) {};

\node[category, anchor=west] (sub6) at ($(pair5.south west)+(0.,-\myspacing)$) {(Informal) caregivers have more personal time};
\node[quote, anchor=west] (q6) at ($(sub6.east)+(0.5cm,0)$) {“The AT also provided carers with additional personal time which was highly valued and, in many instances, helped create the balance between their own personal space and independence with that of staying connected with the person with dementia.” (Sriram et al. 2019 Informal carers experience of assistive technology use in dementia care at home: a systematic review, p. 13)
};
\draw[arrow] (sub6.east) -- (q6.west);
\node[draw=none, fit=(sub6)(q6), inner sep=0pt] (pair6) {};

\node[category, anchor=west] (sub7) at ($(pair6.south west)+(-0.5,-\myspacing)$) {Providing tools to ease fulfilling of complex tasks};
\node[quote, anchor=west] (q7) at ($(sub7.east)+(1.cm,0)$) {“In these therapies, called robotherapies, the interactions and communications between a subject and an artificial complex interactive system are used for evaluation, diagnosis, prognosis or treatment [2, 36, 37].”(Fosch-Villaronga and Albo-Canals 2019 “I’ll take care of you,” said the robot, p. 80)
};
\draw[arrow] (sub7.east) -- (q7.west);
\node[draw=none, fit=(sub7)(q7), inner sep=0pt] (pair7) {};

\node[category, anchor=north west] (sub8) at ($(pair7.south west)+(0.5,0)$) {\textit{Cont. on next page}};
\node[quote, anchor=west] (q8) at ($(sub8.east)+(1.cm,0)$) {};
\node[draw=none, fit=(sub8)(q8), inner sep=0pt] (pair8) {};

\node[category, anchor=north west] (sub9) at ($(pair8.south west)+(-0.5,0)$) {\textit{Cont. on next page}};
\node[quote, anchor=west] (q9) at ($(sub9.east)+(1.cm,0)$) {};
\node[draw=none, fit=(sub9)(q9), inner sep=0pt] (pair9) {};

\node[category, anchor=north west] (sub10) at ($(pair9.south west)+(-0.5,0)$) {\textit{Cont. on next page}};
\node[quote, anchor=west] (q10) at ($(sub10.east)+(1.5cm,0)$) {};
\node[draw=none, fit=(sub10)(q10), inner sep=0pt] (pair10) {};

\draw[connector] (cat1.west) |- ([xshift=-2mm]sub1.west);
\draw[connector] (sub1.west) |- ([xshift=-2mm]sub2.west);
\draw[connector] (cat1.west) |- ([xshift=-2mm]sub3.west);
\draw[connector] (sub3.west) |- ([xshift=-2mm]sub4.west);
\draw[connector] (sub4.west) |- ([xshift=-2mm]sub5.west);
\draw[connector] (sub4.west) |- ([xshift=-2mm]sub6.west);
\draw[connector] (sub3.west) |- ([xshift=-2mm]sub7.west);
\draw[connector] (sub7.west) |- ([xshift=-2mm]sub8.west);
\draw[connector] (sub3.west) |- ([xshift=-2mm]sub9.west);
\draw[connector] (cat1.west) |- ([xshift=-2mm]sub10.west);
\end{tikzpicture}

\begin{tikzpicture}[
  every node/.style={align=left,},
  category/.style={text width=6cm, align=left, inner sep=4pt, font=\small},
  root/.style={text width=10cm, align=left, inner sep=4pt},
  quote/.style={rounded corners, thick, text width=6cm, align=justify, inner sep=4pt, font=\footnotesize},
  arrow/.style={-{Stealth[length=3mm]}, thick},
  connector/.style={thick},
  spacing/.store in=\myspacing, spacing=1.5cm 
]

\node[root] (cat1) at (0, 0) {Ethical opportunities for care providers/facilities (including caring relatives)  (continued)};

\node[category, anchor=west] (sub1) at ($(cat1.south west)+(0.5,-\myspacing+0.5cm)$) {Positively impacting care provider’s health and well-being (continued)};
\node[quote, anchor=west] (q1) at ($(sub1.east)+(1.5cm,0)$) {};
\node[draw=none, fit=(sub1)(q1), inner sep=0pt] (pair1) {};

\node[category, anchor=west] (sub2) at ($(pair1.south west)+(0.5,-\myspacing+0.5cm)$) {Providing tools to ease fulfilling of complex tasks (continued)};
\node[quote, anchor=west] (q2) at ($(sub2.east)+(1.5cm,0)$) {};
\node[draw=none, fit=(sub2)(q2), inner sep=0pt] (pair2) {};

\node[category, anchor=west] (sub3) at ($(pair2.south west)+(0.5,-\myspacing)$) {Providing support for personnel in crisis situations};
\node[quote, anchor=west] (q3) at ($(sub3.east)+(0.5cm,0)$) {“P56 (operator of a care center) sees it similarly and considers robots to be useful support in emergencies: ‘... also to develop emergency skills. This means that if called upon in an emergency, such a system can help a very excited, perhaps still young and fresh employee to deal with a crisis situation, an emergency situation, in a professional manner, so to speak.’” (Translation by the authors)(Carros et al. 2022 Roboter als intelligente Assistenten in Betreuung und Pflege – Grenzen und Perspektiven im Praxiseinsatz, p. 805; in: Pfannstiel 2022 Künstliche Intelligenz im Gesundheitswesen)
};
\draw[arrow] (sub3.east) -- (q3.west);
\node[draw=none, fit=(sub3)(q3), inner sep=0pt] (pair3) {};

\node[category, anchor=west] (sub4) at ($(pair3.south west)+(0.,-\myspacing-0.8cm)$) {Supporting in diagnosis (e. g. testing sight and hearing)};
\node[quote, anchor=west] (q4) at ($(sub4.east)+(0.5cm,0)$) {“Only a few respondents were more specific. For example, P22 (management consultant) noted that robotics could be used very well in the field of AI-supported (early) diagnostics: ‘Or in the broadest sense also a bit of early diagnostics, so if the hearing deteriorates or the eyes deteriorate or something like that, that then either supporting or analyzing tasks are set in the interaction, where it can also go into the care diagnosis or medical diagnosis in the broadest sense.’” (Tranlsation by the authors) (Carros et al. 2022 Roboter als intelligente Assistenten in Betreuung und Pflege – Grenzen und Perspektiven im Praxiseinsatz, p. 805; in: Pfannstiel 2022 Künstliche Intelligenz im Gesundheitswesen)
};
\draw[arrow] (sub4.east) -- (q4.west);
\node[draw=none, fit=(sub4)(q4), inner sep=0pt] (pair4) {};

\node[category, anchor=west] (sub5) at ($(pair4.south west)+(0.,-\myspacing-0.8cm)$) {Providing activities (esp. for cognitive impaired people)};
\node[quote, anchor=west] (q5) at ($(sub5.east)+(0.5cm,0)$) {“The expert interviews conducted on PARO showed that all interviewees initially expected the robotic seal to bring joy to the residents, activate them and improve their general well-being. In this context, one expert from the field of geriatric care emphasized that she saw the robotic seal as a special offer for residents who require a lot of care and are very restricted due to their need for care.” (Translation by the authors)(Scorna 2015 Servicerobotik in der Altenpflege – Eine empirische Untersuchung des Einsatzes der Serviceroboter in der stationären Altenpflege am Beispiel von PARO und Care-O-bot, p. 89; in: Weber et al. 2015 Technisierung des Alltags – Beitrag für ein gutes Leben?)
};
\draw[arrow] (sub5.east) -- (q5.west);
\node[draw=none, fit=(sub5)(q5), inner sep=0pt] (pair5) {};

\node[category, anchor=west] (sub6) at ($(pair5.south west)+(0.,-\myspacing)$) {Supporting therapeutical approaches};
\node[quote, anchor=west] (q6) at ($(sub6.east)+(0.5cm,0)$) {“Furthermore, Paro seems to fulfill its intended purpose. On average, the robot has been used regularly in care facilities for over a year and a half for its intended purpose as a therapeutic aid [...], which speaks for its success in the care context.” (Translation by the authors)(Baisch et al. 2018 Emotionale Roboter im Pflegekontext -Empirische Analyse des bisherigen Einsatzes und der Wirkungen von Paro und Pleo, p. 22)
};
\draw[arrow] (sub6.east) -- (q6.west);
\node[draw=none, fit=(sub6)(q6), inner sep=0pt] (pair6) {};

\node[category, anchor=west] (sub7) at ($(pair6.south west)+(-0.5,-\myspacing)$) {Better networking/coordination with care recipients and informal caregivers};
\node[quote, anchor=west] (q7) at ($(sub7.east)+(1.cm,0)$) {“On the other hand, they should improve the networking of recipients, informal and professional helpers and thus ensure the coordination of needs-based care work in complex structures (cf. Hülsken-Giesler and Krings 2015; Hülsken-Giesler 2015).” (Translation by the authors) (Hülsken-Giesler and Daxberger 2018 Robotik in der Pflege aus pflegewissenschaftlicher Perspektive, p. 127; in: Bendel 2018 Pflegeroboter)
};
\draw[arrow] (sub7.east) -- (q7.west);
\node[draw=none, fit=(sub7)(q7), inner sep=0pt] (pair7) {};

\node[category, anchor=north west] (sub8) at ($(pair7.south west)+(0.5,0)$) {\textit{Cont. on next page}};
\node[quote, anchor=west] (q8) at ($(sub8.east)+(1.cm,0)$) {};
\node[draw=none, fit=(sub8)(q8), inner sep=0pt] (pair8) {};

\node[category, anchor=north west] (sub9) at ($(pair8.south west)+(-0.5,0)$) {\textit{Cont. on next page}};
\node[quote, anchor=west] (q9) at ($(sub9.east)+(1.cm,0)$) {};
\node[draw=none, fit=(sub9)(q9), inner sep=0pt] (pair9) {};

\node[category, anchor=north west] (sub10) at ($(pair9.south west)+(-0.5,0)$) {\textit{Cont. on next page}};
\node[quote, anchor=west] (q10) at ($(sub10.east)+(1.5cm,0)$) {};
\node[draw=none, fit=(sub10)(q10), inner sep=0pt] (pair10) {};

\draw[connector] (cat1.west) |- ([xshift=-2mm]sub1.west);
\draw[connector] (sub1.west) |- ([xshift=-2mm]sub2.west);
\draw[connector] (sub2.west) |- ([xshift=-2mm]sub3.west);
\draw[connector] (sub2.west) |- ([xshift=-2mm]sub4.west);
\draw[connector] (sub2.west) |- ([xshift=-2mm]sub5.west);
\draw[connector] (sub2.west) |- ([xshift=-2mm]sub6.west);
\draw[connector] (sub1.west) |- ([xshift=-2mm]sub7.west);
\draw[connector] (sub7.west) |- ([xshift=-2mm]sub8.west);
\draw[connector] (sub1.west) |- ([xshift=-2mm]sub9.west);
\draw[connector] (cat1.west) |- ([xshift=-2mm]sub10.west);

\end{tikzpicture}

\begin{tikzpicture}[
  every node/.style={align=left,},
  category/.style={text width=6cm, align=left, inner sep=4pt, font=\small},
  root/.style={text width=10cm, align=left, inner sep=4pt},
  quote/.style={rounded corners, thick, text width=6cm, align=justify, inner sep=4pt, font=\footnotesize},
  arrow/.style={-{Stealth[length=3mm]}, thick},
  connector/.style={thick},
  spacing/.store in=\myspacing, spacing=1.5cm 
]

\node[root] (cat1) at (0, 0) {Ethical opportunities for care providers/facilities (including caring relatives)  (continued)};

\node[category, anchor=west] (sub1) at ($(cat1.south west)+(0.5,-\myspacing+0.5cm)$) {Positively impacting care provider’s health and well-being (continued)};
\node[quote, anchor=west] (q1) at ($(sub1.east)+(1.5cm,0)$) {};
\node[draw=none, fit=(sub1)(q1), inner sep=0pt] (pair1) {};

\node[category, anchor=west] (sub2) at ($(pair1.south west)+(0.5,-\myspacing+0.5cm)$) {Better networking/coordination with care recipients and informal caregivers (continued)};
\node[quote, anchor=west] (q2) at ($(sub2.east)+(1.5cm,0)$) {};
\node[draw=none, fit=(sub2)(q2), inner sep=0pt] (pair2) {};

\node[category, anchor=west] (sub3) at ($(pair2.south west)+(0.5,-\myspacing)$) {Virtual visits can save time and energy};
\node[quote, anchor=west] (q3) at ($(sub3.east)+(0.5cm,0)$) {“Telepresence robots, such as Giraff and VGo, have proven effective in providing remote monitoring of adults with dementia and enabling long-distance control or interaction between patients and caregivers, often in combination with telephony and long-range remote control.” (Ienca et al. 2016 Social and Assistive Robotics in Dementia Care: Ethical Recommendations for Research and Practice, pp. 566, 567)
};
\draw[arrow] (sub3.east) -- (q3.west);
\node[draw=none, fit=(sub3)(q3), inner sep=0pt] (pair3) {};

\node[category, anchor=west] (sub4) at ($(pair3.south west)+(-0.5,-\myspacing-1cm)$) {Easing access to relevant patient information (e. g. adherence)};
\node[quote, anchor=west] (q4) at ($(sub4.east)+(1.cm,0)$) {“Two months ago, Ms. Koertig was given a care robot. [...] When the mobile device connects to the care service’s sensor, it recognizes that Stefanie Zimmermann is in the house and travels to her independently to support her. The care planning for Ms. Koertig is stored in the robot along the care process. It provides the necessary information and ensures that Stefanie doesn’t forget anything. This has worked well the last few times. The first step is to measure her blood sugar level, as Ms. Koertig takes medication that needs to be dosed correctly. The robot shows the values from the previous weeks on a display [...].” (Translation bei the authors)(Seefeldt and Hülsken-Giesler 2020 Pflegeethik und Robotik in der Pflege, p. 271)
};
\draw[arrow] (sub4.east) -- (q4.west);
\node[draw=none, fit=(sub4)(q4), inner sep=0pt] (pair4) {};

\node[category, anchor=west] (sub5) at ($(pair4.south west)+(0.,-\myspacing)$) {Reducing difficult behavior of care recipients};
\node[quote, anchor=west] (q5) at ($(sub5.east)+(1.cm,0)$) {“Since social interaction with dementia patients can also be very frustrating for the relatives, some of the burden on the relatives may be alleviated if the interaction with Paro improves the condition of the patient by increasing affective well-being and reducing stress.”(Misselhorn et al. 2013 Ethical Considerations Regarding the Use of Social Robots in the Fourth Age, p. 127)
};
\draw[arrow] (sub5.east) -- (q5.west);
\node[draw=none, fit=(sub5)(q5), inner sep=0pt] (pair5) {};

\node[category, anchor=west] (sub6) at ($(pair5.south west)+(0.,-\myspacing)$) {Easy maintenance of robot pets over real animals};
\node[quote, anchor=west] (q6) at ($(sub6.east)+(1.cm,0)$) {“Paro is helping in the care of dementia sufferers in a much more controlled way than real live companion animals. Paro will not be hurt if his mistress flies into a rage. Paro is easily washed and will not die.”(Bennett 2014 Robot Identity Assurance, p. 11) 
};
\draw[arrow] (sub6.east) -- (q6.west);
\node[draw=none, fit=(sub6)(q6), inner sep=0pt] (pair6) {};

\node[category, anchor=west] (sub7) at ($(pair6.south west)+(-0.5,-\myspacing)$) {Positively implicates relationships};
\node[quote, anchor=west] (q7) at ($(sub7.east)+(1.5cm,0)$) {“The intervention of robot caregivers could improve family unity and other interpersonal relationships because they would not be tainted by our aversion to unpleasant tasks.”(Borenstein and Pearson 2012 Robot Caregivers: Ethical Issues across the Human Lifespan, p. 261; in: Lin et al. 2012 Robot Ethics: The Ethical and Social Implications of Robotics)
};
\draw[arrow] (sub7.east) -- (q7.west);
\node[draw=none, fit=(sub7)(q7), inner sep=0pt] (pair7) {};

\node[category, anchor=west] (sub8) at ($(pair7.south west)+(0.,-\myspacing-0.3cm)$) {Upgrading nursing profession };
\node[quote, anchor=west] (q8) at ($(sub8.east)+(1.5cm,0)$) {“Instead of replacing people, Lio and P-Care enable targeted care for older people and act as an extension of care staff. By engaging with technology and using it in residential homes with the help of robotics, care professions are being upgraded and new occupational fields are being created.” (Translation by the authors)(Früh and Gasser 2018 Erfahrungen aus dem Einsatz von Pflegerobotern für Menschen im Alter, p. 44; in: Bendel 2018 Pflegeroboter)
};
\draw[arrow] (sub8.east) -- (q8.west);
\node[draw=none, fit=(sub8)(q8), inner sep=0pt] (pair8) {};

\node[category, anchor=north west] (sub9) at ($(pair8.south west)+(0.,0)$) {\textit{Cont. on next page}};
\node[quote, anchor=west] (q9) at ($(sub9.east)+(1.cm,0)$) {};
\node[draw=none, fit=(sub9)(q9), inner sep=0pt] (pair9) {};

\draw[connector] (cat1.west) |- ([xshift=-2mm]sub1.west);
\draw[connector] (sub1.west) |- ([xshift=-2mm]sub2.west);
\draw[connector] (sub2.west) |- ([xshift=-2mm]sub3.west);
\draw[connector] (sub1.west) |- ([xshift=-2mm]sub4.west);
\draw[connector] (sub1.west) |- ([xshift=-2mm]sub5.west);
\draw[connector] (sub1.west) |- ([xshift=-2mm]sub6.west);
\draw[connector] (cat1.west) |- ([xshift=-2mm]sub7.west);
\draw[connector] (cat1.west) |- ([xshift=-2mm]sub8.west);
\draw[connector] (cat1.west) |- ([xshift=-2mm]sub9.west);
\end{tikzpicture}

\begin{tikzpicture}[
  every node/.style={align=left,},
  category/.style={text width=6cm, align=left, inner sep=4pt, font=\small},
  root/.style={text width=10cm, align=left, inner sep=4pt},
  quote/.style={rounded corners, thick, text width=6cm, align=justify, inner sep=4pt, font=\footnotesize},
  arrow/.style={-{Stealth[length=3mm]}, thick},
  connector/.style={thick},
  spacing/.store in=\myspacing, spacing=1.5cm 
]

\node[root] (cat1) at (0, 0) {Ethical opportunities for care providers/facilities (including caring relatives)  (continued)};

\node[category, anchor=west] (sub1) at ($(cat1.south west)+(0.5,-\myspacing)$) {Easing management for care facilities};
\node[quote, anchor=west] (q1) at ($(sub1.east)+(1.5cm,0)$) {“The robot was seen as an enabler to increase collaboration with families in facility care and improve family care, both which are general targets in the Finnish elderly care system nowadays.”(Niemelä et al. 2021 Towards Ethical Guidelines of Using Telepresence Robots in Residential Care, p. 436)};
\draw[arrow] (sub1.east) -- (q1.west);
\node[draw=none, fit=(sub1)(q1), inner sep=0pt] (pair1) {};

\node[category, anchor=west] (sub2) at ($(pair1.south west)+(0.5,-\myspacing)$) {Social robots as competitive advantage on the market};
\node[quote, anchor=west] (q2) at ($(sub2.east)+(1.cm,0)$) {“Lio therefore brings additional value for both residents and care staff. At the same time, it enables institutions to stand out from others and secure a place for themselves in the competitive market (Früh and Gasser 2018).” (Translation by the authors)(Bleuler and Caroni 2021 Roboter in der Pflege - Welche Aufgaben können Roboter heute schon übernehmen?, p. 451; in: Bendel 2021 Soziale Roboter)
};
\draw[arrow] (sub2.east) -- (q2.west);
\node[draw=none, fit=(sub2)(q2), inner sep=0pt] (pair2) {};

\node[category, anchor=west] (sub3) at ($(pair2.south west)+(0.,-\myspacing)$) {Social robots can work all day long (without negative behavior change)};
\node[quote, anchor=west] (q3) at ($(sub3.east)+(1.cm,0)$) {“In a study where a humanoid robot Zora was piloted in elder care services, the care personnel noted—among other things—that unlike a human caregiver, the robot does not get tired, it always responds in a friendly way, and it repeats things over and over if needed.”(Pirhonen et al. 2020 Can robots tackle late-life loneliness? Scanning of future opportunities and challenges in assisted living facilities, p. 4)
};
\draw[arrow] (sub3.east) -- (q3.west);
\node[draw=none, fit=(sub3)(q3), inner sep=0pt] (pair3) {};

\node[category, anchor=west] (sub4) at ($(pair3.south west)+(0.,-\myspacing)$) {Controlling adherence of care personnel to treatment standards};
\node[quote, anchor=west] (q4) at ($(sub4.east)+(1.cm,0)$) {“Some formal carers raised the issue of the robot being used to ‘spy’ on them, whilst other formal carers did seem willing to use the robot to check up on, as well as to reinforce, adherence to treatment regimes.”(Draper and Sorell 2017 Ethical values and social care robots for older people: an international qualitative study, p. 67)
};
\draw[arrow] (sub4.east) -- (q4.west);
\node[draw=none, fit=(sub4)(q4), inner sep=0pt] (pair4) {};"

\node[category, anchor=west] (sub5) at ($(pair4.south west)+(0.,-\myspacing-0.5cm)$) {Reducing costs for care facilities };
\node[quote, anchor=west] (q5) at ($(sub5.east)+(1.cm,0)$) {“Care costs can be reduced in the medium term by dividing the support activities into two categories. Demanding management and care functions are carried out by qualified employees, who can then concentrate on meaningful tasks. Repetitive routine and auxiliary functions that do not necessarily require human intervention will be carried out by intelligent personal robots.” (Translation by the authors) (Früh and Gasser 2018 Erfahrungen aus dem Einsatz von Pflegerobotern für Menschen im Alter, p. 45; in: Bendel 2018 Pflegeroboter)
};
\draw[arrow] (sub5.east) -- (q5.west);
\node[draw=none, fit=(sub5)(q5), inner sep=0pt] (pair5) {};

\draw[connector] (cat1.west) |- ([xshift=-2mm]sub1.west);
\draw[connector] (sub1.west) |- ([xshift=-2mm]sub2.west);
\draw[connector] (sub1.west) |- ([xshift=-2mm]sub3.west);
\draw[connector] (sub1.west) |- ([xshift=-2mm]sub4.west);
\draw[connector] (sub1.west) |- ([xshift=-2mm]sub5.west);

\end{tikzpicture}

\begin{tikzpicture}[
  every node/.style={align=left,},
  category/.style={text width=6cm, align=left, inner sep=4pt, font=\small},
  root/.style={text width=10cm, align=left, inner sep=4pt},
  quote/.style={rounded corners, thick, text width=6cm, align=justify, inner sep=4pt, font=\footnotesize},
  arrow/.style={-{Stealth[length=3mm]}, thick},
  connector/.style={thick},
  spacing/.store in=\myspacing, spacing=1.5cm 
]

\node[root] (cat1) at (0, 0) {Ethical hazards for care providers/facilities (including caring relatives) (Main category)};

\node[category, anchor=west] (sub1) at ($(cat1.south west)+(0.5,-\myspacing+0.2)$) {Adverse effect on health/well-being of care providers};
\node[quote, anchor=west] (q1) at ($(sub1.east)+(1.5cm,0)$) {“However, when introducing robotic systems, it must of course be ensured that they do not lead to further strain on care staff.” (Translation by the authors)(Gräb-Schmidt and Stritzelberger 2018 Ethische Herausforderungen durch autonome Systeme und Robotik im Bereich Pflege, pp. 359, 360)};
\draw[arrow] (sub1.east) -- (q1.west);
\node[draw=none, fit=(sub1)(q1), inner sep=0pt] (pair1) {};

\node[category, anchor=west] (sub2) at ($(pair1.south west)+(0.5,-\myspacing)$) {Endangering safety of care providers};
\node[quote, anchor=west] (q2) at ($(sub2.east)+(1.cm,0)$) {“In contrast with robots for factories, assistive technologies such as SAR share space with people and therefore they have the potential to accidentally harm humans and damage facilities.”(Koimizu 2019 Aged Care with Socially Assistive Robotics under Advance Care Planning, p. 34)};
\draw[arrow] (sub2.east) -- (q2.west);
\node[draw=none, fit=(sub2)(q2), inner sep=0pt] (pair2) {};

\node[category, anchor=west] (sub3) at ($(pair2.south west)+(0.5,-\myspacing+0.3cm)$) {Harms caused by malfunction/lack of suitability };
\node[quote, anchor=west] (q3) at ($(sub3.east)+(0.5cm,0)$) {“One rehabilitation staff participant reported the clutter in the hallway and resident room confused the sensors of the robot, hindering optimal movement of the robot.”(Hung et al. 2022 Technological risks and ethical implications of using robots in long-term care, p. 4)};
\node[draw=none, fit=(sub3)(q3), inner sep=0pt] (pair3) {};

\node[category, anchor=west] (sub4) at ($(pair3.south west)+(-0.5,-\myspacing-0.6cm)$) {Feeling uncomfortable with social robot/social robot usage of care recipients };
\node[quote, anchor=west] (q4) at ($(sub4.east)+(1.cm,0)$) {“In addition to the geriatric nurses who were immediately enthusiastic, there were always those who would have rejected care with PARO. The reasons for this are often that there is no money for such investments and that people don’t believe that PARO has a positive effect on the residents, but rather assume that they want to trick those in need of care with the seal.” (Translation by the authors)(Scorna 2015 Servicerobotik in der Altenpflege – Eine empirische Untersuchung des Einsatzes der Serviceroboter in der stationären Altenpflege am Beispiel von PARO und Care-O-bot, p. 91; in: Weber et al. 2015 Technisierung des Alltags – Beitrag für ein gutes Leben?)};
\draw[arrow] (sub4.east) -- (q4.west);
\node[draw=none, fit=(sub4)(q4), inner sep=0pt] (pair4) {};

\node[category, anchor=west] (sub5) at ($(pair4.south west)+(-0.5,-\myspacing-0.1cm)$) {Aggravate work situation for care providers};
\node[quote, anchor=west] (q5) at ($(sub5.east)+(1.5cm,0)$) {“The applications of robotics and autonomous systems in long-term care also have both positive and negative employment implications to the incumbent nursing and health workforce in the future (Khaksar et al., 2016; Stahl and Coeckelbergh 2016; Chou et al., 2018; Metzler et al., 2015; Pfadenhauer and Dukat 2015).”(Tan et al. 2021 Tensions and antagonistic interactions of risks and ethics of using robotics and autonomous systems in long-term care, p. 8)};
\draw[arrow] (sub5.east) -- (q5.west);
\node[draw=none, fit=(sub5)(q5), inner sep=0pt] (pair5) {};

\node[category, anchor=west] (sub6) at ($(pair5.south west)+(0.5,-\myspacing-0.5cm)$) {Investing in new technologies instead of working conditions};
\node[quote, anchor=west] (q6) at ($(sub6.east)+(1.cm,0)$) {“It is a question of societal priorities and demands – and the narratives and imaginaries making these intelligible in the first place – whether resources are spent for the development of technologies like care robots or for structurally making professional and private human care-giving possible, acknowledged and payed properly.”(von Maur 2023 Alice Does not Care – Or: Wy it Matters That Robots “Don’t Give a Damn”, p. 226; in: Loh und Loh 2023 Social Robotics and the Good Life - The Normative Side of Forming Emotional Bonds With Robots)};
\draw[arrow] (sub6.east) -- (q6.west);
\node[draw=none, fit=(sub6)(q6), inner sep=0pt] (pair6) {};

\node[category, anchor=west] (sub7) at ($(pair6.south west)+(0.,-\myspacing)$) {Job loss following replacement through social robot};
\node[quote, anchor=west] (q7) at ($(sub7.east)+(1.cm,0)$) {“From the caregiver's perspective, there is also the risk that they will lose their job due to the robot [5, 17, 18].” (Translation by the authors)(Radic and Vosen 2020 Ethische, rechtliche und soziale Anforderungen an Assistenzroboter in der Pflege – Sicht des Führungspersonals in Kliniken und Pflegeeinrichtungen, p. 630)};
\draw[arrow] (sub7.east) -- (q7.west);
\node[draw=none, fit=(sub7)(q7), inner sep=0pt] (pair7) {};

\node[category, anchor=north west] (sub8) at ($(pair7.south west)+(0.,0)$) {\textit{Cont. on next page}};
\node[quote, anchor=west] (q8) at ($(sub8.east)+(1.cm,0)$) {};
\node[draw=none, fit=(sub8)(q8), inner sep=0pt] (pair8) {};

\node[category, anchor=north west] (sub9) at ($(pair8.south west)+(-0.5,0)$) {\textit{Cont. on next page}};
\node[quote, anchor=west] (q9) at ($(sub9.east)+(1.cm,0)$) {};
\node[draw=none, fit=(sub9)(q9), inner sep=0pt] (pair9) {};

\draw[connector] (cat1.west) |- ([xshift=-2mm]sub1.west);
\draw[connector] (sub1.west) |- ([xshift=-2mm]sub2.west);
\draw[connector] (sub2.west) |- ([xshift=-2mm]sub3.west);
\draw[connector] (sub1.west) |- ([xshift=-2mm]sub4.west);
\draw[connector] (cat1.west) |- ([xshift=-2mm]sub5.west);
\draw[connector] (sub5.west) |- ([xshift=-2mm]sub6.west);
\draw[connector] (sub5.west) |- ([xshift=-2mm]sub7.west);
\draw[connector] (sub5.west) |- ([xshift=-2mm]sub8.west);
\draw[connector] (cat1.west) |- ([xshift=-2mm]sub9.west);

\end{tikzpicture}
\clearpage

\begin{tikzpicture}[
  every node/.style={align=left,},
  category/.style={text width=6cm, align=left, inner sep=4pt, font=\small},
  root/.style={text width=10cm, align=left, inner sep=4pt},
  quote/.style={rounded corners, thick, text width=6cm, align=justify, inner sep=4pt, font=\footnotesize},
  arrow/.style={-{Stealth[length=3mm]}, thick},
  connector/.style={thick},
  spacing/.store in=\myspacing, spacing=1.5cm 
]

\node[root] (cat1) at (0, 0) {Ethical hazards for care providers/facilities (including caring relatives) (continued)};

\node[category, anchor=west] (sub1) at ($(cat1.south west)+(0.5,-\myspacing+0.5cm)$) {Aggravate work situation for care providers
 (continued)};
\node[quote, anchor=west] (q1) at ($(sub1.east)+(1.5cm,0)$) {};
\node[draw=none, fit=(sub1)(q1), inner sep=0pt] (pair1) {};

\node[category, anchor=west] (sub2) at ($(pair1.south west)+(0.5,-\myspacing)$) {Reduction of (the importance of) relationship work};
\node[quote, anchor=west] (q2) at ($(sub2.east)+(1.cm,0)$) {“Nurses from a clinical setting explained they often work together with colleagues and are therefore worried about less teamwork when working with the robot: ‘And there is togetherness and teamwork at the bedside when staff help each other to move patients or make the beds, and that would then all be lost’ (PN 13, FG 1).” (Nielsen et al. 2022 Implementing ethical aspects in the development of a robotic system for nursing care: a qualitative approach, p. 5)};
\draw[arrow] (sub2.east) -- (q2.west);
\node[draw=none, fit=(sub2)(q2), inner sep=0pt] (pair2) {};

\node[category, anchor=west] (sub3) at ($(pair2.south west)+(0.,-\myspacing)$) {Conflict among care providers how to use the social robot};
\node[quote, anchor=west] (q3) at ($(sub3.east)+(1.cm,0)$) {“At the same time, it is important to keep in mind the interests of the carers. Different carers may have different desires or interests regarding how to care for older people and may disagree about how care should be discharged, including the care roles that a robot should assume (Jenkins and Draper 2015). What is beneficial for one type of carer may be detrimental to another.”(Tan et al. 2021 Tensions and antagonistic interactions of risks and ethics of using robotics and autonomous systems in long-term care, p. 9)};
\draw[arrow] (sub3.east) -- (q3.west);
\node[draw=none, fit=(sub3)(q3), inner sep=0pt] (pair3) {};

\node[category, anchor=west] (sub4) at ($(pair3.south west)+(0.,-\myspacing)$) {Loss of trust in care providers by care recipients};
\node[quote, anchor=west] (q4) at ($(sub4.east)+(1.cm,0)$) {“Conversely, however, the presence of a robot to monitor the care delivered by a FC may, if this is known to the older person, undermine the faith that this older person then has in the carer.”(Jenkins and Draper 2015 Care, Monitoring, and Companionship: Views on Care Robots from Older People and Their Carer, p. 678)};
\draw[arrow] (sub4.east) -- (q4.west);
\node[draw=none, fit=(sub4)(q4), inner sep=0pt] (pair4) {};

\node[category, anchor=west] (sub5) at ($(pair4.south west)+(-0.5,-\myspacing)$) {Feeling insecure in handling social robots};
\node[quote, anchor=west] (q5) at ($(sub5.east)+(1.5cm,0)$) {“Reluctance to embrace robots may be due to a variety of reasons such as fear of being replaced by a robot, fear of change, fear of operating highly advanced technology, […] (Broadbent et al. 2010; Verrugio \& Operto, 2008).” (Huschilt et al. 2012 The Use of Socially Assistive Robots for Dementia Care, p. 17)};
\draw[arrow] (sub5.east) -- (q5.west);
\node[draw=none, fit=(sub5)(q5), inner sep=0pt] (pair5) {};

\node[category, anchor=west] (sub6) at ($(pair5.south west)+(0.5,-\myspacing)$) {Lack of competencies for handling social robots};
\node[quote, anchor=west] (q6) at ($(sub6.east)+(1.cm,0)$) {“The staff felt unskilled and that they had no time to learn to use the robot.”(Niemelä et al. 2021Towards Ethical Guidelines of Using Telepresence Robots in Residential Care, p. 433)};
\draw[arrow] (sub6.east) -- (q6.west);
\node[draw=none, fit=(sub6)(q6), inner sep=0pt] (pair6) {};

\node[category, anchor=west] (sub7) at ($(pair6.south west)+(0.,-\myspacing)$) {Need to re-negotiate own role in care};
\node[quote, anchor=west] (q7) at ($(sub7.east)+(1.cm,0)$) {“The introduction of SARs challenges current roles and responsibilities in care practice settings, thus threatening the quality of the practice and the essential element of trust that is constitutive of the relationship between caregivers and care recipients. Which tasks can be responsibly delegated to SARs or not in order to legitimately reshape these roles is a question with deep ethical implications at the organizational level.”(Boada et al. 2021 The ethical issues of social assistive robotics: A critical literature review, p. 9)};
\draw[arrow] (sub7.east) -- (q7.west);
\node[draw=none, fit=(sub7)(q7), inner sep=0pt] (pair7) {};

\node[category, anchor=west] (sub8) at ($(pair7.south west)+(-0.5,-\myspacing)$) {Complicating management for care facilities};
\node[quote, anchor=west] (q8) at ($(sub8.east)+(1.5cm,0)$) {“Purchasing health robots for use at a facility can entail large financial costs, resulting in the discontinuation of other activities.”(The Swedish National Council on Medical Ethics 2015 Robots and Surveillance in Health Care of the Elderly – Ethical Aspects, p. 447)) };
\draw[arrow] (sub8.east) -- (q8.west);
\node[draw=none, fit=(sub8)(q8), inner sep=0pt] (pair8) {};
\node[category, anchor=north west] (sub9) at ($(pair8.south west)+(0.5,0)$) {\textit{Cont. on next page}};
\node[quote, anchor=west] (q9) at ($(sub9.east)+(1.cm,0)$) {};
\node[draw=none, fit=(sub9)(q9), inner sep=0pt] (pair9) {};

\node[category, anchor=north west] (sub10) at ($(pair9.south west)+(-0.5,0)$) {\textit{Cont. on next page}};
\node[quote, anchor=west] (q10) at ($(sub10.east)+(1.5cm,0)$) {};
\node[draw=none, fit=(sub10)(q10), inner sep=0pt] (pair10) {};
\draw[connector] (cat1.west) |- ([xshift=-2mm]sub1.west);
\draw[connector] (sub1.west) |- ([xshift=-2mm]sub2.west);
\draw[connector] (sub1.west) |- ([xshift=-2mm]sub3.west);
\draw[connector] (sub1.west) |- ([xshift=-2mm]sub4.west);
\draw[connector] (cat1.west) |- ([xshift=-2mm]sub5.west);
\draw[connector] (sub5.west) |- ([xshift=-2mm]sub6.west);
\draw[connector] (sub5.west) |- ([xshift=-2mm]sub7.west);
\draw[connector] (cat1.west) |- ([xshift=-2mm]sub8.west);
\draw[connector] (sub5.west) |- ([xshift=-2mm]sub9.west);
\draw[connector] (cat1.west) |- ([xshift=-2mm]sub10.west);

\end{tikzpicture}

\begin{tikzpicture}[
  every node/.style={align=left,},
  category/.style={text width=6cm, align=left, inner sep=4pt, font=\small},
  root/.style={text width=10cm, align=left, inner sep=4pt},
  quote/.style={rounded corners, thick, text width=6cm, align=justify, inner sep=4pt, font=\footnotesize},
  arrow/.style={-{Stealth[length=3mm]}, thick},
  connector/.style={thick},
  spacing/.store in=\myspacing, spacing=1.5cm 
]

\node[root] (cat1) at (0, 0) {Ethical hazards for care providers/facilities (including caring relatives) (continued)};

\node[category, anchor=west] (sub1) at ($(cat1.south west)+(0.5,-\myspacing+0.5cm)$) {Complicating management for care facilities
 (continued)};
\node[quote, anchor=west] (q1) at ($(sub1.east)+(1.5cm,0)$) {};
\node[draw=none, fit=(sub1)(q1), inner sep=0pt] (pair1) {};

\node[category, anchor=west] (sub2) at ($(pair1.south west)+(0.5,-\myspacing)$) {Mismanagement due to complex changes/unclear consequence};
\node[quote, anchor=west] (q2) at ($(sub2.east)+(1.cm,0)$) {"'We started using these systems in order to have a unique selling point. Back then, hardly anyone did that. But it wasn’t economically viable at the time', says Adler. Her care service was one of the first in the region to use autonomous systems. 'We had too many false alarms. That’s expensive.' Autonomous systems in care initially worked with fixed values. For example, the robot requested a blood pressure measurement and transmitted the data directly to the nursing service. 'But not everyone’s blood pressure is the same', says Jana Fiedler." (Translation by the authors)(Seefeldt and Hülsken-Giesler 2020 Pflegeethik und Robotik in der Pflege, p. 272)};
\draw[arrow] (sub2.east) -- (q2.west);
\node[draw=none, fit=(sub2)(q2), inner sep=0pt] (pair2) {};

\node[category, anchor=west] (sub3) at ($(pair2.south west)+(0.5,-\myspacing-0.5cm)$) {Possible loss of nursing staff skills due to social robots};
\node[quote, anchor=west] (q3) at ($(sub3.east)+(0.5cm,0)$) {“Moreover, integration of robots in the healthcare services may also contribute to a deskilling of the personnel [8], i.e., if some of the tasks are carried out by the robot instead of the human caregiver, the human caregiver will not get the necessary hands-on practice in dealing with those types of tasks.”(Saplacan et al. 2021 On Ethical Challenges Raised by Care Robots: A Review of the Existing Regulatory-, Theoretical-, and Research Gaps, p. 223)};
\draw[arrow] (sub3.east) -- (q3.west);
\node[draw=none, fit=(sub3)(q3), inner sep=0pt] (pair3) {};

\node[category, anchor=west] (sub4) at ($(pair3.south west)+(0.,-\myspacing-0.3cm)$) {Investment might turn into waste of resource};
\node[quote, anchor=west] (q4) at ($(sub4.east)+(0.5cm,0)$) {“A Reuters article (21st September 2007) reported that director of a Japanese care home had said that residents ‘liked the ifbot for about a month before they lost interest’. The robot had spent most of the past 2 years languishing alone in a corner, and the director of the facility commented, ‘stuffed animals are more popular’.” (Sharkey and Sharkey 2012 Granny and the robots: ethical issues in robot care for the elderly, pp. 34, 35)};
\draw[arrow] (sub4.east) -- (q4.west);
\node[draw=none, fit=(sub4)(q4), inner sep=0pt] (pair4) {};

\node[category, anchor=west] (sub5) at ($(pair4.south west)+(-0.5,-\myspacing-0.2cm)$) {Social robots turn out strategic disadvantage on the market};
\node[quote, anchor=west] (q5) at ($(sub5.east)+(1.cm,0)$) {“Robots that support care are viewed very critically by senior citizens. Nursing homes with robots do not necessarily enhance the status of an inpatient facility [...].” (Translation by the authors) (Kapitel Meyer 2011 Akzeptanz ausgewählter Anwendungsszenarien, p. 103; in: Meyer 2011 Mein Freund der Roboter - Servicerobotik für ältere Menschen - eine Antwort auf den demographischen Wandel?)};
\draw[arrow] (sub5.east) -- (q5.west);
\node[draw=none, fit=(sub5)(q5), inner sep=0pt] (pair5) {};

\node[category, anchor=west] (sub6) at ($(pair5.south west)+(0.5,-\myspacing-0.7cm)$) {Social robots perceived as indicator for low quality of care};
\node[quote, anchor=west] (q6) at ($(sub6.east)+(0.5cm,0)$) {“The importance of broadening the perspective in this way is shown by the assessment of care facilities in which robots are used. For the interviewees, this would not be a factor that enhances this facility, in contrast to modern furnishings, a cozy atmosphere or appealing lighting. They would assume that the facility does not have enough nursing staff.” (Translation by the authors)(Kapitel Meyer 2011 Akzeptanz ausgewählter Anwendungsszenarien, p. 101; in: Meyer 2011 Mein Freund der Roboter - Servicerobotik für ältere Menschen - eine Antwort auf den demographischen Wandel?)};
\draw[arrow] (sub6.east) -- (q6.west);
\node[draw=none, fit=(sub6)(q6), inner sep=0pt] (pair6) {};

\node[category, anchor=north west] (sub7) at ($(pair6.south west)+(-0.,0)$) {\textit{Cont. on next page}};
\node[quote, anchor=west] (q7) at ($(sub7.east)+(1.cm,0)$) {};
\node[draw=none, fit=(sub7)(q7), inner sep=0pt] (pair7) {};

\node[category, anchor=north west] (sub8) at ($(pair7.south west)+(-0.5,0)$) {\textit{Cont. on next page}};
\node[quote, anchor=west] (q8) at ($(sub8.east)+(1.cm,0)$) {};
\node[draw=none, fit=(sub8)(q8), inner sep=0pt] (pair8) {};

\node[category, anchor=north west] (sub9) at ($(pair8.south west)+(-0.5,0)$) {\textit{Cont. on next page}};
\node[quote, anchor=west] (q9) at ($(sub9.east)+(1.cm,0)$) {};
\node[draw=none, fit=(sub9)(q9), inner sep=0pt] (pair9) {};

\draw[connector] (cat1.west) |- ([xshift=-2mm]sub1.west);
\draw[connector] (sub1.west) |- ([xshift=-2mm]sub2.west);
\draw[connector] (sub2.west) |- ([xshift=-2mm]sub3.west);
\draw[connector] (sub2.west) |- ([xshift=-2mm]sub4.west);
\draw[connector] (sub1.west) |- ([xshift=-2mm]sub5.west);
\draw[connector] (sub5.west) |- ([xshift=-2mm]sub6.west);
\draw[connector] (sub5.west) |- ([xshift=-2mm]sub7.west);
\draw[connector] (sub1.west) |- ([xshift=-2mm]sub8.west);
\draw[connector] (cat1.west) |- ([xshift=-2mm]sub9.west);

\end{tikzpicture}

\begin{tikzpicture}[
  every node/.style={align=left,},
  category/.style={text width=6cm, align=left, inner sep=4pt, font=\small},
  root/.style={text width=10cm, align=left, inner sep=4pt},
  quote/.style={rounded corners, thick, text width=6cm, align=justify, inner sep=4pt, font=\footnotesize},
  arrow/.style={-{Stealth[length=3mm]}, thick},
  connector/.style={thick},
  spacing/.store in=\myspacing, spacing=1.5cm 
]

\node[root] (cat1) at (0, 0) {Ethical hazards for care providers/facilities (including caring relatives) (continued)};

\node[category, anchor=west] (sub1) at ($(cat1.south west)+(0.5,-\myspacing+0.5cm)$) {Complicating management for care facilities
 (continued)};
\node[quote, anchor=west] (q1) at ($(sub1.east)+(1.5cm,0)$) {};
\node[draw=none, fit=(sub1)(q1), inner sep=0pt] (pair1) {};

\node[category, anchor=west] (sub2) at ($(pair1.south west)+(0.5,-\myspacing+0.5cm)$) {Social robots turn out strategic disadvantage on the market (continued)};
\node[quote, anchor=west] (q2) at ($(sub2.east)+(1.5cm,0)$) {};
\node[draw=none, fit=(sub2)(q2), inner sep=0pt] (pair2) {};

\node[category, anchor=west] (sub3) at ($(pair2.south west)+(0.5,-\myspacing)$) {Negative publicity for care institutions when risks realize};
\node[quote, anchor=west] (q3) at ($(sub3.east)+(0.5cm,0)$) {“The data captured by robotics and autonomous systems deployed in long-term care settings cover a significant amount of an old person’s life at the end of the developmental curve. This information is highly sensitive and intimate especially when it comes to data collected towards the end-of-life of the care recipients. If these data are breached for unsolicited purposes that are detrimental and malicious, it will result in strong distrust from the care recipients and their families, as well as negative publicities for the care institutions (Ienca et al., 2016).”
(Tan et al. 2021 Tensions and antagonistic interactions of risks and ethics of using robotics and autonomous systems in long-term care, p. 8)};
\draw[arrow] (sub3.east) -- (q3.west);
\node[draw=none, fit=(sub3)(q2), inner sep=0pt] (pair3) {};

\node[category, anchor=west] (sub3a) at ($(pair3.south west)+(-0.5,-\myspacing-1.5cm)$) {Problems with ongoing maintenance and support};
\node[quote, anchor=west] (q3a) at ($(sub3a.east)+(1.cm,0)$) {“Currently, universal access by fair opportunity to assistive technologies is an ideal but not a reality. A few studies also highlighted ongoing maintenance, cleaning and repair can be an added cost [17, 36].” (Hung et al 2019 The benefits of and barriers to using a social robot PARO in a care setting: a scoping review, p. 6)};
\draw[arrow] (sub3a.east) -- (q3a.west);
\node[draw=none, fit=(sub3a)(q3a), inner sep=0pt] (pair3a) {};

\node[category, anchor=west] (sub4) at ($(pair3a.south west)+(0.,-\myspacing)$) {High training effort};
\node[quote, anchor=west] (q4) at ($(sub4.east)+(1.cm,0)$) {“One researcher participant said, ‘I have seen the hurdle of training. It is a lot of work to motivate people learn about what the robots do and how to use them. Some LTC homes just not have the capacity to adopt new practice and support staff training.’”(Hung et al. 2022 Technological risks and ethical implications of using robots in long-term care, p. 4) };
\draw[arrow] (sub5.east) -- (q5.west);
\node[draw=none, fit=(sub4)(q4), inner sep=0pt] (pair4) {};

\node[category, anchor=west] (sub5) at ($(pair4.south west)+(0.,-\myspacing)$) {Loss of skilled personnel due to replacement};
\node[quote, anchor=west] (q5) at ($(sub5.east)+(1.cm,0)$) {“By substituting robots for human health care professionals, we could lose many dynamic, highly skilled knowledgeable, and compassionate professionals - both regulated and unregulated - who are experts in the care for people with dementia (Butter et al., 2008; Veruggio \& Operto, 2008).” (Huschilt and Clune 2012 The Use of Socially Assistive Robots for Dementia Care, p. 18)};
\draw[arrow] (sub5.east) -- (q5.west);
\node[draw=none, fit=(sub5)(q5), inner sep=0pt] (pair5) {};

\node[category, anchor=west] (sub6) at ($(pair5.south west)+(-0.5,-\myspacing-2cm)$) {Lack of protection due to lack of guidelines/legal regulations};
\node[quote, anchor=west] (q6) at ($(sub6.east)+(1.5cm,0)$) {“Legal concerns exist about the use of SARs for dementia care: Robots are not people and members of a regulatory body. Currently, robots and the performance of robotic tasks are not regulated, nor do they require a registration or license. There is no legally defined scope of practice for robots. Therefore, SARs are not sanctioned to perform interventions that are authorized for licensed nurses. Questions from this premise emerge: What happens if a regulated health care professional delegates a task to the SAR? Who should be held accountable for a robot’s error – the robot, the delegating professional, the designer, the programmer, the manufacturer, or the patient? (Feil-Seifer \& Mataric, 2011; Veruggio \& Operto, 2008). All of these issues must be considered if SARs are to enter mainstream dementia care approaches.” (Huschilt et al. 2012 The Use of Socially Assistive Robots for Dementia care, p. 17)};
\draw[arrow] (sub6.east) -- (q6.west);
\node[draw=none, fit=(sub6)(q6), inner sep=0pt] (pair6) {};

\draw[connector] (cat1.west) |- ([xshift=-2mm]sub1.west);
\draw[connector] (sub1.west) |- ([xshift=-2mm]sub2.west);
\draw[connector] (sub2.west) |- ([xshift=-2mm]sub3.west);
\draw[connector] (sub1.west) |- ([xshift=-2mm]sub3a.west);
\draw[connector] (sub1.west) |- ([xshift=-2mm]sub4.west);
\draw[connector] (sub1.west) |- ([xshift=-2mm]sub5.west);
\draw[connector] (cat1.west) |- ([xshift=-2mm]sub6.west);

\end{tikzpicture}

\begin{tikzpicture}[
  every node/.style={align=left,},
  category/.style={text width=6cm, align=left, inner sep=4pt, font=\small},
  root/.style={text width=10cm, align=left, inner sep=4pt},
  quote/.style={rounded corners, thick, text width=6cm, align=justify, inner sep=4pt, font=\footnotesize},
  arrow/.style={-{Stealth[length=3mm]}, thick},
  connector/.style={thick},
  spacing/.store in=\myspacing, spacing=1.5cm 
]

\node[root] (cat1) at (0, 0) {Ethical opportunities for society/the (healthcare) system (Main category)};

\node[category, anchor=west] (sub1) at ($(cat1.south west)+(0.5,-\myspacing)$) {Supporting state in ensuring comprehensive nursing care for all};
\node[quote, anchor=west] (q1) at ($(sub1.east)+(1.5cm,0)$) {“For the majority of older people and their families there is no reason why buying something that helps with care should not be a normal consumer purchase. While it has not yet been possible for telecare equipment to break into a mass consumer market, similar functionality but packaged in a small robot may have the potential to make it to supermarket shelves.” (Blackman 2013 Care robots for the supermarket shelf: a product gap in assistive technologies, p. 767)};
\draw[arrow] (sub1.east) -- (q1.west);
\node[draw=none, fit=(sub1)(q1), inner sep=0pt] (pair1) {};

"
\node[category, anchor=west] (sub2) at ($(pair1.south west)+(0.5,-\myspacing)$) {Efficient care provision};
\node[quote, anchor=west] (q2) at ($(sub2.east)+(1.cm,0)$) {“They also address the need to offer and provide acceptable care as economically as possible.”(Draper and Sorell 2017 Ethical values and social care robots for older people: an international qualitative study, p. 50)};
\draw[arrow] (sub2.east) -- (q2.west);
\node[draw=none, fit=(sub2)(q2), inner sep=0pt] (pair2) {};"
"
\node[category, anchor=west] (sub3) at ($(pair2.south west)+(0.5,-\myspacing)$) {Reducing costs of care overall};
\node[quote, anchor=west] (q3) at ($(sub3.east)+(0.5cm,0)$) {“Finally, last but not least, among the strongest motivations for introducing care robots is to reduce the cost of healthcare […].”(Salvini 2015 On Ethical, Legal and Social Issues of Care Robots, p. 432; in: Mohammed 2015 Intelligent Assistive Robots - Recent Advances in Assistive Robotics for Everyday Activities) };
\draw[arrow] (sub3.east) -- (q3.west);
\node[draw=none, fit=(sub3)(q3), inner sep=0pt] (pair3) {};"
"
\node[category, anchor=west] (sub4) at ($(pair3.south west)+(-0.5,-\myspacing)$) {Furthering equal access to/provision of care};
\node[quote, anchor=west] (q4) at ($(sub4.east)+(1.cm,0)$) {“SAR has implications for social equality, since depending on how it is developed and implemented, it may either contribute to increase or lessen the equality of care both in terms of access and quality of treatment.”(Boada et al. 2021 The ethical issues of social assistive robotics: A critical literature review, p. 9)};
\draw[arrow] (sub4.east) -- (q4.west);
\node[draw=none, fit=(sub4)(q4), inner sep=0pt] (pair4) {};"
"
\node[category, anchor=west] (sub5) at ($(pair4.south west)+(0.5,-\myspacing)$) {Social robots treat all patients equally };
\node[quote, anchor=west] (q5) at ($(sub5.east)+(0.5cm,0)$) {“Some participants maintained that robots could be more caring than humans and noted that ‘robot care’ would sometimes be preferential. The reasons given for such were that robots treat all patients equally.” (Nyholm 2021 Users ambivalent sense of security with humanoid robots in healthcare, p. 224)};
\draw[arrow] (sub5.east) -- (q5.west);
\node[draw=none, fit=(sub5)(q5), inner sep=0pt] (pair5) {};"
"
\node[category, anchor=west] (sub6) at ($(pair5.south west)+(0.,-\myspacing-1.7cm)$) {Ensuring that care is available in rural areas};
\node[quote, anchor=west] (q6) at ($(sub6.east)+(0.5cm,0)$) {“In particular, technology-supported case and care management and telematics-supported systems in general (tele-nursing) are likely to gain in importance in future in view of the expected increase in the number of very old people living alone and also very old people remaining alone in (rural) regions, and will become an important field of activity for outpatient care services, especially in structurally weak regions.” (Translation by the authors)(Bleses and Dammert 2020 Neue Technologien aus Sicht der Pflegewissenschaft, p. 60; in: Hanika 2020 Künstliche Intelligenz, Robotik und autonome Systeme in der Gesundheitsversorgung)};
\draw[arrow] (sub6.east) -- (q6.west);
\node[draw=none, fit=(sub6)(q6), inner sep=0pt] (pair6) {};"

\node[category, anchor=north west] (sub7) at ($(pair6.south west)+(-0.5,0)$) {\textit{Cont. on next page}};
\node[quote, anchor=west] (q7) at ($(sub7.east)+(1.cm,0)$) {};
\node[draw=none, fit=(sub7)(q7), inner sep=0pt] (pair7) {};

\node[category, anchor=north west] (sub8) at ($(pair7.south west)+(-0.5,0)$) {\textit{Cont. on next page}};
\node[quote, anchor=west] (q8) at ($(sub8.east)+(1.cm,0)$) {};
\node[draw=none, fit=(sub8)(q8), inner sep=0pt] (pair8) {};

\draw[connector] (cat1.west) |- ([xshift=-2mm]sub1.west);
\draw[connector] (sub1.west) |- ([xshift=-2mm]sub2.west);
\draw[connector] (sub2.west) |- ([xshift=-2mm]sub3.west);
\draw[connector] (sub1.west) |- ([xshift=-2mm]sub4.west);
\draw[connector] (sub4.west) |- ([xshift=-2mm]sub5.west);
\draw[connector] (sub4.west) |- ([xshift=-2mm]sub6.west);
\draw[connector] (sub4.west) |- ([xshift=-2mm]sub7.west);
\draw[connector] (cat1.west) |- ([xshift=-2mm]sub8.west);

\end{tikzpicture}

\begin{tikzpicture}[
  every node/.style={align=left,},
  category/.style={text width=6cm, align=left, inner sep=4pt, font=\small},
  root/.style={text width=10cm, align=left, inner sep=4pt},
  quote/.style={rounded corners, thick, text width=6cm, align=justify, inner sep=4pt, font=\footnotesize},
  arrow/.style={-{Stealth[length=3mm]}, thick},
  connector/.style={thick},
  spacing/.store in=\myspacing, spacing=1.5cm 
]

\node[root] (cat1) at (0, 0) {Ethical opportunities for society/the (healthcare) system (continued)};

\node[category, anchor=west] (sub1) at ($(cat1.south west)+(0.5,-\myspacing+0.5cm)$) {Supporting state in ensuring comprehensive nursing care for all (continued)};
\node[quote, anchor=west] (q1) at ($(sub1.east)+(1.5cm,0)$) {};
\node[draw=none, fit=(sub1)(q1), inner sep=0pt] (pair1) {};

"
\node[category, anchor=west] (sub2) at ($(pair1.south west)+(0.5,-\myspacing)$) {Cushioning the nursing shortage/demographic change};
\node[quote, anchor=west] (q2) at ($(sub2.east)+(1.cm,0)$) {"One of the areas of society in which such issues are currently becoming particularly virulent is care for the elderly, which is facing enormous challenges as a result of demographic change: The number of people in need of care has been rising for years, while it is becoming increasingly difficult to recruit nursing staff in sufficient numbers. A massive care gap is already emerging today, which is making the call for technical and, in particular, robotic support louder - innovative assistance technologies for care have been promoted by politicians for years [...]." (Translation by the authors)(Kehl 2018 Entgrenzungen Zwischen Mensch Und Maschine, Oder: Können Roboter Gute Pflege Leisten?, p. 3)};
\draw[arrow] (sub2.east) -- (q2.west);
\node[draw=none, fit=(sub2)(q2), inner sep=0pt] (pair2) {};"
"
\node[category, anchor=west] (sub3) at ($(pair2.south west)+(-0.5,-\myspacing-1.5cm)$) {Reducing discrimination/stigmatization of certain groups};
\node[quote, anchor=west] (q3) at ($(sub3.east)+(1.5cm,0)$) {“According to its proponents, this approach [approach of neutralization] could help to circumvent morally contested issues like the reproduction of gender stereotypes in care ‘because you can probably only get into trouble with that [i.e. gender attributions]’ (I16, 41). One marketing expert argued that a neutral approach was more inclusive and facilitated the establishment of a personal relationship: ‘At the moment everything is gender neutral. Pepper addresses people mostly with ‘you’. That certainly is a door opener because you do not address a person as ‘mister’ or ‘miss’ but directly create a personal connection’ (I4, 27).”(Weßel et al. 2022 Gender and Age Stereotypes in Robotics for Eldercare: Ethical Implications of Stakeholder Perspectives from Technology Development, Industry, and Nursing, p. 8) };
\draw[arrow] (sub3.east) -- (q3.west);
\node[draw=none, fit=(sub3)(q3), inner sep=0pt] (pair3) {};"

"
\node[category, anchor=west] (sub4) at ($(pair3.south west)+(0.5,-\myspacing-0.5cm)$) {Decreasing the taboo around sexuality in old age/disability};
\node[quote, anchor=west] (q4) at ($(sub4.east)+(1.cm,0)$) {“Our aim for doing this was to help recognize that though their common denominator is sex, disability is only partly a framing parameter. It is therefore a case where all these sex gadgets, sex toys and other sex innovations are actually bridging the gaps in thinking on this subject, rather than letting the gaps grow bigger, which is a rather positive development in our view.”(Koumpis and Gees 2020 Sex with Robots: A not-so-niche market for disabled and older persons, p. 229)};
\draw[arrow] (sub4.east) -- (q4.west);
\node[draw=none, fit=(sub4)(q4), inner sep=0pt] (pair4) {};"
"
\node[category, anchor=west] (sub5) at ($(pair4.south west)+(-0.5,-\myspacing-0.4cm)$) {Promoting desired behavior of citizens};
\node[quote, anchor=west] (q5) at ($(sub5.east)+(1.5cm,0)$) {“As we have pointed out, the use of robots may actually even have the potential to alleviate or circumvent some of the pertinent pitfalls and problems, e.g., the sociopsychological mechanisms of stereotype threat that can impede the performance of male nurses in a traditionally female profession.”(Weßel et al. 2021 Gender Stereotyping of Robotic Systems in Eldercare: An Exploratory Analysis of Ethical Problems and Possible Solutions, p. 11)};
\draw[arrow] (sub5.east) -- (q5.west);
\node[draw=none, fit=(sub5)(q5), inner sep=0pt] (pair5) {};"
"
\node[category, anchor=west] (sub6) at ($(pair5.south west)+(0.5,-\myspacing)$) {Developing moral skills of citizens};
\node[quote, anchor=west] (q6) at ($(sub6.east)+(1.cm,0)$) {“On the one hand, because of their ‘interpersonal’ particular kind of interaction with humans, social robots may influence and shape human moral character, by potentially cultivating both vices and virtues.”(Boada et al. 2021 The ethical issues of social assistive robotics: A critical literature review, p. 7)};
\draw[arrow] (sub6.east) -- (q6.west);
\node[draw=none, fit=(sub6)(q6), inner sep=0pt] (pair6) {};"
"
\node[category, anchor=north west] (sub7) at ($(pair6.south west)+(-0.0,0)$) {\textit{Cont. on next page}};
\node[quote, anchor=west] (q7) at ($(sub7.east)+(1.cm,0)$) {};
\node[draw=none, fit=(sub7)(q7), inner sep=0pt] (pair7) {};

\node[category, anchor=north west] (sub8) at ($(pair7.south west)+(-0.5,0)$) {\textit{Cont. on next page}};
\node[quote, anchor=west] (q8) at ($(sub8.east)+(1.cm,0)$) {};
\node[draw=none, fit=(sub8)(q8), inner sep=0pt] (pair8) {};

\draw[connector] (cat1.west) |- ([xshift=-2mm]sub1.west);
\draw[connector] (sub1.west) |- ([xshift=-2mm]sub2.west);
\draw[connector] (cat1.west) |- ([xshift=-2mm]sub3.west);
\draw[connector] (sub3.west) |- ([xshift=-2mm]sub4.west);
\draw[connector] (cat1.west) |- ([xshift=-2mm]sub5.west);
\draw[connector] (sub5.west) |- ([xshift=-2mm]sub6.west);
\draw[connector] (sub5.west) |- ([xshift=-2mm]sub7.west);
\draw[connector] (cat1.west) |- ([xshift=-2mm]sub8.west);

\end{tikzpicture}

\begin{tikzpicture}[
  every node/.style={align=left,},
  category/.style={text width=6cm, align=left, inner sep=4pt, font=\small},
  root/.style={text width=10cm, align=left, inner sep=4pt},
  quote/.style={rounded corners, thick, text width=6cm, align=justify, inner sep=4pt, font=\footnotesize},
  arrow/.style={-{Stealth[length=3mm]}, thick},
  connector/.style={thick},
  spacing/.store in=\myspacing, spacing=1.5cm 
]

\node[root] (cat1) at (0, 0) {Ethical opportunities for society/the (healthcare) system (continued)};

\node[category, anchor=west] (sub1) at ($(cat1.south west)+(0.5,-\myspacing+0.5cm)$) {Promoting desired behavior of citizens (continued)};
\node[quote, anchor=west] (q1) at ($(sub1.east)+(1.5cm,0)$) {};
\node[draw=none, fit=(sub1)(q1), inner sep=0pt] (pair1) {};

\node[category, anchor=west] (sub2) at ($(pair1.south west)+(0.5,-\myspacing)$) {Nudging users to desirable behavior via social robot design};
\node[quote, anchor=west] (q2) at ($(sub2.east)+(1.cm,0)$) {“Beyond, there are also discussions as to whether social robots should be designed with ‘opt-in’, ‘opt-out’ or a ‘no way out’ design features to nudge users towards espousing personal interest or collective interests of the society (Borenstein et al., 2016).”(Tan et al. 2021 Tensions and antagonistic interactions of risks and ethics of using robotics and autonomous systems in long-term care, p. 12)};
\draw[arrow] (sub2.east) -- (q2.west);
\node[draw=none, fit=(sub2)(q2), inner sep=0pt] (pair2) {};"
"
\node[category, anchor=west] (sub3) at ($(pair2.south west)+(-0.5,-\myspacing-1cm)$) {Important field for economic development (location advantage)};
\node[quote, anchor=west] (q3) at ($(sub3.east)+(1.5cm,0)$) {“Various focus technologies are commonly discussed under the heading of Care 4.0, from electronic documentation to AAL systems (Ambient Assisted Living), telecare and telepresence medicine through to robotics. The latter is at the center of this statement because it is considered to be particularly promising for the future in terms of overcoming growing care challenges, but also as a driver of economic development.22” (Translation by the authors)(Deutscher Ethikrat 2020 Robotik für gute Pflege (Stellungnahme), p. 14)};
\draw[arrow] (sub3.east) -- (q3.west);
\node[draw=none, fit=(sub3)(q3), inner sep=0pt] (pair3) {};"

\draw[connector] (cat1.west) |- ([xshift=-2mm]sub1.west);
\draw[connector] (sub1.west) |- ([xshift=-2mm]sub2.west);
\draw[connector] (cat1.west) |- ([xshift=-2mm]sub3.west);

\end{tikzpicture}

\begin{tikzpicture}[
  every node/.style={align=left,},
  category/.style={text width=6cm, align=left, inner sep=4pt, font=\small},
  root/.style={text width=10cm, align=left, inner sep=4pt},
  quote/.style={rounded corners, thick, text width=6cm, align=justify, inner sep=4pt, font=\footnotesize},
  arrow/.style={-{Stealth[length=3mm]}, thick},
  connector/.style={thick},
  spacing/.store in=\myspacing, spacing=1.5cm 
]

\node[root] (cat1) at (0, 0) {Ethical hazards for society/the (healthcare) system (Main Category)};
"
\node[category, anchor=west] (sub1) at ($(cat1.south west)+(0.5,-\myspacing)$) {Moving forward without an adequate knowledge/ discussion base};
\node[quote, anchor=west] (q1) at ($(sub1.east)+(1.5cm,0)$) {“The fact that such dystopias are nevertheless so prominent in ethical and public discourse indicates, on the one hand, that automation efforts in the care sector raise normative questions of a completely new dimension. On the other hand, it also highlights the central deficits of the corresponding debates. As these mainly revolve around speculative scenarios, there is hardly any differentiated discussion of the potential and limits of robotics in care.” (Translation by the authors)(Kehl 2018 Entgrenzungen Zwischen Mensch Und Maschine, Oder: Können Roboter Gute Pflege Leisten?, p. 5)};
\draw[arrow] (sub1.east) -- (q1.west);
\node[draw=none, fit=(sub1)(q1), inner sep=0pt] (pair1) {};"

"
\node[category, anchor=west] (sub2) at ($(pair1.south west)+(0.,-\myspacing-0.6cm)$) {Using a misguided approach to regulation/design of social robots};
\node[quote, anchor=west] (q2) at ($(sub2.east)+(1.5cm,0)$) {“In terms of the decisions regarding the use of robots in care for PwD, the decision-makers (state-funding bodies, owners of private care or nursing homes) are not those who initially bear the costs (those in care). They are, in effect, testing robotic care for older people in ‘living labs’ – care homes. By the time those decision-makers will need that sort of care, many of the problems that initially are presented will have disappeared.” (O’Brolcháin 2017 Robots and people with dementia: Unintended consequences and moral hazard, p. 968)};
\draw[arrow] (sub2.east) -- (q2.west);
\node[draw=none, fit=(sub2)(q2), inner sep=0pt] (pair2) {};"
"
\node[category, anchor=west] (sub3) at ($(pair2.south west)+(0.5,-\myspacing-1.2cm)$) {Prioritizing system/economic advantages over care recipients needs};
\node[quote, anchor=west] (q3) at ($(sub3.east)+(1.cm,0)$) {"Against this backdrop, fears cannot be dismissed that an accelerated use of robots could lead to care work being increasingly subjected to a mechanistic understanding, i.e. narrowed down to purpose-related aspects and the sentient aspects being marginalized accordingly - especially if this can lead to economic savings. This applies all the more as a one-sided understanding of nursing care as a rational problem-solving activity is part of overarching and long-standing standardization and economization efforts [...] which are increasingly subjecting not only nursing care but the healthcare system as a whole to the logic of business management." (Translation by the authors)(Kehl 2018 Entgrenzungen Zwischen Mensch Und Maschine, Oder: Können Roboter Gute Pflege Leisten?, p. 6)};
\draw[arrow] (sub3.east) -- (q3.west);
\node[draw=none, fit=(sub3)(q3), inner sep=0pt] (pair3) {};"
"
\node[category, anchor=west] (sub4) at ($(pair3.south west)+(0.,-\myspacing-2cm)$) {Working with underdeveloped ethical concepts};
\node[quote, anchor=west] (q4) at ($(sub4.east)+(1.cm,0)$) {“For instance, (human) freedom is currently understood as what is philosophically known as ‘negative liberty’ [79]. The ethical implications of SAR regarding this issue, though, could and should be broadened up by approaching the matter from a deeper understanding of this notion in its dimension of ‘positive liberty’ [79]. That is, closer to freedom as autonomy –in which freedom has to do with self-realization, with taking over the own life–. It could also be interesting to think from the perspective of Pettit’s ‘republican conception of freedom as non-domination’ [84]. Delving into the meaning of this notion would definitely allow for a richer normative-oriented reflection on SAR that takes into account the interdependence existing between freedom and the sociopolitical structuring of human life, thus approaching SAR’s power of domination both at an interpersonal and structural level.” (Boada et al. 2021 The ethical issues of social assistive robotics: A critical literature review, p. 10)};
\draw[arrow] (sub4.east) -- (q4.west);
\node[draw=none, fit=(sub4)(q4), inner sep=0pt] (pair4) {};"

\node[category, anchor=north west] (sub5) at ($(pair4.south west)+(-0.0,0)$) {\textit{Cont. on next page}};
\node[quote, anchor=west] (q5) at ($(sub5.east)+(1.cm,0)$) {};
\node[draw=none, fit=(sub5)(q5), inner sep=0pt] (pair5) {};

\node[category, anchor=north west] (sub6) at ($(pair5.south west)+(-0.5,0)$) {\textit{Cont. on next page}};
\node[quote, anchor=west] (q6) at ($(sub6.east)+(1.cm,0)$) {};
\node[draw=none, fit=(sub6)(q6), inner sep=0pt] (pair6) {};

\draw[connector] (cat1.west) |- ([xshift=-2mm]sub1.west);
\draw[connector] (cat1.west) |- ([xshift=-2mm]sub2.west);
\draw[connector] (sub2.west) |- ([xshift=-2mm]sub3.west);
\draw[connector] (sub2.west) |- ([xshift=-2mm]sub4.west);
\draw[connector] (sub2.west) |- ([xshift=-2mm]sub5.west);
\draw[connector] (cat1.west) |- ([xshift=-2mm]sub6.west);

\end{tikzpicture}

\begin{tikzpicture}[
  every node/.style={align=left,},
  category/.style={text width=6cm, align=left, inner sep=4pt, font=\small},
  root/.style={text width=10cm, align=left, inner sep=4pt},
  quote/.style={rounded corners, thick, text width=6cm, align=justify, inner sep=4pt, font=\footnotesize},
  arrow/.style={-{Stealth[length=3mm]}, thick},
  connector/.style={thick},
  spacing/.store in=\myspacing, spacing=1.5cm 
]

\node[root] (cat1) at (0, 0) {Ethical hazards for society/the (healthcare) system (continued)};

\node[category, anchor=west] (sub1) at ($(cat1.south west)+(0.5,-\myspacing+0.5cm)$) {Using a misguided approach to regulation/design of social robots (continued)};
\node[quote, anchor=west] (q1) at ($(sub1.east)+(1.5cm,0)$) {};
\node[draw=none, fit=(sub1)(q1), inner sep=0pt] (pair1) {};

"
\node[category, anchor=west] (sub2) at ($(pair1.south west)+(0.5,-\myspacing-0.3cm)$) {Ignoring relevant aspects of the care situation};
\node[quote, anchor=west] (q2) at ($(sub2.east)+(1.cm,0)$) {“Most of the socially assistive robots have so far been tested only in laboratory settings. Consequently, there is a scarcity of knowledge about their use in real care situations from the point of view of older care-dependent people, especially regarding their use by persons with dementia, and particularly considering the application in domestic environments. Concretely, the testing of robots in laboratory settings cannot capture the complexity and high variability of everyday real-life situations in nursing care (Bioethics Commission, 2018).”(Paletta et al. 2019 AMIGO–A Socially Assistive Robot for Coaching Multimodal Training of Persons with Dementia, p. 267; in: Korn 2019 Social Robots: Technological, Societal and Ethical Aspects of Human-Robot Interaction)};
\draw[arrow] (sub2.east) -- (q2.west);
\node[draw=none, fit=(sub2)(q2), inner sep=0pt] (pair2) {};"
"
\node[category, anchor=west] (sub3) at ($(pair2.south west)+(0.,-\myspacing-2cm)$) {Premature regulation not yet knowing relevant implications};
\node[quote, anchor=west] (q3) at ($(sub3.east)+(1.cm,0)$) {“The source of normative uncertainties and therefore the subject of ethical evaluation is generally not the technology itself, but its role in specific socio-technical contexts of action (Grunwald 2013a, p. 4). This essentially includes the intended and unintended consequences that are associated with the use of technology and whose moral implications need to be clarified ethically. In fields such as care robotics, which are still at a very early stage of development and whose potential applications are therefore still unclear, this consequentialist (consequence-oriented) approach is confronted with fundamental problems - after all, the consequences of technology are still largely in the dark, which means that their ethical assessment is also largely in the dark (which, however, is basically a problem of insufficient prognostic knowledge and not a problem of normative evaluation per se).” (Translation by the authors)(Kehl 2018 Wege zu verantwortungsvoller Forschung und Entwicklung im Bereich der Pfegerobotik: Die ambivalente Rolle der Ethik, p. 148; in: Bendel 2018 Pflegeroboter)};
\draw[arrow] (sub3.east) -- (q3.west);
\node[draw=none, fit=(sub3)(q3), inner sep=0pt] (pair3) {};"
"
\node[category, anchor=west] (sub4) at ($(pair3.south west)+(0.,-\myspacing-0.2cm)$) {Unclear or lacking regulation providing limited orientation};
\node[quote, anchor=west] (q4) at ($(sub4.east)+(1.cm,0)$) {“From a regulatory perspective, robots for dementia are often in a gray zone between the regulation of medical applications and that of general ICT applications. A striking example is Paro, who is classified as Class 2 medical device by the U.S. FDA regulation but not in the EU.”(Ienca et al. 2016 Social and Assistive Robotics in Dementia Care: Ethical Recommendations for Research and Practice, p. 571)};
\draw[arrow] (sub4.east) -- (q4.west);
\node[draw=none, fit=(sub4)(q4), inner sep=0pt] (pair4) {};"
"
\node[category, anchor=west] (sub5) at ($(pair4.south west)+(0.,-\myspacing)$) {Further problematic attitudes/approaches};
\node[quote, anchor=west] (q5) at ($(sub5.east)+(1.cm,0)$) {};
\node[draw=none, fit=(sub5)(q5), inner sep=0pt] (pair5) {};"

"
\node[category, anchor=west] (sub6) at ($(pair5.south west)+(0.5,-\myspacing)$) {Moral decisions are too complex for adequate social robot programming};
\node[quote, anchor=west] (q6) at ($(sub6.east)+(0.5cm,0)$) {“In my opinion, there is little to suggest that Bentham or Kant could return in the form of care robots in the near future. The technical difficulties of implementing a moral-philosophical approach in the form of algorithms are far too great.” (Translation by the authors)(Hilgendorf 2018 Recht und Ethik in der Pflegerobotik – ein Überblick, p. 383)};
\draw[arrow] (sub6.east) -- (q6.west);
\node[draw=none, fit=(sub6)(q6), inner sep=0pt] (pair6) {};"

\node[category, anchor=north west] (sub7) at ($(pair6.south west)+(-0.,0)$) {\textit{Cont. on next page}};
\node[quote, anchor=west] (q7) at ($(sub7.east)+(1.cm,0)$) {};
\node[draw=none, fit=(sub7)(q7), inner sep=0pt] (pair7) {};

\node[category, anchor=north west] (sub8) at ($(pair7.south west)+(-0.5,0)$) {\textit{Cont. on next page}};
\node[quote, anchor=west] (q8) at ($(sub8.east)+(1.cm,0)$) {};
\node[draw=none, fit=(sub8)(q8), inner sep=0pt] (pair8) {};

\node[category, anchor=north west] (sub9) at ($(pair8.south west)+(-0.5,0)$) {\textit{Cont. on next page}};
\node[quote, anchor=west] (q9) at ($(sub9.east)+(1.cm,0)$) {};
\node[draw=none, fit=(sub9)(q9), inner sep=0pt] (pair9) {};

\draw[connector] (cat1.west) |- ([xshift=-2mm]sub1.west);
\draw[connector] (sub1.west) |- ([xshift=-2mm]sub2.west);
\draw[connector] (sub1.west) |- ([xshift=-2mm]sub3.west);
\draw[connector] (sub1.west) |- ([xshift=-2mm]sub4.west);
\draw[connector] (sub1.west) |- ([xshift=-2mm]sub5.west);
\draw[connector] (sub5.west) |- ([xshift=-2mm]sub6.west);
\draw[connector] (sub5.west) |- ([xshift=-2mm]sub7.west);
\draw[connector] (sub1.west) |- ([xshift=-2mm]sub8.west);
\draw[connector] (cat1.west) |- ([xshift=-2mm]sub9.west);

\end{tikzpicture}

\begin{tikzpicture}[
  every node/.style={align=left,},
  category/.style={text width=6cm, align=left, inner sep=4pt, font=\small},
  root/.style={text width=10cm, align=left, inner sep=4pt},
  quote/.style={rounded corners, thick, text width=6cm, align=justify, inner sep=4pt, font=\footnotesize},
  arrow/.style={-{Stealth[length=3mm]}, thick},
  connector/.style={thick},
  spacing/.store in=\myspacing, spacing=1.5cm 
]

\node[root] (cat1) at (0, 0) {Ethical hazards for society/the (healthcare) system (continued)};

\node[category, anchor=west] (sub1) at ($(cat1.south west)+(0.5,-\myspacing+0.5cm)$) {Using a misguided approach to regulation/design of social robots (continued)};
\node[quote, anchor=west] (q1) at ($(sub1.east)+(1.5cm,0)$) {};
\node[draw=none, fit=(sub1)(q1), inner sep=0pt] (pair1) {};

\node[category, anchor=west] (sub2) at ($(sub1.south west)+(0.5,-\myspacing+0.5cm)$) {Further problematic attitudes/approaches (continued)};
\node[quote, anchor=west] (q2) at ($(sub2.east)+(1.5cm,0)$) {};
\node[draw=none, fit=(sub2)(q2), inner sep=0pt] (pair2) {};

"
\node[category, anchor=west] (sub3) at ($(pair2.south west)+(0.5,-\myspacing)$) {Guidelines without processes to hold people accountable};
\node[quote, anchor=west] (q3) at ($(sub3.east)+(0.5cm,0)$) {“To indeed make roboticists ‘accountable for the social, environmental and human health impacts that robotics may impose on the present and future generations,’ as the EP mentions, we encourage central institutions to develop these codes and establish consequences for violations, concerning fines and sanctions. The law could make them binding. Another way to make them binding would be to include them into a private contract, between a robot producer and a user – in this case a healthcare institution [131].”(Fosch-Villaronga and Albo-Canals 2019 “I’ll take care of you,” said the robot, p. 89)};
\draw[arrow] (sub3.east) -- (q3.west);
\node[draw=none, fit=(sub3)(q3), inner sep=0pt] (pair3) {};"
"
\node[category, anchor=west] (sub4) at ($(pair3.south west)+(0.,-\myspacing-1.6cm)$) {Approaches that are not ELSA-informed};
\node[quote, anchor=west] (q4) at ($(sub4.east)+(0.5cm,0)$) {“Coding literacy may be necessary for STEM curricula and may provide students employability in the future [113].  Still, this may entail a greater disconnection from what constitutes to be human in the future. As Bauman highlights, the more we are in the virtual type of proximity, the less time we spend in learning and acquiring the skills needed in the non-virtual type of reality [107]. Not including ethics, philosophy, and history in school curricula may probably prevent future generations from not only solving ethical questions arisen from technology (problem-solving) but what it is more worrying, from identifying such problems and issues in the very first place (problem-finding).”(Fosch-Villaronga and Albo-Canals 2019 “I’ll take care of you,” said the robot, p. 85)};
\draw[arrow] (sub4.east) -- (q4.west);
\node[draw=none, fit=(sub4)(q4), inner sep=0pt] (pair4) {};"
"
\node[category, anchor=west] (sub5) at ($(pair4.south west)+(0.,-\myspacing-2.3cm)$) {Focusing on physical instead of cognitive/emotional harm};
\node[quote, anchor=west] (q5) at ($(sub5.east)+(0.5cm,0)$) {“While this standard is unique in addressing service robot related physical safety issues [53], robots used in therapeutic settings interact with the user mostly socially [54, 55]. Social robots express and perceive emotions, communicate in high-level dialogue, learn/recognize models of other agents, establish and maintain social relationships, use natural cues such as gaze or gestures, exhibit distinctive personality and character, and might learn or even develop social competencies [56]. Those social robots that assist users through social interaction have been called socially assistive robots (SAR) [57]. Socially-interactive robots raise the question of whether available service robot safety standards suffice to mitigate hazards that mostly relate to cognitive aspects. Since the European Parliament (EP) on its latest resolution concerning the legal aspects of robot technology mentions, ‘human dignity and autonomy – both physical and psychological – is always to be respected,’ in this subsection we address some points that challenge this cognitive side.” (Fosch-Villaronga and Albo-Canals 2019 “I’ll take care of you,” said the robot, p. 82)};
\draw[arrow] (sub5.east) -- (q5.west);
\node[draw=none, fit=(sub5)(q5), inner sep=0pt] (pair5) {};"
"

\node[category, anchor=north west] (sub6) at ($(pair5.south west)+(0.5,0)$) {\textit{Cont. on next page}};
\node[quote, anchor=west] (q6) at ($(sub6.east)+(1.cm,0)$) {};
\node[draw=none, fit=(sub6)(q6), inner sep=0pt] (pair6) {};

\node[category, anchor=north west] (sub7) at ($(pair6.south west)+(-0.5,0)$) {\textit{Cont. on next page}};
\node[quote, anchor=west] (q7) at ($(sub7.east)+(1.cm,0)$) {};
\node[draw=none, fit=(sub7)(q7), inner sep=0pt] (pair7) {};

\node[category, anchor=north west] (sub8) at ($(pair7.south west)+(-0.5,0)$) {\textit{Cont. on next page}};
\node[quote, anchor=west] (q8) at ($(sub8.east)+(1.cm,0)$) {};
\node[draw=none, fit=(sub8)(q8), inner sep=0pt] (pair8) {};

\draw[connector] (cat1.west) |- ([xshift=-2mm]sub1.west);
\draw[connector] (sub1.west) |- ([xshift=-2mm]sub2.west);
\draw[connector] (sub2.west) |- ([xshift=-2mm]sub3.west);
\draw[connector] (sub2.west) |- ([xshift=-2mm]sub4.west);
\draw[connector] (sub2.west) |- ([xshift=-2mm]sub5.west);
\draw[connector] (sub2.west) |- ([xshift=-2mm]sub6.west);
\draw[connector] (sub1.west) |- ([xshift=-2mm]sub7.west);
\draw[connector] (cat1.west) |- ([xshift=-2mm]sub8.west);

\end{tikzpicture}

\clearpage

\begin{tikzpicture}[
  every node/.style={align=left,},
  category/.style={text width=6cm, align=left, inner sep=4pt, font=\small},
  root/.style={text width=10cm, align=left, inner sep=4pt},
  quote/.style={rounded corners, thick, text width=6cm, align=justify, inner sep=4pt, font=\footnotesize},
  arrow/.style={-{Stealth[length=3mm]}, thick},
  connector/.style={thick},
  spacing/.store in=\myspacing, spacing=1.5cm 
]

\node[root] (cat1) at (0, 0) {Ethical hazards for society/the (healthcare) system (continued)};

\node[category, anchor=west] (sub1) at ($(cat1.south west)+(0.5,-\myspacing+0.5cm)$) {Using a misguided approach to regulation/design of social robots (continued)};
\node[quote, anchor=west] (q1) at ($(sub1.east)+(1.5cm,0)$) {};
\node[draw=none, fit=(sub1)(q1), inner sep=0pt] (pair1) {};

\node[category, anchor=west] (sub2) at ($(pair1.south west)+(0.5,-\myspacing+0.5cm)$) {Further problematic attitudes/approaches (continued)};
\node[quote, anchor=west] (q2) at ($(sub2.east)+(1.5cm,0)$) {};
\node[draw=none, fit=(sub2)(q2), inner sep=0pt] (pair2) {};

\node[category, anchor=west] (sub3) at ($(pair2.south west)+(0.5,-\myspacing)$) {Ignoring consequences of social robots in the long run };
\node[quote, anchor=west] (q3) at ($(sub3.east)+(0.5cm,0)$) {“There are few, if any, mechanisms that protect users from broader and long-term consequences of a particular technology. Indeed, designers may correctly specify an objective function, their system may perform a task in the environment, but they may ignore the fact that their technology may have broader consequences in the long run.”(Fosch-Villaronga and Albo Canals 2019 “I’ll take care of you,” said the robot - Reflecting upon the legal and ethical aspects of the use and development of social robots for therapy, p. 84)};
\draw[arrow] (sub3.east) -- (q3.west);
\node[draw=none, fit=(sub3)(q3), inner sep=0pt] (pair3) {};"
"
\node[category, anchor=west] (sub4) at ($(pair3.south west)+(0.0,-\myspacing-1cm)$) {Focusing too much on implications for the individual};
\node[quote, anchor=west] (q4) at ($(sub4.east)+(0.5cm,0)$) {“The individual-centered perspective comes with a disproportionately fewer attention to SAR implications from the perspective of the (care) practice in which its artifacts are used, as well as from the macro sociopolitical level of justice. Descriptively, several reasons could be found behind this tendency.11 However, from a normative-oriented point of view, what matters the most is that this is an important deficiency of the current ethical approach to SAR, which shows a continuation of the individualist assumptions and the ‘neglect of the political’ underlying the mainstream philosophy of technology and ethics of technology [82].”(Boada et al. 2021 The ethical issues of social assistive robotics: A critical literature review, p. 10)};
\draw[arrow] (sub4.east) -- (q4.west);
\node[draw=none, fit=(sub4)(q4), inner sep=0pt] (pair4) {};"
"
\node[category, anchor=west] (sub4a) at ($(pair4.south west)+(0.,-\myspacing)$) {Seeing digital sexuality as pathological instead of atypical};
\node[quote, anchor=west] (q4a) at ($(sub4a.east)+(0.5cm,0)$) {“Those who favour the advent of sex with robots are often perceived as being sexually deviant, even perverted.”(Koumpis and Gees 2020 Sex with Robots: A not-so-niche market for disabled and older persons, p. 228)};
\draw[arrow] (sub4a.east) -- (q4a.west);
\node[draw=none, fit=(sub4a)(q4a), inner sep=0pt] (pair4a) {};"
"
\node[category, anchor=west] (sub5) at ($(pair4a.south west)+(0.,-\myspacing)$) {Understanding vulnerability solely as annoyance/capacity problem};
\node[quote, anchor=west] (q5) at ($(sub5.east)+(0.5cm,0)$) {“The idea that individual nursing activities could be performed by technology or robots harbors the risk of falling back into a deficit-oriented view of patients.” (Translation by the authors)(Manzeschke and Petersen 2020 Digitalisierung und Roboterisierung in der Pflege – Ethisch-anthropologische Überlegungen, pp. 169, 170)};
\draw[arrow] (sub5.east) -- (q5.west);
\node[draw=none, fit=(sub5)(q5), inner sep=0pt] (pair5) {};"
"
\node[category, anchor=west] (sub6) at ($(pair5.south west)+(0.,-\myspacing-0.5cm)$) {Raising false hopes without really assessing technical feasibility};
\node[quote, anchor=west] (q6) at ($(sub6.east)+(0.5cm,0)$) {“Visions of technical futures are painted and unfolded in a quasi-deterministic inevitability, the relevance of which appears at least questionable on closer inspection. Real challenges, which primarily relate to the shaping of technological development, are lost from view. Not least, this raises false expectations among the public and paints a picture of care robotics and its capabilities that does not correspond to reality.” (Translation by the authors)(Kehl 2018 Wege zu verantwortungsvoller Forschung und Entwicklung im Bereich der Pfegerobotik: Die ambivalente Rolle der Ethik, p. 155; in: Bendel 2018 Pflegeroboter)};
\draw[arrow] (sub6.east) -- (q6.west);
\node[draw=none, fit=(sub6)(q6), inner sep=0pt] (pair6) {};"

\node[category, anchor=north west] (sub7) at ($(pair6.south west)+(-0.0,0)$) {\textit{Cont. on next page}};
\node[quote, anchor=west] (q7) at ($(sub7.east)+(1.cm,0)$) {};
\node[draw=none, fit=(sub7)(q7), inner sep=0pt] (pair7) {};

\node[category, anchor=north west] (sub8) at ($(pair7.south west)+(-0.5,0)$) {\textit{Cont. on next page}};
\node[quote, anchor=west] (q8) at ($(sub8.east)+(1.cm,0)$) {};
\node[draw=none, fit=(sub8)(q8), inner sep=0pt] (pair8) {};

\node[category, anchor=north west] (sub9) at ($(pair8.south west)+(-0.5,0)$) {\textit{Cont. on next page}};
\node[quote, anchor=west] (q9) at ($(sub9.east)+(1.cm,0)$) {};
\node[draw=none, fit=(sub9)(q9), inner sep=0pt] (pair9) {};

\draw[connector] (cat1.west) |- ([xshift=-2mm]sub1.west);
\draw[connector] (sub1.west) |- ([xshift=-2mm]sub2.west);
\draw[connector] (sub2.west) |- ([xshift=-2mm]sub3.west);
\draw[connector] (sub2.west) |- ([xshift=-2mm]sub4.west);
\draw[connector] (sub2.west) |- ([xshift=-2mm]sub4a.west);
\draw[connector] (sub2.west) |- ([xshift=-2mm]sub5.west);
\draw[connector] (sub2.west) |- ([xshift=-2mm]sub6.west);
\draw[connector] (sub2.west) |- ([xshift=-2mm]sub7.west);
\draw[connector] (sub1.west) |- ([xshift=-2mm]sub8.west);
\draw[connector] (cat1.west) |- ([xshift=-2mm]sub9.west);

\end{tikzpicture}
\clearpage

\begin{tikzpicture}[
  every node/.style={align=left,},
  category/.style={text width=6cm, align=left, inner sep=4pt, font=\small},
  root/.style={text width=10cm, align=left, inner sep=4pt},
  quote/.style={rounded corners, thick, text width=6cm, align=justify, inner sep=4pt, font=\footnotesize},
  arrow/.style={-{Stealth[length=3mm]}, thick},
  connector/.style={thick},
  spacing/.store in=\myspacing, spacing=1.5cm 
]

\node[root] (cat1) at (0, 0) {Ethical hazards for society/the (healthcare) system (continued)};

\node[category, anchor=west] (sub1) at ($(cat1.south west)+(0.5,-\myspacing-0.5cm)$) {Beneficial social robots not used to due diverse barriers };
\node[quote, anchor=west] (q1) at ($(sub1.east)+(1.5cm,0)$) {“Finally, in addition to the lack of functionality of the device and a lack of knowledge on the part of employees, structural barriers to the use of the device are also recognized. The key issues here are cost coverage and access to the internet. In addition to the lack of cost efficiency and doubts about effectiveness, this aspect is considered to be the reason for the generally still low spread of technical assistance systems (Weinberger \& Decker, 2015).” (Translation by the authors)(Geier et al. 2020 Wie Pflegekräfte im ambulanten Bereich den Einsatz von Telepräsenzsystemen einschätzen – eine qualitative Studie, p. 49)};
\draw[arrow] (sub1.east) -- (q1.west);
\node[draw=none, fit=(sub1)(q1), inner sep=0pt] (pair1) {};

\node[category, anchor=west] (sub2) at ($(sub1.south west)+(0.5,-\myspacing-1.9cm)$) {Lack of acceptance/fear of beneficial social robots};
\node[quote, anchor=west] (q2) at ($(sub2.east)+(1cm,0)$) {“They state that one major problem that impedes the implementation of robots in elderly care is the population’s fear towards robots in care in general and older people’s fear in particular (see also Nomura, Kanda and Suzuki 2006).”(Hoppe et al. 2020 Assistive robots in care: Expectations and perceptions of older people, p. 140; in: Haltaufderheide et al. 2020 Aging between Participation and Simulation - Ethical Dimensions of Social Assistive Technologies)};
\draw[arrow] (sub2.east) -- (q2.west);
\node[draw=none, fit=(sub2)(q2), inner sep=0pt] (pair2) {};

"
\node[category, anchor=west] (sub3) at ($(pair2.south west)+(0.5,-\myspacing)$) {Inadequate design for target groups leads to skepticism};
\node[quote, anchor=west] (q3) at ($(sub3.east)+(0.5cm,0)$) {“As unmet users’ expectations are a major indicator of suboptimal adoption, the lack of user-oriented approaches in product design risks to generate a vicious circle where unmet user needs cause lower-than-expected uptake which, in turn, perpetuates unmet user needs.”(Ienca et al. 2016 Social and Assistive Robotics in Dementia Care: Ethical Recommendations for Research and Practice, p. 567)};
\draw[arrow] (sub3.east) -- (q3.west);
\node[draw=none, fit=(sub3)(q3), inner sep=0pt] (pair3) {};"
"
\node[category, anchor=west] (sub4) at ($(pair3.south west)+(0.,-\myspacing)$) {Lack of knowledge about social robots leads to skepticism};
\node[quote, anchor=west] (q4) at ($(sub4.east)+(0.5cm,0)$) {“However, it has been found that public knowledge about robots seems vague, as the public tends to lack information on how care robots can be used in elder care [35]. Previous studies report that when participants did not have experiences with the robot in question, more negative attitudes were prevalent [56, 57].”(Johansson-Pajala et al. 2020 Care Robot Orientation: What, Who and How? Potential Users’ Perceptions, p. 1105)};
\draw[arrow] (sub4.east) -- (q4.west);
\node[draw=none, fit=(sub4)(q4), inner sep=0pt] (pair4) {};"
"
\node[category, anchor=west] (sub5) at ($(pair4.south west)+(-0.5,-\myspacing-1.6cm)$) {High costs prevent usage/development of social robots};
\node[quote, anchor=west] (q5) at ($(sub5.east)+(1.cm,0)$) {“Despite many years of development efforts and countless prototypes, only a handful of systems have made it into practical use so far - all of them special applications (in addition to the Paro seal, these include some eating aids as well as a few exoskeletons and telepresence robots), which have no gripper arms and only limited autonomy and therefore deviate significantly from the model of the multifunctional everyday assistant. This is not only due to the high technical hurdles that need to be overcome. There is also a difficult market environment: as the care sector is under high economic pressure and only a few service providers have sufficient investment potential, it is unclear whether the high development costs can be amortized.” (Translation by the authors) (Kehl 2018 Entgrenzungen Zwischen Mensch Und Maschine, Oder: Können Roboter Gute Pflege Leisten?, pp. 4, 5) };
\draw[arrow] (sub5.east) -- (q5.west);
\node[draw=none, fit=(sub5)(q5), inner sep=0pt] (pair5) {};"
"
\node[category, anchor=west] (sub6) at ($(pair5.south west)+(0.,-\myspacing)$) {Further barriers};
\node[quote, anchor=west] (q6) at ($(sub6.east)+(1.cm,0)$) {“The robotic device not being adapted to the intended environment (eg, it was too large, or movement pattern did not fit in the setting) and the technology failing in certain locations were seen as barriers as well.” (Servaty et al. 2020 Implementation of robotic devices in nursing care. Barriers and facilitators: an integrative review, p. 9)};
\draw[arrow] (sub6.east) -- (q6.west);
\node[draw=none, fit=(sub6)(q6), inner sep=0pt] (pair6) {};"

\node[category, anchor=north west] (sub7) at ($(pair6.south west)+(-0.5,0)$) {\textit{Cont. on next page}};
\node[quote, anchor=west] (q7) at ($(sub7.east)+(1.cm,0)$) {};
\node[draw=none, fit=(sub7)(q7), inner sep=0pt] (pair7) {};

\draw[connector] (cat1.west) |- ([xshift=-2mm]sub1.west);
\draw[connector] (sub1.west) |- ([xshift=-2mm]sub2.west);
\draw[connector] (sub2.west) |- ([xshift=-2mm]sub3.west);
\draw[connector] (sub2.west) |- ([xshift=-2mm]sub4.west);
\draw[connector] (sub1.west) |- ([xshift=-2mm]sub5.west);
\draw[connector] (sub1.west) |- ([xshift=-2mm]sub6.west);
\draw[connector] (cat1.west) |- ([xshift=-2mm]sub7.west);

\end{tikzpicture}

\begin{tikzpicture}[
  every node/.style={align=left,},
  category/.style={text width=6cm, align=left, inner sep=4pt, font=\small},
  root/.style={text width=10cm, align=left, inner sep=4pt},
  quote/.style={rounded corners, thick, text width=6cm, align=justify, inner sep=4pt, font=\footnotesize},
  arrow/.style={-{Stealth[length=3mm]}, thick},
  connector/.style={thick},
  spacing/.store in=\myspacing, spacing=1.5cm 
]

\node[root] (cat1) at (0, 0) {Ethical hazards for society/the (healthcare) system (continued)};

"
\node[category, anchor=west] (sub1) at ($(cat1.south west)+(0.5,-\myspacing-0.4cm)$) {Problematic change in (under)standing of care};
\node[quote, anchor=west] (q1) at ($(sub1.east)+(1.5cm,0)$) {“One problem that is currently arising for nursing as a profession, but which could possibly change our society as a whole with regard to our understanding of help and care, is the question of whether and how our human understanding of help and care will be reshaped if robots take over a not inconsiderable part of these activities and a 'feeling' for the interpersonal dimension of help is lost because the functional element is so strongly in the foreground.” (Translation by the authors)(Manzeschke 2019 Roboter in der Pflege - Von Menschen, Maschinen und anderen hilfreichen Wesen, p. 5)};
\draw[arrow] (sub1.east) -- (q1.west);
\node[draw=none, fit=(sub1)(q1), inner sep=0pt] (pair1) {};"
"
\node[category, anchor=west] (sub2) at ($(pair1.south west)+(0.5,-\myspacing)$) {Reduction in significance of care};
\node[quote, anchor=west] (q2) at ($(sub2.east)+(1.cm,0)$) {“[…] the central ethical question is how to reshape the traditional roles and functions of the professionals in the practice, i.e., which are the tasks that should be delegated to robots and why. Human substitution by robots could not only impact the practice quality, but also the meaning of care.”(Boada et al. 2021 The ethical issues of social assistive robotics: A critical literature review, p. 8)};
\draw[arrow] (sub2.east) -- (q2.west);
\node[draw=none, fit=(sub2)(q2), inner sep=0pt] (pair2) {};"
"
\node[category, anchor=west] (sub3) at ($(pair2.south west)+(0.,-\myspacing-0.5cm)$) {Dehumanization/mechanization of care};
\node[quote, anchor=west] (q3) at ($(sub3.east)+(1.cm,0)$) {“However, there are also fundamental objections that are generally directed against the use of robots in care and to which the systems presented here in more detail are also exposed. One particularly daunting idea is that robots could one day completely replace human interaction in care. The elderly would then live isolated from the rest of the world in care facilities where they are fed, washed and monitored by robots that are also responsible for entertainment and attention.” (Translation by the authors)(Misselhorn 2018 Pflegesysteme, p. 152; in: Misselhorn 2018 Grundfragen der Maschinenethik)};
\draw[arrow] (sub3.east) -- (q3.west);
\node[draw=none, fit=(sub3)(q3), inner sep=0pt] (pair3) {};"
"
\node[category, anchor=west] (sub4) at ($(pair3.south west)+(0.5,-\myspacing-1.2cm)$) {Reductionist view of care as input-output process};
\node[quote, anchor=west] (q4) at ($(sub4.east)+(0.5cm,0)$) {“Reducing the complex process of caring into single sequences of input and output, does harm to this phenomenon and carries the risk to change it ultimately. The normalization of implementing care robots to fight loneliness bears the danger of a structural and even institutionalized (acceptance of) objectification and that we may not even expect essential features of care from humans anymore, if we get more and more used to encounters where our opposite does not care about us. This is to be phrased as a hypothetical state of affairs, as a potential outcome of the implementation and acceptance – the normalization – of care robots in societies.”(von Maur 2023 Alice Does not Care – Or: Wy it Matters That Robots “Don’t Give a Damn”, p. 226; in: Loh und Loh 2023 Social Robotics and the Good Life - The Normative Side of Forming Emotional Bonds With Robots)};
\draw[arrow] (sub4.east) -- (q4.west);
\node[draw=none, fit=(sub4)(q4), inner sep=0pt] (pair4) {};"
"
\node[category, anchor=west] (sub5) at ($(pair4.south west)+(-1.,-\myspacing+0.2cm)$) {Problematic change in understanding of being human};
\node[quote, anchor=west] (q5) at ($(sub5.east)+(1.5cm,0)$) {“It is argued that the robot degrades the person in need of help to a problem that can be solved by technology.” (Translation by the authors)(Radic and Vosen 2020 Ethische, rechtliche und soziale Anforderungen an Assistenzroboter in der Pflege – Sicht des Führungspersonals in Kliniken und Pflegeeinrichtungen, p. 630) };
\draw[arrow] (sub5.east) -- (q5.west);
\node[draw=none, fit=(sub5)(q5), inner sep=0pt] (pair5) {};"
"
\node[category, anchor=west] (sub6) at ($(pair5.south west)+(0.5,-\myspacing-0.5cm)$) {Humanizing social robots and objectifying humans};
\node[quote, anchor=west] (q6) at ($(sub6.east)+(1.cm,0)$) {“But if we – we, as the society being responsible for the narratives and imaginaries which make certain solutions to societal problems visible while obscuring others – do not see the robot as an object, but rather endorse the illusion that a robot cares about us, we humanize it and objectify ourselves: We objectify ourselves by humanizing an object.17”(Von Maur 2023 Alice Does not Care – Or: Wy it Matters That Robots “Don’t Give a Damn”, p. 225; in: Loh und Loh 2023 Social Robotics and the Good Life - The Normative Side of Forming Emotional Bonds With Robots)};
\draw[arrow] (sub6.east) -- (q6.west);
\node[draw=none, fit=(sub6)(q6), inner sep=0pt] (pair6) {};"
"
\node[category, anchor=north west] (sub7) at ($(pair6.south west)+(-0.5,0)$) {\textit{Cont. on next page}};
\node[quote, anchor=west] (q7) at ($(sub7.east)+(1.cm,0)$) {};
\node[draw=none, fit=(sub7)(q7), inner sep=0pt] (pair7) {};

\draw[connector] (cat1.west) |- ([xshift=-2mm]sub1.west);
\draw[connector] (sub1.west) |- ([xshift=-2mm]sub2.west);
\draw[connector] (sub1.west) |- ([xshift=-2mm]sub3.west);
\draw[connector] (sub3.west) |- ([xshift=-2mm]sub4.west);
\draw[connector] (cat1.west) |- ([xshift=-2mm]sub5.west);
\draw[connector] (sub5.west) |- ([xshift=-2mm]sub6.west);
\draw[connector] (cat1.west) |- ([xshift=-2mm]sub7.west);

\end{tikzpicture}

\begin{tikzpicture}[
  every node/.style={align=left,},
  category/.style={text width=6cm, align=left, inner sep=4pt, font=\small},
  root/.style={text width=10cm, align=left, inner sep=4pt},
  quote/.style={rounded corners, thick, text width=6cm, align=justify, inner sep=4pt, font=\footnotesize},
  arrow/.style={-{Stealth[length=3mm]}, thick},
  connector/.style={thick},
  spacing/.store in=\myspacing, spacing=1.5cm 
]

\node[root] (cat1) at (0, 0) {Ethical hazards for society/the (healthcare) system (continued)};

"
\node[category, anchor=west] (sub1) at ($(cat1.south west)+(0.5,-\myspacing)$) {Damaging interpersonal and societal connectedness};
\node[quote, anchor=west] (q1) at ($(sub1.east)+(1.5cm,0)$) {“The use of social robotics in care for PwD will diminish the social relationship between healthy independent adults and PwD, by moving the latter further out of the mainstream of society and rendering them socially invisible “(O’Brolchain 2017 Robots and people with dementia: Unintended consequences and moral hazard, p. 969)};
\draw[arrow] (sub1.east) -- (q1.west);
\node[draw=none, fit=(sub1)(q1), inner sep=0pt] (pair1) {};"
"
\node[category, anchor=west] (sub2) at ($(pair1.south west)+(0.5,-\myspacing-0.3cm)$) {Normalization of robotic care changes expectation in care/interaction};
\node[quote, anchor=west] (q2) at ($(sub2.east)+(1.cm,0)$) {“One structural effect of normalizing the implementation of care robots could be that humans will not only stay lonely while interacting with robots, but that they will give up expectations of real care and true relationships in the first place.”(von Maur 2023 Alice Does not Care – Or: Wy it Matters That Robots “Don’t Give a Damn”, p. 210; in: Loh und Loh 2023 Social Robotics and the Good Life - The Normative Side of Forming Emotional Bonds With Robots)};
\draw[arrow] (sub2.east) -- (q2.west);
\node[draw=none, fit=(sub2)(q2), inner sep=0pt] (pair2) {};"
"
\node[category, anchor=west] (sub3) at ($(pair2.south west)+(0.5,-\myspacing-0.4cm)$) {Fostering an instrumentalist view of relationships};
\node[quote, anchor=west] (q3) at ($(sub3.east)+(0.5cm,0)$) {“Second, because the unauthentic intersubjectivity involved in HRI may affect the proper development or exercising of human moral faculties (empathy, care …), as well as foster an instrumentalist view of relationships, in which the ‘other’ is not a ‘you’ but rather an object. This endangers relations and forms of life that are intrinsically valuable and define us.”(Boada et al. 2021 The ethical issues of social assistive robotics: A critical literature review, p. 7)};
\draw[arrow] (sub3.east) -- (q3.west);
\node[draw=none, fit=(sub3)(q3), inner sep=0pt] (pair3) {};"
"
\node[category, anchor=west] (sub4) at ($(pair3.south west)+(-0.5,-\myspacing-0.7cm)$) {No exchange/understanding between generations};
\node[quote, anchor=west] (q4) at ($(sub4.east)+(1.cm,0)$) {“The implication to society as a whole could be dire in the long-term as this would mean that the society, especially the younger generation, will be less cognizant about the challenges one faces at old age, including having less exposure to the vulnerability and infirmity that may inflict the older generation. The changing nature of social relationships might also erode opportunities to extend virtues such as care and empathy to functionally dependant [sic] older people (O’Brolchain 2017).”(Tan et al. 2021 Tensions and antagonistic interactions of risks and ethics of using robotics and autonomous systems in long-term care, p. 10)};
\draw[arrow] (sub4.east) -- (q4.west);
\node[draw=none, fit=(sub4)(q4), inner sep=0pt] (pair4) {};"
"
\node[category, anchor=west] (sub5) at ($(pair4.south west)+(-0.5,-\myspacing-1.5cm)$) {Moral de-skilling};
\node[quote, anchor=west] (q5) at ($(sub5.east)+(1.5cm,0)$) {“SAR may disrupt human moral practices that are constitutive of our societies and culture and, in turn, endanger both the internal goods of these practices, and certain human moral capacities that can only be developed and exercised through these forms of activity. SAR may erode care as a central practice of human moral life because it may reduce our engagement in such activity, thus lessening the cultivation of its associated moral skills and leading to a moral and professional deskilling. This has implications for the organizational sphere of human life, given that it challenges the core values of the exercised practice of care, which calls into question those of the whole institutional context within which is held.”(Boada et al. 2021 The ethical issues of social assistive robotics: A critical literature review, p. 8)};
\draw[arrow] (sub5.east) -- (q5.west);
\node[draw=none, fit=(sub5)(q5), inner sep=0pt] (pair5) {};"
"

"
\node[category, anchor=north west] (sub6) at ($(pair5.south west)+(0.5,0)$) {\textit{Cont. on next page}};
\node[quote, anchor=west] (q6) at ($(sub6.east)+(1.cm,0)$) {};
\node[draw=none, fit=(sub6)(q6), inner sep=0pt] (pair6) {};

\node[category, anchor=north west] (sub7) at ($(pair6.south west)+(-0.5,0)$) {\textit{Cont. on next page}};
\node[quote, anchor=west] (q7) at ($(sub7.east)+(1.cm,0)$) {};
\node[draw=none, fit=(sub7)(q7), inner sep=0pt] (pair7) {};

\draw[connector] (cat1.west) |- ([xshift=-2mm]sub1.west);
\draw[connector] (sub1.west) |- ([xshift=-2mm]sub2.west);
\draw[connector] (sub2.west) |- ([xshift=-2mm]sub3.west);
\draw[connector] (sub1.west) |- ([xshift=-2mm]sub4.west);
\draw[connector] (cat1.west) |- ([xshift=-2mm]sub5.west);
\draw[connector] (sub5.west) |- ([xshift=-2mm]sub6.west);
\draw[connector] (cat1.west) |- ([xshift=-2mm]sub7.west);
\end{tikzpicture}

\begin{tikzpicture}[
  every node/.style={align=left,},
  category/.style={text width=6cm, align=left, inner sep=4pt, font=\small},
  root/.style={text width=10cm, align=left, inner sep=4pt},
  quote/.style={rounded corners, thick, text width=6cm, align=justify, inner sep=4pt, font=\footnotesize},
  arrow/.style={-{Stealth[length=3mm]}, thick},
  connector/.style={thick},
  spacing/.store in=\myspacing, spacing=1.5cm 
]

\node[root] (cat1) at (0, 0) {Ethical hazards for society/the (healthcare) system (continued)};

\node[category, anchor=west] (sub1) at ($(cat1.south west)+(0.5,-\myspacing+0.5cm)$) {Moral de-skilling (continued)};
\node[quote, anchor=west] (q1) at ($(sub1.east)+(1.5cm,0)$) {};
\node[draw=none, fit=(sub1)(q1), inner sep=0pt] (pair1) {};

"
\node[category, anchor=west] (sub2) at ($(pair1.south west)+(0.5,-\myspacing)$) {Reducing opportunities to develop moral skills};
\node[quote, anchor=west] (q2) at ($(sub2.east)+(1.cm,0)$) {“On the other hand, the adoption of SARs in care, by outsourcing practices central to human existence to non-human actors, could blind us from the awareness of the constitutive vulnerability and (inter)dependance of human life, thus threatening the cultivation of virtues essential to a flourishing society. More specifically, the new technological practices could reduce the opportunities for cultivating moral skills regarding human caregiving.”(Boada et al. 2021 The ethical issues of social assistive robotics: A critical literature review, p. 7)};
\draw[arrow] (sub2.east) -- (q2.west);
\node[draw=none, fit=(sub2)(q2), inner sep=0pt] (pair2) {};"
"
\node[category, anchor=west] (sub3) at ($(pair2.south west)+(0.5,-\myspacing-0.5cm)$) {By social robot never challenging personal biases and prejudices};
\node[quote, anchor=west] (q3) at ($(sub3.east)+(0.5cm,0)$) {“Eventually, the constant repetition of such everyday life effects of stereotyping in robotic systems for eldercare might even influence the users’ personality itself by generating, promoting, or reinforcing biased perspectives and prejudiced attitudes. In the end, this could undermine their fundamental capability or willingness to critically reflect and overcome gender stereotypes and consequently cultivate a sexist and narrow-minded character.” (Weßel et al. 2021 Gender Stereotyping of Robotic Systems in Elderly Care An Exploratory Analysis of Ethical Problems and Possible Solutions, p. 5) };
\draw[arrow] (sub3.east) -- (q3.west);
\node[draw=none, fit=(sub3)(q3), inner sep=0pt] (pair3) {};"
"
\node[category, anchor=west] (sub4) at ($(pair3.south west)+(0.,-\myspacing)$) {Constrained communication/dialogue system of social robots reduces capacities for moral reasoning};
\node[quote, anchor=west] (q4) at ($(sub4.east)+(0.5cm,0)$) {“A partly adjacent concern related to human moral deskilling is the influence that language-capable robots, because of their acting as social moral agents (and given their constrained dialogue systems), may have on human moral reasoning.”(Boada et al. 2021 The ethical issues of social assistive robotics: A critical literature review, p. 8)};
\draw[arrow] (sub4.east) -- (q4.west);
\node[draw=none, fit=(sub4)(q4), inner sep=0pt] (pair4) {};"
"
\node[category, anchor=west] (sub5) at ($(pair4.south west)+(-0.5,-\myspacing-0.5cm)$) {Improperly extending treatment of social robot to humans};
\node[quote, anchor=west] (q5) at ($(sub5.east)+(1.cm,0)$) {“Human improper behavior and interaction with robots could foster a human’s moral corruption. Even without explicit or intentional abusive behavior, the very same simulation of unconditional recognition carried out by robots could lead to a moral deskilling (for instance, it could normalize the experience of exercising control and power over what is seen as an autonomous agent with recognitive abilities).”(Boada et al. 2021 The ethical issues of social assistive robotics: A critical literature review, p. 7)};
\draw[arrow] (sub5.east) -- (q5.west);
\node[draw=none, fit=(sub5)(q5), inner sep=0pt] (pair5) {};"
"
\node[category, anchor=west] (sub6) at ($(pair5.south west)+(0.5,-\myspacing-0.5cm)$) {Sex robots further improper treatment of women (or men)};
\node[quote, anchor=west] (q6) at ($(sub6.east)+(0.5cm,0)$) {“Professor Kathleen Richardson has published on this topic (see, for example, [14] and also her Campaign Against Sex Robots [13]) and offers several good reasons why sex robots are not a good idea at all. Among her arguments are the facts that ‘sex robots further sexually objectify women and children’. One might of course think that sex robots similarly objectify men as well, but her argument is, of course, quite correct and valid.”(Koumpis and Gees 2020 Sex with Robots: A not-so-niche market for disabled and older persons, p. 231)};
\draw[arrow] (sub6.east) -- (q6.west);
\node[draw=none, fit=(sub6)(q6), inner sep=0pt] (pair6) {};"

\node[category, anchor=north west] (sub7) at ($(pair6.south west)+(-1.0,0)$) {\textit{Cont. on next page}};
\node[quote, anchor=west] (q7) at ($(sub7.east)+(1.cm,0)$) {};
\node[draw=none, fit=(sub7)(q7), inner sep=0pt] (pair7) {};

\draw[connector] (cat1.west) |- ([xshift=-2mm]sub1.west);
\draw[connector] (sub1.west) |- ([xshift=-2mm]sub2.west);
\draw[connector] (sub2.west) |- ([xshift=-2mm]sub3.west);
\draw[connector] (sub2.west) |- ([xshift=-2mm]sub4.west);
\draw[connector] (sub1.west) |- ([xshift=-2mm]sub5.west);
\draw[connector] (sub5.west) |- ([xshift=-2mm]sub6.west);
\draw[connector] (cat1.west) |- ([xshift=-2mm]sub7.west);

\end{tikzpicture}

\begin{tikzpicture}[
  every node/.style={align=left,},
  category/.style={text width=6cm, align=left, inner sep=4pt, font=\small},
  root/.style={text width=10cm, align=left, inner sep=4pt},
  quote/.style={rounded corners, thick, text width=6cm, align=justify, inner sep=4pt, font=\footnotesize},
  arrow/.style={-{Stealth[length=3mm]}, thick},
  connector/.style={thick},
  spacing/.store in=\myspacing, spacing=1.5cm 
]

\node[root] (cat1) at (0, 0) {Ethical hazards for society/the (healthcare) system (continued)};

\node[category, anchor=west] (sub1) at ($(cat1.south west)+(0.5,-\myspacing-0.5cm)$) {Fueling discrimination/stigmatization of certain groups };
\node[quote, anchor=west] (q1) at ($(sub1.east)+(1.5cm,0)$) {“These considerations already point to the fundamental concern that the very idea of stereotyping in and of itself contravenes basic egalitarian principles of justice in modern moral thought that call for equal respect and mutual recognition of all individuals. Thus, one could argue that stereotyping of care and social robots implies the affirmation and reinforcement of traditional gender stereotypes and therefore amounts to a technological promotion of existing societal bias, injustice, and discrimination.”(Weßel et al. 2021 Gender Stereotyping of Robotic Systems in Elderly Care An Exploratory Analysis of Ethical Problems and Possible Solutions, p. 6) };
\draw[arrow] (sub1.east) -- (q1.west);
\node[draw=none, fit=(sub1)(q1), inner sep=0pt] (pair1) {};
"
\node[category, anchor=west] (sub2) at ($(pair1.south west)+(0.5,-\myspacing-3cm)$) {Social robot programming reinforces (gender) participation gaps/inequality};
\node[quote, anchor=west] (q2) at ($(sub2.east)+(1.cm,0)$) {“As far as the implementation of stereotyping strategies in the design or programming of eldercare robots is concerned, the question is how gender stereotypes affect the possibilities of participation and inclusion of older men and women in different areas of social life. Thus, it would appear rather objectionable if certain assistive functions were neglected or not promoted because they were connected to corresponding social practices that are considered as not gender-relevant or gender-inappropriate, e.g., needlework for older men or technical activities for older women. In particular, it would be highly problematic if stereotyping strategies in care robots for older people would simply adopt and reproduce norms that lead to a gender participation gap [56] and thus either systematically inhibit or impede the social and political participation of older women or consequently favor or even enforce the social or political participation of older men. Especially options for political participation should not be coupled to gender assignments but rather should be left to the individual citizen’s own preferences and self-determined decisions.”(Weßel et al. 2021 Gender Stereotyping of Robotic Systems in Elderly Care An Exploratory Analysis of Ethical Problems and Possible Solutions, p. 6) };
\draw[arrow] (sub2.east) -- (q2.west);
\node[draw=none, fit=(sub2)(q2), inner sep=0pt] (pair2) {};"
"
\node[category, anchor=west] (sub3) at ($(pair2.south west)+(0.,-\myspacing)$) {Social robot design reinforces ageist stereotypes/deficit-orientation};
\node[quote, anchor=west] (q3) at ($(sub3.east)+(1.cm,0)$) {“Neven analysed the interaction between robots, older test users and robot designers and found that the designers were influenced by stereotypes of older people as lonely, fragile, in need of care and company [53].”(Frennert and Östlund 2014 Review: Seven Matters of Concern of Social Robots and Older People, p. 303)};
\draw[arrow] (sub3.east) -- (q3.west);
\node[draw=none, fit=(sub3)(q3), inner sep=0pt] (pair3) {};"
"
\node[category, anchor=west] (sub4) at ($(pair3.south west)+(0.,-\myspacing)$) {Social robot appearance reinforces racist/sexist/ethnic stereotypes};
\node[quote, anchor=west] (q4) at ($(sub4.east)+(1.cm,0)$) {“Jecker notes that one criticism of using sex robots is that it may endorse an extreme form of objectification—sex robots are primarily designed to look like young attractive women, which may bolster society’s propensity to objectify women for the purpose of serving heterosexual men.” (Bianchi 2021 Considering sex robots for older adults with cognitive impairments, p. 37)};
\draw[arrow] (sub4.east) -- (q4.west);
\node[draw=none, fit=(sub4)(q4), inner sep=0pt] (pair4) {};"
"
\node[category, anchor=west] (sub5) at ($(pair4.south west)+(-0.5,-\myspacing)$) {Unfair distribution of benefits and burdens of social robot introduction};
\node[quote, anchor=west] (q5) at ($(sub5.east)+(1.5cm,0)$) {“Indeed, one’s economic, social, physical, psychological, cultural-religious, educational, etc. capabilities and functionings should not lead to unfair advantages or disadvantages regarding the possibilities of SAR use (d’Avack et al., 2017; Kornfield-Matte, 2017; Palmerini et al., 2014).” (Vandemeulebroucke et al. 2021 Socially Assistive Robots in Aged Care: Ethical Orientations Beyond the Care-Romantic and Technology-Deterministic Gaze, p. 17)};
\draw[arrow] (sub5.east) -- (q5.west);
\node[draw=none, fit=(sub5)(q5), inner sep=0pt] (pair5) {};"

\node[category, anchor=north west] (sub6) at ($(pair5.south west)+(0.5,0)$) {\textit{Cont. on next page}};
\node[quote, anchor=west] (q6) at ($(sub6.east)+(1.cm,0)$) {};
\node[draw=none, fit=(sub6)(q6), inner sep=0pt] (pair6) {};

\node[category, anchor=north west] (sub7) at ($(pair6.south west)+(-0.5,0)$) {\textit{Cont. on next page}};
\node[quote, anchor=west] (q7) at ($(sub7.east)+(1.cm,0)$) {};
\node[draw=none, fit=(sub7)(q7), inner sep=0pt] (pair7) {};

\draw[connector] (cat1.west) |- ([xshift=-2mm]sub1.west);
\draw[connector] (sub1.west) |- ([xshift=-2mm]sub2.west);
\draw[connector] (sub1.west) |- ([xshift=-2mm]sub3.west);
\draw[connector] (sub1.west) |- ([xshift=-2mm]sub4.west);
\draw[connector] (cat1.west) |- ([xshift=-2mm]sub5.west);
\draw[connector] (sub5.west) |- ([xshift=-2mm]sub6.west);
\draw[connector] (cat1.west) |- ([xshift=-2mm]sub7.west);

\end{tikzpicture}

\begin{tikzpicture}[
  every node/.style={align=left,},
  category/.style={text width=6cm, align=left, inner sep=4pt, font=\small},
  root/.style={text width=10cm, align=left, inner sep=4pt},
  quote/.style={rounded corners, thick, text width=6cm, align=justify, inner sep=4pt, font=\footnotesize},
  arrow/.style={-{Stealth[length=3mm]}, thick},
  connector/.style={thick},
  spacing/.store in=\myspacing, spacing=1.5cm 
]

\node[root] (cat1) at (0, 0) {Ethical hazards for society/the (healthcare) system (continued)};

\node[category, anchor=west] (sub1) at ($(cat1.south west)+(0.5,-\myspacing+0.5cm)$) {Unfair distribution of benefits and burdens of social robot introduction (continued)};
\node[quote, anchor=west] (q1) at ($(sub1.east)+(1.5cm,0)$) {};
\node[draw=none, fit=(sub1)(q1), inner sep=0pt] (pair1) {};

"
\node[category, anchor=west] (sub2) at ($(pair1.south west)+(0.5,-\myspacing)$) {Burdens are not shouldered by those reaping the benefits};
\node[quote, anchor=west] (q2) at ($(sub2.east)+(1.cm,0)$) {“The question of whether only the affluent older people will be able to reap the benefits of robotics and autonomous systems and whether the less privileged older people will be disproportionately subjected to the unintended consequences of robotics and autonomous systems has been raised (Laitinen et al., 2016).”(Tan et al. 2021 Tensions and antagonistic interactions of risks and ethics of using robotics and autonomous systems in long-term care, p. 10)};
\draw[arrow] (sub2.east) -- (q2.west);
\node[draw=none, fit=(sub2)(q2), inner sep=0pt] (pair2) {};"
"
\node[category, anchor=west] (sub3) at ($(pair2.south west)+(0.,-\myspacing-0.3cm)$) {Only rich can afford expensive human care (robot care is cheaper)};
\node[quote, anchor=west] (q3) at ($(sub3.east)+(1.cm,0)$) {“Risk potentials correspond primarily with the fear that ‘people in need of care will receive less human attention through the use of care robots’ (y=0.48***), and secondly with the expectation that ‘the use of robots will ultimately lead to only the wealthy being cared for by humans’ (y=0.28***).” (Translation by the authors)(Zwick and Hampel 2019 Cui bono? Zum Für und Wider von Robotik in der Pflege – Ergebnisse einer Repräsentativbefragung, p. 54)};
\draw[arrow] (sub3.east) -- (q3.west);
\node[draw=none, fit=(sub3)(q3), inner sep=0pt] (pair3) {};"
"
\node[category, anchor=west] (sub4) at ($(pair3.south west)+(0.,-\myspacing-0.2cm)$) {Unequal access to effective robot technology};
\node[quote, anchor=west] (q4) at ($(sub4.east)+(1.cm,0)$) {“It is found that one of the noticeable issues in robot ethics is having equal access to assistive medical robots.” (Mansouri et al. 2017 Ethical framework of assistive devices: review and reflection, p. 6)};
\draw[arrow] (sub4.east) -- (q4.west);
\node[draw=none, fit=(sub4)(q4), inner sep=0pt] (pair4) {};"
"
\node[category, anchor=west] (sub5) at ($(pair4.south west)+(0.5,-\myspacing-1.4cm)$) {Due to lack of customization options};
\node[quote, anchor=west] (q5) at ($(sub5.east)+(0.5cm,0)$) {“[…] the proportion of devices selectively designed to support specific stages of dementia is currently low. This lack of specificity could represent a significant obstacle toward the massive adoption of IATs for dementia and could add an additional reason to the limited uptake of IATs. In fact, patients suffering from different stages of AD may present specific needs and limitations that may be qualitatively and/or quantitatively different than those of people with other age-related cognitive disturbances or disabilities [50] . Also, people with mild AD often present different needs and limitations than people with moderate or advanced dementia and vice versa [51].” (Ienca 2016 Intelligent Assistive Technology for Alzheimer’s Disease and Other Dementias: A Systematic Review, p. 1336)};
\draw[arrow] (sub5.east) -- (q5.west);
\node[draw=none, fit=(sub5)(q5), inner sep=0pt] (pair5) {};"
"
\node[category, anchor=west] (sub6) at ($(pair5.south west)+(0.,-\myspacing-0.9cm)$) {Due to high costs and limited public finances};
\node[quote, anchor=west] (q6) at ($(sub6.east)+(0.5cm,0)$) {“Distributive justice is complicated by the fact that the more complex the assistive technology, the higher the risk that financial factors limit access for many individuals suffering from dementia and other disabilities (Alzheimer Europe 2010b). In view of these financial aspects, it will be necessary to avoid a digital divide and to find ways so that existing socioeconomic inequalities will not be increased.”(Hildt 2019 Shaping the Development and Use of Intelligent Assistive Technologies in Dementia – Some Thoughts, pp. 133, 134; in: Jotterand et al. 2019 Intelligent Assistive Technologies for Dementia: Clinical, Ethical, Social and Regulatory Implications) };
\draw[arrow] (sub6.east) -- (q6.west);
\node[draw=none, fit=(sub6)(q6), inner sep=0pt] (pair6) {};"
"

\node[category, anchor=north west] (sub7) at ($(pair6.south west)+(-0.,0)$) {\textit{Cont. on next page}};
\node[quote, anchor=west] (q7) at ($(sub7.east)+(1.cm,0)$) {};
\node[draw=none, fit=(sub7)(q7), inner sep=0pt] (pair7) {};

\node[category, anchor=north west] (sub8) at ($(pair7.south west)+(-0.5,0)$) {\textit{Cont. on next page}};
\node[quote, anchor=west] (q8) at ($(sub8.east)+(1.cm,0)$) {};
\node[draw=none, fit=(sub8)(q8), inner sep=0pt] (pair8) {};

\node[category, anchor=north west] (sub9) at ($(pair8.south west)+(-0.5,0)$) {\textit{Cont. on next page}};
\node[quote, anchor=west] (q9) at ($(sub9.east)+(1.cm,0)$) {};
\node[draw=none, fit=(sub9)(q9), inner sep=0pt] (pair9) {};

\draw[connector] (cat1.west) |- ([xshift=-2mm]sub1.west);
\draw[connector] (sub1.west) |- ([xshift=-2mm]sub2.west);
\draw[connector] (sub1.west) |- ([xshift=-2mm]sub3.west);
\draw[connector] (sub1.west) |- ([xshift=-2mm]sub4.west);
\draw[connector] (sub4.west) |- ([xshift=-2mm]sub5.west);
\draw[connector] (sub4.west) |- ([xshift=-2mm]sub6.west);
\draw[connector] (sub4.west) |- ([xshift=-2mm]sub7.west);
\draw[connector] (sub1.west) |- ([xshift=-2mm]sub8.west);
\draw[connector] (cat1.west) |- ([xshift=-2mm]sub9.west);

\end{tikzpicture}

\begin{tikzpicture}[
  every node/.style={align=left,},
  category/.style={text width=6cm, align=left, inner sep=4pt, font=\small},
  root/.style={text width=10cm, align=left, inner sep=4pt},
  quote/.style={rounded corners, thick, text width=6cm, align=justify, inner sep=4pt, font=\footnotesize},
  arrow/.style={-{Stealth[length=3mm]}, thick},
  connector/.style={thick},
  spacing/.store in=\myspacing, spacing=1.5cm 
]

\node[root] (cat1) at (0, 0) {Ethical hazards for society/the (healthcare) system (continued)};

\node[category, anchor=west] (sub1) at ($(cat1.south west)+(0.5,-\myspacing+0.5cm)$) {Unfair distribution of benefits and burdens of social robot introduction (continued)};
\node[quote, anchor=west] (q1) at ($(sub1.east)+(1.5cm,0)$) {};
\node[draw=none, fit=(sub1)(q1), inner sep=0pt] (pair1) {};

\node[category, anchor=west] (sub2) at ($(pair1.south west)+(0.5,-\myspacing+0.5cm)$) {Unequal access to effective robot technology (continued)};
\node[quote, anchor=west] (q2) at ($(sub2.east)+(1.5cm,0)$) {};
\node[draw=none, fit=(sub2)(q2), inner sep=0pt] (pair2) {};

"
\node[category, anchor=west] (sub3) at ($(pair2.south west)+(0.5,-\myspacing)$) {Due to limited information among certain groups};
\node[quote, anchor=west] (q3) at ($(sub3.east)+(0.5cm,0)$) {“Clinicians and other health professionals are often unaware of new IATs and their applicability to dementia care, as little cooperation has occurred between technology development and medical implementation.” (Ienca 2016 Intelligent Assistive Technology for Alzheimer’s Disease and Other Dementias: A Systematic Review, p. 1302)};
\draw[arrow] (sub3.east) -- (q3.west);
\node[draw=none, fit=(sub3)(q3), inner sep=0pt] (pair3) {};"
"
\node[category, anchor=west] (sub4) at ($(pair3.south west)+(0.,-\myspacing-0.5cm)$) {Social robots meet only needs of those groups participating in design};
\node[quote, anchor=west] (q4) at ($(sub4.east)+(0.5cm,0)$) {“This becomes even more important as many robots tested or actually used in the care sector were not originally and specifically designed for care purposes but, for example, as industrial robots like Kuka’s LBR robotic arms, or as companion robots in general like SoftBank’s Pepper.”(Weßel et al. 2021 Gender Stereotyping of Robotic Systems in Elderly Care An Exploratory Analysis of Ethical Problems and Possible Solutions, p. 3)};
\draw[arrow] (sub4.east) -- (q4.west);
\node[draw=none, fit=(sub4)(q4), inner sep=0pt] (pair4) {};"
"
\node[category, anchor=west] (sub5) at ($(pair4.south west)+(0.,-\myspacing-0.5cm)$) {Social robots meet only needs subset of population trained on};
\node[quote, anchor=west] (q5) at ($(sub5.east)+(0.5cm,0)$) {“According to the informants, it is vital that the potential users of the technology, such as older persons, families, and care professionals, participate in the development process, but this seem to be lacking for several reasons. Manufacturers and companies develop products without consulting the users, or they involve the users too late in the process, only when it is time to test the product.” (Johansson-Pajala and Gustafsson 2022 Significant challenges when introducing care roots in Swedish elder care, p. 171)};
\draw[arrow] (sub5.east) -- (q5.west);
\node[draw=none, fit=(sub5)(q5), inner sep=0pt] (pair5) {};"
"
\node[category, anchor=west] (sub6) at ($(pair5.south west)+(0.,-\myspacing)$) {Sex robots not produced for elderly due to ageism};
\node[quote, anchor=west] (q6) at ($(sub6.east)+(0.5cm,0)$) {“Jecker is right to call out sexual ageism. Older people often do have sexual needs, and this should not be stigmatised or ignored. But we worry that a focus on sex robots may inadvertently strengthen the very ageism that Jecker decries.” (Earp and Grunt-Mejer 2021 Robots and sexual ethics, p. 2)};
\draw[arrow] (sub6.east) -- (q6.west);
\node[draw=none, fit=(sub6)(q6), inner sep=0pt] (pair6) {};"
"
\node[category, anchor=west] (sub7) at ($(pair6.south west)+(-1.,-\myspacing)$) {Erosion of responsibility/gaps in the responsibility structure};
\node[quote, anchor=west] (q7) at ($(sub7.east)+(1.5cm,0)$) {“The erosion of responsibility in the development, approval and use of robotic technologies should be counteracted by establishing transparent responsibility structures.” (Translation by the authors)(Deutscher Ethikrat 2020 Robotik für gute Pflege (Stellungnahme), p. 50)};
\draw[arrow] (sub7.east) -- (q7.west);
\node[draw=none, fit=(sub7)(q7), inner sep=0pt] (pair7) {};"
"

\node[category, anchor=north west] (sub8) at ($(pair7.south west)+(0.5,0)$) {\textit{Cont. on next page}};
\node[quote, anchor=west] (q8) at ($(sub8.east)+(1.cm,0)$) {};
\node[draw=none, fit=(sub8)(q8), inner sep=0pt] (pair8) {};

\node[category, anchor=north west] (sub9) at ($(pair8.south west)+(-0.5,0)$) {\textit{Cont. on next page}};
\node[quote, anchor=west] (q9) at ($(sub9.east)+(1.cm,0)$) {};
\node[draw=none, fit=(sub9)(q9), inner sep=0pt] (pair9) {};

draw[connector] (cat1.west) |- ([xshift=-2mm]sub1.west);
\draw[connector] (sub1.west) |- ([xshift=-2mm]sub2.west);
\draw[connector] (sub2.west) |- ([xshift=-2mm]sub3.west);
\draw[connector] (sub2.west) |- ([xshift=-2mm]sub4.west);
\draw[connector] (sub2.west) |- ([xshift=-2mm]sub5.west);
\draw[connector] (sub2.west) |- ([xshift=-2mm]sub6.west);
\draw[connector] (cat1.west) |- ([xshift=-2mm]sub7.west);
\draw[connector] (sub7.west) |- ([xshift=-2mm]sub8.west);
\draw[connector] (cat1.west) |- ([xshift=-2mm]sub9.west);
\end{tikzpicture}

\begin{tikzpicture}[
  every node/.style={align=left,},
  category/.style={text width=6cm, align=left, inner sep=4pt, font=\small},
  root/.style={text width=10cm, align=left, inner sep=4pt},
  quote/.style={rounded corners, thick, text width=6cm, align=justify, inner sep=4pt, font=\footnotesize},
  arrow/.style={-{Stealth[length=3mm]}, thick},
  connector/.style={thick},
  spacing/.store in=\myspacing, spacing=1.5cm 
]

\node[root] (cat1) at (0, 0) {Ethical hazards for society/the (healthcare) system (continued)};
\node[category, anchor=west] (sub1) at ($(cat1.south west)+(0.5,-\myspacing+0.5cm)$) {Erosion of responsibility/gaps in the responsibility structure (continued)};
\node[quote, anchor=west] (q1) at ($(sub1.east)+(1.5cm,0)$) {};
\node[draw=none, fit=(sub1)(q1), inner sep=0pt] (pair1) {};

"
\node[category, anchor=west] (sub2) at ($(pair1.south west)+(0.5,-\myspacing-1cm)$) {Underdefined liability where social robots make mistakes};
\node[quote, anchor=west] (q2) at ($(sub2.east)+(1.cm,0)$) {As to current issues concerning liability, scholars point out that in case of damages caused by an autonomous robot it might be more difficult to identify a responsible due to the increased complexity of the causal chain 130]. As a matter of fact, to turn a robot into a market product involves many actors: manufacturer, importers, wholesalers, retailers, repairers, installers, inspectors, and the users. Moreover, as pointed out by Lehman-Wilzig, the manufacturer has now doubled into the hardware and the software manufacturers. Therefore, it could be more difficult to attribute or share the responsibility for a defect or a fault and in many cases there might be ‘no one at fault!’ [30]. The issue of the indeterminacy of liability is related to robot autonomy, but also to share control robot […].”(Salvini 2015 On Ethical, Legal and Social Issues of Care Robots, p. 441; in: Mohammed 2015 Intelligent Assistive Robots - Recent Advances in Assistive Robotics for Everyday Activities)};
\draw[arrow] (sub2.east) -- (q2.west);
\node[draw=none, fit=(sub2)(q2), inner sep=0pt] (pair2) {};"
"
\node[category, anchor=west] (sub3) at ($(pair2.south west)+(-0.5,-\myspacing-0.6cm)$) {Inefficient use of social robots};
\node[quote, anchor=west] (q3) at ($(sub3.east)+(1.5cm,0)$) {“Although IATs open up promising prospects for the future of dementia care, their adoption is still lower than expected [8]. This has been attributed to suboptimal information transfer and dissemination across technology development und medical implementation [9] as well as to the lack of solid and highly generalizable clinical validation of many IATs [10].” (Ienca 2016 Intelligent Assistive Technology for Alzheimer’s Disease and Other Dementias: A Systematic Review, p. 1302)};
\draw[arrow] (sub3.east) -- (q3.west);
\node[draw=none, fit=(sub3)(q3), inner sep=0pt] (pair3) {};"
"
\node[category, anchor=west] (sub4) at ($(pair3.south west)+(0.5,-\myspacing-1.5cm)$) {Difficulty in continuing the use of social robot after projects end};
\node[quote, anchor=west] (q4) at ($(sub4.east)+(1.cm,0)$) {“Implementation of technology was perceived as a time-consuming and complex process. One concern was the ‘project problem’, meaning that, when a project ends, it all ends, rather than the project being firmly anchored in the organisation, so that the technology can be implemented after the project ends. Another challenge is how to launch the products on the market. Often, for a single product and small businesses, there is a shortage of capital, financing, and investment, and the unattainable market even seems to have reduced the inflow of innovation.”(Johansson-Pajala and Gustafsson 2022 Significant challenges when introducing care robots in Swedish elder care, p. 171)};
\draw[arrow] (sub4.east) -- (q4.west);
\node[draw=none, fit=(sub4)(q4), inner sep=0pt] (pair4) {};"
"
\node[category, anchor=west] (sub5) at ($(pair4.south west)+(0.,-\myspacing)$) {Positive benefits of social robot use cannot sustain long-term};
\node[quote, anchor=west] (q5) at ($(sub5.east)+(1.cm,0)$) {“However, reported positive effects could only be shown for a short time [20] and undesirable effects as well as obstacles in the application of SAR are supposed.”(Lehmann et al. 2021 Using a Socially Assistive Robot in a Nursing Home, p. 149)};
\draw[arrow] (sub5.east) -- (q5.west);
\node[draw=none, fit=(sub5)(q5), inner sep=0pt] (pair5) {};"
"
\node[category, anchor=north west] (sub6) at ($(pair5.south west)+(0.5,0)$) {\textit{Cont. on next page}};
\node[quote, anchor=west] (q6) at ($(sub6.east)+(1.cm,0)$) {};
\node[draw=none, fit=(sub6)(q6), inner sep=0pt] (pair6) {};

\node[category, anchor=north west] (sub7) at ($(pair6.south west)+(-0.5,0)$) {\textit{Cont. on next page}};
\node[quote, anchor=west] (q7) at ($(sub7.east)+(1.cm,0)$) {};
\node[draw=none, fit=(sub7)(q7), inner sep=0pt] (pair7) {};

\node[category, anchor=north west] (sub8) at ($(pair7.south west)+(-0.5,0)$) {\textit{Cont. on next page}};
\node[quote, anchor=west] (q8) at ($(sub8.east)+(1.cm,0)$) {};
\node[draw=none, fit=(sub8)(q8), inner sep=0pt] (pair8) {};

\draw[connector] (cat1.west) |- ([xshift=-2mm]sub1.west);
\draw[connector] (sub1.west) |- ([xshift=-2mm]sub2.west);
\draw[connector] (cat1.west) |- ([xshift=-2mm]sub3.west);
\draw[connector] (sub3.west) |- ([xshift=-2mm]sub4.west);
\draw[connector] (sub3.west) |- ([xshift=-2mm]sub5.west);
\draw[connector] (sub5.west) |- ([xshift=-2mm]sub6.west);
\draw[connector] (sub3.west) |- ([xshift=-2mm]sub7.west);
\draw[connector] (cat1.west) |- ([xshift=-2mm]sub8.west);
\end{tikzpicture}

\begin{tikzpicture}[
  every node/.style={align=left,},
  category/.style={text width=6cm, align=left, inner sep=4pt, font=\small},
  root/.style={text width=10cm, align=left, inner sep=4pt},
  quote/.style={rounded corners, thick, text width=6cm, align=justify, inner sep=4pt, font=\footnotesize},
  arrow/.style={-{Stealth[length=3mm]}, thick},
  connector/.style={thick},
  spacing/.store in=\myspacing, spacing=1.5cm 
]

\node[root] (cat1) at (0, 0) {Ethical hazards for society/the (healthcare) system (continued)};

\node[category, anchor=west] (sub1) at ($(cat1.south west)+(0.5,-\myspacing+0.5cm)$) {Inefficient use of social robots (continued)};
\node[quote, anchor=west] (q1) at ($(sub1.east)+(1.5cm,0)$) {};
\node[draw=none, fit=(sub1)(q1), inner sep=0pt] (pair1) {};

"
\node[category, anchor=west] (sub2) at ($(pair1.south west)+(0.5,-\myspacing)$) {High costs of social robots and their implementation};
\node[quote, anchor=west] (q2) at ($(sub2.east)+(1.cm,0)$) {“One major factor likely to influence the actual take-up of robots in elder care is cost. The high costs of current robots are likely to limit their adoption.” (Sharkey and Sharkey 2012 The eldercare factory, p. 283)};
\draw[arrow] (sub2.east) -- (q2.west);
\node[draw=none, fit=(sub2)(q2), inner sep=0pt] (pair2) {};"
"
\node[category, anchor=west] (sub3) at ($(pair2.south west)+(0.5,-\myspacing)$) {Difficult to develop robots in a divided/small market };
\node[quote, anchor=west] (q3) at ($(sub3.east)+(0.5cm,0)$) {“This caring matrix points out four different target-groups for effective digital assisting tools in, as each sector of the matrix has other claims for assistance. This is one reason why it is difficult to develop effective ‘CareSAR’-tools, as the market is divided in small sectors.”(Hasenauer et al. 2022 Living Labs in Social Service Institutions: An Effective Method to Improve the Ethical, Reliable Use of Digital Assistive Robots to Support Social Services, p. 2)};
\draw[arrow] (sub3.east) -- (q3.west);
\node[draw=none, fit=(sub3)(q3), inner sep=0pt] (pair3) {};"
"
\node[category, anchor=west] (sub4) at ($(pair3.south west)+(-1.,-\myspacing)$) {Ecologically and socially not sustainable };
\node[quote, anchor=west] (q4) at ($(sub4.east)+(1.5cm,0)$) {“Seemingly, it is the local distribution of benefits and costs that is under ethical reflection (with the exception of the question of ecological sustainability, which is presented as both a local and global matter of concern).”(Boada et al. 2021 The ethical issues of social assistive robotics: A critical literature review, p. 9)};
\node[draw=none, fit=(sub4)(q4), inner sep=0pt] (pair4) {};"
"
\node[category, anchor=west] (sub5) at ($(pair4.south west)+(0.5,-\myspacing)$) {Using of minerals that fuel conflict};
\node[quote, anchor=west] (q5) at ($(sub5.east)+(1.cm,0)$) {“SAR use in aged-care settings forces societal stakeholders to tackle the problem of what has been called ‘conflict minerals’. The sources of production materials and minerals to produce SARs need to be questioned.”(Vandemeulebroucke et al. 2021 Socially Assistive Robots in Aged Care: Ethical Orientations Beyond the Care-Romantic and Technology-Deterministic Gaze, p. 17)};
\draw[arrow] (sub5.east) -- (q5.west);
\node[draw=none, fit=(sub5)(q5), inner sep=0pt] (pair5) {};"
"
\node[category, anchor=west] (sub6) at ($(pair5.south west)+(0.,-\myspacing)$) {Working conditions in production not safe/healthy};
\node[quote, anchor=west] (q6) at ($(sub6.east)+(1.cm,0)$) {“SAR use in aged-care settings forces societal stakeholders to tackle the problem of what has been called ‘conflict minerals’. The sources of production materials and minerals to produce SARs need to be questioned. Efforts need to be made to improve working and social conditions for workers that mine, collect, and process minerals and materials in unhealthy and unsafe circumstances.”(Vandemeulebroucke et al. 2021 Socially Assistive Robots in Aged Care: Ethical Orientations Beyond the Care-Romantic and Technology-Deterministic Gaze, p. 17)};
\draw[arrow] (sub6.east) -- (q6.west);
\node[draw=none, fit=(sub6)(q6), inner sep=0pt] (pair6) {};"
"
\node[category, anchor=west] (sub7) at ($(pair6.south west)+(0.,-\myspacing)$) {Social robot production/usage/dumping too resource intensive};
\node[quote, anchor=west] (q7) at ($(sub7.east)+(1.cm,0)$) {“However, the supply of raw materials for robots, the energy consumption they require, and the dumping waste that these new care technologies generate are important ethical challenges of SAR.”(Boada et al. 2021 The ethical issues of social assistive robotics: A critical literature review, p. 9)};
\draw[arrow] (sub7.east) -- (q7.west);
\node[draw=none, fit=(sub7)(q7), inner sep=0pt] (pair7) {};"
"
\node[category, anchor=west] (sub8) at ($(pair7.south west)+(-0.5,-\myspacing-0.6cm)$) {Social robots get too much power and endanger humans/human society};
\node[quote, anchor=west] (q8) at ($(sub8.east)+(1.5cm,0)$) {“In this mode of rule, which proceeds by transforming its commands into programs and its agents into automatons (or in the case of the care robots: which delegates its socio-political tasks, made necessary by demographic change, to automatons), power, which was already far away in the past, now becomes completely intangible.” (Translation by the authors)(Andreas 2018 Autonomous Lethality. Lebenskritische Entscheidungen in der Roboterethik, p. 151; in: Andreas et al. 2018 Unterwachen und Schlafen. Anthrophile Medien nach dem Interface)};
\draw[arrow] (sub8.east) -- (q8.west);
\node[draw=none, fit=(sub8)(q8), inner sep=0pt] (pair8) {};"

\draw[connector] (cat1.west) |- ([xshift=-2mm]sub1.west);
\draw[connector] (sub1.west) |- ([xshift=-2mm]sub2.west);
\draw[connector] (sub2.west) |- ([xshift=-2mm]sub3.west);
\draw[connector] (cat1.west) |- ([xshift=-2mm]sub4.west);
\draw[connector] (sub4.west) |- ([xshift=-2mm]sub5.west);
\draw[connector] (sub4.west) |- ([xshift=-2mm]sub6.west);
\draw[connector] (sub4.west) |- ([xshift=-2mm]sub7.west);
\draw[connector] (cat1.west) |- ([xshift=-2mm]sub8.west);
\end{tikzpicture}

\begin{tikzpicture}[
  every node/.style={align=left,},
  category/.style={text width=6cm, align=left, inner sep=4pt, font=\small},
  root/.style={text width=10cm, align=left, inner sep=4pt},
  quote/.style={rounded corners, thick, text width=6cm, align=justify, inner sep=4pt, font=\footnotesize},
  arrow/.style={-{Stealth[length=3mm]}, thick},
  connector/.style={thick},
  spacing/.store in=\myspacing, spacing=1.5cm 
]

\node[root] (cat1) at (0, 0) {Unsettled questions (Main category)};

"
\node[category, anchor=west] (sub1) at ($(cat1.south west)+(0.5,-\myspacing-1.5cm)$) {Do the benefits of social robots outweigh the costs?};
\node[quote, anchor=west] (q1) at ($(sub1.east)+(1.5cm,0)$) {“The principle of justice governs the fair distribution of scarce resources. This can be a very difficult topic when discussing experimental treatments such as SAR. The authors know of no SAR systems that are currently used outside the research setting, so discussion of the actual cases in the field is premature. However, we can presume that for the foreseeable future, robots will be somewhat expensive. Thus, a question that should be asked is: Do the benefits of SAR outweigh the costs? Like other proposed therapies, quality of life surveys or other methods for assessing medical economy can be used to assess relative benefits, and costs can be weighed against improvements observed [1], [24]. There does not seem to be a significant difference between calculating the costs and benefits of robots compared to other assistive devices.” (Feil-Seifer and Mataric 2011 Socially Assistive Robotics - Ethical Issues Related to Technology, p. 30)};
\draw[arrow] (sub1.east) -- (q1.west);
\node[draw=none, fit=(sub1)(q1), inner sep=0pt] (pair1) {};"
"
\node[category, anchor=west] (sub2) at ($(pair1.south west)+(0.,-\myspacing)$) {Who should receive/how to allocate scarce (robotic) resources?};
\node[quote, anchor=west] (q2) at ($(sub2.east)+(1.5cm,0)$) {“Carebots are likely to be a relatively expensive resource, and decisions will have to be made to who gets what and why.”(Gallagher et al. 2016 Robots in elder care: Some ethical questions, p. 370) };
\draw[arrow] (sub2.east) -- (q2.west);
\node[draw=none, fit=(sub2)(q2), inner sep=0pt] (pair2) {};"
"
\node[category, anchor=west] (sub3) at ($(pair2.south west)+(0.,-\myspacing-0.5cm)$) {Whose interests regarding social robot usage should be prioritized? };
\node[quote, anchor=west] (q3) at ($(sub3.east)+(1.5cm,0)$) {“One ethical question is only mentioned here in passing, but it is anything but trivial to answer: Should a robot be unconditionally aligned with the interests or preferences of the person in need of support? And if not, with what other interests and according to what criteria?” (Translation by the authors)(Manzeschke 2019 Roboter in der Pflege - Von Menschen, Maschinen und anderen hilfreichen Wesen, p. 7)};
\draw[arrow] (sub3.east) -- (q3.west);
\node[draw=none, fit=(sub3)(q3), inner sep=0pt] (pair3) {};"
"
\node[category, anchor=west] (sub4) at ($(pair3.south west)+(0.5,-\myspacing-0.6cm)$) {Should care benefit be realized if staff is negatively affected?};
\node[quote, anchor=west] (q4) at ($(sub4.east)+(1.cm,0)$) {“One could argue that the robot should primarily help to care for the cared-for persons so that their wishes should have priority. However, these wishes might conflict with the aim of the care robot so that the care task cannot be executed as intended. The respective wishes might also be discriminating against other members of the care setting and thus collide with the principle of justice.” (Weßel et al. 2021 Gender Stereotyping of Robotic Systems in Elderly Care An Exploratory Analysis of Ethical Problems and Possible Solutions, p. 10.)};
\draw[arrow] (sub4.east) -- (q4.west);
\node[draw=none, fit=(sub4)(q4), inner sep=0pt] (pair4) {};"
"
\node[category, anchor=west] (sub5) at ($(pair4.south west)+(0.,-\myspacing-0.3cm)$) {To whom among care home residents should social robots be customized?};
\node[quote, anchor=west] (q5) at ($(sub5.east)+(1.cm,0)$) {“With regard to autonomy, queering only seems to offer a solution where the robot is merely used by one person, or all users have common preferences about the design. Otherwise, conflicts can arise from the individualization of the design. This raises the question of whose preferences should be prioritized.”(Weßel et al. 2021 Gender Stereotyping of Robotic Systems in Elderly Care An Exploratory Analysis of Ethical Problems and Possible Solutions, p. 10.)};
\draw[arrow] (sub5.east) -- (q5.west);
\node[draw=none, fit=(sub5)(q5), inner sep=0pt] (pair5) {};"
"
\node[category, anchor=north west] (sub6) at ($(pair5.south west)+(-0.5,0)$) {\textit{Cont. on next page}};
\node[quote, anchor=west] (q6) at ($(sub6.east)+(1.cm,0)$) {};
\node[draw=none, fit=(sub6)(q6), inner sep=0pt] (pair6) {};

\draw[connector] (cat1.west) |- ([xshift=-2mm]sub1.west);
\draw[connector] (cat1.west) |- ([xshift=-2mm]sub2.west);
\draw[connector] (cat1.west) |- ([xshift=-2mm]sub3.west);
\draw[connector] (sub3.west) |- ([xshift=-2mm]sub4.west);
\draw[connector] (sub3.west) |- ([xshift=-2mm]sub5.west);
\draw[connector] (cat1.west) |- ([xshift=-2mm]sub6.west);

\end{tikzpicture}

\begin{tikzpicture}[
  every node/.style={align=left,},
  category/.style={text width=6cm, align=left, inner sep=4pt, font=\small},
  root/.style={text width=10cm, align=left, inner sep=4pt},
  quote/.style={rounded corners, thick, text width=6cm, align=justify, inner sep=4pt, font=\footnotesize},
  arrow/.style={-{Stealth[length=3mm]}, thick},
  connector/.style={thick},
  spacing/.store in=\myspacing, spacing=1.5cm 
]

\node[root] (cat1) at (0, 0) {Unsettled questions (continued)};

"
\node[category, anchor=west] (sub1) at ($(cat1.south west)+(0.5,-\myspacing)$) {Design by stereotype to foster wellbeing/compliance while reproducing harmful effects/reducing acceptance of diverse ways of life? };
\node[quote, anchor=west] (q1) at ($(sub1.east)+(1.5cm,0)$) {“[…] we then map possible moral problems and conflicts such as the tension between a potentially beneficial impact of stereotyping on user compliance and wellbeing on the one hand, and its implications for modern pluralistic values such as diversity and mutual recognition on the other.” (Weßel et al. 2021 Gender Stereotyping of Robotic Systems in Eldercare: An Exploratory Analysis of Ethical Problems and Possible Solutions, p. 2)};
\draw[arrow] (sub1.east) -- (q1.west);
\node[draw=none, fit=(sub1)(q1), inner sep=0pt] (pair1) {};"
"
\node[category, anchor=west] (sub2) at ($(pair1.south west)+(0.,-\myspacing-1.3cm)$) {What privacy risks are acceptable for security gains?};
\node[quote, anchor=west] (q2) at ($(sub2.east)+(1.5cm,0)$) {“It would therefore be much more interesting to find out how to deal with the various ambivalences that can arise when using autonomous assistance systems - important questions that arise particularly with dementia patients concern, for example, the possible deceptive nature of social therapy robots or the conflict between the protection of privacy on the one hand and safety-related welfare requirements on the other (for example in the course of technical monitoring of cognitively impaired people) (Remmers 2016).” (Translation by the authors)(Kehl 2018 Wege zu verantwortungsvoller Forschung und Entwicklung im Bereich der Pfegerobotik: Die ambivalente Rolle der Ethik, p. 149; in: Bendel 2018 Pflegeroboter)};
\draw[arrow] (sub2.east) -- (q2.west);
\node[draw=none, fit=(sub2)(q2), inner sep=0pt] (pair2) {};"
"
\node[category, anchor=west] (sub3) at ($(pair2.south west)+(0.,-\myspacing-3cm)$) {Is deception about relationships acceptable to realize benefits?};
\node[quote, anchor=west] (q3) at ($(sub3.east)+(1.5cm,0)$) {“In fact, it cannot be ruled out that people with dementia in particular, who suffer from an advanced loss of mental capacity, are deceived about the nature of robots. In principle - as Nozick's thought experiment shows - it can also be assumed that people would prefer their existing reality to a perfect fictional world. However, stepping out of everyday reality for a short time in order to be fooled can be appealing. We often enjoy the variety of fictional parallel worlds with virtual reality glasses or in 4D cinemas and then return to our normal lives. Similarly, interacting with a robot does not mean permanently leaving the real world in favor of a preferred illusory existence. Taking into account the strong intuition that most people do not want to remain in a fictitious, albeit perfect, world, the exchange between humans and machines must also be limited to regulated and frequented time windows. Under these conditions, robots are very well able to maintain the interest of the person in need of care and improve the quality of life of those affected, contrary to what critics such as Sparrow/Sparrow say.” (Translation by the authors)(Kreis 2018 Umsorgen, überwachen, unterhalten - sind Pflegeroboter ethisch vertretbar?, p. 221; in: Bendel 2018 Pflegeroboter)};
\draw[arrow] (sub3.east) -- (q3.west);
\node[draw=none, fit=(sub3)(q3), inner sep=0pt] (pair3) {};"
"
\node[category, anchor=north west] (sub6) at ($(pair3.south west)+(-0.,0)$) {\textit{Cont. on next page}};
\node[quote, anchor=west] (q6) at ($(sub6.east)+(1.cm,0)$) {};
\node[draw=none, fit=(sub6)(q6), inner sep=0pt] (pair6) {};

\draw[connector] (cat1.west) |- ([xshift=-2mm]sub1.west);
\draw[connector] (cat1.west) |- ([xshift=-2mm]sub2.west);
\draw[connector] (cat1.west) |- ([xshift=-2mm]sub3.west);
\draw[connector] (cat1.west) |- ([xshift=-2mm]sub6.west);

\end{tikzpicture}

\begin{tikzpicture}[
  every node/.style={align=left,},
  category/.style={text width=6cm, align=left, inner sep=4pt, font=\small},
  root/.style={text width=10cm, align=left, inner sep=4pt},
  quote/.style={rounded corners, thick, text width=6cm, align=justify, inner sep=4pt, font=\footnotesize},
  arrow/.style={-{Stealth[length=3mm]}, thick},
  connector/.style={thick},
  spacing/.store in=\myspacing, spacing=1.5cm 
]

\node[root] (cat1) at (0, 0) {Unsettled questions (continued)};

\node[category, anchor=west] (sub1) at ($(cat1.south west)+(0.5,-\myspacing+0.5cm)$) {Is deception about relationships acceptable to realize benefits? (continued)};
\node[quote, anchor=west] (q1) at ($(sub1.east)+(1.5cm,0)$) {};
\node[draw=none, fit=(sub1)(q1), inner sep=0pt] (pair1) {};

"
\node[category, anchor=west] (sub2) at ($(pair1.south west)+(0.5,-\myspacing-1cm)$) {Do the benefits of anthropomorphization outweigh negative effects?};
\node[quote, anchor=west] (q2) at ($(sub2.east)+(1.cm,0)$) {“Yet, the anthropomorphization of the robot does not necessarily always have to create harm but might also increase wellbeing. In fact, this possibility is explicitly used in technology development to support a robot’s performance in its setting. For example, Matilda is designed as child-like in order not to appear threatening to older people with cognitive impairments [62]. In consequence, explaining that such a robot is only a machine could have detrimental effects on its utilization and functioning and the wellbeing of the users. Furthermore, it is possible that users have already formed an emotional attachment to a robot before it is explained to them, or an emotional attachment is crucial for the users to feel comfortable and to interact with the robot in an effective way. If the robot is ‘de-humanized’ by an explanation, this might ultimately affect these users’ wellbeing because they do not want to use ‘just’ a machine but need an emotional bond to increase their wellbeing.” (Weßel et al. 2021 Gender Stereotyping of Robotic Systems in Eldercare: An Exploratory Analysis of Ethical Problems and Possible Solutions, p. 8.)};
\draw[arrow] (sub2.east) -- (q2.west);
\node[draw=none, fit=(sub2)(q2), inner sep=0pt] (pair2) {};"
"
\node[category, anchor=west] (sub3) at ($(pair2.south west)+(-0.5,-\myspacing-0.5cm)$) {Should control of/access to social robots be restricted to reduce risk of harm?};
\node[quote, anchor=west] (q3) at ($(sub3.east)+(1.5cm,0)$) {“We also suggest that the use of the robot should reflect the autonomy of patients where possible. Our participants, however, were open to both formal and ICs limiting autonomy to ensure the safety of older people, especially where protecting safety might secure autonomy in the longer-term.” (Jenkins and Draper 2015 Care, Monitoring, and Companionship: Views on Care Robots from Older People and Their Carer, p. 682)};
\draw[arrow] (sub3.east) -- (q3.west);
\node[draw=none, fit=(sub3)(q3), inner sep=0pt] (pair3) {};"
"
\node[category, anchor=west] (sub4) at ($(pair3.south west)+(0.5,-\myspacing)$) {Should customization be possible risking abuse by other actors?};
\node[quote, anchor=west] (q4) at ($(sub4.east)+(1.cm,0)$) {“What if several people can instruct the robot, say, me, my children living remotely and my doctor. Who can instruct the robot to do what actions? If the instructions conflict, who’s instructions take precedence?”(Bennett 2014 Robot Identity Assurance, p. 11)};
\draw[arrow] (sub4.east) -- (q4.west);
\node[draw=none, fit=(sub4)(q4), inner sep=0pt] (pair4) {};"
"
\node[category, anchor=west] (sub5) at ($(pair4.south west)+(-0.5,-\myspacing-1cm)$) {Who should/how to decide when people are cognitively impaired? };
\node[quote, anchor=west] (q5) at ($(sub5.east)+(1.5cm,0)$) {“Let us suppose that a person, throughout her entire life-time, has established a strong intellectual self-identity. She has seen people in nursing homes play with Paro and concluded that ‘I am not the kind of person who is ever going to play with such a thing!’ She informs her family about this decision and puts it – together with other things – down in her advance healthcare directive. She then develops late-stage dementia and moves into a residence. One day she sees a group of people interact with Paro and desperately wants to join in. Is it right to involve her in the group, or should we prevent her from playing with Paro?”(Misselhorn et al. 2013 Ethical Considerations Regarding the Use of Social Robots in the Fourth Age, pp. 130, 131)};
\draw[arrow] (sub5.east) -- (q5.west);
\node[draw=none, fit=(sub5)(q5), inner sep=0pt] (pair5) {};"
"
\node[category, anchor=north west] (sub6) at ($(pair5.south west)+(-0.,0)$) {\textit{Cont. on next page}};
\node[quote, anchor=west] (q6) at ($(sub6.east)+(1.cm,0)$) {};
\node[draw=none, fit=(sub6)(q6), inner sep=0pt] (pair6) {};

\draw[connector] (cat1.west) |- ([xshift=-2mm]sub1.west);
\draw[connector] (sub1.west) |- ([xshift=-2mm]sub2.west);
\draw[connector] (cat1.west) |- ([xshift=-2mm]sub3.west);
\draw[connector] (sub3.west) |- ([xshift=-2mm]sub4.west);
\draw[connector] (cat1.west) |- ([xshift=-2mm]sub5.west);
\draw[connector] (cat1.west) |- ([xshift=-2mm]sub6.west);

\end{tikzpicture}

\begin{tikzpicture}[
  every node/.style={align=left,},
  category/.style={text width=6cm, align=left, inner sep=4pt, font=\small},
  root/.style={text width=10cm, align=left, inner sep=4pt},
  quote/.style={rounded corners, thick, text width=6cm, align=justify, inner sep=4pt, font=\footnotesize},
  arrow/.style={-{Stealth[length=3mm]}, thick},
  connector/.style={thick},
  spacing/.store in=\myspacing, spacing=1.5cm 
]

\node[root] (cat1) at (0, 0) {Unsettled questions (continued)};

"
\node[category, anchor=west] (sub1) at ($(cat1.south west)+(0.5,-\myspacing)$) {Who bears responsibility for the social robot and its actions?};
\node[quote, anchor=west] (q1) at ($(sub1.east)+(1.5cm,0)$) {“Another justice-related issue when discussing robotics in socially assistive settings is the notion of responsibility: Who is responsible when things go wrong?” (Feil-Seifer und Mataric 2011 Socially Assistive Robotics - Ethical Issues Related to Technology, p. 30) };
\draw[arrow] (sub1.east) -- (q1.west);
\node[draw=none, fit=(sub1)(q1), inner sep=0pt] (pair1) {};"
"
\node[category, anchor=west] (sub2) at ($(pair1.south west)+(0.,-\myspacing-0.5cm)$) {(When) should social robots be allowed to act without consent to protect? };
\node[quote, anchor=west] (q2) at ($(sub2.east)+(1.5cm,0)$) {“The actions of robots can be advisable concerning consequentialism or utilitarian approaches. For instance, if an older person wants to walk on the high-traffic road and a robot can stop him, then what should the robot do? Herein, the problem would be based on the robot’s classification algorithm about the dangerous activity. In another scenario, Should the robot stop a person from eating much food? Would it be fine to restrict an individual’s liberty?” (Noori et al. 2019 Robot-Care for the Older People: Ethically Justified or Not?, p. 45)};
\draw[arrow] (sub2.east) -- (q2.west);
\node[draw=none, fit=(sub2)(q2), inner sep=0pt] (pair2) {};"
"
\node[category, anchor=west] (sub3) at ($(pair2.south west)+(0.5,-\myspacing)$) {Should social robots motivate/remind while possibly perceived as controlling?};
\node[quote, anchor=west] (q3) at ($(sub3.east)+(1.cm,0)$) {“The use of care assistant robots also raises questions. How pushy may a robot become, for example, in reminding someone to take medication?” (Van Est et al. 2016 Robotisation as Rationalisation - In Search for a Human Robot Future, p. 45)};
\draw[arrow] (sub3.east) -- (q3.west);
\node[draw=none, fit=(sub3)(q3), inner sep=0pt] (pair3) {};"
"
\node[category, anchor=west] (sub4) at ($(pair3.south west)+(-0.5,-\myspacing-0.5cm)$) {Should autonomy be bestowed onto social robots (at all/to what extent)?};
\node[quote, anchor=west] (q4) at ($(sub4.east)+(1.5cm,0)$) {“In general, the use of social robots in therapy and care raises a number of ethical questions: How autonomous should a therapy robot be? How much control should the caregivers and therapists have?” (Translation by the authors)(Lange and Bauer 2021 - Eine Robbe für Oma - Die zukünftige Dauerausstellung Robotik im Deutschen Museum, S. 236; in: Inthorn und Seising 2021 Digitale Patientenversorgung - Zur Computerisierung von Diagnostik, Therapie und Pflege)};
\draw[arrow] (sub4.east) -- (q4.west);
\node[draw=none, fit=(sub4)(q4), inner sep=0pt] (pair4) {};"

"
\node[category, anchor=west] (sub5) at ($(pair4.south west)+(0.5,-\myspacing)$) {Should social robots have moral capabilities (and which ones)?};
\node[quote, anchor=west] (q5) at ($(sub5.east)+(1.cm,0)$) {“Should a robot assistant have a moral reasoning system that tells it what is ‘good care’?” (Kemenade et al. 2019 Do You Care for Robots That Care? Exploring the Opinions of Vocational Care Students on the Use of Healthcare Robots, p. 8)};
\draw[arrow] (sub5.east) -- (q5.west);
\node[draw=none, fit=(sub5)(q5), inner sep=0pt] (pair5) {};"

"
\node[category, anchor=west] (sub6) at ($(pair5.south west)+(-0.5,-\myspacing-0.3cm)$) {What normative status should we assign social robots?};
\node[quote, anchor=west] (q6) at ($(sub6.east)+(1.5cm,0)$) {“Robots, as they are currently being developed for care, are complex technical counterparts that enter into social interaction with humans, although it is not yet clear what social and consequently what normative status we should assign to these manifestations.” (Translation by the authors) (Manzeschke 2019 Roboter in der Pflege - Von Menschen, Maschinen und anderen hilfreichen Wesen, p. 1)};
\draw[arrow] (sub6.east) -- (q6.west);
\node[draw=none, fit=(sub6)(q6), inner sep=0pt] (pair6) {};"

"
\node[category, anchor=north west] (sub7) at ($(pair6.south west)+(-0.,0)$) {\textit{Cont. on next page}};
\node[quote, anchor=west] (q7) at ($(sub7.east)+(1.cm,0)$) {};
\node[draw=none, fit=(sub7)(q7), inner sep=0pt] (pair7) {};

\draw[connector] (cat1.west) |- ([xshift=-2mm]sub1.west);
\draw[connector] (cat1.west) |- ([xshift=-2mm]sub2.west);
\draw[connector] (sub2.west) |- ([xshift=-2mm]sub3.west);
\draw[connector] (cat1.west) |- ([xshift=-2mm]sub4.west);
\draw[connector] (sub4.west) |- ([xshift=-2mm]sub5.west);
\draw[connector] (cat1.west) |- ([xshift=-2mm]sub6.west);
\draw[connector] (cat1.west) |- ([xshift=-2mm]sub7.west);

\end{tikzpicture}

\begin{tikzpicture}[
  every node/.style={align=left,},
  category/.style={text width=6cm, align=left, inner sep=4pt, font=\small},
  root/.style={text width=10cm, align=left, inner sep=4pt},
  quote/.style={rounded corners, thick, text width=6cm, align=justify, inner sep=4pt, font=\footnotesize},
  arrow/.style={-{Stealth[length=3mm]}, thick},
  connector/.style={thick},
  spacing/.store in=\myspacing, spacing=1.5cm 
]

\node[root] (cat1) at (0, 0) {Unsettled questions (continued)};

"
\node[category, anchor=west] (sub1) at ($(cat1.south west)+(0.5,-\myspacing-1.5cm)$) {How do we understand the concept of care?};
\node[quote, anchor=west] (q1) at ($(sub1.east)+(1.5cm,0)$) {“Lifting patients out of their beds can be described as both goal-oriented and practice-oriented. From the goal-oriented perspective, it is simply a matter of changing the patient's position. In the practice-oriented sense, however, lifting is also about touch and empathic interaction between the caregiver and the patient. Depending on the perspective from which an action is viewed, the use of a care robot or assistance system in this case may or may not appear morally problematic: if the aim is simply to achieve the goal, it is fine to use the support of a machine. If the focus is on touch and empathic interaction, then this is questionable. There is no right or wrong perspective, but it depends largely on what meaning an action has for the person in need of care.” (Translation by the authors)(Misselhorn 2018 Pflegesysteme, p. 154; in: Misselhorn 2018 Grundfragen der Maschinenethik)};
\draw[arrow] (sub1.east) -- (q1.west);
\node[draw=none, fit=(sub1)(q1), inner sep=0pt] (pair1) {};"
"
\node[category, anchor=west] (sub2) at ($(pair1.south west)+(0.5,-\myspacing-2cm)$) {Do social robots per se fail to care?};
\node[quote, anchor=west] (q2) at ($(sub2.east)+(1.cm,0)$) {“For him, deep care means something like deep human emotions when looking at the person in need of care. This criterion is somewhat unclear, but according to my explanation it could be understood as follows: in the sense of deep care, the caregiver not only cares for the patient, but is concerned about the patient. He is existentially involved and offers this existential relationship to the patient. Robots are only suitable for superficial care. Of course, human nurses do not always provide deep care either. They sometimes carry out work like robots. But patients perceive them as human beings and they are always the objects of a human relationship for patients.” (Translation by the authors)(Kovács 2021 Anthropologische und ethische Aspekte des Einsatzes von Robotern im Gesundheitssektor, p. 32; in: Inthorn und Seising 2021 Digitale Patientenversorgung – Zur Computerisierung von Diagnostik, Therapie und Pflege) };
\draw[arrow] (sub2.east) -- (q2.west);
\node[draw=none, fit=(sub2)(q2), inner sep=0pt] (pair2) {};"

\draw[connector] (cat1.west) |- ([xshift=-2mm]sub1.west);
\draw[connector] (sub1.west) |- ([xshift=-2mm]sub2.west);

\end{tikzpicture}
\clearpage
\section*{Appendix B: References of included fulltexts}\label{secA2}

\begin{enumerate}
\item Akalin N, Kristoffersson A, Loutfi A (2019) {Evaluating the Sense of Safety and
  Security in Human–Robot Interaction with Older People}. In: Korn O (ed)
  {Social Robots: Technological, Societal and Ethical Aspects of Human-Robot
  Interaction}. {Human-Computer Interaction Series}, Springer, Cham, p 237--264

\item {Aldinhas Ferreira} MI, Sequeira JS (2017) {Robots in Ageing Societies}. In:
  {Aldinhas Ferreira} MI, {Silva Sequeira} J, Tokhi MO, et~al (eds) {A World
  with Robots: International Conference on Robot Ethics: ICRE 2015}, vol~84.
  {Springer International Publishing}, Cham, p 217--223,
  \doi{10.1007/978-3-319-46667-5_17}

\item Amirabdollahian F, den Akker Ro, Bedaf S, et~al (2013{\natexlab{a}}) {Assistive
  technology design and development for acceptable robotics companions for
  ageing years}. {Paladyn, Journal of Behavioral Robotics} 4(2).
  \doi{10.2478/pjbr-2013-0007}

\item Amirabdollahian F, {op den Akker} R, Bedaf S, et~al (2013{\natexlab{b}})
  {Accompany: Acceptable robotiCs COMPanions for AgeiNG Years —
  Multidimensional aspects of human-system interactions}. In: {2013 6th
  International Conference on Human System Interactions (HSI)}. IEEE, pp
  570--577, \doi{10.1109/HSI.2013.6577882}

\item {Ammicht Quinn} R (2019) {Zwischen Fürsorge und Kontrolle}. {EthikJournal}
  5:1--20

\item Anderson M (2020) {Ein wertegesteuerter Roboter}. {Spektrum der Wissenschaft 20
  Spezial Biologie Medizin Hirnforschung} pp 3--5

\item Anderson SL, Anderson M (2018) {Ethische Roboter für die Altenpflege}. In:
  {3TH1CS}. {Bundeszentrale für Politische Bildung, 2018}, Bonn

\item Andreas M (2018) {Autonomous Lethality. Lebenskritische Entscheidungen in der
  Roboterethik}. {meson press}, \doi{10.25969/MEDIAREP/1278}

\item Baisch S, Kolling T, Rühl S, et~al (2018) {Emotionale Roboter im
  Pflegekontext}. {Zeitschrift für Gerontologie und Geriatrie} 51(1):16--24.
  \doi{10.1007/s00391-017-1346-8}

\item Battistuzzi L, Papadopoulos C, Hill T, et~al (2021) {Socially Assistive Robots,
  Older Adults and Research Ethics: The Case for Case-Based Ethics Training}.
  {International Journal of Social Robotics} 13(4):647--659.
  \doi{10.1007/s12369-020-00652-x}

\item Battistuzzi L, Sgorbissa A, Papadopoulos C, et~al (2018) {Embedding Ethics in
  the Design of Culturally Competent Socially Assistive Robots}. In: {2018
  IEEE/RSJ International Conference on Intelligent Robots and Systems (IROS)}.
  IEEE, pp 1996--2001, \doi{10.1109/IROS.2018.8594361}

\item Bauberger S (2020) {Welche KI?} {Carl Hanser Verlag GmbH {\&} Co. KG},
  München, \doi{10.3139/9783446465527}

\item Bedaf S, Marti P, de~Witte L (2019) {What are the preferred characteristics of
  a service robot for the elderly? A multi-country focus group study with older
  adults and caregivers}. {Assistive Technology} 31(3):147--157.
  \doi{10.1080/10400435.2017.1402390}

\item Bendel O (2018) {Roboter im Gesundheitsbereich: Operations-, Therapie-und
  Pflegeroboter aus ethischer Sicht}. In: Bendel O (ed) {Pflegeroboter}.
  {Springer Fachmedien Wiesbaden}, Wiesbaden, p 195--212

\item Bendel O (2020) {Care Robots with Sexual Assistance Functions}. In: {AAAI 2020
  Spring Symposium "Applied AI in Healthcare: Safety, Community, and the
  Environment"(Stanford University)}, vol abs/2004.04428. CoRR,
  \doi{10.48550/arXiv.2004.04428}

\item Bennett B (2019) {Technology, ageing and human rights: Challenges for an ageing
  world}. {International Journal of Law and Psychiatry} 66:101449.
  \doi{10.1016/j.ijlp.2019.101449}

\item Bennett B, McDonald F, Beattie E, et~al (2017) {Assistive technologies for
  people with dementia: ethical considerations}. {Bulletin of the World Health
  Organization} 95(11):749--755. \doi{10.2471/BLT.16.187484}

\item Bennett L (2014) {Robot Identity Assurance}. {ITNOW} 56(3):10--11.
  \doi{10.1093/itnow/bwu064}

\item Bertolini A, Arian S (2020) {Do Robots Care?: Towards an Anthropocentric
  Framework in the Caring of Frail Individuals through Assistive Technology}.
  In: Haltaufderheide J, Hovemann J, Vollmann J (eds) {Aging between
  Participation and Simulation: Ethical Dimensions of Socially Assistive
  Technologies in Elderly Care}. {De Gruyter}, p 35--52,
  \doi{10.1515/9783110677485-003}

\item Bianchi A (2021) {Considering sex robots for older adults with cognitive
  impairments}. {Journal of Medical Ethics} 47(1):37--38.
  \doi{10.1136/medethics-2020-106927}

\item {Bioethikkommission beim Bundeskanzleramt Österreich} (2019) {Roboter in der
  Betreuung alter Menschen: Stellungnahme der Bioethikkommission}. {Jahrbuch
  für Wissenschaft und Ethik} 24(1):355--384. \doi{10.1515/jwiet-2019-0015}

\item Blackman T (2013) {Care robots for the supermarket shelf: a product gap in
  assistive technologies}. {Ageing and Society} 33(5):763--781.
  \doi{10.1017/S0144686X1200027X}

\item Bleses H, Dammert M (2020) {Neue Technologien aus Sicht der
  Pflegewissenschaft}. In: Hanika H (ed) {Künstliche Intelligenz, Robotik und
  autonome Systeme in der Gesundheitsversorgung}. {Verlag Wissenschaft {\&}
  Praxis}, Sternenfels, p 55--84

\item Bleuler T, Caroni P (2021) {Roboter in der Pflege: Welche Aufgaben können
  Roboter heute schon übernehmen?} In: Bendel O (ed) {Soziale Roboter:
  Technikwissenschaftliche, wirtschaftswissenschaftliche, philosophische,
  psychologische und soziologische Grundlagen}. {Springer Gabler}, Wiesbaden, p
  441--457, \doi{10.1007/978-3-658-31114-8}

\item Boada JP, Maestre BR, Genís CT (2021) {The Ethical Issues of Social Assistive
  Robotics: A Critical Literature Review}. {Technology in Society} 67:1--13.
  \doi{10.1016/j.techsoc.2021.101726}

\item Bogue R (2013) {Robots to aid the disabled and the elderly}. {Industrial Robot:
  An International Journal} 40(6):519--524. \doi{10.1108/IR-07-2013-372}

\item Boni-Saenz AA (2021) {Are sex robots enough?} {Journal of Medical Ethics}
  47(1):35. \doi{10.1136/medethics-2020-106928}

\item Borenstein J, Pearson Y (2014) {Robot Caregivers: Ethical Issues across the
  Human Lifespan}. In: Lin P, Abney K, Bekey G (eds) {Robot Ethics: The Ethical
  and Social Implications of Robotics}. {MIT Press}, Cambridge, p 251--265

\item Bradwell HL, Winnington R, Thill S, et~al (2020) {Ethical perceptions towards
  real-world use of companion robots with older people and people with
  dementia: survey opinions among younger adults}. {BMC Geriatrics} 20(1).
  \doi{10.1186/s12877-020-01641-5}

\item {Büro für Technikfolgen-Abschätzung beim Deutschen Bundestag} (2018)
  {Robotik in Der Pflege ‐ Gesellschaftliche Herausforderungen}. Berlin

\item Byers P, Matthews S, Kennett J (2021) {Truthfulness in Dementia Care}.
  {Bioethics} 35(9):839--841. \doi{10.1111/bioe.12970}

\item Carnevale A (2017) {“I Tech Care”: How Healthcare Robotics Can Change the
  Future of Love, Solidarity, and Responsibility}. In: Hakli R, Seibt J (eds)
  {Sociality and Normativity for Robots}. {Studies in the Philosophy of
  Sociality}, {Springer International Publishing}, Cham, p 217--232,
  \doi{10.1007/978-3-319-53133-5_11}

\item Carros F, Eilers H, Langendorf J, et~al (2022) {Roboter als intelligente
  Assistenten in Betreuung und Pflege - Grenzen und Perspektiven im
  Praxiseinsatz}. In: {Künstliche Intelligenz im Gesundheitswesen}. {Springer
  Gabler, 2022}, Wiesbaden

\item Casey D, Felzmann H, Pegman G, et~al (2016) {What People with Dementia Want:
  Designing MARIO an Acceptable Robot Companion}. In: Miesenberger K, Bühler
  C, Penaz P (eds) {Computers Helping People with Special Needs}, {Lecture
  Notes in Computer Science}, vol 9758. {Springer International Publishing},
  Cham, p 318--325, \doi{10.1007/978-3-319-41264-1_44}

\item Coeckelbergh M (2015) {Care Robots and the Future of ICT-mediated Elderly Care:
  A Response to Doom Scenarios}. {AI {\&} SOCIETY} 31(4):455--462.
  \doi{10.1007/s00146-015-0626-3}

\item Coghlan S (2022) {Robots and the Possibility of Humanistic Care}.
  {International Journal of Social Robotics} 14(10):2095--2108.
  \doi{10.1007/s12369-021-00804-7}

\item Coghlan S, Waycott J, Lazar A, et~al (2021) {Dignity, Autonomy, and Style of
  Company}. {Proceedings of the ACM on Human-Computer Interaction}
  5(CSCW1):1--25. \doi{10.1145/3449178}

\item Conti D, {Di Nuovo} S, {Di Nuovo} A (2021) {A Brief Review of Robotics
  Technologies to Support Social Interventions for Older Users}. In: Zimmermann
  A, Howlett RJ, Jain LC (eds) {Human Centred Intelligent Systems}, {Smart
  Innovation, Systems and Technologies}, vol 189. {Springer Singapore},
  Singapore, p 221--232, \doi{10.1007/978-981-15-5784-2_18}

\item Decker M (2008) {Caregiving robots and ethical reflection: the perspective of
  interdisciplinary technology assessment}. {AI {\&} SOCIETY} 22(3):315--330.
  \doi{10.1007/s00146-007-0151-0}

\item Depner D, Hülsken-Giesler M (2017) {Robotik in der Pflege - Eckpunkte für
  eine prospektive ethische Bewertung in der Langzeitpflege}. {Personzentrierte
  Langzeitpflege} \doi{10.14623/zfme.2017.1.51-62}

\item {Deutscher Ethikrat} (10. März 2020) {Robotik für gute Pflege:
  Stellungnahme}. {Deutscher Ethikrat}, Berlin

\item Diaz-Orueta U, Hopper L, Konstantinidis E (2020) {Shaping technologies for
  older adults with and without dementia: Reflections on ethics and
  preferences}. {Health Informatics Journal} 26(4):3215--3230.
  \doi{10.1177/1460458219899590}

\item Döring N (2018) {Sollten Pflegeroboter auch sexuelle Assistenzfunktionen
  bieten?} In: Bendel O (ed) {Pflegeroboter}. {Springer Fachmedien Wiesbaden},
  Wiesbaden, p 249--267, \doi{10.1007/978-3-658-22698-5_14}

\item Dosso JA, Bandari E, Malhotra A, et~al (2022) {User perspectives on emotionally
  aligned social robots for older adults and persons living with dementia}.
  {Journal of rehabilitation and assistive technologies engineering} 9.
  \doi{10.1177/20556683221108364}

\item Draper H, Sorell T (2017) {Ethical Values and Social Care Robots for Older
  People: An International Qualitative Study}. {Ethics and Information
  Technology} 19(1):49--68. \doi{10.1007/s10676-016-9413-1}

\item Earp BD, Grunt-Mejer K (2021) {Robots and sexual ethics}. {Journal of Medical
  Ethics} 47(1):1--2. \doi{10.1136/medethics-2020-107153}

\item {van Est} R, Royakkers L (2016) {Robotisation as Rationalisation – In Search
  for a Human Robot Future}. {Amsterdam University Press},
  \doi{10.25969/MEDIAREP/13393}

\item Espingardeiro A (2014) {A Roboethics Framework for the Development and
  Introduction of Social Assistive Robots in Elderly Care}. PhD thesis,
  {University of Salford (United Kingdom)}

\item Feil-Seifer D, Matarić M (2011) {Socially Assistive Robotics}. {IEEE Robotics
  {\&} Automation Magazine} 18(1):24--31. \doi{10.1109/MRA.2010.940150}

\item Felber NA, Pageau F, McLean A, et~al (2022) {The Concept of Social Dignity as a
  Yardstick to Delimit Ethical Use of Robotic Assistance in the Care of Older
  Persons}. {Medicine, Health Care {\&} Philosophy} 25(1):99--110.
  \doi{10.1007/s11019-021-10054-z}

\item Fiorini L, Rovini E, Russo S, et~al (2022) {On the Use of Assistive Technology
  during the COVID-19 Outbreak: Results and Lessons Learned from Pilot
  Studies}. {Sensors} 22(17):6631. \doi{10.3390/s22176631}

\item Fosch-Villaronga E, Albo-Canals J (2019) {“I’ll take care of you,” said
  the robot}. {Paladyn, Journal of Behavioral Robotics} 10(1):77--93.
  \doi{10.1515/pjbr-2019-0006}

\item Fosch-Villaronga E, Poulsen A (2020) {Sex care robots}. {Paladyn, Journal of
  Behavioral Robotics} 11(1):1--18. \doi{10.1515/pjbr-2020-0001}

\item Frebel L (2015) {Roboter gegen das Vergessen?: Technische Assistenz bei
  Altersdemenz Im Speilfilm aus medizinethischer Sicht}. In: Weber K, Frommeld
  D, Manzeschke A, et~al (eds) {Technisierung Des Alltags: Beitrag Für Ein
  Gutes Leben?} {Franz Steiner}, Stuttgart, p 99--116

\item Frennert S, Östlund B (2014) {Review: Seven Matters of Concern of Social
  Robots and Older People}. {International Journal of Social Robotics}
  6(2):299--310. \doi{10.1007/s12369-013-0225-8}

\item Früh M, Gasser A (2018) {Erfahrungen aus dem Einsatz von Pflegerobotern für
  Menschen im Alter}. In: Bendel O (ed) {Pflegeroboter}. {Springer Fachmedien
  Wiesbaden}, Wiesbaden, p 37--62, \doi{10.1007/978-3-658-22698-5_3}

\item Gallagher A, Nåden D, Karterud D (2016) {Robots in elder care}. {Nursing
  Ethics} 23(4):369--371. \doi{10.1177/0969733016647297}

\item Geier J, Mauch M, Patsch M, et~al (2020) {Wie Pflegekräfte im ambulanten
  Bereich den Einsatz von Telepräsenzsystemen einschätzen - Eine qualitative
  Studie}. {Pflege} 33(1):43--51. \doi{10.1024/1012-5302/a000709}

\item Gelin R (2017) {The Domestic Robot: Ethical and Technical Concerns}. In:
  {Aldinhas Ferreira} MI, {Silva Sequeira} J, Tokhi MO, et~al (eds) {A World
  with Robots: International Conference on Robot Ethics: ICRE 2015}. {Springer
  International Publishing}, Cham, p 207--216

\item Giansanti D (2021) {The Social Robot in Rehabilitation and Assistance: What Is
  the Future?} {Healthcare} 9(3):244. \doi{10.3390/healthcare9030244}

\item Gisinger C (2018) {Pflegeroboter aus Sicht der Geriatrie}. In: Bendel O (ed)
  {Pflegeroboter}. {Springer Fachmedien Wiesbaden}, Wiesbaden, p 113--124,
  \doi{10.1007/978-3-658-22698-5_6}

\item Glende S, Conrad I, Krezdorn L, et~al (2016) {Increasing the Acceptance of
  Assistive Robots for Older People Through Marketing Strategies Based on
  Stakeholder Needs}. {International Journal of Social Robotics} 8(3):355--369.
  \doi{10.1007/s12369-015-0328-5}

\item Gochoo M, Alnajjar F, Tan TH, et~al (2021) {Towards Privacy-Preserved Aging in
  Place: A Systematic Review}. {Sensors} 21(9):3082. \doi{10.3390/s21093082}

\item Gräb-Schmidt E, Stritzelberger CP (2018) {Ethische Herausforderungen durch
  autonome Systeme und Robotik im Bereich der Pflege}. {Zeitschrift für
  medizinische Ethik} \doi{10.14623/zfme.2018.4.357-372}

\item Grunwald A, Kehl C (2020) {Mit Robotern gegen den Pflegenotstand}. {Spektrum
  der Wissenschaft 20 Spezial Biologie Medizin Hirnforschung} pp 16--17

\item Haltaufderheide J, Lucht A, Strünck C, et~al (2023) {Socially Assistive
  Devices in Healthcare–a Systematic Review of Empirical Evidence from an
  Ethical Perspective}. {Science and Engineering Ethics} 29(1).
  \doi{10.1007/s11948-022-00419-9}

\item Hasenauer R, Ehrenmueller I, Belviso C (2022) {Living Labs in Social Service
  Institutions: An Effective Method to Improve the Ethical, Reliable Use of
  Digital Assistive Robots to Support Social Services}. In: {2022 Portland
  International Conference on Management of Engineering and Technology
  (PICMET)}. IEEE, pp 1--9, \doi{10.23919/PICMET53225.2022.9882746}

\item Heeser AC (2020) {Schöne digitale Welt}. {Pflegezeitschrift} 73(9):10--12.
  \doi{10.1007/s41906-020-0776-x}

\item Hildt E (2019) {Shaping the Development and Use of Intelligent Assistive
  Technologies in Dementia}. In: Jotterand F, Ienca M, Wangmo T, et~al (eds)
  {Intelligent Assistive Technologies for Dementia}. {Oxford University
  PressNew York}, p 130--144, \doi{10.1093/med/9780190459802.003.0008}

\item Hilgendorf E (2018) {Recht und Ethik in der Pflegerobotik: Ein Überblick}.
  {Zeitschrift für medizinische Ethik} \doi{10.14623/zfme.2018.4.373-385}

\item Honekamp I, Sauer L, Wache T, et~al (2019) {Akzeptanz von Pflegerobotern Im
  Krankenhaus: Eine Quantitative Studie}. {TATuP - Zeitschrift für
  Technikfolgenabschätzung in Theorie und Praxis} 28(2):58--63.
  \doi{10.14512/tatup.28.2.s58}

\item Hoppe JA, Johansson-Pajala RM, Gustafsson C, et~al (2020) {Assistive robots in
  care: Expectations and perceptions of older people.} In: Haltaufderheide J,
  Hovemann J, Vollmann J (eds) {Aging between Participation and Simulation:
  Ethical Dimensions of Socially Assistive Technologies in Elderly Care}. {De
  Gruyter}, p 139--156

\item Hübner G, Müller S (2020) {Roboter in der Pflege}. {Spektrum der Wissenschaft
  20 Spezial Biologie Medizin Hirnforschung} pp 6--8

\item Hülsken-Giesler M, Daxberger S (2018) {Robotik in der Pflege aus
  pflegewissenschaftlicher Perspektive}. In: Bendel O (ed) {Pflegeroboter}.
  {Springer Fachmedien Wiesbaden}, Wiesbaden, p 125--139,
  \doi{10.1007/978-3-658-22698-5_7}

\item Hung L, Gregorio M, Mann J, et~al (2021) {Exploring the perceptions of people
  with dementia about the social robot PARO in a hospital setting}. {Dementia}
  20(2):485--504. \doi{10.1177/1471301219894141}

\item Hung L, Liu C, Woldum E, et~al (2019) {The Benefits of and Barriers to Using a
  Social Robot PARO in Care Settings: A Scoping Review}. {BMC geriatrics}
  19(1):1--10. \doi{10.1186/s12877-019-1244-6}

\item Hung L, Mann J, Perry J, et~al (2022) {Technological Risks and Ethical
  Implications of Using Robots in Long-Term Care}. {Journal of rehabilitation
  and assistive technologies engineering} 9:1--10.
  \doi{10.1177/20556683221106917}

\item Huschilt J, Clune L (2012) {The Use of Socially Assistive Robots for Dementia
  Care}. {Journal of Gerontological Nursing} 38(10):15--19.
  \doi{10.3928/00989134-20120911-02}

\item Ienca M, Fabrice J, Elger B, et~al (2017) {Intelligent Assistive Technology for
  Alzheimer’s Disease and Other Dementias: A Systematic Review}. {Journal of
  Alzheimer’s Disease} 56(4):1301--1340. \doi{10.3233/JAD-161037}

\item Ienca M, Jotterand F, {Vic\u a} C, et~al (2016) {Social and Assistive Robotics
  in Dementia Care: Ethical Recommendations for Research and Practice}.
  {International Journal of Social Robotics} 8(4):565--573.
  \doi{10.1007/s12369-016-0366-7}

\item Ienca M, Villaronga EF (2019) {Privacy and Security Issues in Assistive
  Technologies for Dementia}. In: Jotterand F, Ienca M, Wangmo T, et~al (eds)
  {Intelligent Assistive Technologies for Dementia}. {Oxford University
  PressNew York}, p 221--239, \doi{10.1093/med/9780190459802.003.0013}

\item Isabet B, Pino M, Lewis M, et~al (2021) {Social Telepresence Robots: A
  Narrative Review of Experiments Involving Older Adults before and during the
  COVID-19 Pandemic}. {International Journal of Environmental Research and
  Public Health} 18(7):3597. \doi{10.3390/ijerph18073597}

\item Jecker NS (2021{\natexlab{a}}) {Nothing to be ashamed of: sex robots for older
  adults with disabilities}. {Journal of Medical Ethics} 47(1):26--32.
  \doi{10.1136/medethics-2020-106645}

\item Jecker NS (2021{\natexlab{b}}) {Sex robots for older adults with disabilities:
  reply to critics}. {Journal of Medical Ethics} 47(2):113--114.
  \doi{10.1136/medethics-2020-107148}

\item Jecker N (2021{\natexlab{c}}) {You’ve Got a Friend in Me: Sociable Robots for
  Older Adults in an Age of Global Pandemics}. {Ethics and Information
  Technology} 23(S1):35--43. \doi{10.1007/s10676-020-09546-y}

\item Jenkins S, Draper H (2015) {Care, Monitoring, and Companionship: Views on Care
  Robots from Older People and Their Carers}. {International Journal of Social
  Robotics} 7(5):673--683. \doi{10.1007/s12369-015-0322-y}

\item Johansson-Pajala RM, Gustafsson C (2022) {Significant challenges when
  introducing care robots in Swedish elder care}. {Disability and
  Rehabilitation: Assistive Technology} 17(2):166--176.
  \doi{10.1080/17483107.2020.1773549}

\item Johansson-Pajala RM, Thommes K, Hoppe JA, et~al (2020) {Care Robot Orientation:
  What, Who and How? Potential Users’ Perceptions}. {International Journal of
  Social Robotics} 12(5):1103--1117. \doi{10.1007/s12369-020-00619-y}

\item Johnston C (2022) {Ethical Design and Use of Robotic Care of the Elderly}.
  {Journal of Bioethical Inquiry} 19(1):11--14.
  \doi{10.1007/s11673-022-10181-z}

\item Kamphof I (2015) {In the Company of Robots: Health Care and the Identity of
  People with Dementia}. In: Swinnen A, Schweda M (eds) {Popularizing Dementia:
  Public Expressions and Representations of Forgetfulness}. Transcript,
  Bielefeld, p 359--376, \doi{10.1515/9783839427101-017}

\item Kayser D, Gasser A, Früh M (2022) {F{\&}P Robotics, Assistenzroboter Lio. Ein
  Erfahrungsbericht}. In: Stronegger W, Platzer J (eds) {Technisierung der
  Pflege: 4. Goldegger Dialogforum Mensch und Endlichkeit}. {Bioethik in
  Wissenschaft und Gesellschaft}, Nomos, Baden-Baden, p 67--78

\item Kehl C (2018{\natexlab{a}}) {Robotik und assistive Neurotechnologien in der
  Pflege - gesellschaftliche Herausforderungen. Vertiefung des Projekts
  Mensch-Maschine-Entgrenzungen: TAB Arbeitsbericht}.
  \doi{10.5445/IR/1000094095}

\item Kehl C (2018{\natexlab{b}}) {Wege zu verantwortungsvoller Forschung und
  Entwicklung im Bereich der Pflegerobotik: Die ambivalente Rolle der Ethik}.
  In: Bendel O (ed) {Pflegeroboter}. {Springer Fachmedien Wiesbaden},
  Wiesbaden, p 141--160, \doi{10.1007/978-3-658-22698-5_1}

\item Kehl C (2018{\natexlab{c}}) {Entgrenzungen zwischen Mensch und Maschine, Oder:
  Können Roboter zu guter Pflege beitragen?} {Aus Politik und Zeitgeschichte}
  68(6-8):22--28

\item Keibel A (2020) {Warum tut sich die Pflegerobotik so schwer?} {Spektrum der
  Wissenschaft 20 Spezial Biologie Medizin Hirnforschung} pp 12--15

\item {van Kemenade} MAM, Hoorn JF, Konijn EA (2018) {Healthcare Students’ Ethical
  Considerations of Care Robots in The Netherlands}. {Applied Sciences}
  8(10):1712. \doi{10.3390/app8101712}

\item {van Kemenade} MAM, Hoorn JF, Konijn EA (2019) {Do You Care for Robots That
  Care? Exploring the Opinions of Vocational Care Students on the Use of
  Healthcare Robots}. {Robotics} 8(1):22. \doi{10.3390/robotics8010022}

\item Kenigsberg PA, Aquino JP, Bérard A, et~al (2019) {Assistive Technologies to
  Address Capabilities of People with Dementia: From Research to Practice}.
  {Dementia} 18(4):1568--1595. \doi{10.1177/1471301217714093}

\item Khaksar W, Saplacan D, Bygrave LA, et~al (2023) {Robotics in Elderly
  Healthcare: A Review of 20 Recent Research Projects}.
  \doi{10.48550/arXiv.2302.04478}

\item Kim JW, Choi YL, Jeong SH, et~al (2022) {A Care Robot with Ethical Sensing
  System for Older Adults at Home}. {Sensors} 22(19):7515.
  \doi{10.3390/s22197515}

\item Klebbe R, Klüber K, Dahms R, et~al (2023) {Caregivers’ Perspectives on
  Human–Robot Collaboration in Inpatient Elderly Care Settings}. {Machines}
  11(1):34. \doi{10.3390/machines11010034}

\item Kodate N, Maeda Y, Hauray B, et~al (2022{\natexlab{a}}) {Hopes and fears
  regarding care robots: Content analysis of newspapers in East Asia and
  Western Europe, 2001–2020}. {Frontiers in Rehabilitation Sciences} 3.
  \doi{10.3389/fresc.2022.1019089}

\item Kodate N, Donnelly S, Suwa S, et~al (2022{\natexlab{b}}) {Home‐care robots
  – Attitudes and perceptions among older people, carers and care
  professionals in Ireland: A questionnaire study}. {Health {\&} Social Care in
  the Community} 30(3):1086--1096. \doi{10.1111/hsc.13327}

\item Koeszegi S, {Weiss Astrid} (2021) {Mein neuer Teamkollege ist ein Roboter! Wie
  soziale Roboter die Zukunft der Arbeit verändern können}. In: Altenburger
  R, Schmidpeter R (eds) {CSR und Künstliche Intelligenz}. {Springer Gabler},
  Berlin and Heidelberg, p 279--303

\item Koh WQ, Ang FXH, Casey D (2021) {Impacts of Low-cost Robotic Pets for Older
  Adults and People With Dementia: Scoping Review}. {JMIR Rehabilitation and
  Assistive Technologies} 8(1):e25340. \doi{10.2196/25340}

\item Kohlen H (2021) {Sorge-Ethik als menschliche Praxis im Unterschied zu Technik
  und Robotik in der Pflege}. {Imago Hominis} 28:129--135

\item Koimizu J (2019) {Aged Care with Socially Assistive Robotics under Advance Care
  Planning}. In: {2019 IEEE International Conference on Advanced Robotics and
  its Social Impacts (ARSO)}. IEEE, pp 34--38,
  \doi{10.1109/ARSO46408.2019.8948742}

\item Körtner T (2016) {Ethical challenges in the use of social service robots for
  elderly people}. {Zeitschrift für Gerontologie und Geriatrie}
  49(4):303--307. \doi{10.1007/s00391-016-1066-5}

\item Koumpis A, Gees T (2020) {Sex with Robots: A Not-so-Niche Market for Disabled
  and Older Persons}. {Paladyn: Journal of Behavioral Robotics} 11(1):228--232.
  \doi{10.1515/pjbr-2020-0009}

\item Kovács L (2021) {Anthropologische Und Ethische Aspekte Des Einsatzes von
  Robotern Im Gesundheitssektor}. In: Inthorn J, Seising R (eds) {Digitale
  Patientenversorgung: Zur Computerisierung von Diagnostik, Therapie Und
  Pflege}, {Medical Humanities}, vol~3. {transcript Verlag}, Bielefeld, p
  19--34

\item Kreis J (2018) {Umsorgen, überwachen, unterhalten ‐ sind Pflegeroboter
  ethisch vertretbar?} In: Bendel O (ed) {Pflegeroboter}. {Springer Fachmedien
  Wiesbaden}, Wiesbaden, p 213--228, \doi{10.1007/978-3-658-22698-5_12}

\item Lancaster K (2023) {Granny and the Sexbots: An ethical appraisal of the use of
  sexbots in residential care institutions for elderly people}. In: Loh J, Loh
  W (eds) {Social robotics and the good life}. {Philosophy}, {transcript
  Verlag}, Bielefeld, p 181--207

\item Lange N, Bauer L (2021) {Eine Robbe für Oma - Die zukünftige Dauerausstellung
  Robotik im Deutschen Museum}. In: Inthorn J, Seising R (eds) {Digitale
  Patientenversorgung: Zur Computerisierung von Diagnostik, Therapie Und
  Pflege}. {Medical Humanities}, {transcript Verlag}, Bielefeld, p 221--240

\item Lau YY, {van't Hof} C, {van Est} R (2009{\natexlab{a}}) {R{\&}D in healthcare
  robotics}. In: Lau YY, {van't Hof} C, {van Est} R (eds) {Beyond the surface}.
  {Technology assessment}, {Rathenau Institute}, The Hague, p 17--34

\item Lau YY, {van't Hof} C, {van Est} R (2009{\natexlab{b}}) {Robotics for an aging
  Japanese society}. In: Lau YY, {van't Hof} C, {van Est} R (eds) {Beyond the
  surface}. {Technology assessment}, {Rathenau Institute}, The Hague, p 9--16

\item Lee H, Chung MA, Kim H, et~al (2022) {The Effect of Cognitive Function Health
  Care Using Artificial Intelligence Robots for Older Adults: Systematic Review
  and Meta-analysis}. {JMIR Aging} 5(2):e38896. \doi{10.2196/38896}

\item Lehmann S, Ruf E, Misoch S (2020) {Robot Use for Older Adults – Attitudes,
  Wishes and Concerns. First Results from Switzerland}. In: Stephanidis C,
  Antona M (eds) {HCI International 2020 - Posters}, {Communications in
  Computer and Information Science}, vol 1226. {Springer International
  Publishing}, Cham, p 64--70, \doi{10.1007/978-3-030-50732-9_9}

\item Lehmann S, Ruf E, Misoch S (2021) {Using a Socially Assistive Robot in a
  Nursing Home: Caregivers’ Expectations and Concerns}. In: Stephanidis C,
  Antona M, Ntoa S (eds) {HCI International 2021 - Posters}, {Communications in
  Computer and Information Science}, vol 1420. {Springer International
  Publishing}, Cham, p 148--155, \doi{10.1007/978-3-030-78642-7_20}

\item Lehoux P, Grimard D (2018) {When robots care: Public deliberations on how
  technology and humans may support independent living for older adults}.
  {Social Science {\&} Medicine} 211:330--337.
  \doi{10.1016/j.socscimed.2018.06.038}

\item Li S, {van Wynsberghe} A, Roeser S (2020) {The Complexity of Autonomy: A
  Consideration of the Impacts of Care Robots on the Autonomy of Elderly Care
  Receivers}. In: Nørskov M, Seibt J, Quick OS (eds) {Culturally Sustainable
  Social Robotics: Proceedings of Robophilosophy 2020}. {IOS Press}, Amsterdam,
  Berlin, Washington, DC, p 316--325, \doi{10.3233/FAIA200928}

\item Liang A, Piroth I, Robinson H, et~al (2017) {A Pilot Randomized Trial of a
  Companion Robot for People With Dementia Living in the Community}. {Journal
  of the American Medical Directors Association} 18(10):871--878.
  \doi{10.1016/j.jamda.2017.05.019}

\item Mansouri N, Goher K, Hosseini SE (2017) {Ethical framework of assistive
  devices: review and reflection}. {Robotics and Biomimetics} 4(1).
  \doi{10.1186/s40638-017-0074-2}

\item Manzeschke A (2019) {Roboter in der Pflege: Von Menschen, Maschinen und anderen
  hilfreichen Wesen}. {EthikJournal} 2019(1):1--11

\item Manzeschke A (2022) {Robotik in der Pflege? Ethische Merkposten für ihren
  Einsatz}. {Public Health Forum} 30(1):41--43. \doi{10.1515/pubhef-2021-0123}

\item Manzeschke A, Petersen J (2020) {Digitalisierung und Roboterisierung in der
  Pflege: Ethisch-anthropologische Überlegungen}. In: Dibelius O,
  Piechotta-Henze G (eds) {Menschenrechtsbasierte Pflege}. Hogrefe, Bern, p
  161--173

\item Marchang J, {Di Nuovo} A (2022) {Assistive Multimodal Robotic System (AMRSys):
  Security and Privacy Issues, Challenges, and Possible Solutions}. {Applied
  Sciences} 12(4):1--29. \doi{10.3390/app12042174}

\item {van Maris} A, Zook N, Caleb-Solly P, et~al (2020) {Designing Ethical Social
  Robots‐-A Longitudinal Field Study With Older Adults}. {Frontiers in
  Robotics and AI} 7:1--14. \doi{10.3389/frobt.2020.00001}

\item {van Maris} A, Zook N, Dogramadzi S, et~al (2021) {A New Perspective on Robot
  Ethics through Investigating Human–Robot Interactions with Older Adults}.
  {Applied Sciences} 11(21):10136. \doi{10.3390/app112110136}

\item Martens R, Hildebrand C (2021) {Dementia care, robot pets, and aliefs}.
  {Bioethics} 35(9):870--876. \doi{10.1111/bioe.12952}

\item Matarić M, Scassellati B (2016) {Socially Assistive Robotics}. In: Siciliano
  B, Khatib O (eds) {Springer Handbook of Robotics}. {Springer Handbooks},
  Springer, \doi{10.1007/978-3-319-32552-1}

\item Matthias A (2015) {Robot Lies in Health Care: When Is Deception Morally
  Permissible?} {Kennedy Institute of Ethics Journal} 25(2):169--162.
  \doi{10.1353/ken.2015.0007}

\item von Maur I (2023) {Alice Does Not Care: Or: Why It Matters That Robots
  “Don’t Give a Damn”}. In: Loh J, Loh W (eds) {Social Robotics and the
  Good Life: The Normative Side of Forming Emotional Bonds With Robots}.
  {transcript Verlag}, Bielefeld, p 209--231

\item Meacham D, Studley M (2017) {Could a Robot Care? It’s All in the Movement}.
  In: Abney K, Lin P, Jenkins R (eds) {Robot ethics 2.0}. {Oxford University
  Press}, New York, NY, p 97--113

\item Meiland F, Innes A, Mountain G, et~al (2017) {Technologies to Support
  Community-Dwelling Persons With Dementia: A Position Paper on Issues
  Regarding Development, Usability, Effectiveness and Cost-Effectiveness,
  Deployment, and Ethics}. {JMIR Rehabilitation and Assistive Technologies}
  4(1):e1. \doi{10.2196/rehab.6376}

\item Metzler TA, Barnes SJ (2014) {Three dialogues concerning robots in elder care}.
  {Nursing Philosophy} 15(1):4--13. \doi{10.1111/nup.12027}

\item Metzler TA, Lewis LM, Pope LC (2016) {Could robots become authentic companions
  in nursing care?} {Nursing Philosophy} 17(1):36--48. \doi{10.1111/nup.12101}

\item Meyer S (2011{\natexlab{a}}) {Akzeptanz ausgewählter Anwendungsszenarien}. In:
  {Mein Freund der Roboter: Servicerobotik für ältere Menschen ‐ eine
  Antwort auf den demografischen Wandel?} VDE-Verlag, Berlin, Offenbach, p
  59--103

\item Meyer S (2011{\natexlab{b}}) {Akzeptanzbedingungen für Roboter-Assistenten}.
  In: {Mein Freund der Roboter: Servicerobotik für ältere Menschen ‐ eine
  Antwort auf den demografischen Wandel?} VDE-Verlag, Berlin, Offenbach, p
  104--113

\item Mishra N, Tulsulkar G, Li H, et~al (2021) {Does Elderly Enjoy Playing Bingo
  with a Robot? A Case Study with the Humanoid Robot Nadine}. In:
  Magnenat-Thalmann N, Interrante V, Thalmann D, et~al (eds) {Advances in
  Computer Graphics}, {Lecture Notes in Computer Science}, vol 13002. {Springer
  International Publishing}, Cham, p 491--503,
  \doi{10.1007/978-3-030-89029-2_38}

\item Misselhorn C (2018) {Pflegesysteme}. In: {Grundfragen Der Maschinenethik}.
  {Reclams Universal Bibliothek}, Reclam, Ditzingen, p 136--154

\item Misselhorn C (2019) {Moralische Maschinen in der Pflege? Grundlagen und eine
  Roadmap für ein moralisch lernfähiges Altenpflegesystem}. In: Woopen C,
  Jannes M (eds) {Roboter in der Gesellschaft}. {Springer Berlin Heidelberg},
  Berlin, Heidelberg, p 53--68, \doi{10.1007/978-3-662-57765-3_4}

\item Misselhorn C (2021) {Von empathischen Pflegerobotern und virtuellen
  Seelenklempnern}. In: Misselhorn C (ed) {Künstliche Intelligenz und
  Empathie}. {/Was bedeutet das alles?]}, Reclam, Ditzingen, p 61--74

\item Misselhorn C, Pompe U, Stapleton M (2013) {Ethical Considerations Regarding the
  Use of Social Robots in the Fourth Age}. {GeroPsych} 26(2):121--133.
  \doi{10.1024/1662-9647/a000088}

\item Moyle W, Jones C, Murfield J, et~al (2019) {Using a therapeutic companion robot
  for dementia symptoms in long-term care: reflections from a cluster-RCT}.
  {Aging {\&} Mental Health} 23(3):329--336.
  \doi{10.1080/13607863.2017.1421617}

\item Nass E, Schneider M (2022) {Maschinen mit Moral für eine gute Pflege der
  Zukunft?} In: Pfannstiel M (ed) {Künstliche Intelligenz im Gesundheitswesen:
  Entwicklungen, Beispiele und Perspektiven}. {Springer Gabler}, Wiesbaden, p
  311--323

\item Navon M (2021) {The Virtuous Servant Owner—A Paradigm Whose Time has Come
  (Again)}. {Frontiers in Robotics and AI} 8. \doi{10.3389/frobt.2021.715849}

\item Nestorov N, Stone E, Lehane P, et~al (2014) {Aspects of Socially Assistive
  Robots Design for Dementia Care}. In: {2014 IEEE 27th International Symposium
  on Computer-Based Medical Systems}. IEEE, pp 396--400,
  \doi{10.1109/CBMS.2014.16}

\item Nielsen S, Langensiepen S, Madi M, et~al (2022) {Implementing ethical aspects
  in the development of a robotic system for nursing care: a qualitative
  approach}. {BMC Nursing} 21(1). \doi{10.1186/s12912-022-00959-2}

\item Niemelä M, {van Aerschot} L, Tammela A, et~al (2021) {Towards Ethical
  Guidelines of Using Telepresence Robots in Residential Care}. {International
  Journal of Social Robotics} 13:431--439. \doi{10.1007/s12369-019-00529-8}

\item Noori FM, Uddin Z, Torresen J (2019) {Robot-Care for the Older People:
  Ethically Justified or Not?} In: {2019 Joint IEEE 9th International
  Conference on Development and Learning and Epigenetic Robotics
  (ICDL-EpiRob)}. IEEE, pp 43--47, \doi{10.1109/DEVLRN.2019.8850706}

\item Nordgren A (2018) {How to respond to resistiveness towards assistive
  technologies among persons with dementia}. {Medicine, Health Care and
  Philosophy} 21(3):411--421. \doi{10.1007/s11019-017-9816-8}

\item Nyholm L, Santamäki-Fischer R, Fagerström L (2021) {Users’ ambivalent sense
  of security with humanoid robots in healthcare}. {Informatics for Health and
  Social Care} 46(2):218--226. \doi{10.1080/17538157.2021.1883027}

\item O’Brolcháin F (2019) {Robots and people with dementia: Unintended
  consequences and moral hazard}. {Nursing Ethics} 26(4):962--972.
  \doi{10.1177/0969733017742960}

\item Paganini C (2022) {Mit Kranken, Hochbetagten und Sterbenden kommunizieren.
  Sollen Pflegeroboter immer die Wahrheit sagen?} In: Stronegger W, Platzer J
  (eds) {Technisierung der Pflege: 4. Goldegger Dialogforum Mensch und
  Endlichkeit}. {Bioethik in Wissenschaft und Gesellschaft}, Nomos,
  Baden-Baden, p 91--106

\item Paletta L, Schüssler S, Zuschnegg J, et~al (2019) {AMIGO‐-A Socially
  Assistive Robot for Coaching Multimodal Training of Persons with Dementia}.
  In: Korn O (ed) {Social Robots: Technological, Societal and Ethical Aspects
  of Human-Robot Interaction}. {Human-Computer Interaction Series}, Springer,
  Cham, p 265--284, \doi{10.1007/978-3-030-17107-0}

\item Parviainen J, Turja T, {van Aerschot} L (2019) {Social Robots and Human Touch
  in Care: The Perceived Usefulness of Robot Assistance Among Healthcare
  Professionals}. In: Korn O (ed) {Social Robots: Technological, Societal and
  Ethical Aspects of Human-Robot Interaction}. {Human-Computer Interaction
  Series}, Springer, Cham, p 187--204, \doi{10.1007/978-3-030-17107-0}

\item {van Patten} R, Keller AV, Maye JE, et~al (2020) {Home-Based Cognitively
  Assistive Robots: Maximizing Cognitive Functioning and Maintaining
  Independence in Older Adults Without Dementia}. {Clinical interventions in
  aging} 15:1129--1139. \doi{10.2147/CIA.S253236}

\item Pilotto A, Boi R, Petermans J (2018) {Technology in geriatrics}. {Age and
  Ageing} 47(6):771--774. \doi{10.1093/ageing/afy026}

\item Pino M, Boulay M, Jouen F, et~al (2015) {"Are we ready for robots that care for
  us?" Attitudes and opinions of older adults toward socially assistive
  robots}. {Frontiers in aging neuroscience} 7:141.
  \doi{10.3389/fnagi.2015.00141}

\item Pirhonen J, Melkas H, Laitinen A, et~al (2020{\natexlab{a}}) {Could robots
  strengthen the sense of autonomy of older people residing in assisted living
  facilities?—A future-oriented study}. {Ethics and Information Technology}
  22(2):151--162. \doi{10.1007/s10676-019-09524-z}

\item Pirhonen J, Tiilikainen E, Pekkarinen S, et~al (2020{\natexlab{b}}) {Can Robots
  Tackle Late-Life Loneliness? Scanning of Future Opportunities and Challenges
  in Assisted Living Facilities}. {Futures} 124:1--12.
  \doi{10.1016/j.futures.2020.102640}

\item Poulsen A, Burmeister OK (2019) {Overcoming carer shortages with care robots:
  Dynamic value trade-offs in run-time}. {Australasian Journal of Information
  Systems} 23. \doi{10.3127/ajis.v23i0.1688}

\item Poulsen A, Fosch-Villaronga E, Burmeister OK (2020) {Cybersecurity, value
  sensing robots for LGBTIQ+ elderly, and the need for revised codes of
  conduct}. {Australasian Journal of Information Systems} 24.
  \doi{10.3127/ajis.v24i0.2789}

\item Preuß D, Legal F (2017) {Living with the animals: animal or robotic companions
  for the elderly in smart homes?} {Journal of Medical Ethics} 43(6):407--410.
  \doi{10.1136/medethics-2016-103603}

\item Radic M, Vosen A (2020) {Ethische, rechtliche und soziale Anforderungen an
  Assistenzroboter in der Pflege}. {Zeitschrift für Gerontologie und
  Geriatrie} 53(7):630--636. \doi{10.1007/s00391-020-01791-6}

\item Rantanen T, Lehto P, Vuorinen P, et~al (2018) {The adoption of care robots in
  home care—A survey on the attitudes of Finnish home care personnel}.
  {Journal of Clinical Nursing} 27(9-10):1846--1859. \doi{10.1111/jocn.14355}

\item Reiß T (2019) {Editorial}. {EthikJournal} 5(1):1--5

\item Remmers H (2018) {Pflegeroboter: Analyse und Bewertung aus Sicht pflegerischen
  Handelns und ethischer Anforderungen}. In: Bendel O (ed) {Pflegeroboter}.
  {Springer Fachmedien Wiesbaden}, Wiesbaden, p 161--179,
  \doi{10.1007/978-3-658-22698-5_9}

\item Remmers H (2019) {Pflege und Technik. Stand der Diskussion und zentrale
  ethische Fragen}. {Ethik in der Medizin} 31(4):407--430.
  \doi{10.1007/s00481-019-00545-2}

\item Robillard JM, Kabacińska K (2020) {Realizing the Potential of Robotics for
  Aged Care Through Co-Creation}. {Journal of Alzheimer's Disease}
  76(2):461--466. \doi{10.3233/JAD-200214}

\item Ruf E, Lehmann S, Misoch S (2021) {Ethical Concerns of the General Public
  regarding the Use of Robots for Older Adults}. In: {Proceedings of the 7th
  International Conference on Information and Communication Technologies for
  Ageing Well and e-Health}. {SCITEPRESS - Science and Technology
  Publications}, pp 221--227, \doi{10.5220/0010478202210227}

\item Ruf E, Lehmann S, Pauli C, et~al (2020) {Roboter Zur Unterstützung Im Alter}.
  {HMD Praxis der Wirtschaftsinformatik} 57(6):1251--1270.
  \doi{10.1365/s40702-020-00681-0}

\item Sahm S (2019) {Digitale Anthropologie: Ethische Probleme der Anwendung
  künstlicher Intelligenz und Robotik in der Pflege und Medizin}.
  {Medizinrecht} 37(12):927--933. \doi{10.1007/s00350-019-5395-4}

\item Salvini P (2015) {On Ethical, Legal and Social Issues of Care Robots}. In:
  Mohammed S, Moreno J, Kong K, et~al (eds) {Intelligent Assistive Robots:
  Recent Advances in Assistive Robotics for Everyday Activities}. {Springer
  Tracts in Advanced Robotics}, {Springer International Publishing}, Cham, p
  431--445, \doi{10.1007/978-3-319-12922-8}

\item Saplacan D, Khaksar W, Torresen J (2021) {On Ethical Challenges Raised by Care
  Robots: A Review of the Existing Regulatory-, Theoretical-, and Research
  Gaps}. In: {2021 IEEE International Conference on Advanced Robotics and Its
  Social Impacts (ARSO)}. IEEE, pp 219--226,
  \doi{10.1109/ARSO51874.2021.9542844}

\item Schicktanz S, Schweda M (2021) {Aging 4.0? Rethinking the ethical framing of
  technology-assisted eldercare}. {History and Philosophy of the Life Sciences}
  43(3). \doi{10.1007/s40656-021-00447-x}

\item Schmidhuber M (2022) {Werden Roboter Menschen in der Pflege ersetzen? Ethische
  Überlegungen}. In: Stronegger W, Platzer J (eds) {Technisierung der Pflege:
  4. Goldegger Dialogforum Mensch und Endlichkeit}. {Bioethik in Wissenschaft
  und Gesellschaft}, Nomos, Baden-Baden, p 167--174,
  \doi{10.5771/9783748928720}

\item Schmidhuber M, Stöger K (2021) {Ethisches und Rechtliches zur Zukunft der
  Robotik in der Pflege: Grundfragen für österreichische und deutsche
  Debatten}. In: {Die Zukunft von Medizin und Gesundheitswesen}. {Königshausen
  {\&} Neumann, 2021}, Würzburg

\item Schmietow, Bettina (2020) {Reconfigurations of autonomy in digital health and
  the ethics of (socially) assistive technologies}. In: Haltaufderheide J,
  Hovemann J, Vollmann J (eds) {Aging between Participation and Simulation:
  Ethical Dimensions of Socially Assistive Technologies in Elderly Care}. {De
  Gruyter}, p 171--184

\item Schwaninger I (2020) {Practice-Based Trust Research: Towards Situated
  Human-Robot Interaction in Older People’s Living Spaces}. In: Nørskov M,
  Seibt J, Quick OS (eds) {Culturally Sustainable Social Robotics}. {Frontiers
  in Artificial Intelligence and Applications}, {IOS Press},
  \doi{10.3233/FAIA200975}

\item Scorna U (2015) {Servicerobotik in der Altenpflege: Eine empirische
  Untersuchung des Einsatzes der Serviceroboter in der stationären Altenpflege
  am Beispiel von PARO und Care-O-bot}. In: {Technisierung des Alltags}. {Franz
  Steiner Verlag, 2015}, Stuttgart

\item Sedenberg E, Chuang J, Mulligan D (2016) {Designing Commercial Therapeutic
  Robots for Privacy Preserving Systems and Ethical Research Practices Within
  the Home}. {International Journal of Social Robotics} 8(4):575--587.
  \doi{10.1007/s12369-016-0362-y}

\item Seefeldt D, Hülsken-Giesler M (2020) {Pflegeethik und Robotik in der Pflege}.
  In: Monteverde S (ed) {Handbuch Pflegeethik}. {Pflegepraxis}, {Verlag W.
  Kohlhammer}, Stuttgart, p 271--284

\item Segers S (2022) {Robot Technology for the Elderly and the Value of Veracity:
  Disruptive Technology or Reinvigorating Entrenched Principles?} {Science and
  Engineering Ethics} 28(6). \doi{10.1007/s11948-022-00420-2}

\item Servaty R, Kersten A, Brukamp K, et~al (2020) {Implementation of robotic
  devices in nursing care. Barriers and facilitators: an integrative review}.
  {BMJ Open} 10(9):e038650. \doi{10.1136/bmjopen-2020-038650}

\item Sharkey A (2014) {Robots and human dignity: a consideration of the effects of
  robot care on the dignity of older people}. {Ethics and Information
  Technology} 16(1):63--75. \doi{10.1007/s10676-014-9338-5}

\item Sharkey A, Sharkey N (2012) {Granny and the robots: ethical issues in robot
  care for the elderly}. {Ethics and Information Technology} 14(1):27--40.
  \doi{10.1007/s10676-010-9234-6}

\item Sharkey A, Sharkey N (2011) {Children, the Elderly, and Interactive Robots}.
  {IEEE Robotics {\&} Automation Magazine} 18(1):32--38.
  \doi{10.1109/MRA.2010.940151}

\item Sharkey N (2008) {The Ethical Frontiers of Robotics}. {Science}
  322(5909):1800--1801. \doi{10.1126/science.1164582}

\item Sharkey N, Sharkey A (2014) {The Rights and Wrongs of Robot Care}. In: Lin P,
  Abney K, Bekey G (eds) {Robot Ethics: The Ethical and Social Implications of
  Robotics}. {MIT Press}, Cambridge, p 267--282

\item Sharkey N, Sharkey A (2012) {The Eldercare Factory}. {Gerontology}
  58(3):282--288. \doi{10.1159/000329483}

\item Sharts-Hopko NC (2014) {The Coming Revolution in Personal Care Robotics}.
  {Nursing Administration Quarterly} 38(1):5--12.
  \doi{10.1097/NAQ.0000000000000000}

\item Shelton BE, Uz C (2015) {Immersive Technology and the Elderly: A Mini-Review}.
  {Gerontology} 61(2):175--185. \doi{10.1159/000365754}

\item Shim J, Arkin RC (2016) {Other-Oriented Robot Deception: How Can a Robot’s
  Deceptive Feedback Help Humans in HRI?} In: Agah A, Cabibihan JJ, Howard AM,
  et~al (eds) {Social Robotics}, {Lecture Notes in Computer Science}, vol 9979.
  {Springer International Publishing}, Cham, p 222--232,
  \doi{10.1007/978-3-319-47437-3_22}

\item Sorell T, Draper H (2014) {Robot carers, ethics, and older people}. {Ethics and
  Information Technology} 16(3):183--195. \doi{10.1007/s10676-014-9344-7}

\item Sparrow R (2002) {The March of the Robot Dogs}. {Ethics and Information
  Technology} 4(4):305--318. \doi{10.1023/A:1021386708994}

\item Sparrow R (2016) {Robots in Aged Care: A Dystopian Future?} {AI {\&} SOCIETY}
  31(4):445--454. \doi{10.1007/s00146-015-0625-4}

\item Sparrow R (2021) {Sex robot fantasies}. {Journal of Medical Ethics}
  47(1):33--34. \doi{10.1136/medethics-2020-106932}

\item Sriram V, Jenkinson C, Peters M (2019) {Informal carers’ experience of
  assistive technology use in dementia care at home: a systematic review}. {BMC
  Geriatrics} 19(1). \doi{10.1186/s12877-019-1169-0}

\item Steinrötter B (2020) {Personal Robots in Der Pflege}. In: Ebers M, Heinze C,
  Krügel T, et~al (eds) {Künstliche Intelligenz Und Robotik: Rechtshandbuch}.
  {Verlag C.H.BECK}, München, p 789--827

\item Street J, Barrie H, Eliott J, et~al (2022) {Older Adults’ Perspectives of
  Smart Technologies to Support Aging at Home: Insights from Five World Café
  Forums}. {International Journal of Environmental Research and Public Health}
  19(13):7817. \doi{10.3390/ijerph19137817}

\item Strünck C, Reuter V, Gerling V, et~al (2022) {Socially assistive robots on the
  market}. {Zeitschrift für Gerontologie und Geriatrie} 55(5):376--380.
  \doi{10.1007/s00391-022-02087-7}

\item Suwa S, Tsujimura M, Ide H, et~al (2020) {Home-Care Professionals’ Ethical
  Perceptions of the Development and Use of Home-care Robots for Older Adults
  in Japan}. {International Journal of Human‐Computer Interaction}
  36(14):1295--1303. \doi{10.1080/10447318.2020.1736809}

\item The Swedish National Council on Medical Ethics (2017) { Robots
  and Surveillance in Health Care of the Elderly – Ethical Aspects}.
  {Jahrbuch für Wissenschaft und Ethik} 21(1):445--452.
  \doi{10.1515/jwiet-2017-0125}
\item Tan SY, Taeihagh A, Tripathi A (2021) {Tensions and Antagonistic Interactions
  of Risks and Ethics of Using Robotics and Autonomous Systems in Long-Term
  Care}. {Technological Forecasting and Social Change} 167:1--15.
  \doi{10.1016/j.techfore.2021.120686}

\item Tanioka T (2019) {Nursing and Rehabilitative Care of the Elderly Using Humanoid
  Robots}. {The Journal of Medical Investigation} 66(1.2):19--23.
  \doi{10.2152/jmi.66.19}

\item Teo Y (2021) {Recognition, collaboration and community: science fiction
  representations of robot carers in Robot {\&} Frank , Big Hero 6 and Humans}.
  {Medical Humanities} 47(1):95--102. \doi{10.1136/medhum-2019-011744}

\item Thalmann NM (2022) {Social Robots: Their History and What They Can Do for Us}.
  In: Werthner H, Prem E, Lee E, et~al (eds) {Perspectives on Digital
  Humanism}. {Springer International Publishing}, Cham, p 9--17,
  \doi{10.1007/978-3-030-86144-5_2}

\item Tørresen J (2021) {Undertaking Research with Humans within Artificial
  Intelligence and Robotics: Multimodal Elderly Care Systems}.
  {Technology|Architecture + Design} 5(2):141--145.
  \doi{10.1080/24751448.2021.1967052}

\item Tørresen J, Kurazume R, Prestes E (2020) {Special Issue on Elderly Care
  Robotics – Technology and Ethics}. {Journal of Intelligent {\&} Robotic
  Systems} 98(1):3--4. \doi{10.1007/s10846-020-01148-6}

\item Turja T, Taipale S, Niemelä M, et~al (2022) {Positive Turn in Elder-Care
  Workers’ Views Toward Telecare Robots}. {International Journal of Social
  Robotics} 14(4):931--944. \doi{10.1007/s12369-021-00841-2}

\item Tzafestas SG (2016) {Roboethics}, vol~79. {Springer International Publishing},
  Cham, \doi{10.1007/978-3-319-21714-7}

\item Umbrello S, Capasso M, Balistreri M, et~al (2021) {Value Sensitive Design to
  Achieve the UN SDGs with AI: A Case of Elderly Care Robots}. {Minds and
  Machines} 31(3):395--419. \doi{10.1007/s11023-021-09561-y}

\item Vandemeulebroucke T, de~Casterlé BD, Gastmans C (2018{\natexlab{a}}) {How do
  older adults experience and perceive socially assistive robots in aged care:
  a systematic review of qualitative evidence}. {Aging {\&} Mental Health}
  22(2):149--167. \doi{10.1080/13607863.2017.1286455}

\item Vandemeulebroucke T, {Dierckx de Casterlé} B, Gastmans C (2018{\natexlab{b}})
  {The Use of Care Robots in Aged Care: A Systematic Review of Argument-Based
  Ethics Literature}. {Archives of Gerontology and Geriatrics} 74:15--25.
  \doi{10.1016/j.archger.2017.08.014}

\item Vandemeulebroucke T, {Dierckx de Casterlé} B, Gastmans C (2020) {Ethics of
  socially assistive robots in aged-care settings: a socio-historical
  contextualisation}. {Journal of Medical Ethics} 46(2):128--136.
  \doi{10.1136/medethics-2019-105615}

\item Vandemeulebroucke T, {Dierckx de Casterlé} B, Gastmans C (2021) {Socially
  Assistive Robots in Aged Care: Ethical Orientations Beyond the Care-Romantic
  and Technology-Deterministic Gaze}. {Science and Engineering Ethics} 27(2).
  \doi{10.1007/s11948-021-00296-8}

\item Vandemeulebroucke T, {Dierckx de Casterlé} B, Welbergen L, et~al (2020) {The
  Ethics of Socially Assistive Robots in Aged Care. A Focus Group Study With
  Older Adults in Flanders, Belgium}. {The Journals of Gerontology: Series B}
  75(9):1996--2007. \doi{10.1093/geronb/gbz070}

\item Vandemeulebroucke T, Dzi K, Gastmans C (2021) {Older adults’ experiences with
  and perceptions of the use of socially assistive robots in aged care: A
  systematic review of quantitative evidence}. {Archives of Gerontology and
  Geriatrics} 95:104399. \doi{10.1016/j.archger.2021.104399}

\item Vercelli A, Rainero I, Ciferri L, et~al (2018) {Robots in Elderly Care}.
  {DigitCult@Scientific Journal on Digital Cultures} 2:37--50

\item {Vollmer Dahlke} D, Ory MG (2020) {Emerging Issues of Intelligent Assistive
  Technology Use Among People With Dementia and Their Caregivers: A U.S.
  Perspective}. {Frontiers in Public Health} 8. \doi{10.3389/fpubh.2020.00191}

\item Wachsmuth I (2018) {Robots Like Me: Challenges and Ethical Issues in Aged
  Care}. {Frontiers in Psychology} 9(432):1--3. \doi{10.3389/fpsyg.2018.00432}

\item Wada K, Shibata T (2007) {Living With Seal Robots‐-Its Sociopsychological and
  Physiological Influences on the Elderly at a Care House}. {IEEE Transactions
  on Robotics} 23(5):972--980. \doi{10.1109/TRO.2007.906261}

\item Wagner E, Borycki EM (2022) {The Use of Robotics in Dementia Care: An Ethical
  Perspective}. In: Mantas J, Hasman A, Househ MS, et~al (eds) {Informatics and
  Technology in Clinical Care and Public Health}. {Studies in Health Technology
  and Informatics}, {IOS Press}, \doi{10.3233/SHTI210934}

\item Wahl HW, Mombaur K, Schubert A (2021) {Robotik und Altenpflege: Freund oder
  Feind?} {Pflegezeitschrift} 74(11):62--66. \doi{10.1007/s41906-021-1156-x}

\item Walters ML, Koay KL, Syrdal DS, et~al (2013) {Companion robots for elderly
  people: Using theatre to investigate potential users' views}. In: {2013 IEEE
  RO-MAN}. IEEE, pp 691--696, \doi{10.1109/ROMAN.2013.6628393}

\item Wang J, Liu T, Liu Z, et~al (2019) {Affective Interaction Technology of
  Companion Robots for the Elderly: A Review}. In: {El Rhalibi} A, Pan Z, Jin
  H, et~al (eds) {E-Learning and Games}, {Lecture Notes in Computer Science},
  vol 11462. {Springer International Publishing}, Cham, p 79--83,
  \doi{10.1007/978-3-030-23712-7_11}

\item Wangmo T, Lipps M, Kressig RW, et~al (2019) {Ethical concerns with the use of
  intelligent assistive technology: findings from a qualitative study with
  professional stakeholders}. {BMC Medical Ethics} 20(1).
  \doi{10.1186/s12910-019-0437-z}

\item Wayne K (2019) {How Can Ethics Support Innovative Health Care for an Aging
  Population?} {Ethics {\&} Behavior} 29(3):227--253.
  \doi{10.1080/10508422.2018.1526087}

\item Weber K (2015) {MEESTAR: ein Modell zur ethischen Evaluierung sozio-technischer
  Arrangements in der Pflege- und Gesundheitsversorgung}. In: {Technisierung
  des Alltags}. {Franz Steiner Verlag, 2015}, Stuttgart

\item Weber K (2017) {Demografie, Technik, Ethik: Methoden der normativen Gestaltung
  technisch gestützter Pflege}. {Pflege {\&} Gesellschaft}

\item Weßel M, Ellerich-Groppe N, Koppelin F, et~al (2022) {Gender and Age
  Stereotypes in Robotics for Eldercare: Ethical Implications of Stakeholder
  Perspectives from Technology Development, Industry, and Nursing}. {Science
  and engineering ethics} 28(4):34. \doi{10.1007/s11948-022-00394-1}

\item Weßel M, Ellerich-Groppe N, Schweda M (2021) {Gender Stereotyping of Robotic
  Systems in Eldercare: An Exploratory Analysis of Ethical Problems and
  Possible Solutions}. {International Journal of Social Robotics}
  15:1963--1976. \doi{10.1007/s12369-021-00854-x}

\item Wiczorek R, Bayles M, Rogers W (2020) {Domestic Robots for Older Adults: Design
  Approaches and Recommendations}. In: Woodcock A, Moody L, McDonagh D, et~al
  (eds) {Design of assistive technology for ageing populations}. {Intelligent
  Systems Reference Library}, Springer, Cham, p 203--219

\item Wiertz S (2020) {Trusting Robots?: On the Concept of Trust and on Forms of
  Human Vulnerability}. In: Haltaufderheide J, Hovemann J, Vollmann J (eds)
  {Aging between Participation and Simulation: Ethical Dimensions of Socially
  Assistive Technologies in Elderly Care}. {De Gruyter}, p 53--68

\item Wirth L, Siebenmann J, Gasser A (2020) {Erfahrungen aus dem Einsatz von
  Assistenzrobotern für Menschen im Alter}. In: Buxbaum HJ (ed)
  {Mensch-Roboter-Kollaboration}. {Mensch-Roboter-Kollaboration}, {Springer
  Gabler}, Wiesbaden, p 257--279, \doi{10.1007/978-3-658-28307-0_17}

\item Wu YH, Fassert C, Rigaud AS (2012) {Designing robots for the elderly:
  Appearance issue and beyond}. {Archives of Gerontology and Geriatrics}
  54(1):121--126. \doi{10.1016/j.archger.2011.02.003}

\item Wu YH, Wrobel J, Cornuet M, et~al (2014) {Acceptance of an assistive robot in
  older adults: a mixed-method study of human–robot interaction over a
  1-month period in the Living Lab setting}. {Clinical Interventions in Aging}
  p 801. \doi{10.2147/CIA.S56435}

\item {van Wynsberghe} A (2016{\natexlab{a}}) {Designing Care Robots with Care}. In:
  {van Wynsberghe} A (ed) {Healthcare Robots: Ethics, Design and
  Implementation}. Routledge, p 9--20, \doi{10.4324/9781315586397}

\item {van Wynsberghe} A (2016{\natexlab{b}}) {What is a Care Robot?} In: {van
  Wynsberghe} A (ed) {Healthcare Robots: Ethics, Design and Implementation}.
  Routledge, p 61--68

\item Yasuhara Y, Tanioka R, Tanioka T, et~al (2019) {Ethico-Legal Issues With
  Humanoid Caring Robots and Older Adults in Japan}. {International Journal for
  Human Caring} 23(2):141--148. \doi{10.20467/1091-5710.23.2.141}

\item Yew GCK (2021) {Trust in and Ethical Design of Carebots: The Case for Ethics of
  Care}. {International Journal of Social Robotics} 13(4):629--645.
  \doi{10.1007/s12369-020-00653-w}

\item Zardiashvili L, Fosch-Villaronga E (2020) {“Oh, Dignity too?” Said the
  Robot: Human Dignity as the Basis for the Governance of Robotics}. {Minds and
  Machines} 30(1):121--143. \doi{10.1007/s11023-019-09514-6}

\item Zhang Z, Zhang C, Li X (2022) {The Ethical Governance for the Vulnerability of
  Care Robots: Interactive-Distance-Oriented Flexible Design}. {Sustainability}
  14(4):2303. \doi{10.3390/su14042303}

\item Zöllick J, Kuhlmey A, Nordheim J, et~al (2020) {Technik und Pflege - eine
  ambivalente Beziehung}. {Pflegezeitschrift} 73(3):50--53.
  \doi{10.1007/s41906-019-0653-7}

\item Zöllick J, Kuhlmey A, Nordheim J, et~al (2022{\natexlab{a}}) {Robotik in der
  Pflege – Potenziale und Grenzen}. {Der Hautarzt} 73(5):405--407.
  \doi{10.1007/s00105-022-04965-y}

\item Zöllick JC, Rössle S, Kluy L, et~al (2022{\natexlab{b}}) {Potenziale und
  Herausforderungen von sozialen Robotern für Beziehungen älterer Menschen:
  eine Bestandsaufnahme mittels "rapid review"}. {Zeitschrift für Gerontologie
  und Geriatrie} 55(4):298--304. \doi{10.1007/s00391-021-01932-5}

\item Zwick MM, Hampel J (2019) {Cui bono? Zum Für und Wider von Robotik in der
  Pflege}. {TATuP - Zeitschrift für Technikfolgenabschätzung in Theorie und
  Praxis} 28(2):52--57. \doi{10.14512/tatup.28.2.s52}

\end{enumerate}

\clearpage
\section*{Appendix C: Further information on extracted data}\label{secA3}

\begin{figure}[ht]
\centering
\begin{tikzpicture}
\begin{axis}[
    xbar,
    width=8cm,
    bar width=12pt,
    y dir=reverse,
    xlabel={Number of publications},
    symbolic y coords={
        {Social Sciences and Humanities (incl. Philosophy, Law, Politics)},
        {Health Science and Medicine (incl. Psychology)},
        {Computer Sciences and Engineering},
        {Other (e.g. Natural Sciences)},
        {Unclear}
    },
    ytick=data,
    nodes near coords,
    nodes near coords style={font=\footnotesize},
    yticklabel style={
        text width=6cm,
        align=right,
        font=\footnotesize
    },
    enlarge y limits=0.15,
    xmin=0,
    xmax=130,
    title={Discipline(s) of the first authors\textsuperscript{*}},
    title style={font=\normalsize},
    tick label style={font=\footnotesize},
    label style={font=\footnotesize},
    after end axis/.code={
     \node[anchor=north west, font=\footnotesize, text width=8cm, align=left, inner sep=4pt, draw=none] at (rel axis cs:0, -0.15) 
     {\textsuperscript{*}\textit{One author was assigned to two disciplines.}};
    },
]
\addplot[fill=gray!50] coordinates {
    (103,{Social Sciences and Humanities (incl. Philosophy, Law, Politics)})
    (54,{Health Science and Medicine (incl. Psychology)})
    (50,{Computer Sciences and Engineering})
    (26,{Other (e.g. Natural Sciences)})
    (16,{Unclear})
};
\end{axis}
\end{tikzpicture}
\caption{Distribution of first authors' disciplines.}
\end{figure}
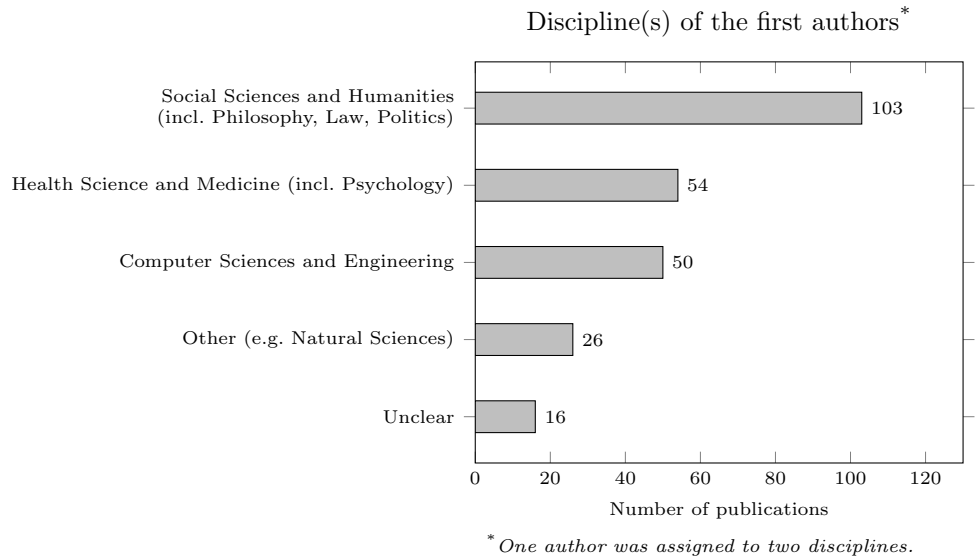

\begin{figure}[ht]
\centering
\begin{tikzpicture}
\begin{axis}[
    xbar,
    width=8cm,
    bar width=12pt,
    y dir=reverse,
    xlabel={Number of publications},
    symbolic y coords={
        {Ethical/Philosophical Analysis},
        {Other (academic)},
        {Empirical Study},
        {Literature Review},
        {Other (non-academic)},
        {Policy Report}
    },
    ytick=data,
    nodes near coords,
    nodes near coords style={font=\footnotesize},
    yticklabel style={
        text width=6cm,
        align=right,
        font=\footnotesize
    },
    enlarge y limits=0.15,
    xmin=0,
    xmax=130,
    title={Types of Articles},
    title style={font=\normalsize},
    tick label style={font=\footnotesize},
    label style={font=\footnotesize},
 after end axis/.code={
     \node[anchor=north west, font=\footnotesize, text width=8cm, align=left, inner sep=4pt, draw=none] at (rel axis cs:0, -0.15) 
     {\textit{}};
    },
]

\addplot[fill=gray!50] coordinates {
    (78,{Ethical/Philosophical Analysis})
    (67,{Other (academic)})
    (66,{Empirical Study})
    (28,{Literature Review})
    (6,{Other (non-academic)})
    (5,{Policy Report})
};

\end{axis}
\end{tikzpicture}
\caption{Distribution of article types.}
\end{figure}
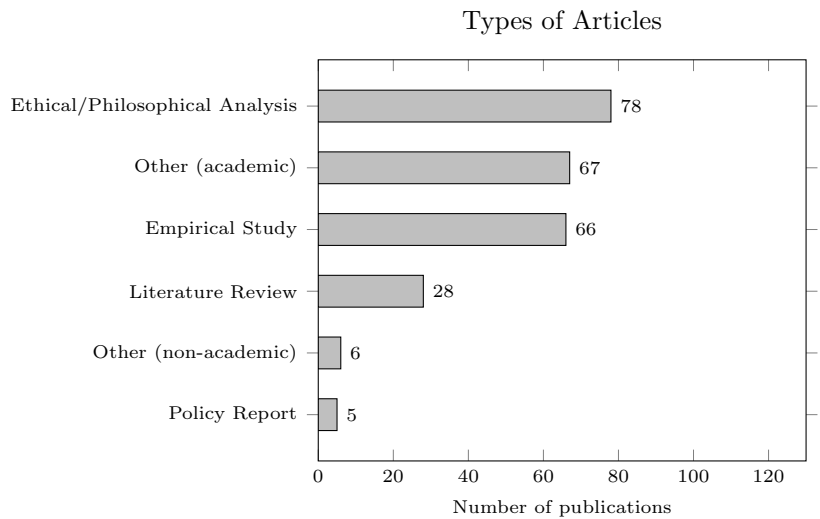

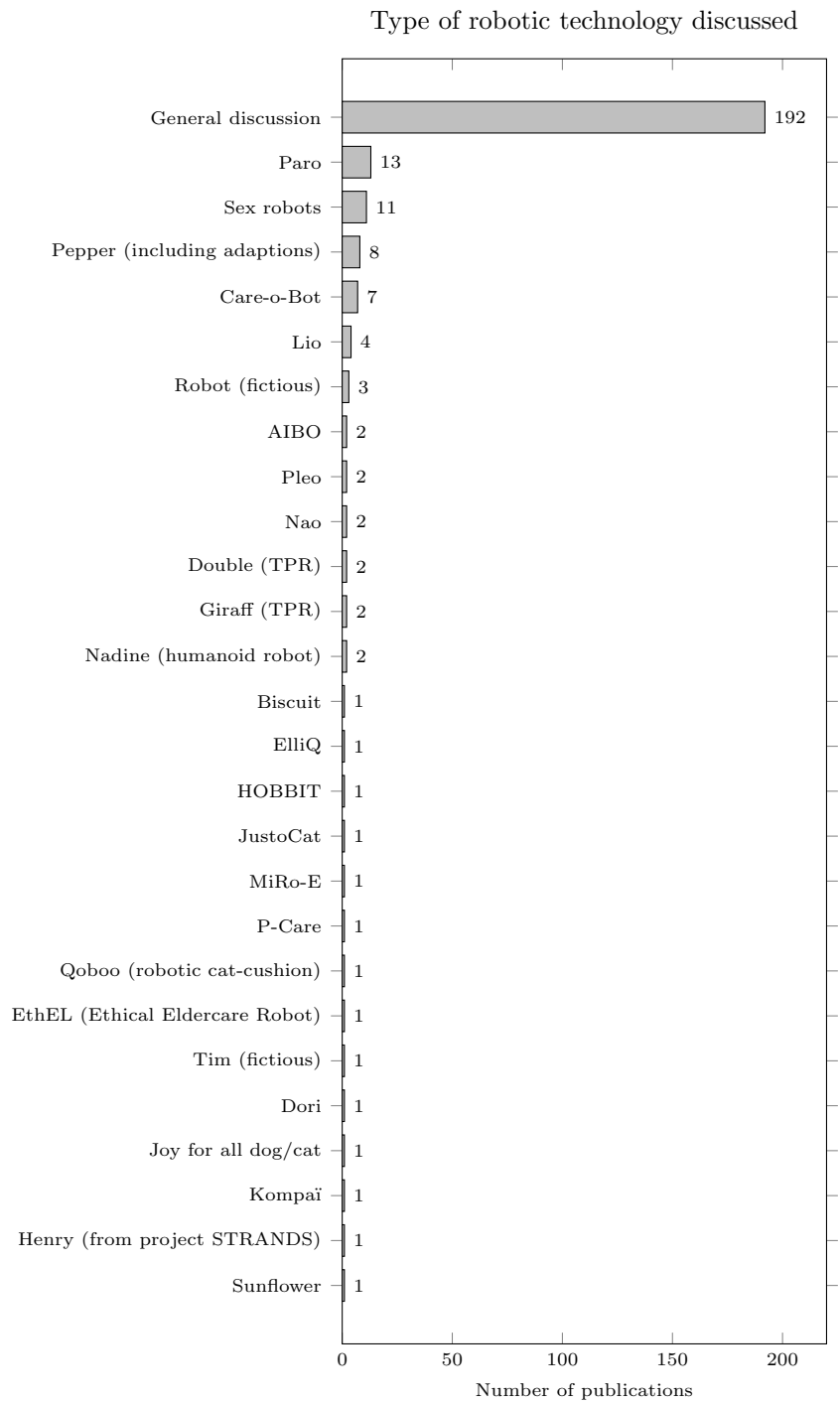
\begin{figure}[ht]
\centering
\begin{tikzpicture}
\begin{axis}[
    xbar,
    width=8cm,
    bar width=12pt,
    xlabel={Number of publications},
    symbolic y coords={
        {Sunflower},
        {Henry (from project STRANDS)},
        {Kompaï},
        {Joy for all dog/cat},
        {Dori},
        {Tim (fictious)},
        {EthEL (Ethical Eldercare Robot)},
        {Qoboo (robotic cat-cushion)},
        {P-Care},
        {MiRo-E},
        {JustoCat},
        {HOBBIT},
        {ElliQ},
        {Biscuit},
        {Nadine (humanoid robot)},
        {Giraff (TPR)},
        {Double (TPR)},
        {Nao},
        {Pleo},
        {AIBO},
        {Robot (fictious)},
        {Lio},
        {Care-o-Bot},
        {Pepper (including adaptions)},
        {Sex robots},
        {Paro},
        {General discussion}
    },
     ytick=data,
     y=0.6cm,
    nodes near coords,
    nodes near coords style={font=\footnotesize},
    yticklabel style={
        text width=6cm,
        align=right,
        font=\footnotesize
    },
    enlarge y limits=0.05,
    xmin=0,
    xmax=220,
    title={Type of robotic technology discussed},
    title style={font=\normalsize},
    tick label style={font=\footnotesize},
    label style={font=\footnotesize},
 after end axis/.code={
     \node[anchor=north west, font=\footnotesize, text width=8cm, align=left, inner sep=4pt, draw=none] at (rel axis cs:0, -0.05) 
     {\textit{ }};
    },
]

\addplot[fill=gray!50] coordinates {
        (1,{Sunflower})
        (1,{Henry (from project STRANDS)})
        (1,{Kompaï})
        (1,{Joy for all dog/cat})
        (1,{Dori})
        (1,{Tim (fictious)})
        (1,{EthEL (Ethical Eldercare Robot)})
        (1,{Qoboo (robotic cat-cushion)})
        (1,{P-Care})
        (1,{MiRo-E})
        (1,{JustoCat})
        (1,{HOBBIT})
        (1,{ElliQ})
        (1,{Biscuit})
        (2,{Nadine (humanoid robot)})
        (2,{Giraff (TPR)})
        (2,{Double (TPR)})
        (2,{Nao})
        (2,{Pleo})
        (2,{AIBO})
        (3,{Robot (fictious)})
        (4,{Lio})
        (7,{Care-o-Bot})
        (8,{Pepper (including adaptions)})
        (11,{Sex robots})
        (13,{Paro})
        (192,{General discussion})
};

\end{axis}
\end{tikzpicture}
\caption{Distribution of type of robotic technology discussed.}
\end{figure}

\begin{figure}[ht]
\centering
\begin{tikzpicture}
\begin{axis}[
    xbar,
    width=8cm,
    bar width=12pt,
    y dir=reverse,
    xlabel={Number of publications},
    symbolic y coords={
        {Hospital},
        {At home/Oupatient},
        {Nursing Home},
        {Not specified/general applicability implied}
    },
    ytick=data,
    nodes near coords,
    nodes near coords style={font=\footnotesize},
    yticklabel style={
        text width=6cm,
        align=right,
        font=\footnotesize
    },
   enlarge y limits=0.15,
    xmin=0,
    xmax=130,
    title={Types of care contexts discussed\textsuperscript{*}},
    title style={font=\normalsize},
    tick label style={font=\footnotesize},
    label style={font=\footnotesize},
 after end axis/.code={
     \node[anchor=north west, font=\footnotesize, text width=8cm, align=left, inner sep=4pt, draw=none] at (rel axis cs:0, -0.15) 
     {\textsuperscript{*}\textit{Settings with hybrid characteristics were assigned to both categories.”}};
    },
]

\addplot[fill=gray!50] coordinates {
    (78,{Hospital})
    (68,{At home/Oupatient})
    (67,{Nursing Home})
    (28,{Not specified/general applicability implied})
};

\end{axis}
\end{tikzpicture}
\caption{Distribution of types of care context discussed.}
\end{figure}

\begin{figure}[ht]
\centering
\begin{tikzpicture}
\begin{axis}[
    xbar,
    width=8cm,
    bar width=12pt,
    xlabel={Type of Devices},
    symbolic y coords={
        {Shared human values (Friedman and Kahn)},
        {Concept of Social Dignity},
        {Kant's formula of humanity (from the categorical imperative)},
        {Ethics of Care},
        {AEG (Alzheimer Europe's Guidelines on the Ethical Use of Assistive Technologies for/by People with Dementia)},
        {MEESTAR},
        {Virtue Ethics},
        {Orientation at standards of "good care"},
        {Value Sensitive DesignCare-Centred Value Sensitive Design},
        {Human rights framework},
        {Capabilities approach},
        {Four principle approach (Beauchamp and Childress)},
        {Adaptions of four principle approach},
        {Self-defined framework (no existing framework used)}
    },
     ytick=data,
     y=1.2cm,
    nodes near coords,
    nodes near coords style={font=\footnotesize},
    yticklabel style={
        text width=6cm,
        align=right,
        font=\footnotesize
    },
    enlarge y limits=0.05,
    xmin=0,
    xmax=40,
    title={Ethics-Framework used (if any)\textsuperscript{*}},
    title style={font=\normalsize},
    tick label style={font=\footnotesize},
    label style={font=\footnotesize},
 after end axis/.code={
     \node[anchor=north west, font=\footnotesize, text width=8cm, align=left, inner sep=4pt, draw=none] at (rel axis cs:0, -0.05) 
     {\textsuperscript{*}\textit{Identification of frameworks was based on the asessment of the authors' arguments and whether a substantive portion of arguments made reference to a framework.}};
    },
]

\addplot[fill=gray!50] coordinates {
        (1,{Shared human values (Friedman and Kahn)})
        (1,{Concept of Social Dignity})
        (1,{Kant's formula of humanity (from the categorical imperative)})
        (1,{Ethics of Care})
        (1,{AEG (Alzheimer Europe's Guidelines on the Ethical Use of Assistive Technologies for/by People with Dementia)})
        (2,{MEESTAR})
        (3,{Virtue Ethics})
        (4,{Orientation at standards of "good care"})
        (4,{Value Sensitive DesignCare-Centred Value Sensitive Design})
        (5,{Human rights framework})
        (6,{Capabilities approach})
        (6,{Four principle approach (Beauchamp and Childress)})
        (11,{Adaptions of four principle approach})
        (30,{Self-defined framework (no existing framework used)})
};

\end{axis}
\end{tikzpicture}
\caption{Distribution of Ethics-Framework used (if any)}
\end{figure}

\clearpage
\section*{Appendix D: Search strings}\label{secA4}

\subsection*{CINAHL}
{\raggedright
((MH "Robotics+") OR (MH "Assistive Technology") OR TX robot* OR (TX social* AND TX assistiv*) OR (TX sozial* and TX assistiv*) OR (TX social* AND TX interactiv*) OR (TX sozial* AND interaktiv*)) AND ((MH "Aged+") OR (MH "Geriatrics+") OR (MH "Dementia+") OR (MH "Dementia Patients") OR (MH "Nursing Home Patients") OR TX elderl* OR TX Pflege OR TX Altenpflege OR TX old* OR TX aged OR TX senior* OR TX geriatr* OR TX dement* OR TX Demenz*) AND ((MH "Ethics+") OR (MH "Ethics, Medical") OR (MH "Ethics, Nursing") OR (MH "Morale") OR (MH "Morals") OR TX ethic* OR TX ethis* OR TX Ethik OR (TX moral* NOT TX "Morales“))
}

\subsection*{TIB}
{\raggedright
(title:(Robot*) OR "socially assistive" OR "sozial assistiv" OR "sozial assistive" OR "socially interactive" OR "sozial interaktiv" OR "sozial interaktive" OR "companion robot" OR "companion robots" OR "Begleitroboter" OR "care robots" OR "care robot" OR "Pflegeroboter" OR "service robot" OR "service robots" OR "Serviceroboter") AND (Pflege OR Altenpflege OR alt* OR old* OR aged OR elderl* OR senior* OR geriatr* OR dement* OR Demenz*) AND (ethic* OR ethis* OR Ethik OR moral* NOT "Morales")
}

\subsection*{BELIT}
{\raggedright
(Robot* OR (social* AND assistiv*) OR (sozial* AND assistiv*) OR (social* AND interactiv*) OR (sozial* AND interaktiv*)) AND (Pflege OR Altenpflege OR alt* OR old* OR aged OR elderl* OR senior* OR geriatr* OR dement* OR Demenz*)

}




\end{appendices}

\end{document}